\documentclass{aa}  

\usepackage{graphicx}
\usepackage{xcolor}
\usepackage{natbib}
\usepackage{pdflscape}
\usepackage{longtable}
\usepackage{adjustbox}
\usepackage{gensymb}
\usepackage{txfonts}
\usepackage[colorlinks=true,
  urlcolor=blue,
  citecolor= blue,
  linkcolor = blue]{hyperref}

\begin{document}

   \title{Gaia early DR3 systemic motions of Local Group dwarf galaxies and orbital properties with a massive Large Magellanic Cloud.}
   
   \titlerunning{eGDR3 bulk motions of Local Group dwarf galaxies}


   \author{G. Battaglia
          \inst{1,2}\fnmsep\email{gbattaglia@iac.es}
          \and
          S. Taibi\inst{1,2}
          \and
          G. F. Thomas\inst{1,2}
          \and
          T. K. Fritz\inst{1,2}
          }

   \institute{
Instituto de Astrofísica de Canarias, Calle Vía Láctea s/n, E-38206 La Laguna, Tenerife, Spain\\
\and 
Universidad de La Laguna, Avda. Astrofísico Fco. Sánchez, E-38205 La Laguna, Tenerife, Spain\\
             }

   \date{Received: ; accepted: }

 
  \abstract
   {}
   {We perform a comprehensive determination of the systemic proper motions of 74 dwarf galaxies and dwarf galaxy candidates in the Local Group based on Gaia early data release 3. The outputs of the analysis for each galaxy, including probabilities of membership, will be made publicly available. The analysis is augmented by a determination of the orbital properties of galaxies within 500 kpc.
   }
   {
   We adopt the flexible Bayesian methodology presented by McConnachie \& Venn (2020), which takes into account the location of the stars on the sky, on the colour-magnitude diagram and on the proper motion plane. We apply some modifications, in particular to the way the colour-magnitude diagram and spectroscopic information are factored in, e.g. by including stars in several evolution phases. The bulk motions are integrated in three gravitational potentials: two where the Milky Way is treated in isolation and has a mass 0.9 \& 1.6 $\times 10^{12}$M$_{\odot}$ and the time-varying potential by Vasiliev et al. (2021), which includes the infall of a massive Large Magellanic Cloud (LMC).
   }
   {
   We are able to determine bulk proper motions for 73 systems, and we consider reliable 66 of these measurements. For the first time, systemic motions are presented for galaxies out to a distance of 1.4 Mpc, in the NGC~3109 association. The inclusion of the infall of a massive LMC significantly modifies the orbital trajectories of the objects, with respect to orbit integration in static Milky Way-only potentials, and leads to 6 galaxies being likely associated to the LMC, 3 possibly associated and 1 recently captured object. We discuss the results of the orbit integration in the context of the relation of the galaxies to the system of Milky Way satellites, implications for the too-big-to-fail problem, impact on star formation histories, and tidal disruption.
   }
   {}


   \maketitle
%

\section{Introduction}
Knowledge of the bulk motions of galaxies residing in the Local Group (LG) is a precious 
resource for a wealth of galaxy evolution and near-field cosmology investigations, e.g. inferences of the mass, barycenter position and velocity of the LG \citep[e.g.][to mention a few]{Kahn_59, Peebles_01, Li_08,  Marel_12a, Gonzalez_14, Penarrubia_14}, studies of the possible history of past interactions between the Milky Way (MW) and the M31 system and its future fate \citep[e.g.][and references therein]{Loeb_05,Marel_12b, Salomon_20}; determinations of the mass of the MW through dynamical modelling of tracers of its gravitational potential, such as its satellite galaxies \citep[e.g.][see references in Fritz et al. 2020 for an overview on the works on this topic]{Wilkinson_99, Battaglia_05,  Boylan-Kolchin_13, Patel_18, Callingham_19, Fritz_20}; group infall as well as the significance and stability of the Vast Polar Structure \citep[e.g.][]{Metz_08,Pawlowski_13,Fritz_18, Kallivayalil_18, Li_EDR3}; considerations on the missing satellite problem \citep[e.g.][]{Simon_18, Fritz_18}. The orbital history of MW satellite galaxies is also very likely to influence several aspects of their evolution, through e.g. the impact of ram-pressure stripping and tidal effects onto their gas content, star formation history (SFH), morphology and dark matter (DM) halo properties \citep[e.g.][and references therein]{Mayer_06, Munoz_08, Kazantzidis_11, Battaglia_15, Hausammann_19, Iorio_19, Miyoshi_20, Ruiz-Lara_21, Rusakov_21, DiCintio_21, Genina_20}.

Before the second data release of the Gaia mission (GDR2) \citep{Gaiamission_16, GaiaDR2_18}, measurements of the systemic proper motions (PMs) of galaxies in the LG were essentially limited to the Magellanic Clouds, the so-called classical MW dwarf spheroidal galaxies (dSphs), one "ultra faint dwarf" (UFD), M31, M33 and IC~10, mostly from HST observations and a few VLBI observations of OH masers (see references in Sect.~\ref{sec:literature}); some of them having become available surprisingly recently \citep[e.g. in the case of the Sextans MW dSph the first such measurement was published only in 2018 by][]{Casetti-Dinescu_18}. 

Since April 2018, the situation has seen a dramatic improvement, starting with the Gaia science verification article \citep{Helmi_18}. Multiple determinations of the systemic PM of a large number of MW satellite galaxies and galaxy candidates blossomed in a matter of weeks after GDR2, led by several groups in the community, and using a variety of techniques, e.g. focusing only on stars with prior spectroscopic information \citep{Simon_18, Fritz_18} or including the full set of stars with astrometric information \citep{Kallivayalil_18, Massari_18}. It is now becoming routine to use Gaia astrometric data also to remove contaminants, as well as to attempt systemic proper motions determinations along with the study of other properties of the systems \citep[e.g.][]{Longeard_18a, Torrealba_19}. Surveys of the MW stellar halo and sub-structures within are and will make plentiful use of Gaia astrometry to boost the success rate in target selection \citep[e.g.][but also the surveys to be carried out with 4MOST and WEAVE, to mention some of the upcoming ones]{Conroy_19, Li_19_S5, AllendePrieto_20_DESI}. 

The methodologies applied in the early GDR2 works mentioned above were rather simple ones, being based on iterative cleanings of the data-sets via $\sigma$-clipping and no statistical treatment of the foreground/background contamination. Later on, more sophisticated methods were used, e.g. with simultaneous statistical modelling of the properties of the dwarf galaxy and the contamination. For instance, \citet{Pace_19} used the spatial and PM  information of stars preselected to have magnitude and color lying on an isochrone and adopted a multi-variate Gaussian in proper motion for both the dwarf galaxy and the MW, while \citet{McConnachie_DR2} used all the  observables at once and adopted the empirical distribution of the contaminant stars in the PM and the colour-magnitude (CM) planes. 

Interestingly, the use of GDR2 data has been also pushed beyond the MW system, with determinations of the tangential motions of M31 \& M33 \citep{Marel_19}, as well as of a few LG dwarf galaxies such as NGC~6822, IC~1613, WLM and Leo~A  \citep{McConnachie_SOLO}. 

The early third data release of Gaia data, hereafter eGDR3,  \citep{GaiaeDR3_21} has implied more precise and accurate astrometric measurements; in particular, for PMs the precision has increased of a  factor of two and systematic errors decreased by a factor $\sim$2.5 \citep{Lindegren_20a}. \citet{McConnachie_EDR3} provides updated systemic PMs for the 58 MW satellites previously considered by the same team with DR2 data, with the improved astrometry now allowing to detect the systemic PM of Bo\"{o}tes~IV, Cetus~III, Pegasus~III and Virgo~I. Recently, \citet{Li_EDR3} provide an independent determination of systemic PMs for 46 MW satellites, and integrate the 3D motions in four isolated MW potential models, with a  total mass from $2.8 \times 10^{11}$ to $15 \times 10^{11}$ M$_{\odot}$. \citet{Martinez-Garcia_21} combined the astrometric and spectroscopic information available for 14 MW satellites to study their internal kinematics and quantify the presence of velocity gradients.

In this work we aim at providing a comprehensive determination of systemic proper motions based on eGDR3 not only for MW satellites, but for LG dwarf galaxies in general\footnote{For brevity, hereafter we refer to all the systems in the sample as "dwarf galaxies", including when they are only "candidates" and even in the case of larger galaxies like M~33 and NGC~3109.}, and to push for the first time these determinations to even larger distances, i.e. reaching out to the NGC~3109 association, at $\sim$ 1.4~Mpc. We are making use of the best techniques in the literature, i.e. those by \citet{McConnachie_DR2, McConnachie_SOLO}, inspired by \citet{Pace_19}, since they take full advantage of the observables available for the largest number of stars with full astrometric and photometric information (location on the sky with respect to the dwarf galaxy centre, and on the colour-magnitude diagram and on the proper motion planes), and model  them in a Bayesian way with a mixture model accounting for contaminant sources. We have introduced a few modifications to these techniques, mainly aimed at making an even more realistic treatment of the stellar population content of the dwarf galaxies and its distribution on the colour-magnitude plane accounting also for the photometric completeness of eGDR3 data; this allows us also to determine probability of memberships for stars in different evolutionary phases, which we will make available to the community \footnote{http://research.iac.es/proyecto/GaiaDR3LocalGroup/pages/en/home.php}, together with several other outputs of our analysis. 

Finally, we study the orbital properties of the galaxies surroundings of the MW by integrating their bulk motions in a set of gravitational potentials, bracketing a 0.9-1.6$\times 10^{12}$~M$_{\odot}$ range for the mass of the MW. Motivated by the work by \citet{Patel_20}, who showed that the orbits of MW satellites can differ significantly between a gravitational potential including only the MW and one where the gravitational influence of the LMC (and SMC) are taken into account (and all galaxies are free to move in response), we integrate the bulk motions also in the triaxial time-varying MW potential made available by \citet{Vasiliev_21}, where the infall of a massive LMC and the response of the MW to this infall are modelled. In this context, we also revisit the association of dwarfs surrounding the MW to the LMC system. 

In Sect.~\ref{sec:sample} we introduce the sample of galaxies analysed and in Sect.~\ref{sec:data} the data-sets used and the quality selection criteria applied. In Sect.~\ref{sec:propmot} we present the methodology for the systemic proper motion determinations, and discuss the output and the tests performed to tackle the robustness of the method; in Sect.~\ref{sec:systematics} we complement the results with a determination of the zero-points and additional errors due to systematics in the Gaia eDR3 data, using QSO. Our systemic proper motions are compared to those in the literature in Sect.~\ref{sec:literature}. In Sect.~\ref{sec:orbits} the bulk motions are integrated in the three gravitational potentials and the resulting orbital trajectories and parameters are then used to address the impact of the LMC on the reconstructed orbital history, make considerations on the too-big-to-fail problem, on the system of MW and LMC satellites, and on observed properties such as SFHs.  We discuss other potential applications of our work in Sect.~\ref{sec:applications} and present our conclusions and summary in Sect.~\ref{sec:conclusions}. 


\section{Sample} \label{sec:sample}

  \begin{figure}
  \includegraphics[width=\columnwidth]{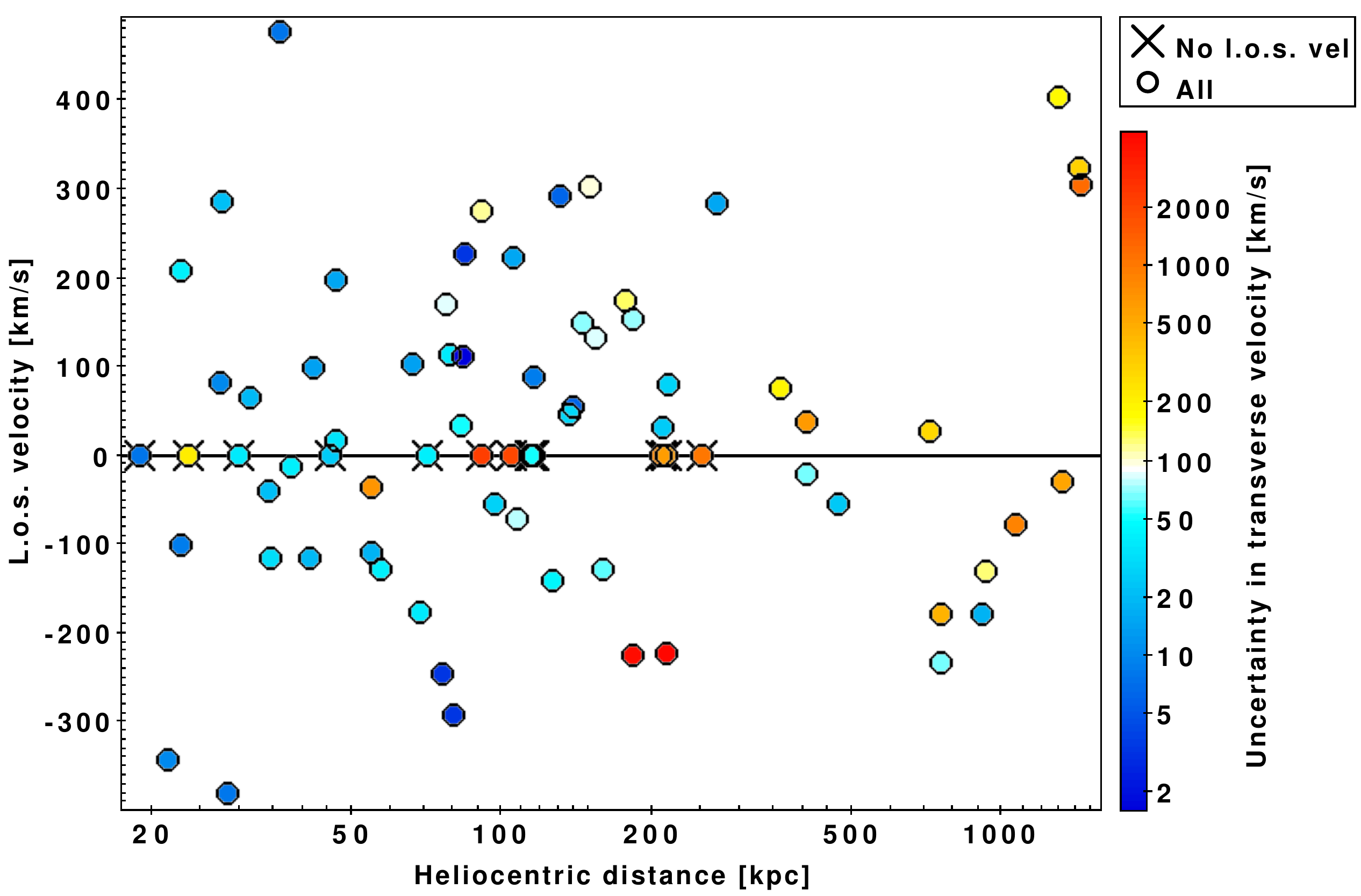}
      \caption{L.o.s. velocity as a function of heliocentric distance for the galaxies in the sample (circles). The systems without a literature measurement of the l.o.s. velocity are assigned a null value, exclusively for the purpose of this figure, and are indicated as crosses. The color-coding is based on the uncertainty in heliocentric transverse velocity, derived from the statistical uncertainties in the systemic proper motions and distance modulus (see Sect.~\ref{sec:propmot}). 
      }
         \label{fig:sample}
   \end{figure}
   
The 74 systems considered in this work are listed in Tab.~\ref{tab:sample}, together with their main global properties (see also Fig.~\ref{fig:sample}). The sample is the union of the dwarf galaxies studied in \citet{Fritz_18}, \citet{Fritz_19}, and some other recently discovered satellites of the MW. We also included isolated LG dwarf galaxies within $\sim$1.4$\,$Mpc.

Due to the large distance of the isolated LG dwarf galaxies, in terms of resolved or partially resolved sources, only H~II regions, young main sequence, blue/red super-giants and asymptotic giant branch (AGB) stars brighter than the tip of the red giant branch (RGB) are potentially detected above the magnitude limit for {\it Gaia} astrometric measurements. Therefore not all isolated dwarf galaxies will have enough (or any) eGDR3 sources  with full astrometric solution. 

In order to select which systems to consider on a first pass, we gave priority to those galaxies within 1.5Mpc with an HI detection, as listed in \citet{McConnachie_12},  and hence likely to host young stars and/or HII regions. The systems that showed to have a clear enough detection in {\it Gaia} eDR3 of centrally concentrated sources (after a first, rough, spatial, parallax and proper motion selection) were retained. We exclude IC~10 because, apart from being only marginally detected, its analysis is complicated by the very high and patchy extinction  in its direction. The borderline systems like LGS3, Antlia, NGC~205, NGC~185 were excluded but the closer galaxy Leo~T was retained. 

We exclude the Magellanic Clouds, M~31 and the Sagittarius (Sgr) dwarf galaxy, because we are neglecting internal motions in our analysis, while they might be relevant for the systemic PM  determination of these systems. These galaxies have anyway already been the subject of very detailed analyses based on {\it Gaia} data \citep[see][the latter using GDR2]{Salomon_20,Luri_20,DelPino_21}. Note that in their GDR2-based study, \citet{Marel_19} found no difference in the centre-of-mass proper motion when explicitly modelling M33 rotation or neglecting it. This means we include the following galaxies outside the virial radius of the MW: Eridanus~II, Leo~T, Phoenix, NGC~6822, WLM, IC~1613, Leo~A, M33, Peg-dIrr, UGC~4879, Sgr-dIrr, Sextans~A, Sextans~B, NGC~3109.

We caution the reader that, even for those galaxies for which a determination of the systemic PM is possible, depending on the distance, the error in the transverse velocity is still too large for scientific applications (see Fig.~\ref{fig:sample}). We refer the reader to Sect.~\ref{sec:systematics} to learn about which galaxies have their error budget dominated by random or statistical errors.

\section{Data-sets} \label{sec:data}

{\it Gaia} eDR3 astrometric and photometric measurements for stars with full astrometric solutions constitute the bulk of the data over which the analysis is performed. 
For the distant systems (i.e., those beyond 400~kpc) we complement {\it Gaia} photometry with the deeper data-set from the Pan-STARRS1 Surveys \citep[PS1,][]{Chambers_16}; this will be used to identify the regions in the colour-magnitude plane where to select candidate massive blue stars and red super giants stars and to determine their spatial distribution, in some cases. 
We refer to Sect.~\ref{sec:method} for a detailed explanation of the methodology.  

We download eGDR3 and PS1 data over an area centered on the systems under consideration, with a radius such to guarantee at least 2000 objects where we expect contaminants to be clearly dominant over stars from the galaxy under consideration, i.e. beyond 5 half-light radii, $R_h$\footnote{The colour-magnitude diagrams were inspected for all the systems, to confirm the dominance by contaminant sources in those regions.}. 

We concentrate exclusively on those eGDR3 detected objects that are not flagged as duplicated source, have a full astrometric solution (astrometric\_params\_solved $\ge$ 31),  high-quality  astrometry  \citep[renormalised unit-weighted error, ruwe, $<$1.4, see e.g.][]{Lindegren_18} and reliable photometric measurements. For this last factor, we retain the measurements with an absolute value of the corrected excess factor within 5 times the standard deviation expected at the corresponding G-mag   \citep[see Eqs. 6, 18 and Table 2 in][]{Riello_21}. In order to exclude sources seen as extended or not isolated by {\it Gaia}, we retain objects with ipd\_frac\_multi\_peak $\le$ 2 and   ipd\_gof\_harmonic\_amplitude $<$ 0.2; this latter cut is less restrictive with respect to what adopted in \citet{Fabricius_21} but adjusts better to the distribution of values seen for the sources around the  galaxies under consideration. Objects with source\_id with a match in the {\it Gaia} AGN catalogue are excluded.

We exclude clear foreground stars by requiring the parallax of each individual source to be consistent with the parallax expected at the dwarf distance modulus within 3$\sigma_\pi$; here $\sigma_\pi$ is the sum in quadrature of the parallax error on the individual measurements and that due to the uncertainty of the galaxy distance modulus. We apply a global zero-point offset of $-0.017$~mas to the parallax measurements of the individual stars \citep{Lindegren_20b}. We do not correct for the 
{\it Gaia} parallax zero-point as a function of location, magnitude and colour, because of its negligible effect on our analysis: even at the brightest magnitudes considered here ($m_G \sim 13.4$~mag for the tip of the RGB of Delve~1, the closest system in the sample), the maximum difference between the zero-point applied and that expected at $m_G \sim 13.4$~mag would be $\sim$0.03~mas \citep{Lindegren_20b}, smaller than the $3\times \sigma_\pi$ range under consideration. 

The apparent $G$-mag for the sources with 6-parameters solutions are corrected as in \citet{Riello_21}, with the python code presented in \citet{Brown_21}. Finally, the apparent $G-$mag and the $BP-RP$ color are corrected for extinction using the \citet{Schlegel_11} maps interpolated at the position of the stars and using the  \citet{Marigo_2008} coefficients for the Gaia filters, based on \citet[][]{Evans_2018} \citep[see][]{sestito2019}.

\section{Determination of systemic proper motions} \label{sec:propmot}

\subsection{Method} \label{sec:method}

Rather than relying only on eGDR3 sources with previous spectroscopic observations, we adopt the flexible methology by \citet[][hereafter, MV20a]{McConnachie_DR2}; \citet[][hereafter, Mc21]{McConnachie_SOLO}, inspired by \citet{Pace_19},
which allows to make use of all the stars with {\it Gaia} astrometric and photometric measurements (aside from the quality cuts detailed in the previous section). We refer the reader to the original source for a detailed explanation of the methodology. Here suffices to say that it is based on a maximum likelihood procedure, which has three  free-parameters: the systemic proper motion of the galaxy $\mu_{\alpha,*, sys}$ and $\mu_{\delta, sys}$ and the fraction of stars in the dwarf galaxy under consideration over the total, $f_{gal}$. It is assumed that  the intrinsic dispersion is negligible in the distribution of PM measurements. 

The likelihood of a star to belong to the system/object  under consideration, $L_{\rm gal}$,  or to the contamination, $L_{\rm c}$, is estimated taking into account a spatial, colour-magnitude and proper motion likelihood term. 

After having determined $\mu_{\alpha,*, sys}$, $\mu_{\delta, sys}$ and $f_{gal}$, these, together with the likelihoods, can be used to obtain the probability of membership of each star to a given galaxy (Eq.~5 in MV20a):
\begin{equation}
    P_{\rm gal} = \frac{f_{\rm gal} L_{\rm gal}}{f_{\rm gal} L_{\rm gal} + (1 - f_{\rm gal}) L_{\rm c}}
\end{equation}

Below we describe the methodology followed to determine the various terms of the likelihood function, and provide an overview of how each galaxy was treated in Tab.~\ref{tab:method}.

\subsubsection{Spatial Distribution} 

The contaminants are assumed to be uniformly distributed over the areas considered around each system.

The spatial term of the likelihood function for the stars belonging to the dwarf galaxy is based on the 2D structure of the dwarf galaxy stellar component. In order to evaluate it, we adopt two main approaches, depending on the system under consideration: either we parametrize it to have an elliptical shape and an exponentially declining surface number density profile, as done by MV20a, or determine it empirically, as done by Mc21 ("Exp" and "Emp" in column "Spatial" of Tab.~\ref{tab:method}). 

We refer to MV20a and Mc21 for the possible caveats concerning this approach. 
    
For the "Exp" case, a 2D look-up map is created by co-adding and then normalising $N=1000$ MonteCarlo realizations of the expected 2D surface number density at a given position on the sky; in each realization, values for the ellipticity, position angle and half-light radii\footnote{The ellipticity is defined as $1 - b/a$, where $b$ and $a$ are the projected minor and major axes; the position angle increases from North to East; the half-light radius here refers to the projected one on the sky, along the major axis.} are randomly extracted from a Gaussian distribution centred on the values listed in Tab.~\ref{tab:sample} and with dispersion given by the average of the lower and upper 1-$\sigma$ uncertainties listed in the same table. For those galaxies where a determination of the ellipticity is missing or only an upper limit is available, we assume the spatial distribution to be circular. This approach is applied essentially to all the systems for which the spatial distribution of the majority of sources detected by {\it Gaia} and that we will use in the analysis have a smooth, spheroidal-looking morphology (i.e. the "classical" dSphs and the UFDs) or for some late-type systems for which there is not enough statistics for an empirical determination (see below). 

For the distant late-type systems, the majority of the sources detected by {\it Gaia} will be blue massive stars, red super-giants (RSGs) and AGB stars. Among these, we will concentrate on the blue and RSGs as more easily identified on the CMD (see also Mc21). These young stars are those that give an "irregular" morphology to some of these galaxies, due to asymmetries in their spatial distribution. Therefore, the approach to follow for the late-type systems is decided after a visual analysis of the spatial distribution of stars with colors consistent with being young main-sequence or blue super-giant stars in the PS1 photometry: if their spatial distribution is well defined, then the probability distribution of the spatial term is determined empirically from these stars as a normalized 2D histogram within 3 half-light radii (apart from M33, that is missing this quantity, for which we consider 0.5~deg). For those cases where the statistics of blue stars are not sufficient for an empirical determination, we resort to modeling the spatial term as an exponentially declining profile. In all cases, the spatial distribution of the RSGs is assumed to follow that of the blue stars, which is true to a good approximation.

\subsubsection {Distribution on the colour-magnitude plane} 

\begin{table}[]
    \centering
    \begin{tabular}{c|c|l}
    \hline\hline
  Spatial & CM & Galaxies \\
  \hline
Exp  & Emp & Carina, Draco, Fornax, Leo~I, Leo~II, \\
  &  & Sculptor, Sextans, Ursa~Minor \\
  & & \\
Exp & Syn &  All those not listed in the other rows \\
& & \\
Emp & Box & IC~1613, NGC~6822, WLM \\
 & & NGC~3109, Sextans~A, Sextans~B, M33 \\
 & & \\
Exp & Box & Leo~A, Peg-dIrr, Sg-dIrr, UGC~4879 \\
    \hline\hline
    \end{tabular}
    \caption{Methodology used for the spatial and CM  term of the likelihood for stars in the dwarf galaxy, when determining the systemic PM (Sect.~\ref{sec:method}. In the column "Spatial", "Exp" indicates that an exponential profile was used for the surface number density distribution and "Emp" that the spatial distribution was determined empirically. In the column, "CM", "Emp" indicates that the probability distribution on the CM plane was determined empirically (from within 1 half-light radius), "Syn" that the synthetic CMD was used and "Box" that a uniform probability was given to stars within regions of the CMD compatible with hosting massive blue stars and red super-giants.}
    \label{tab:method}
\end{table}

We concentrate on sources with $-1.0<$(BP-RP)$_0 <2.5$, 
apart for the late-types, for which we adopt $-1.5<$(BP-RP)$_0 < 2.5$.
We have verified that these color cuts works also for metal-rich systems such as Fornax, WLM, NGC~6822 etc.

As in MV20a, the distribution of contaminants onto the colour-magnitude plane is determined empirically, from eGDR3 sources at semi-major axis radii larger than $5\times R_{h}$, with the exception of those systems for which this limit exceeded the spatial extension of the catalogue downloaded (e.g. for Antlia~II), in which case we assume radii larger than $3\times R_{h}$.
For M33 we do not use a value of the half-light radius but define the region by eye, beyond 1deg. 

For the CM probability distribution of stars belonging to the systems of interest, we introduce a few changes with respect to the method of MV20a:

\begin{enumerate}
    \item {\it Empirical determination} ("Emp" in column "CM" of Tab.~\ref{tab:method}) For well populated and nearby systems, such as the "classical dSphs", the distribution on the CM-plane is determined empirically, from the region within one half-light radii. The dwarf galaxy's stellar population within this region dominates over that of the contaminants and the increase in statistics over considering a smaller area is worth the introduction of a few contaminants.  While we do not expect the choice of using an empirical determination of the dwarf's CM likelihood term to cause a significant difference in the determination of the systemic proper motion over e.g. using an isochrone, we wish to factor in the CMD information in the estimate of the probability of membership for stars in different evolutionary phases, since this is one of the products that we make available. 
    
    Classical dSphs are well-known to display stellar population gradients; however a complete modelling of the CMD as a function of distance from the dwarf centre is outside of the scope of this work, and the CMD of the central regions contains all the features present also in the outer parts.
    
    \item {\it Synthetic CMD} ("Syn" in column "CM" of Tab.~\ref{tab:method}) All those galaxies closer than 440~kpc that are not included in the category above do not have enough signal inside their half-light radius for an empirical determination of the CM probability distribution. In addition, many of these systems are very faint and sparsely populated;  therefore we wish to adopt an approach that includes all the relevant evolutionary phases and does not exclude a priori possible members, e.g. if they were not to fall on the locus of an isochrone of a given age and metallicity. 
    
    To this aim, rather than an isochrone as done in MV20a, we use Basti-IAC to create a synthetic CMD in the eGDR3 photometric filters\footnote{The synthetic CMD was courtesy of S. Cassisi; the website of Basti-IAC is  http://basti-iac.oa-abruzzo.inaf.it/index.html.}, based on the stellar evolutionary models presented in \citet{Hidalgo_18}. These models include also the He-burning phase, which can be precious in faint systems, since in particular the blue part of the horizontal branch is a region of rather reduced contamination.  An advantage of using a synthetic CMD over using an isochrone is that stars are distributed on the magnitude and color plane in the correct proportion  (for a given SFH, chemical enrichment law, initial mass function,..). 
    
    Given that the systems we are applying this method to are either completely or mostly dominated by ancient stellar populations, we adopt a constant star formation rate between 12 and 13 Gyr ago, and a metallicity centered around [Fe/H] $=-2.3$ with a spread of 0.5~dex; this is representative of the metallicity distribution function of stars in UFDs \citep[see review by][]{Simon_19}. 
    
    One hundred realizations are carried out, where the synthetic CMD is shifted in distance modulus, drawing from Gaussian distributions centered on the values listed in Tab.~\ref{tab:sample} and with dispersion given by the average of the upper and lower 1-$\sigma$  errors. At the same time, the photometric errors are introduced by scattering the BP-RP colours of the stars in the synthetic CMD according to the photometric errors derived from the eGDR3 catalogue corresponding to each given object, at the appropriate apparent G-mag. 
    
     {\it Correction for photometric completeness} The synthetic CMD of course does not suffer from photometric incompleteness; on the other hand, it should be considered that the CMD of the contaminants does suffer from this issue, since it is derived empirically from the eGDR3 data, and that the completeness varies depending on the number of transits in a given region of the sky. This might alter the relative probabilities of dwarf galaxy's stars versus contaminant stars in some regions of the CMD, in favour of the former.
     
     In order to introduce an (approximate) corretion to take this effect into account, we resort to Gaia Universe Model Snapshot \citep[GUMS][]{Robin_12}. We download GUMS models from the {\it Gaia} archive around the position of each system, and calculate the ratio of the luminosity function of contaminants stars in the observed eGDR3 catalogues (which will be mostly MW stars) and that of model stars, in a representative color range, 0.5 $<$ (BP-RP)$_0$ $<$ 2.5. In the assumption that the model is a reasonable approximation of the data, at G-mag where {\it Gaia} should not suffer from completeness issues, this ratio should be around unity. In practise, there are some deviations; therefore, we normalize the ratio to the median value in the range 17 $<$ G-mag $<$ 20. After this step, the ratio of the two luminosity functions oscillates around one, except at faint magnitudes, where a decline towards zero is observed, which can be assumed to be due to the {\it Gaia} photometric completeness. This will be the factor by which we multiply the counts in the CMD look-up map as a function of magnitude\footnote{In practise, to avoid introducing noise, we only multiply the counts by this correction factor at G-mag $>$19 and if the ratio is $<$0.75.}. As we will see in Sect.~\ref{sec:tests}, this correction has a minor effect, but this will be adopted for our baseline results, listed in  Tab~\ref{tab:sysmotions}.

    Even though Phoenix and Leo~T do contain a sprinkle of young, blue stars detected in eGDR3, since the majoritarian population in eGDR3 data is by far represented by RGB stars, these systems are treated with the synthetic CMD. This choice does not impact the determination of the systemic PM but it implies that these young stars will be missing from our list of probable members. 
    
    \item {\it Box} ("Box" in column "CM" of Tab.~\ref{tab:method}) For the distant ($>$ 440~kpc) and well populated galaxies, we follow closely the approach by Mc21, and focus on blue sources and RSGs. The color and magnitude limits are chosen by visual inspection of the PS1 photometry, and transferred to the {\it Gaia} eGDR3 bands, using the stars in common between the data-sets for each galaxy. We assign a uniform probability inside these boxes.  
    
\end{enumerate}

For both the dwarfs and the contaminants, we construct the CM  look-up map in bins of magnitude and color, smooth it with a boxcar kernel and then proceed to normalizing it.

\subsubsection{Proper motion} 
For the distribution on the PM plane, we adopt the same approach as MV20a, i.e. an empirical determination for the contamination, while assuming a multi-variate Gaussian distribution for the dwarf galaxy, taking into account the correlation terms between the $\mu_\alpha,*$ and $\mu_\delta$ of the individual stars.

For the MW classical dSphs and UFDs, we restrict the range of the analysis to within $\pm$5 mas yr$^{-1}$, corresponding to a generous tangential velocity cut of $>$470 km s$^{-1}$ at a heliocentric distance larger than 20~kpc ($\pm$3 mas yr$^{-1}$ for Antlia~II, to reduce the overwhelmingly large contamination). 

For the other systems we filter out sources whose proper motion in each component at the distance of the galaxy would imply tangential velocities 3x in excess of a given velocity dispersion (around the reflex proper motion at the galaxy's sky location). As dispersion, we consider the square-root of the quadratic sum of the uncertainty given by the proper motion measurements and 200~km\,s$^{-1}$, where the latter is the observed scatter in l.o.s. velocities for the whole sample of galaxies in Tab.~\ref{tab:sample} (assuming that the scatter in tangential velocities is the same)\footnote{We tested the performance when relaxing the cut in tangential velocity and the resulting systemic PMs are always within 1 or at most 2-$\sigma$ from each other, showing that the method is robust also when removing this condition.}.

\subsection{Results}

The outcome of the analysis is summarized in Tab.~\ref{tab:sysmotions}. 

For the majority of the systems in the sample, in output there are sizeable numbers of stars with large membership probability (42 and 52 systems with $>$10 stars with P$>0.95$ and P$>0.5$, respectively). In fact, some galaxies are extremely well populated. For example, we obtain $>$1000 stars with membership probabilities P$>$0.95 in each of the classical MW dSphs (except Leo~II) as well as in M~33 ($>$20000 for Fornax). Ten systems have between 100-1000 $P>0.95$ members, among which several of the distant galaxies (Phoenix, NGC~6822, IC~1613, WLM, NGC~3109).  

On the other end of the spectrum, there are systems with only an handful of probable members or none at all. The analysis does not lead to a systemic PM determination for Pisces~II and Virgo~I, with PDFs that are essentially flat. Other systems with clearly problematic PDFs are DESJ0225+0304, Pegasus~III, Tucana~V, with strong lopsideness and/or very extended wings of high amplitude, and Indus~I, with a double peaked PDF. For Pisces~II and Tucana~V, however, we are able to obtain a systemic PM when including the information about l.o.s. velocities for the stars observed spectroscopically (see Sect.~\ref{sec:tests}). We would advice against using the systemic PMs for all these cases (whose names are highlighted in red in Tab.~\ref{tab:sysmotions}) and advice to use the motions obtained when considering the spectroscopic information for Pisces~II and Tucana~V. 

In addition to the above, the shape of the PDFs and of the distribution of probable member stars on the plane of the sky, proper motion and colour-magnitude leads us to advise exerting caution when considering the motions of Cetus~III, Indus~II, Aquarius~II, Delve~1, Reticulum~III, Bootes~IV (see Sect.~\ref{sec:individual} for more details; these are highlighted in orange in Tab.\ref{tab:sysmotions}). 

\begin{figure*}
   \centering
  \includegraphics[width=0.8\textwidth]{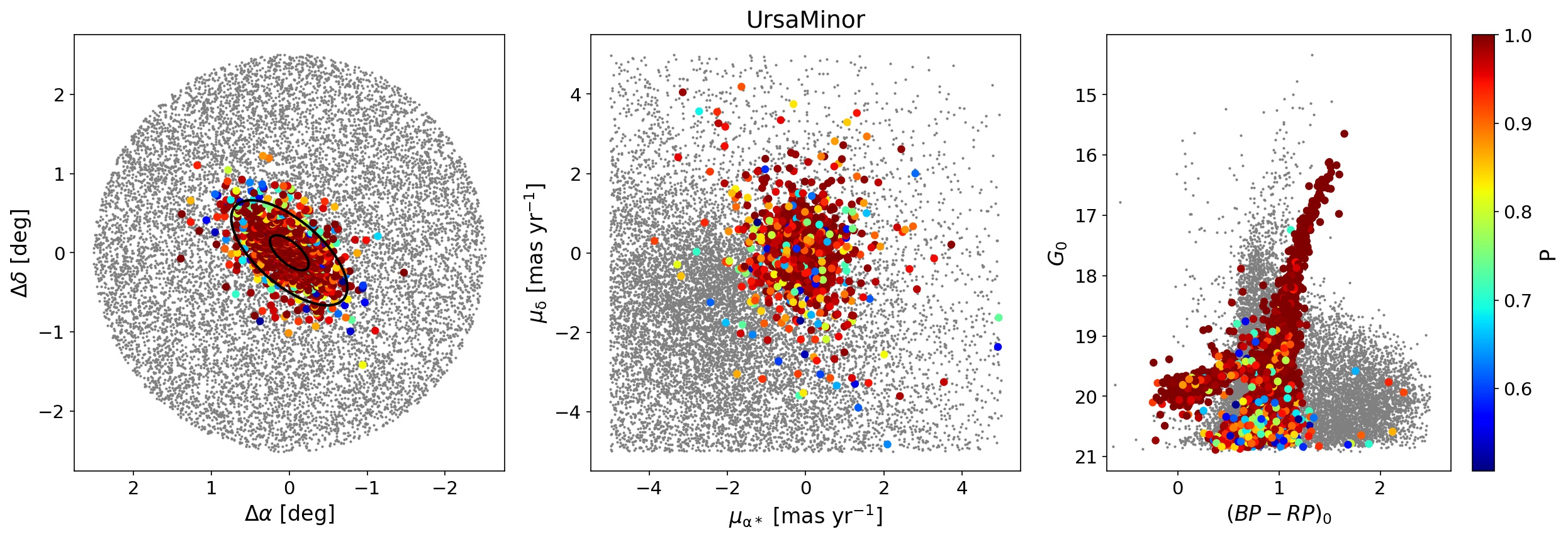}
  \includegraphics[width=0.8\textwidth]{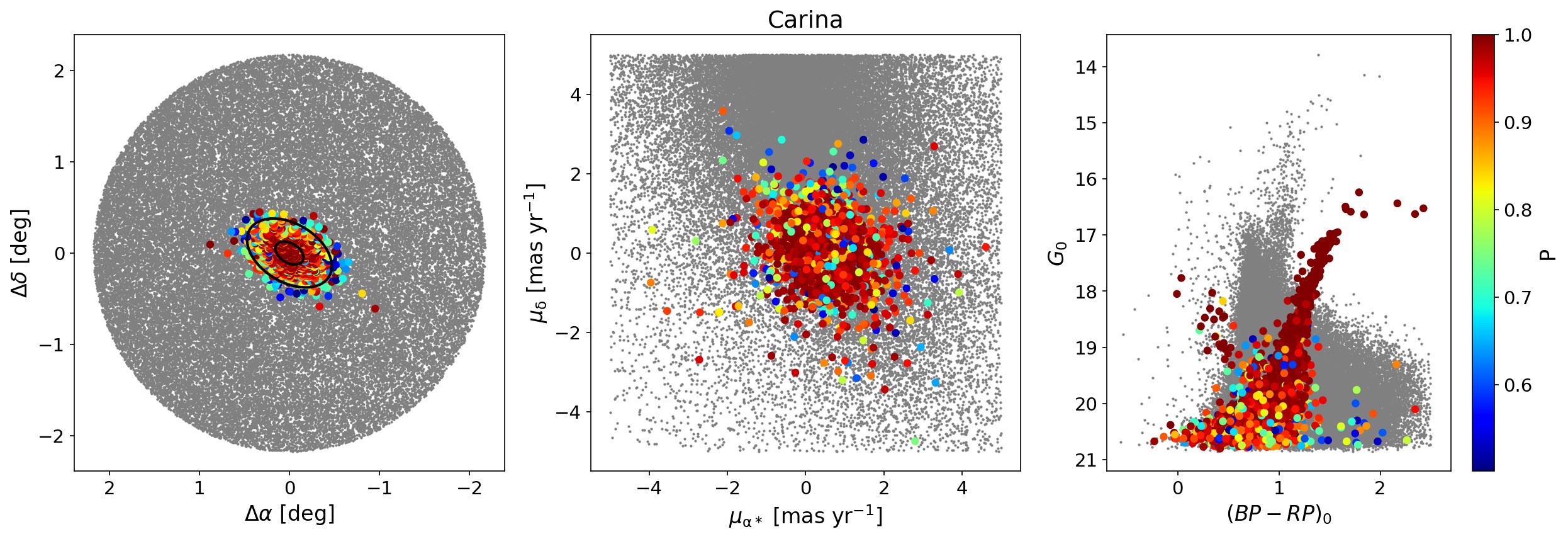}
  \includegraphics[width=0.8\textwidth]{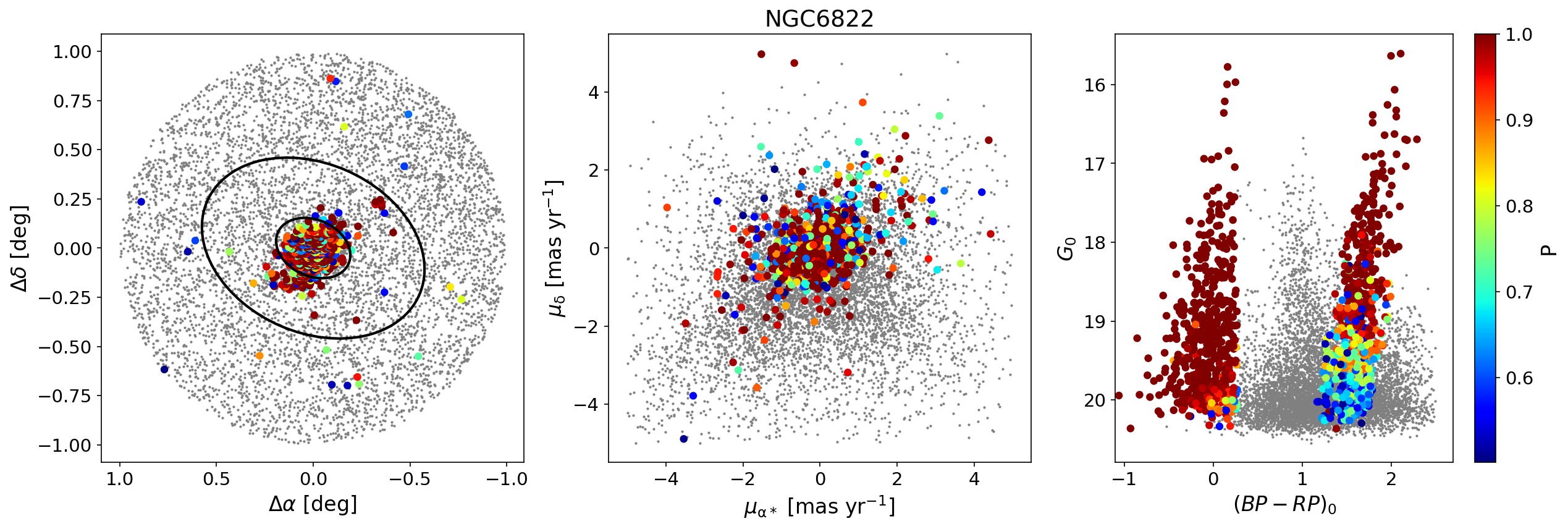}
  \includegraphics[width=0.8\textwidth]{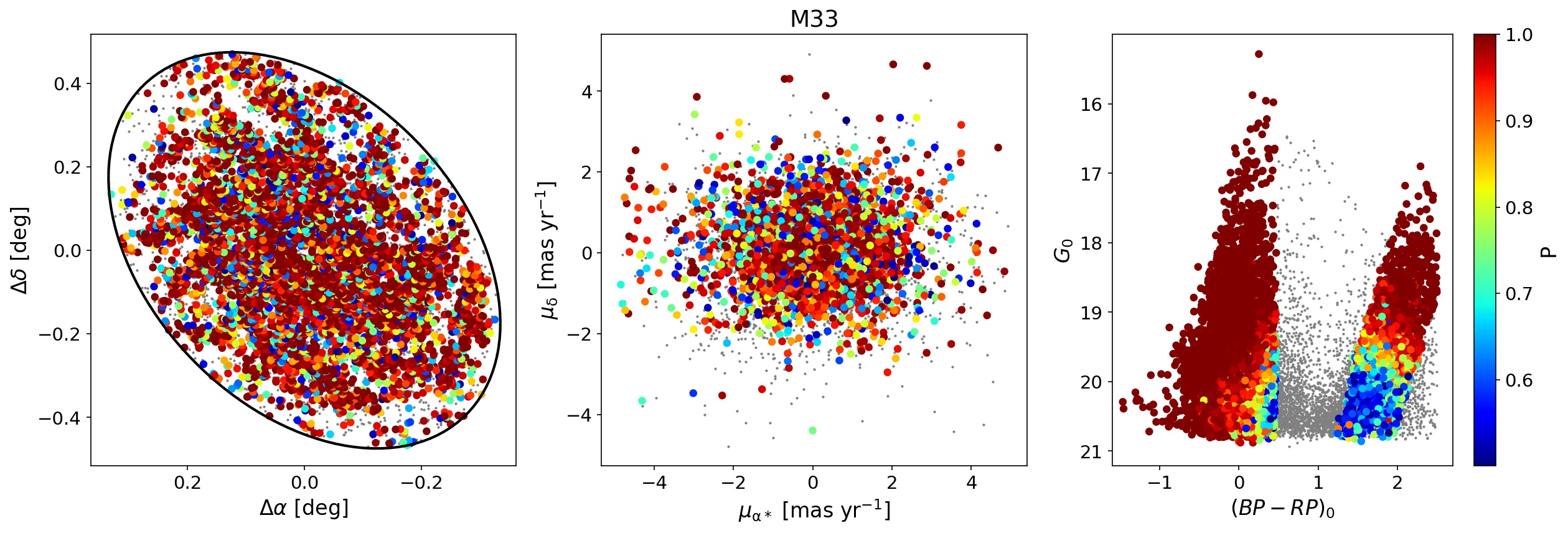}
      \caption{Distribution of member stars (large circles) projected on the tangent plane passing through the galaxy centre (left), and on the proper motion (middle) and colour-magnitude plane (right), for systems in the regime of $>$500 stars with P$>0.95$, in increasing order of distance from top to bottom. The galaxy names are indicated in the figure titles. The colour-coding indicates the probability of membership (only when above $>0.5$; the stars with P $<0.5$ are shown as grey dots. The ellipses in the left panel have semi-major axes equal to 1x and 3x the half-light radii in Tab.~\ref{tab:method} (apart from M33, that is missing this quantity, for which we consider 0.5~deg), and ellipticity and position angle taken from the same table.}
         \label{fig:out_1}
   \end{figure*}

\begin{figure*}
   \centering
  \includegraphics[width=0.8\textwidth]{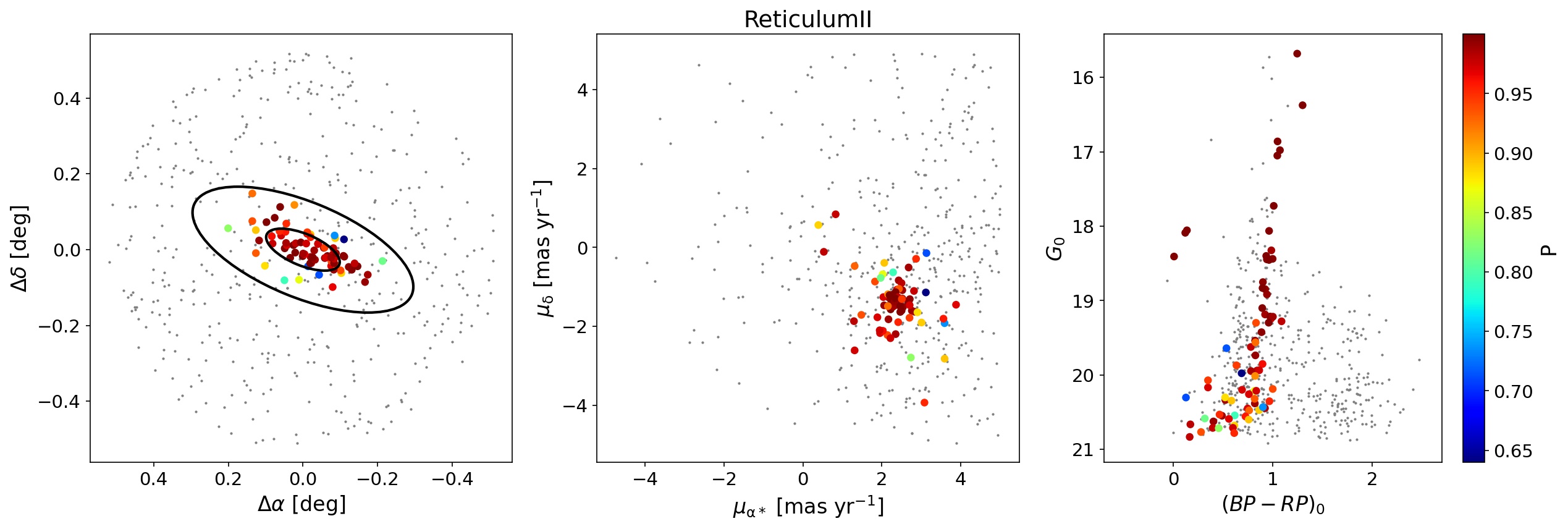}
  \includegraphics[width=0.8\textwidth]{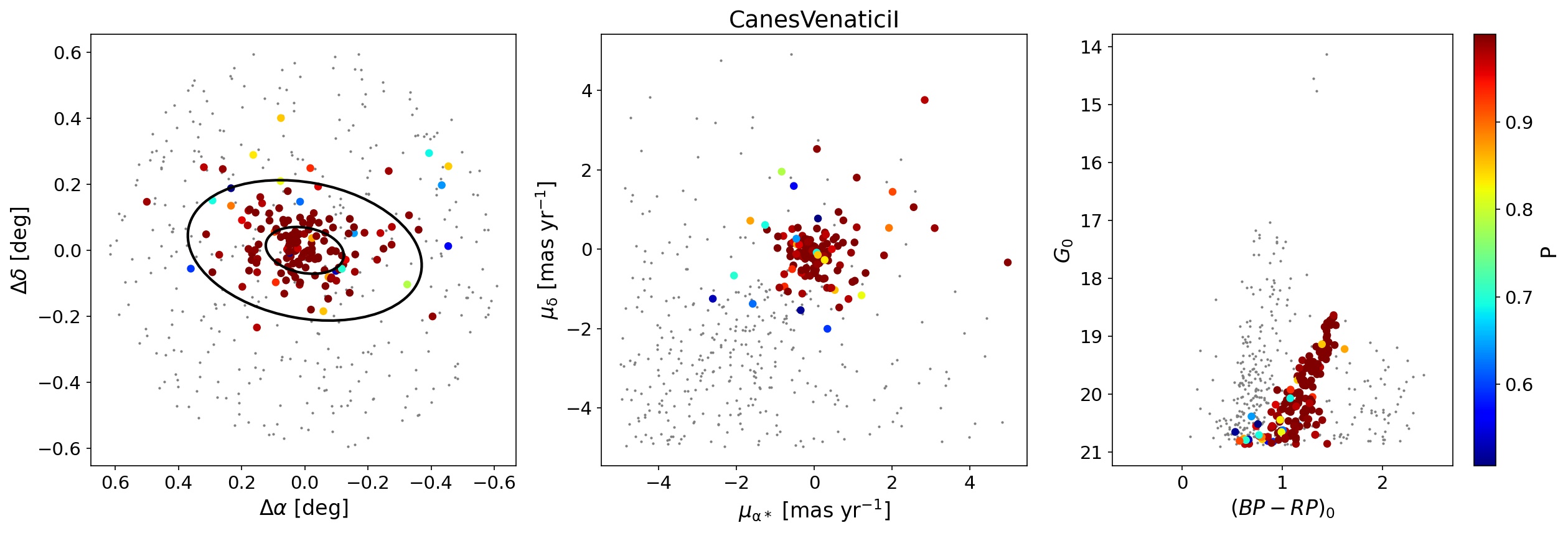}
  \includegraphics[width=0.8\textwidth]{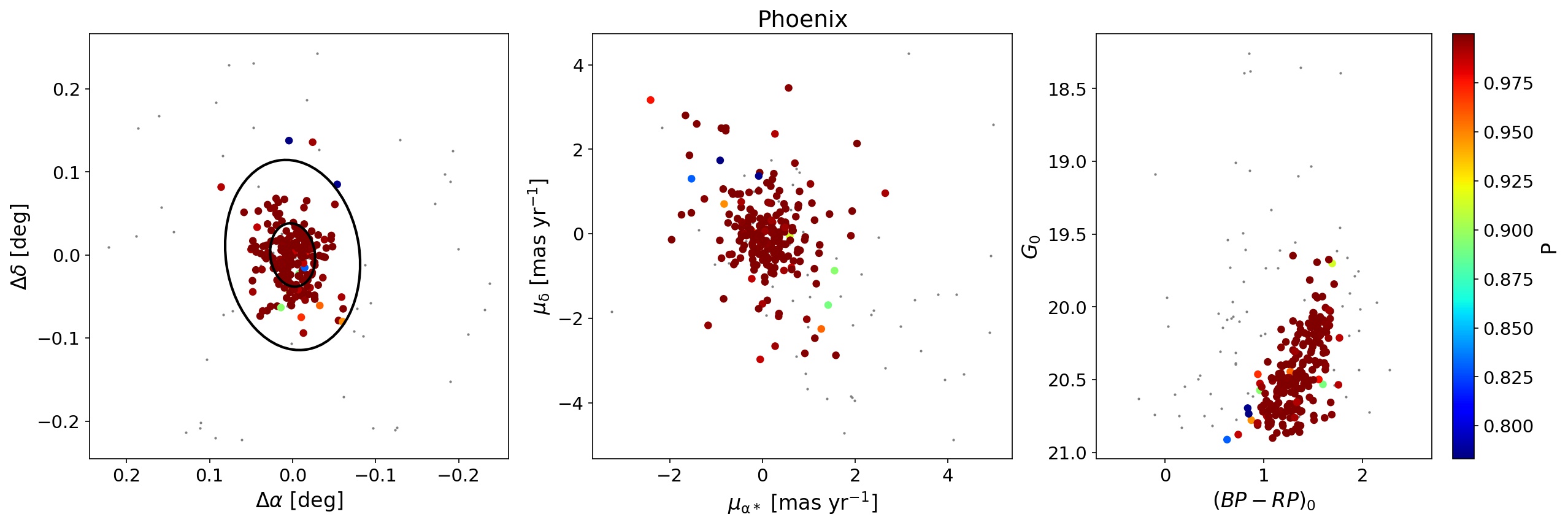}
  \includegraphics[width=0.8\textwidth]{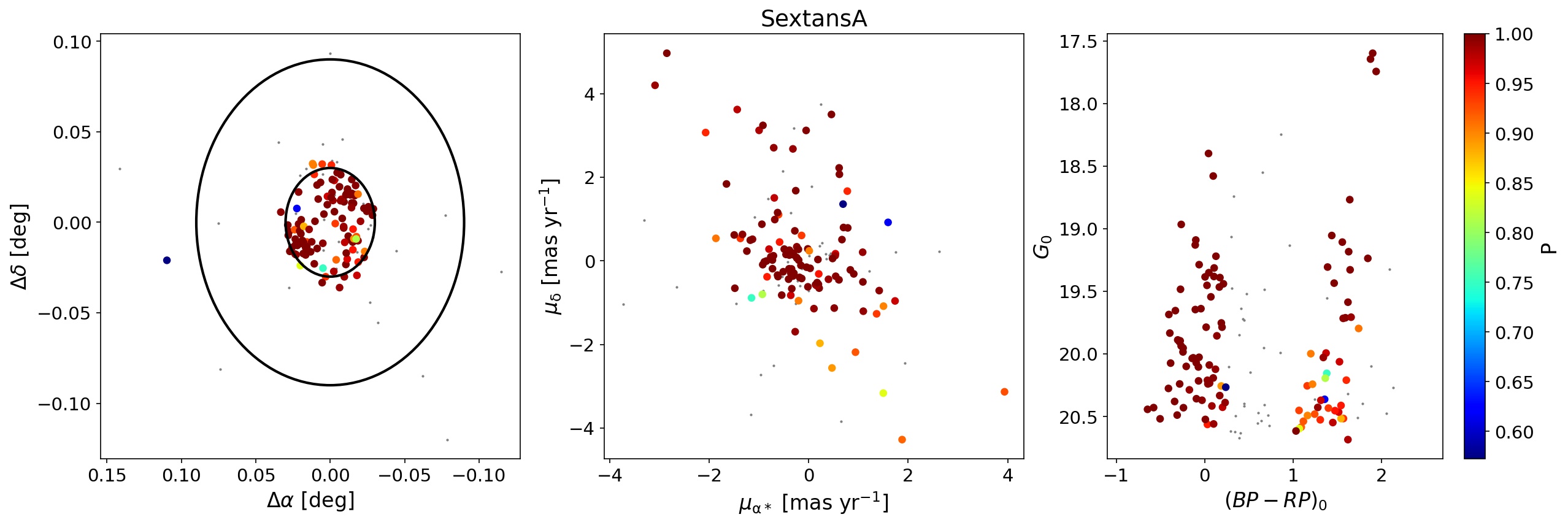}
      \caption{As in Fig.~\ref{fig:out_1} but for the regime of 50-200 stars with P$>0.95$.
              }
         \label{fig:out_2}
   \end{figure*}

\begin{figure*}
   \centering
  \includegraphics[width=0.8\textwidth]{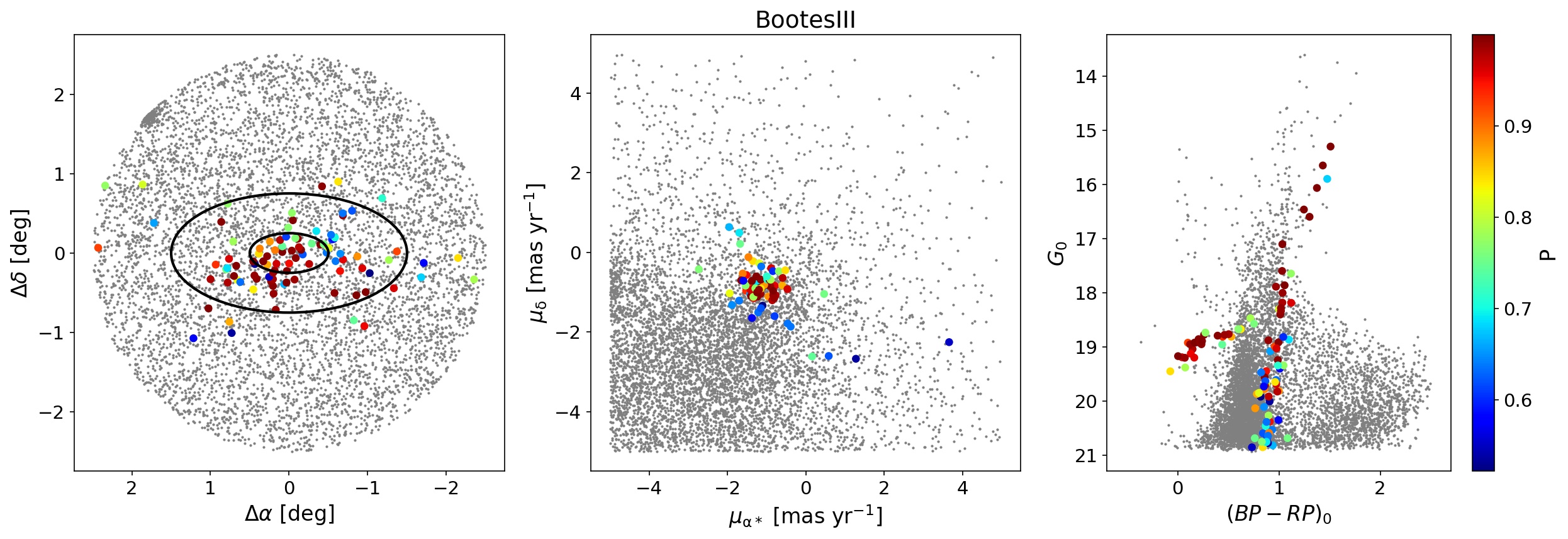}
  \includegraphics[width=0.8\textwidth]{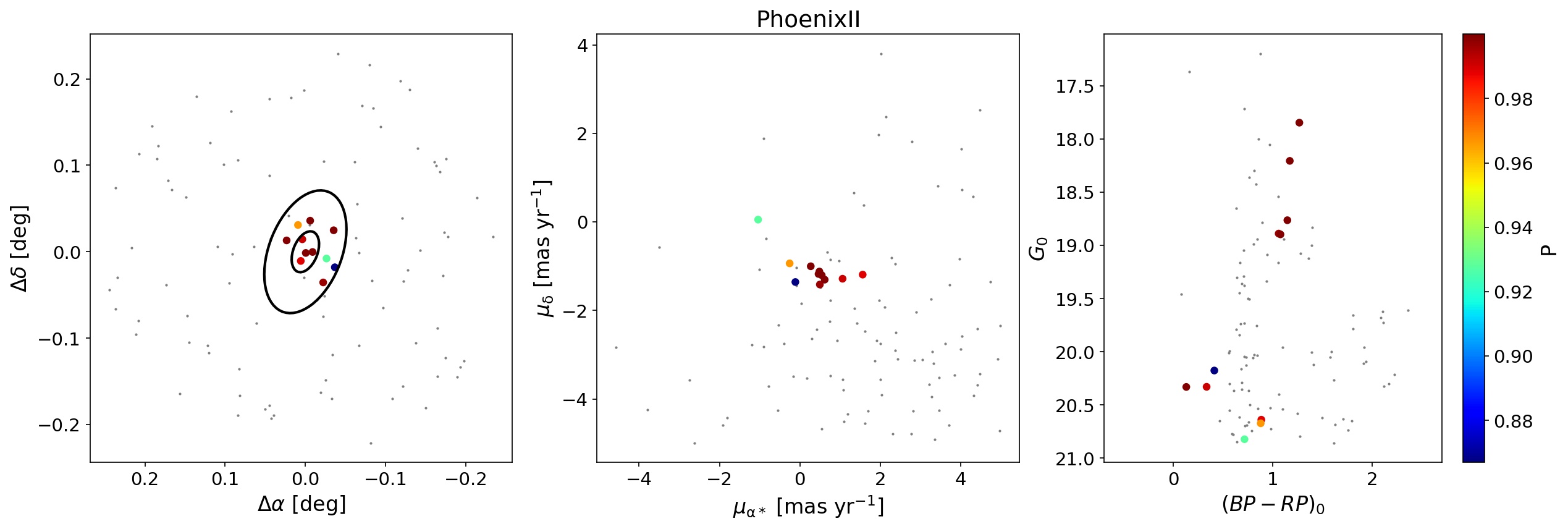}
  \includegraphics[width=0.8\textwidth]{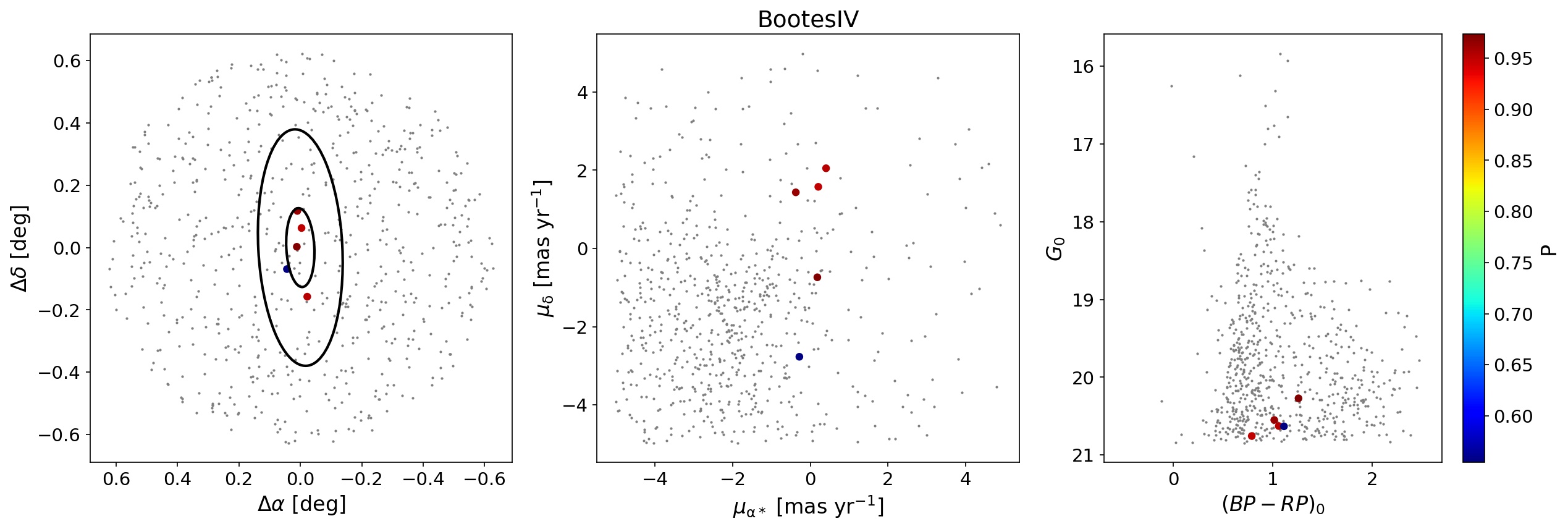}
  \includegraphics[width=0.8\textwidth]{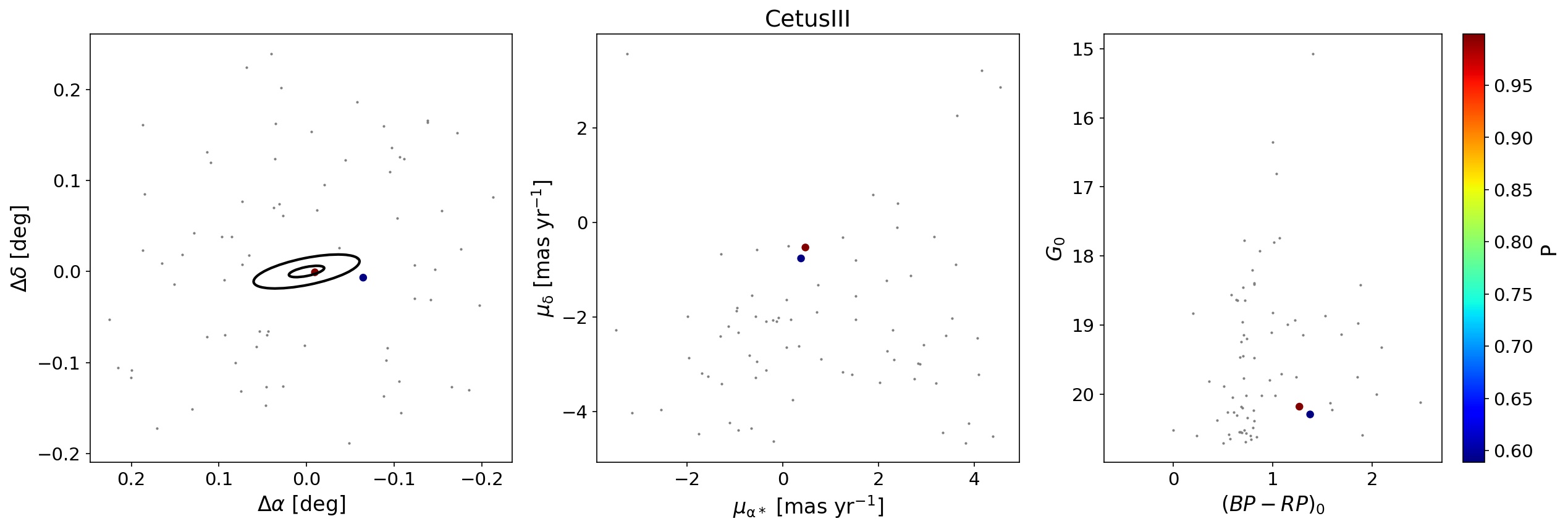}
      \caption{As in Fig.~\ref{fig:out_1} but for the regime of $<$50 stars with P$>0.95$. Cetus~III and Bootes~IV are two cases in which the uncertainties in \citet{McConnachie_EDR3} are much smaller than in our determination.}
         \label{fig:out_3}
   \end{figure*}

Example of plots of the distribution of member stars projected on the tangent plane passing through the galaxy centre, and on the proper motion and colour-magnitude plane are given in Figs.\ref{fig:out_1}-\ref{fig:out_3}, for systems in various regimes, in terms of number of member stars, heliocentric distances, and morphological types.

In summary, out of 74 systems analysed, we are able to determine systemic proper motions for 72 systems without considering complementary spectroscopic information (73 when considering the spectroscopic information), and consider certainly reliable 62 (64 with spectroscopy). The majority of these 64 systems are found in the vicinity of the MW, within 300~kpc, but the galaxies for which we provide systemic PMs are as distant as 1.4~Mpc (Sextans~A and Sextans~B). This is the largest, and most extended in volume, set of systemic proper motions for galaxies and galaxy candidates. Of course, it should be kept in mind that the same uncertainty in proper motion will translate into an uncertainty in transverse velocity 10x larger for a galaxy at 1~Mpc than for one at 100~kpc! 

%
   
\subsubsection{Tests and validations} \label{sec:tests}
$\bullet$ {\it Photometric completeness} As discussed in Sect.~\ref{sec:method}, a correction for the photometric completeness of eGDR3 data was applied to the CM probability distribution of the galaxies treated with the synthetic CMD (see Tab.~\ref{tab:method}). Fig.~\ref{fig:compl} shows the comparison of the systemic proper motions determined with and without applying this correction: the determinations are always in very good agreement, well within the 1-$\sigma$ errors, apart from Antlia~II, for which a larger difference is seen, but still within 2-$\sigma$. The good agreement between the two determinations is likely due to the correction kicking in at faint magnitudes, there were the individual proper motions are less accurate and therefore have a lower weight in the global determination.

\noindent $\bullet$ {\it Inclusion of spectroscopy information} MV20a analyzed the galaxies in their sample with and without considering additional information on the stars's membership from spectroscopic observations. Specifically, when including information from spectroscopy, they modified the prior on the systemic proper motion by multiplying it by a bi-variate Gaussian with mean and dispersion given by the weighted mean proper motion and associated uncertainty of the stars with {\it Gaia} astrometric measurements that are also probable spectroscopic members. The authors concluded that information from stars with spectroscopic follow-up was not required to obtain reasonable estimates of the systemic PMs; in fact, in only a few systems, the inclusion of this information played an important role: Carina~III, Segue~1,  Triangulum~II  and  Tucana~IV, for which the spectroscopy allowed to go from a bi-modal to a uni-modal PDF (which in the case of Carina~III was due to the presence of Carina~II in the field-of-view). In general, the inclusion of this prior had the effect of reducing the size of the error-bars, even though in some cases only slightly. The authors also warn about the dangers of including this information for systems with only an handful of spectroscopic members, since interlopers could of course lurk  among them too. 

We explore the possible improvements due to the use of spectroscopic information in a different way: rather than using the iteratively derived mean proper motion of spectroscopic member stars to modify the prior, we introduce another term in the likelihood, in which the l.o.s. velocity distribution is modelled as the sum of two 1D Gaussians, one for the dwarf galaxy and one for the MW. Since not all the stars with astrometric information do have a l.o.s. velocity measurement, we assign a l.o.s. velocity equal to 0 km\,s$^{-1}$ and a l.o.s. velocity uncertainty of 10000 km\,s$^{-1}$ to the stars in the astrometric sample that do not have a spectroscopic match; these arbitrarily large velocity uncertainties have the effect of giving these stars no weight in the estimate, but do allow us to treat the spectroscopic information as a further likelihood term.  The value of the peak l.o.s. velocity and l.o.s. velocity dispersion for the dwarf galaxies are fixed to the values in Tab.~\ref{tab:sample}, fixing the velocity dispersion to 5~km\,s$^{-1}$ when only an upper limit is available.
At the same time we solve for the peak l.o.s. velocity and l.o.s. velocity dispersion of the MW component (we have tried also keeping them fixed to 0 km\,s$^{-1}$ and 200 km\,s$^{-1}$, respectively, and the results do not vary).  
Fig.~\ref{fig:spec} shows the comparison of our baseline case with the determinations using a spectroscopic prior for the category for which the largest differences could in principle be expected, i.e. the ultra faint dwarfs. 
The only two systems with significant differences are Pisces~II and Tucana~V, where the size of the error-bars reduces  drastically. This is due to two stars with high probability of membership found for Pisces~II and a PDF with wings of much lower level for Tucana~V in the run with spectroscopic information with respect to that without. For the great majority of the other cases, the differences are minor both in terms of systemic motions and associated uncertainties. In Appendix~\ref{sec:individual}, we comment on those cases where the difference between the systemic PMs with and without the spectroscopic information is larger than 0.5 $\sigma$. 

As MV20a concluded, it is re-assuring that spectroscopic follow-up is not a necessary condition for systemic PM determinations. 
Overall we see a lower degree of improvement than that found by  MV20a on GDR2 data. Likely, the main reason is that eGDR3 data, in particular the PMs, have become more precise, which makes it easier for the algorithm to find galaxies even if they have only a few stars above the \textit{Gaia} magnitude limit. This might make it potentially easier to find galaxies using only \textit{Gaia} data in the future data releases, but perhaps also already in eGDR3 \citep[see also][]{Darragh-Ford_20}.

\noindent $\bullet$ {\it RRLyrae} As an additional check of the robustness of the results, we compared the systemic PMs with the individual measurements for RRLyrae found at projected distances within 5$\times$ half-light radii and with magnitudes approximately compatible with the horizontal branch of each system. As catalogue of RRLyrae stars, we use the union of the {\it Gaia} DR2 SOS gaiadr2.vari\_rrlyrae \citep{GaiaDR2_Holl_18, GaiaDR2_19_variables, GaiaDR2_Clementini_19}, the stars classified as RRLyrae of ab, c, d type in the general variability catalogues gaiadr2.vari\_classifier\_result, and the PS1 RRLyrae of ab or cd type by \citep{Sesar_17}\footnote{With classification score above 0.6.}. The comparison is very good. The only system in which some outliers in the RRLyrae PMs are found is Hydrus~I, but this is not a cause for concern, because a closest examination shows the presence of RRLyrae compatible with belonging to the SMC-LMC system in the background and we made no attempt of statistically account for contamination in the RRLyrae variables data-set.

\section{Systematic errors and distance errors} \label{sec:systematics}

It is known that \textit{Gaia} PM measurements are affected by systematic errors, which can be thought of as a component on small angular scales, $\lesssim$ 1 deg, and a component on large scales, with a scale-length of $\sim$16~deg for eGDR3 \citep[see][]{Lindegren_20a}.  Given the spatial scales typically involved in our analysis, the effect of the large-scale component should be to act as a zero-point in the observed systemic motions. On the other hand, the small-scale component will average out for some of the systems with the largest angular size, but not for a significant number of them. 

Therefore we follow two routes: we treat the bias on small-scales as an additional source of noise, while we determine the zero-point from the large-scale component for each galaxy separately from QSOs. 

For the small-scale error, we use the determination by \citet{Vasiliev_EDR3}\footnote{Their equation 2, but with 400/(1+$\theta$/3), not 400 x (1+$\theta$/3), which was a typo (Vasiliev, private communication).}, rather than that by \citet{Lindegren_20a}, since the former was derived on globular cluster stars, in which there are more close neighbours on smaller scales than among quasars, used by the latter work. As typical scale of our systems, we use the "circularized"  half-light radius, $\theta_{\rm half}$ (for M33 we use the radius containing half of the member stars). This leads to errors between 13 and 23 $\mu$as yr$^{_1}$ for both dimensions, $\sigma_{\rm vas} (\theta_{\rm half})$. 

For the other component, we calculate the weighted average of the PMs of QSOs (from the table agn\_cross\_id provided within {\it Gaia} eDR3) within 7 deg around each galaxy. We found this scale to be a good choice in terms of overall error and scatter among the galaxies, among the explored scales of 3-10deg with 1deg steps. We concentrate on QSOs with 5p solutions, since they are known to have more precise measurements \citep{Lindegren_20a, Fabricius_21}, and retain those with $G < 19$, to reduce statistical errors, and ruwe $<$1.4,  ipd\_gof\_harmonic\_amplitude $<$ 0.2 and  ipd\_frac\_multi\_peak $\le$ 2 for ensuring good astrometric measurements.  The zero-point, to be subtracted to the systemic PM, is calculated as a weighted mean (and its error as error in the weighted mean) after two iterations. This yields a minimum of 50 QSO, with the median being $\sim$900. 

Since the \citet{Vasiliev_EDR3} formula includes the effect from the large-scale component, we subtract from $\sigma_{\rm vas} (\theta_{\rm half}$) the corresponding value from the same formula on the scales of the determination from QSO and account for the error on the weighted mean of QSO PMs. 

Both the zero-points and additional error per PM component are given in Tab.~\ref{tab:systematics} and are used for the orbit integration analysis in Sect.~\ref{sec:orbits}) (the PMs and errors in Tab.~\ref{tab:sysmotions} do not include these additional errors/corrections). 

In general, we find that the dominant\footnote{Here defined as being at least 1.2$\times$ larger.} source of error is the random one for all the galaxies, apart from 
 Fornax, Sculptor, Ursa~Minor, Draco, Carina, NGC~6822, Leo~I, Sextans, Antlia~II, Bootes~I, Hydrus~I, Reticulum~II, Carina~II, IC~1613, Crater~II, M33. Since \citet{Lindegren_20a} find that the systematic PM error decreases with a similar factor with time as the random error, this is not expected to change.
 
In the great majority of applications, systemic PMs need to be converted into a velocity, and uncertainties in the distance modulus will contribute to the uncertainties in the physical transverse velocity. Therefore, it is interesting to known in which cases that is the largest source of error (see also Tab.~\ref{tab:systematics}); these are: Bootes~I, Bootes~II, Bootes~III, Carina, Carina~II, Carina~III, Cetus~II, Coma~Berenices, Delve~1, Fornax, Grus~II, Horologium~I, Hydrus~I, Phoenix~II, Pictor~II, Reticulum~II, Sagittarius~II, Segue~1, Segue~2, Tucana~II, Tucana~III, Tucana~IV, Tucana~V, Ursa~Major~I, Ursa~Major~II, Willman~1. It turns out that the majority of galaxies within 100kpc have their uncertainty in transverse velocity dominated by distance errors, as compared to that due to the statistical and systematic errors in the systemic PM. 
Note that we are including an additional 0.1mag error in the distance modulus of galaxies whose published uncertainties are lower than that value; this in order to mimic the typical mismatch between values of distance modulus found from different techniques. If we drop this additional factor, the situation changes only for Fornax and Bootes~II, which become dominated by the PM component. Sticking to the published uncertainties, there are systems where the distance factor can be as large as 3 to 7 times the systemic PM one (going from 10-40 km s$^{-1}$ the former, while the latter is within 3-12 km s$^{-1}$), like Carina~III, Cetus~II, Hydrus~I, Reticulum~II, Sagittarius~II, Segue~1, Tucana~II, Ursa~Major~II. These are all very faint systems, and it will be hard to improve on their distance estimates, but it might be worth the trouble.

\section{Comparison with the literature} \label{sec:literature}

\begin{figure*}
   \centering
   \includegraphics[width=0.30\textwidth,angle=0]{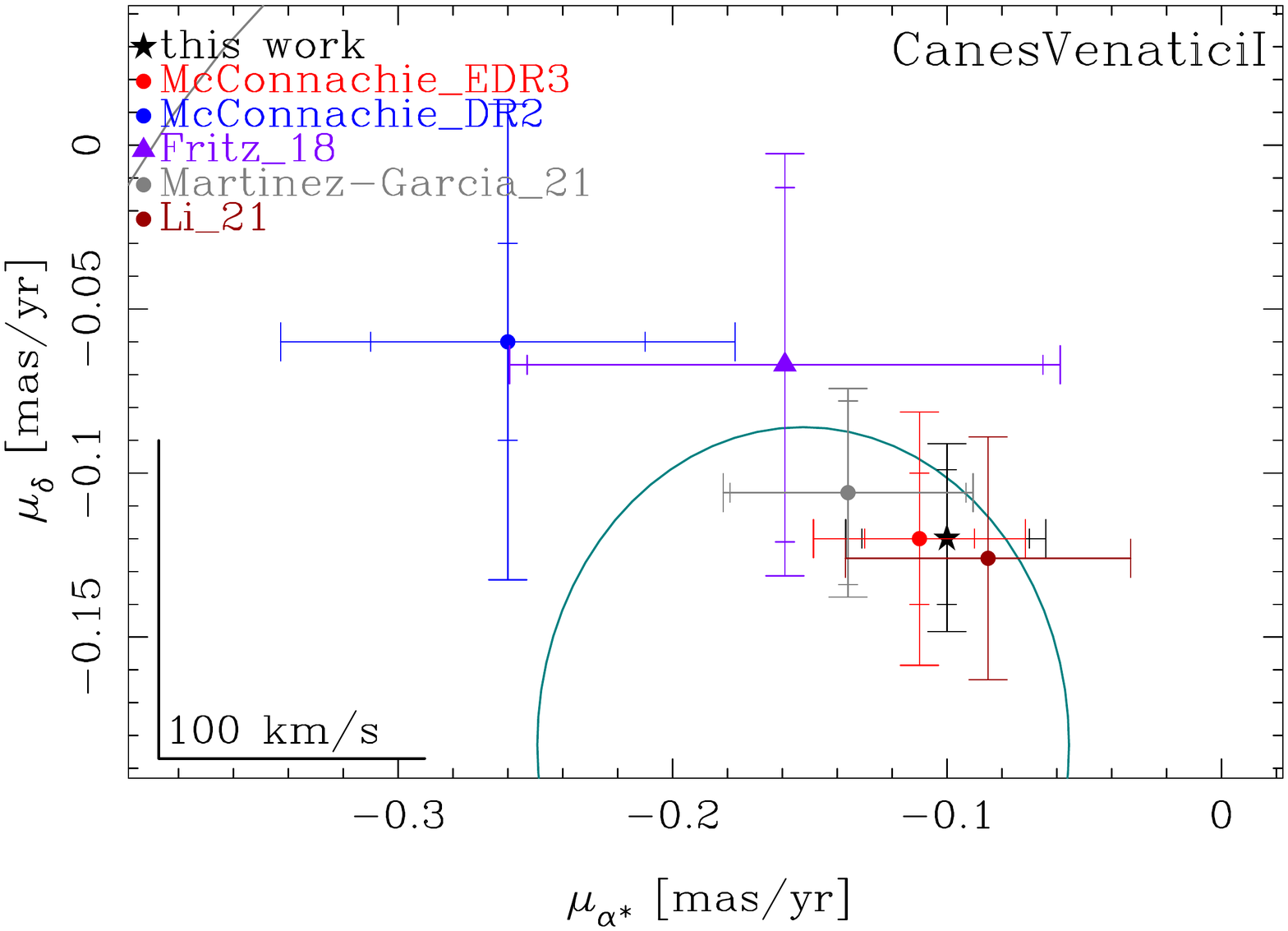}   
   \includegraphics[width=0.30\textwidth,angle=0]{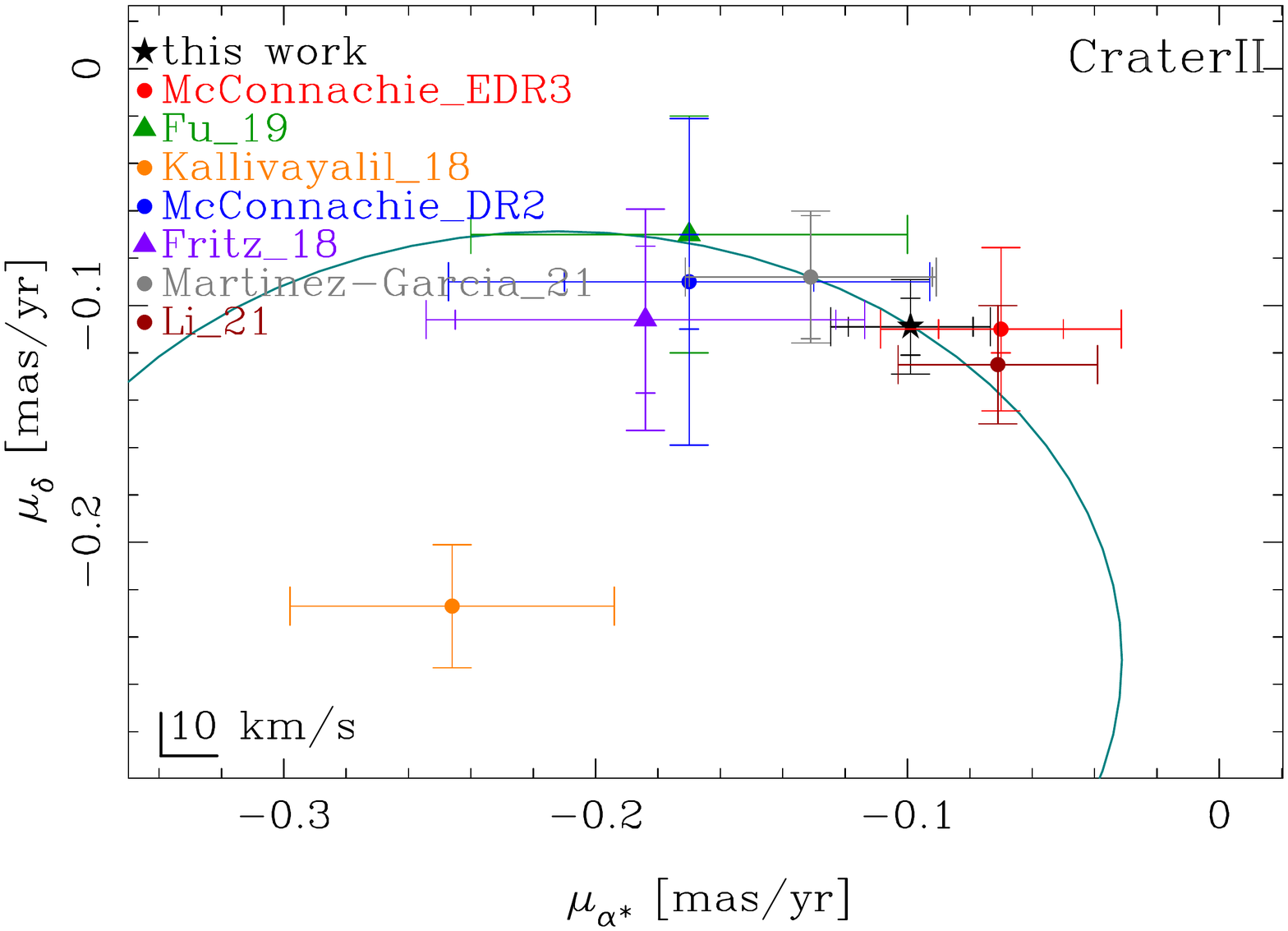}
     \includegraphics[width=0.30\textwidth,angle=0]{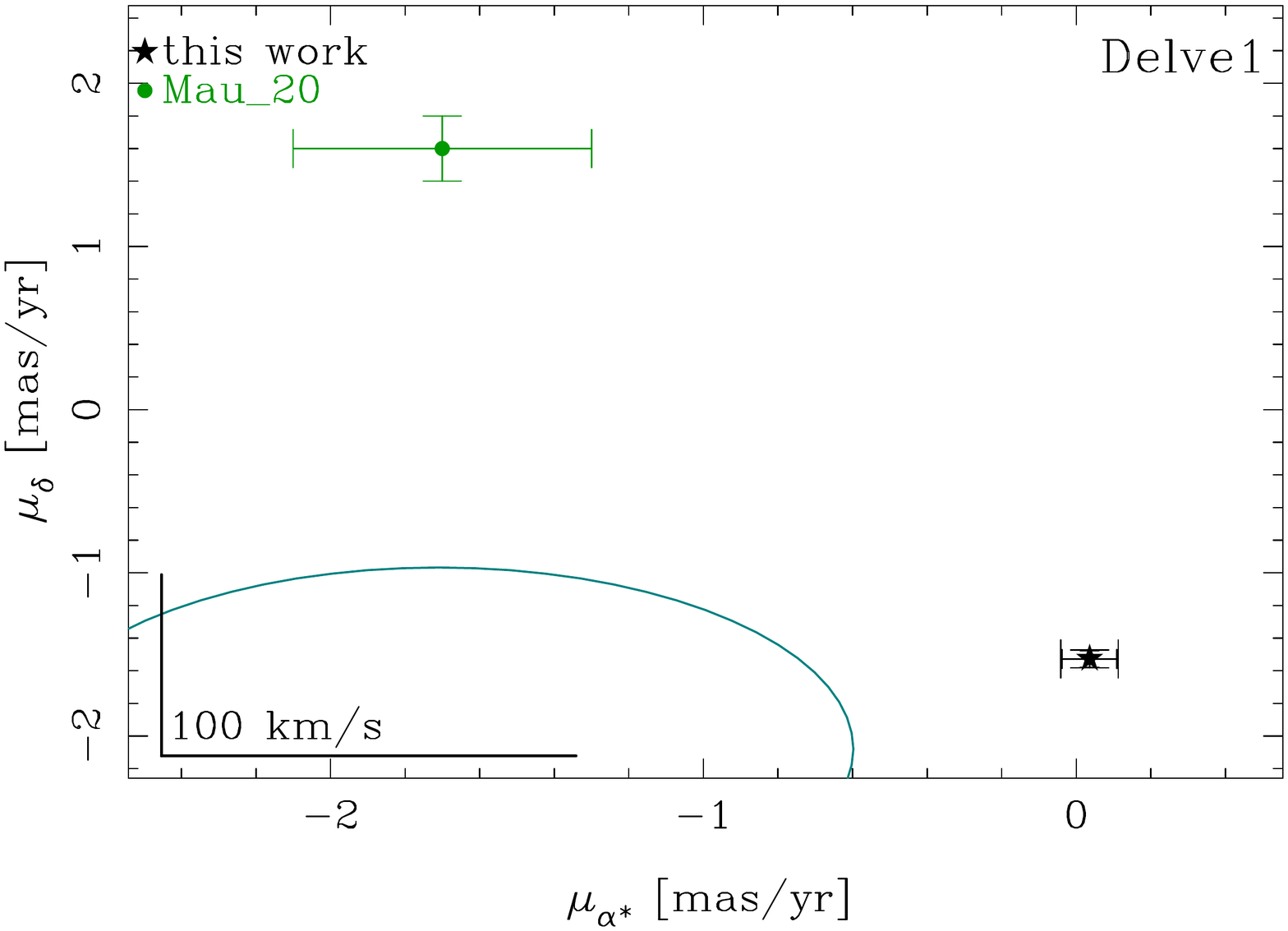}
     \includegraphics[width=0.30\textwidth,angle=0]{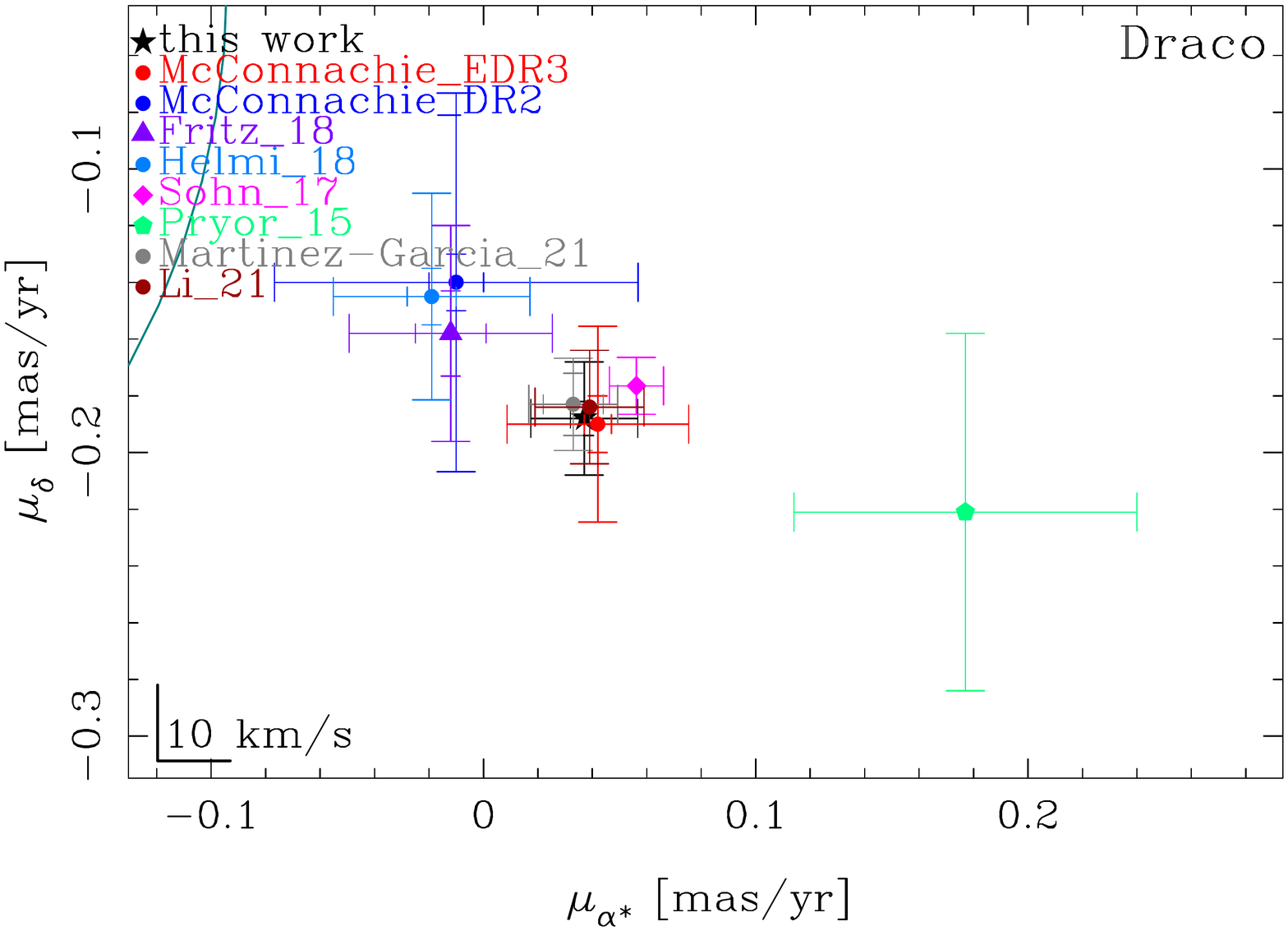}
         \includegraphics[width=0.30\textwidth,angle=0]{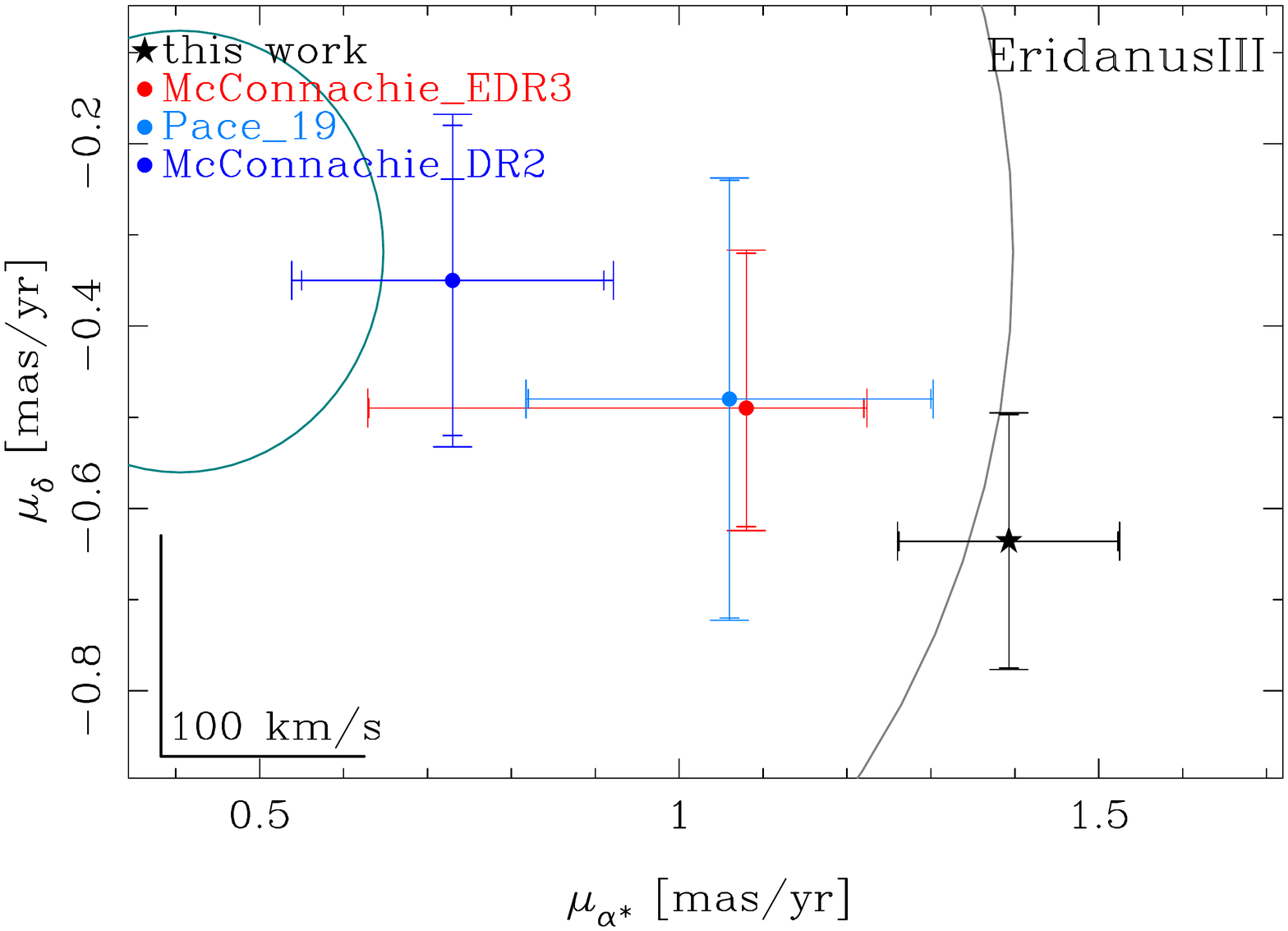}  
         \includegraphics[width=0.30\textwidth,angle=0]{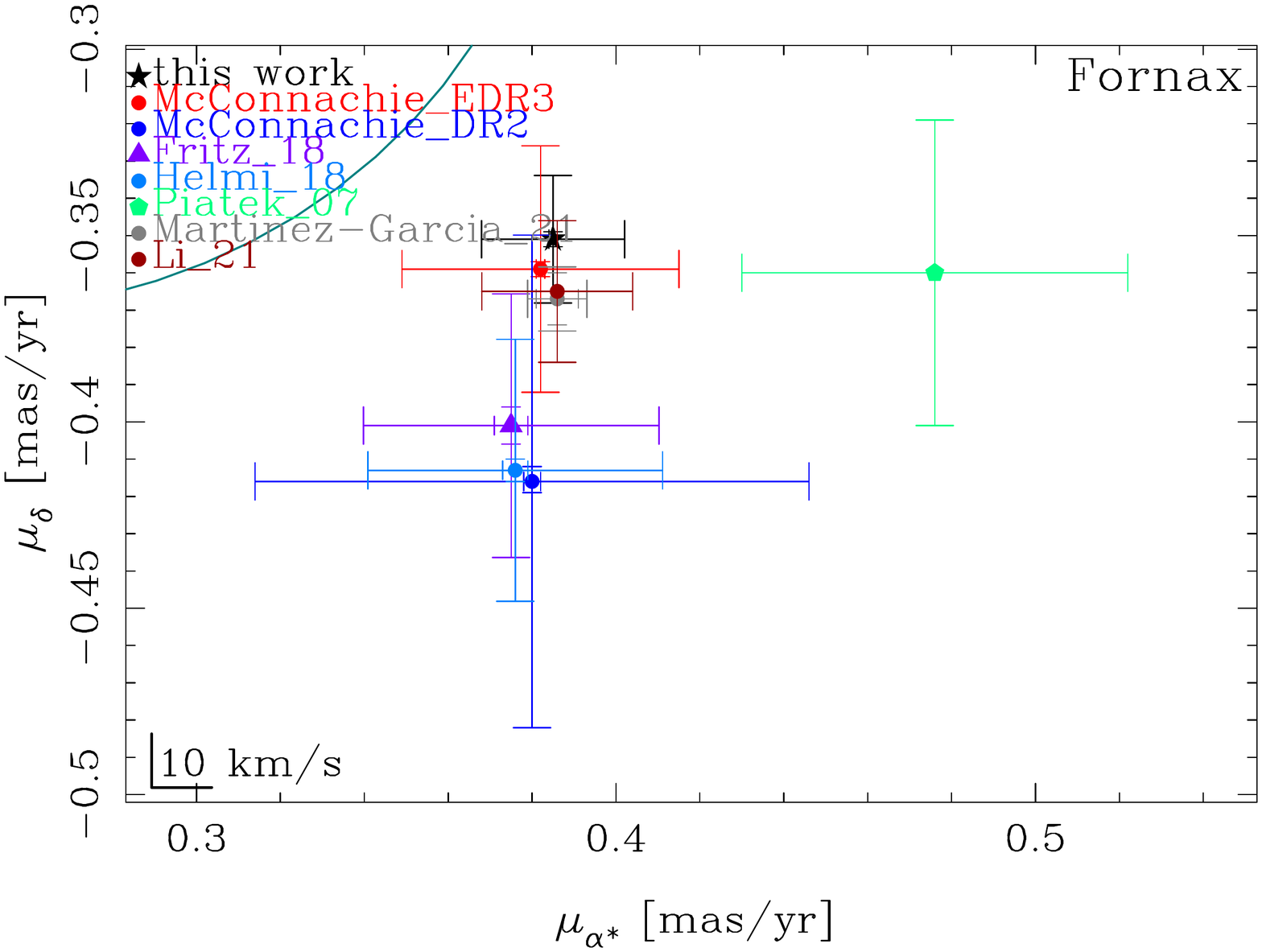}
         \includegraphics[width=0.30\textwidth,angle=0]{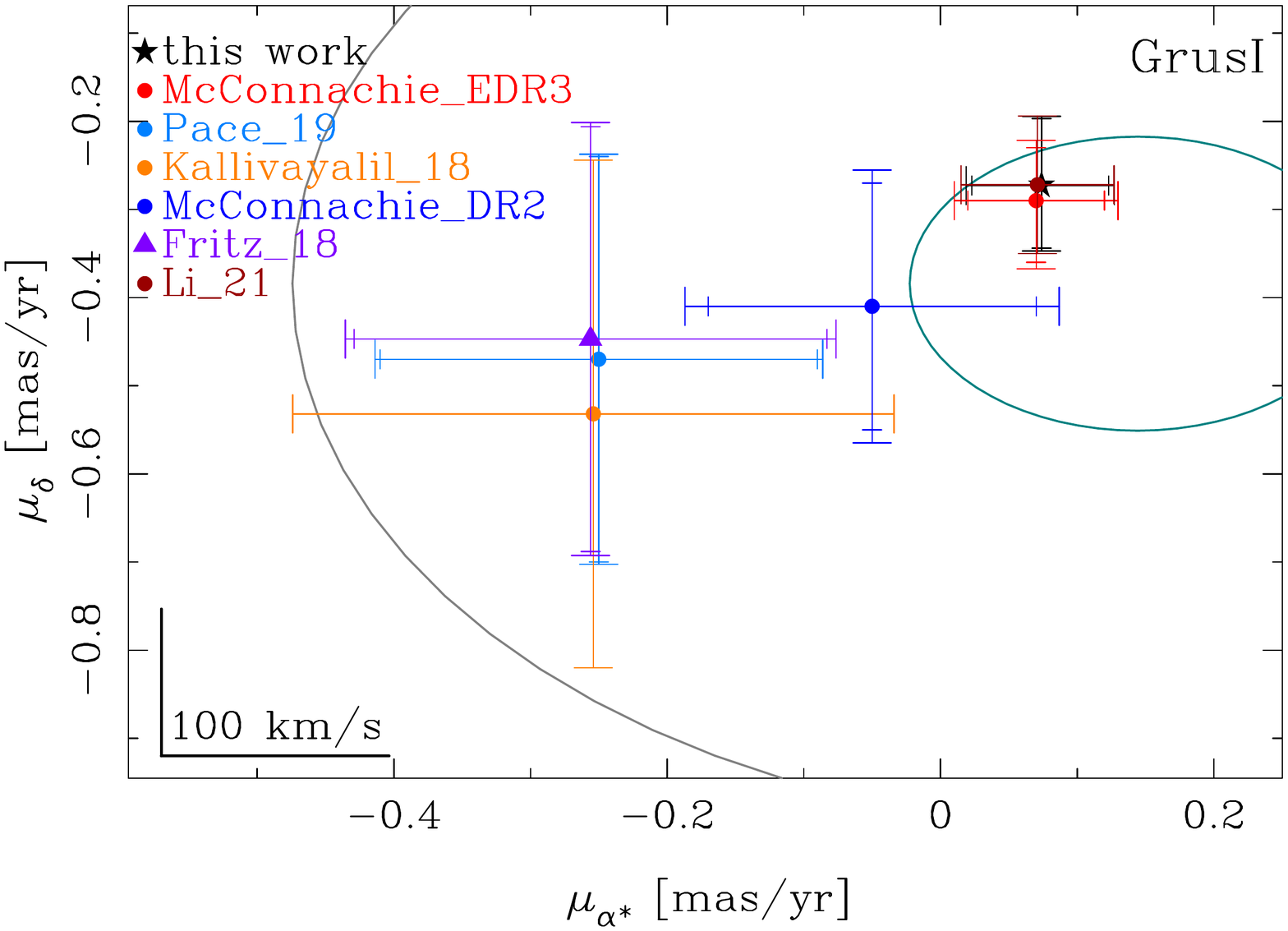}
         \includegraphics[width=0.30\textwidth,angle=0]{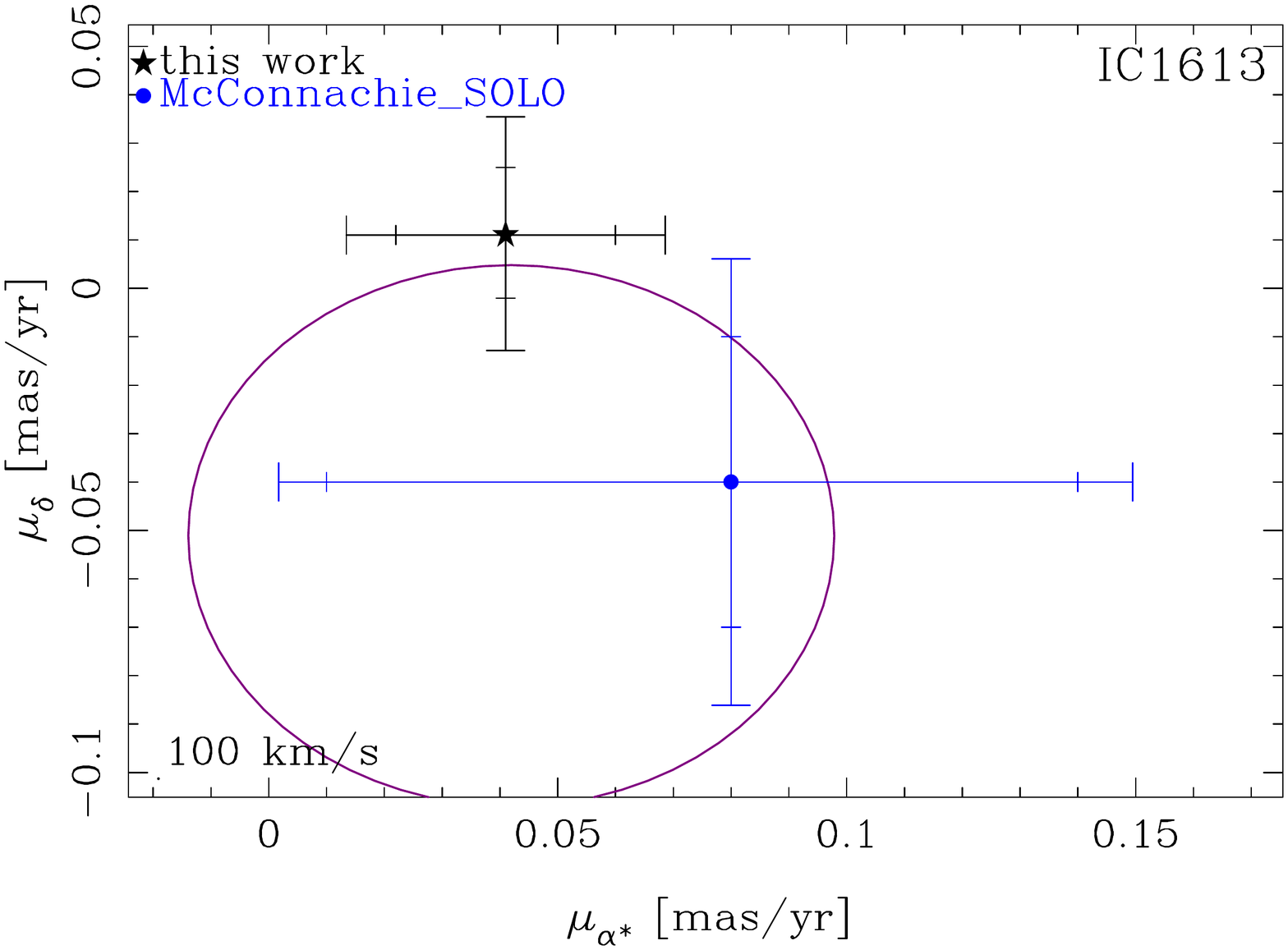}
        \includegraphics[width=0.30\textwidth,angle=0]{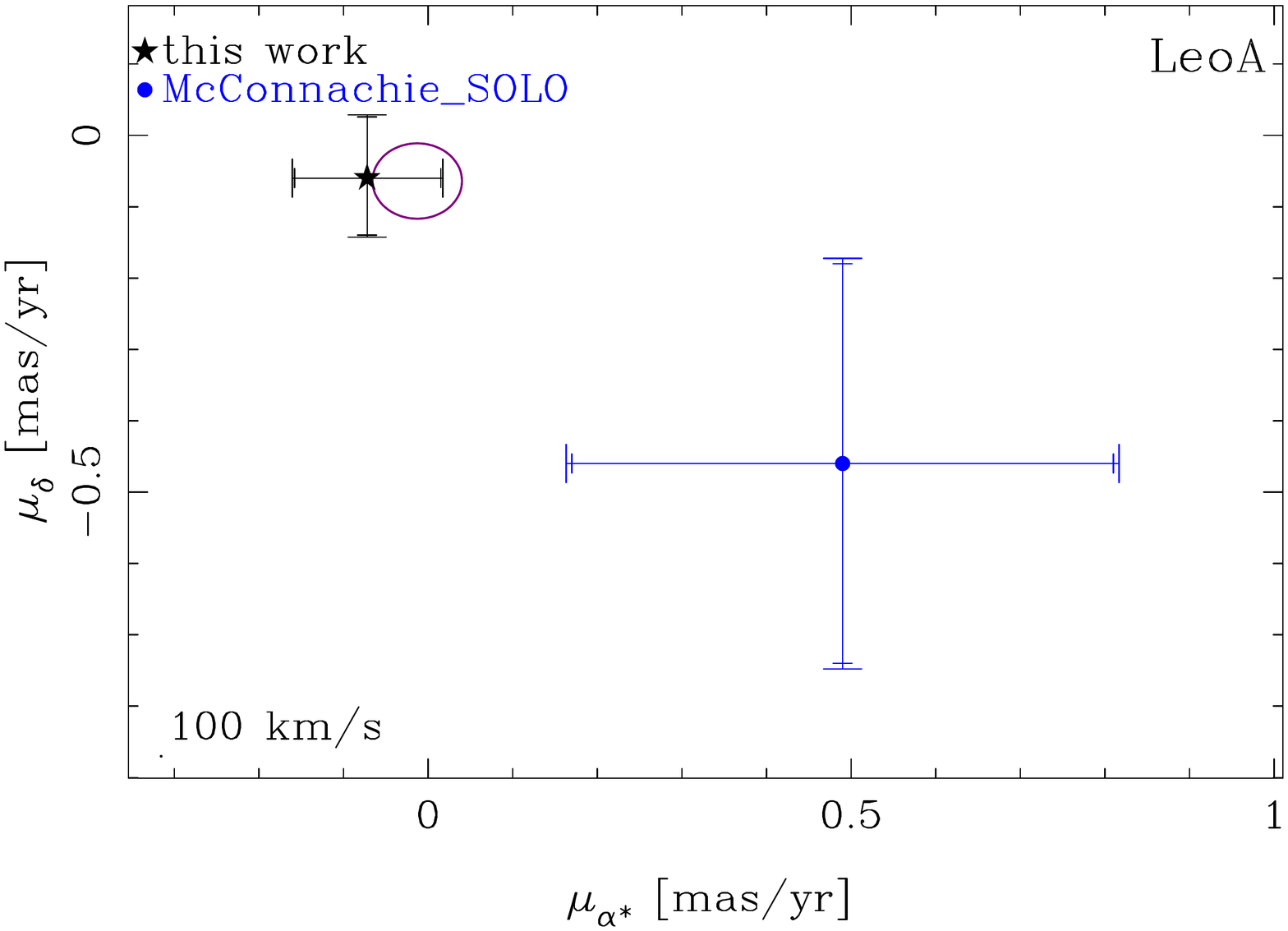}
        \includegraphics[width=0.30\textwidth,angle=0]{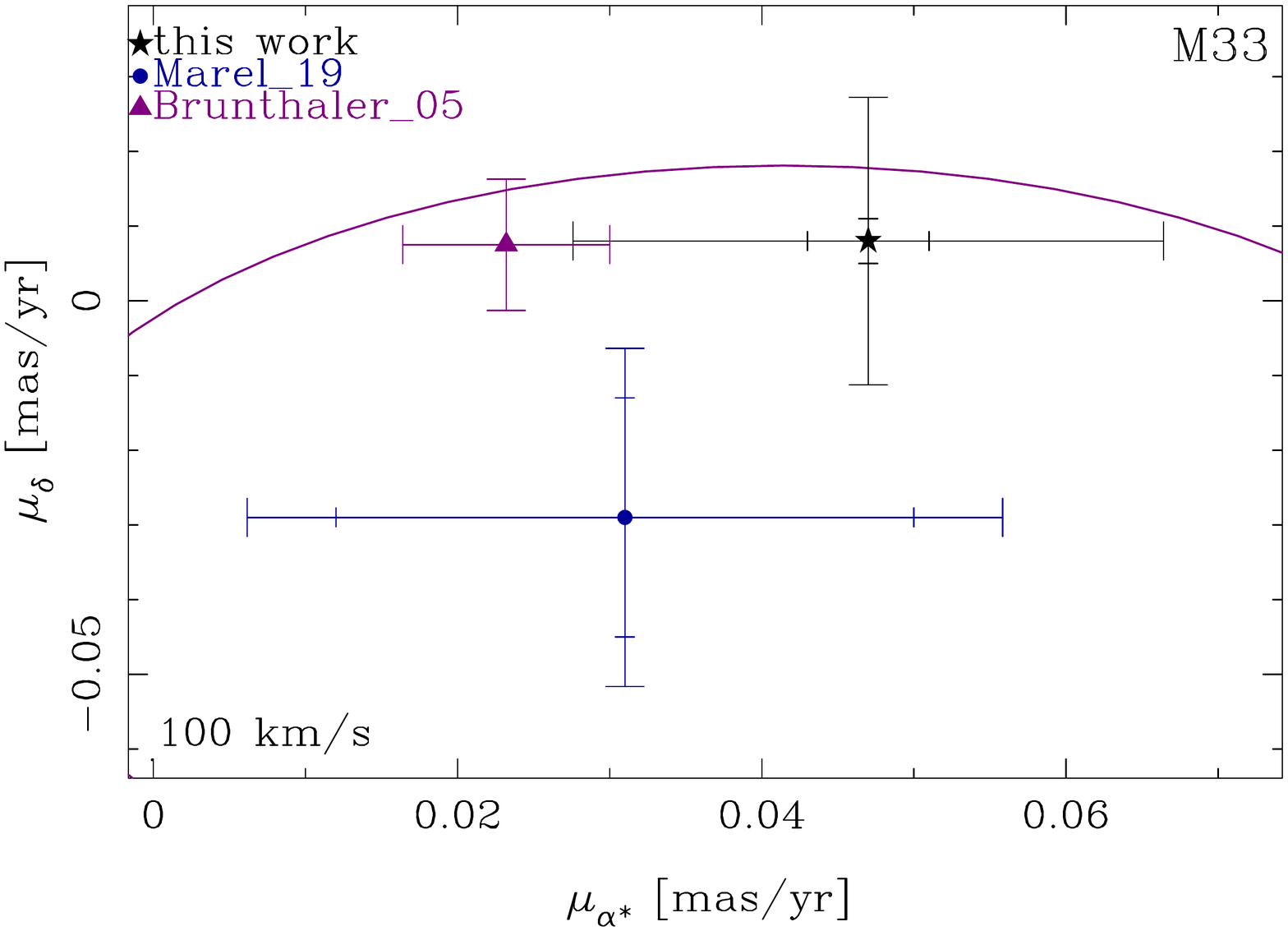}
        \includegraphics[width=0.30\textwidth,angle=0]{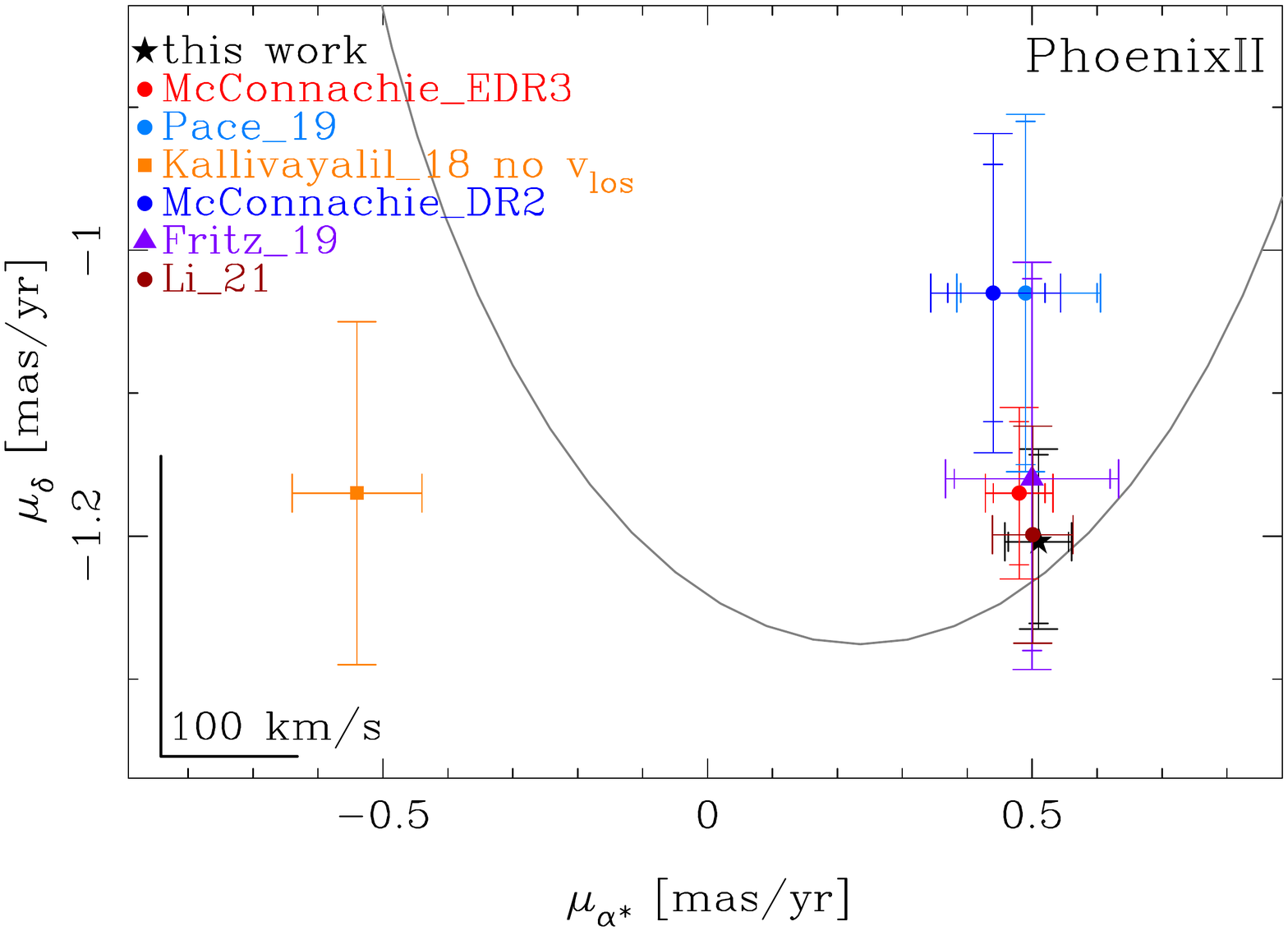}
        \includegraphics[width=0.30\textwidth,angle=0]{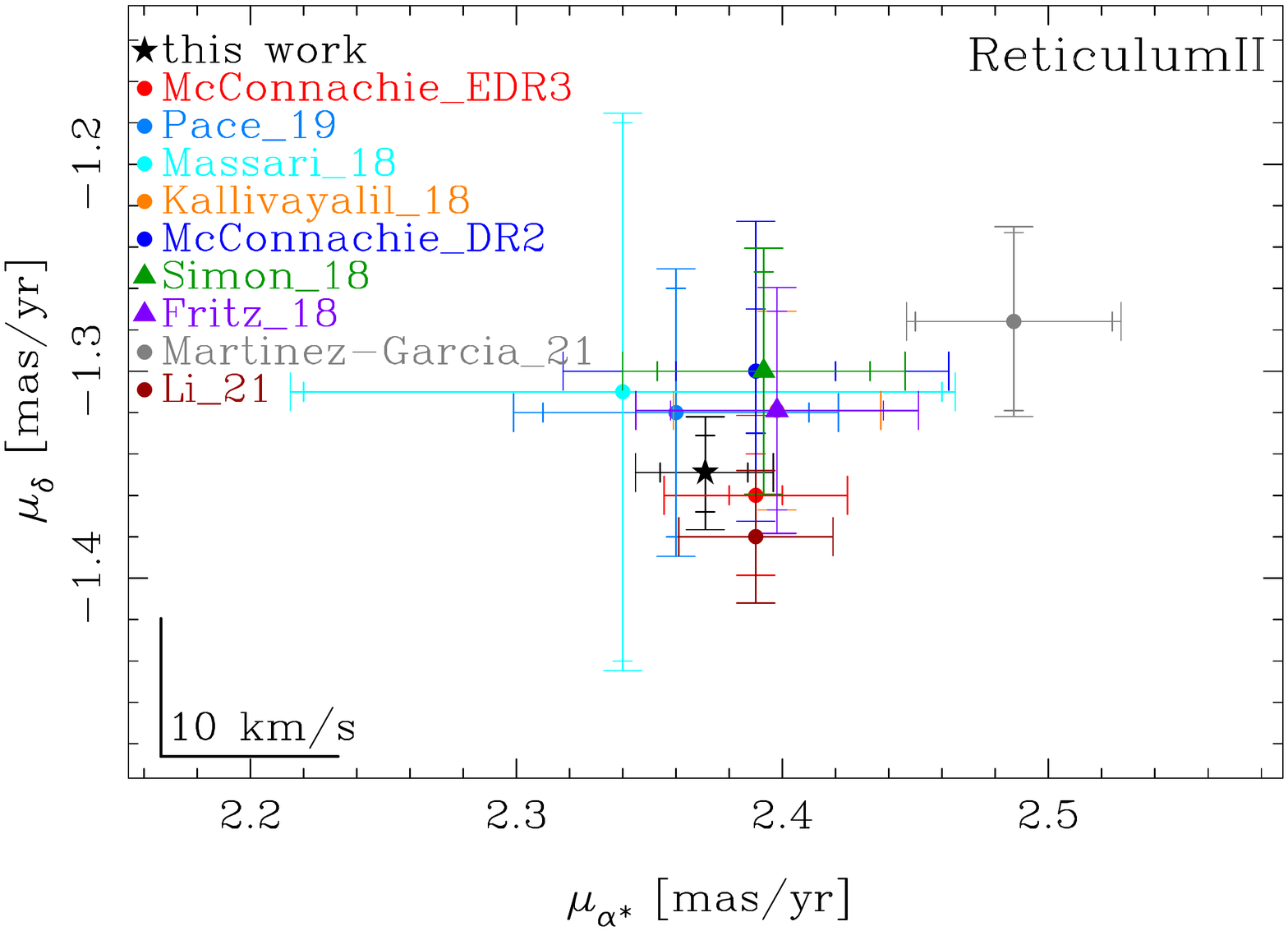}
        \includegraphics[width=0.30\textwidth,angle=0]{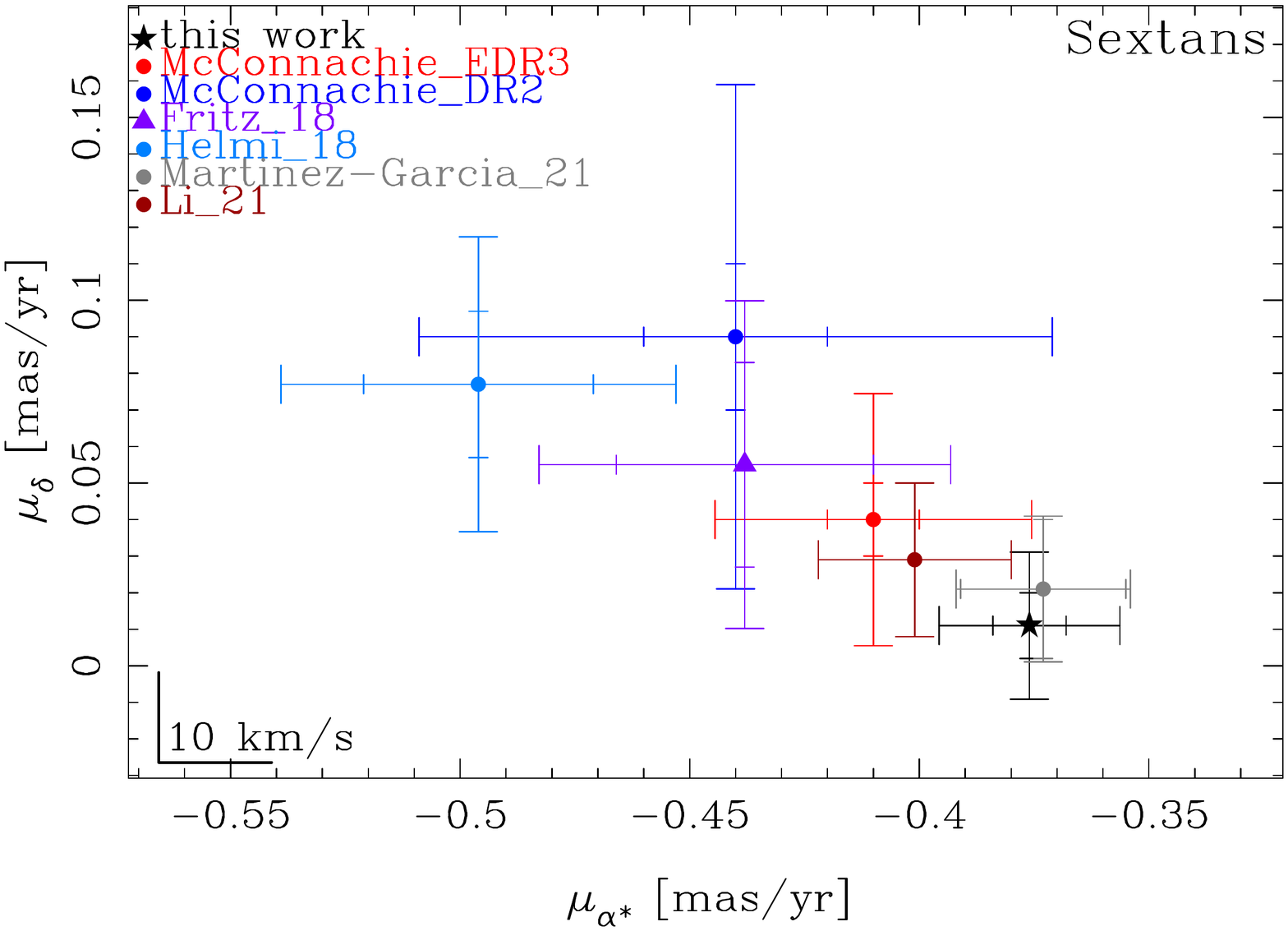}
        \includegraphics[width=0.30\textwidth,angle=0]{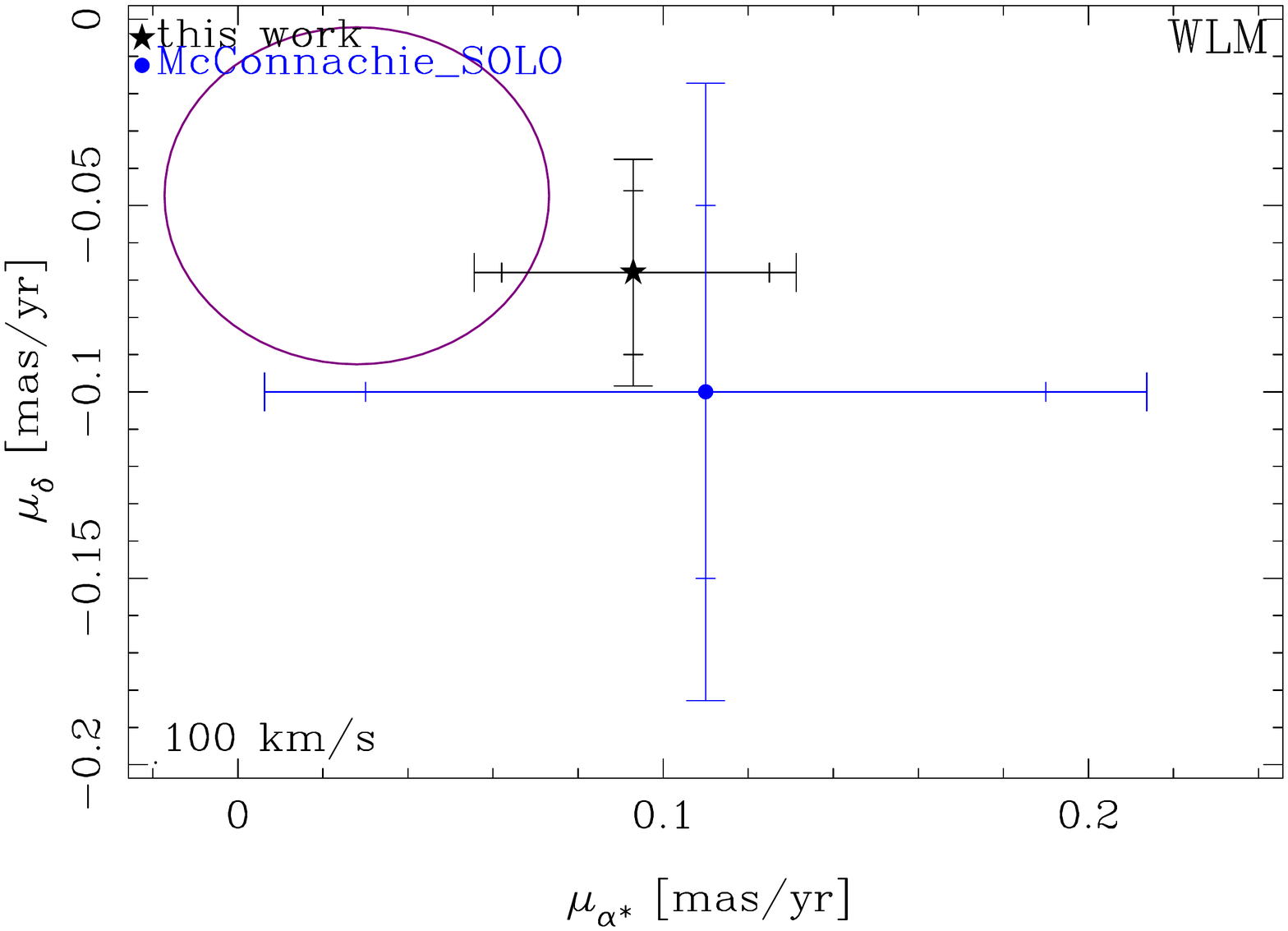}
        \includegraphics[width=0.30\textwidth,angle=0]{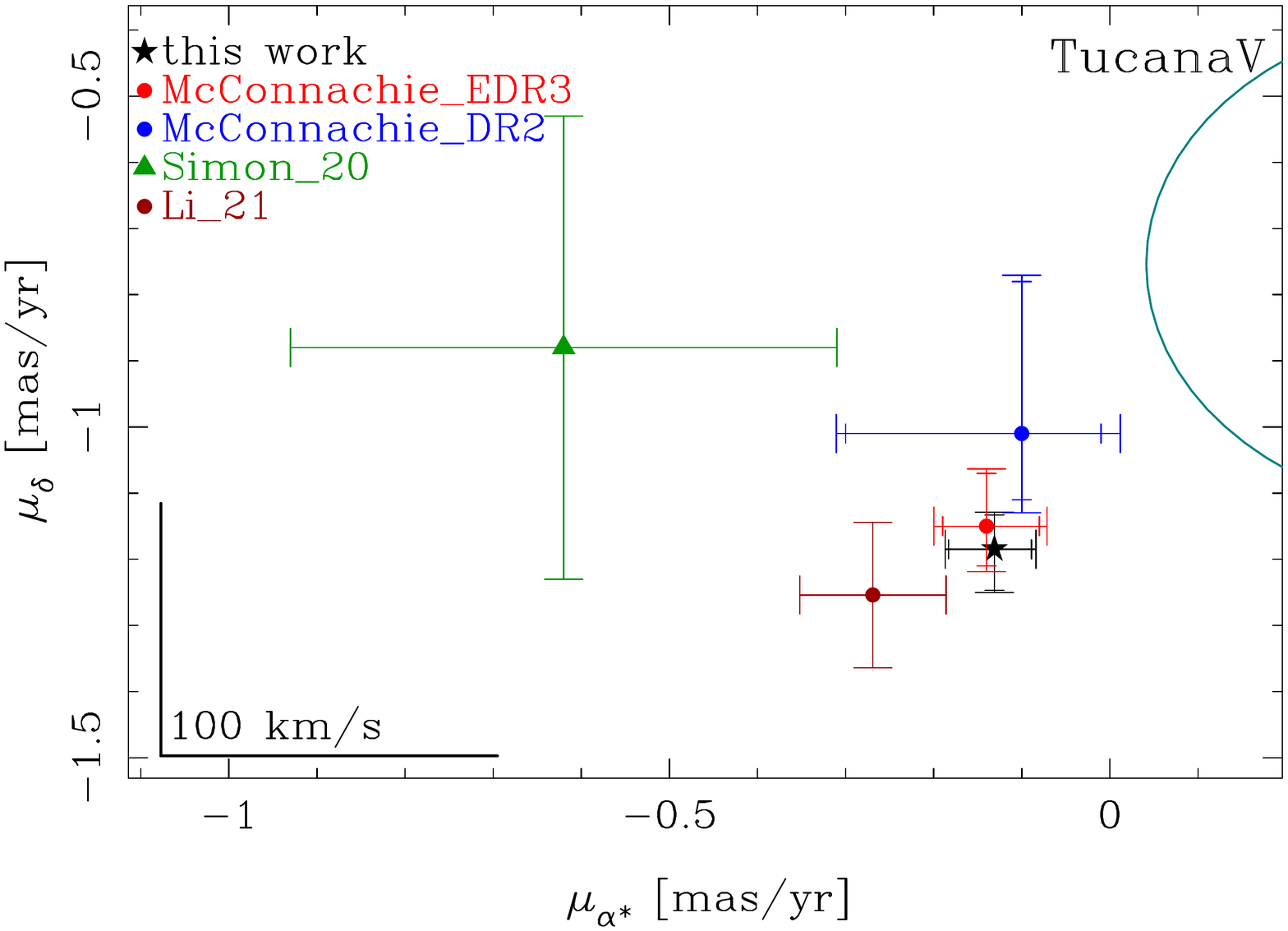}
      \caption{Comparison of our systemic PM measurements (labelled "This work", shown as a black star) with literature measurements. The \textit{Gaia} measurements are from \citet{Helmi_18,Simon_18,Simon_20,Fritz_18,Fritz_19,Carlin_18,Massari_18,Kallivayalil_18,Pace_19,Pace_20,Fu_19,McConnachie_DR2,McConnachie_EDR3,Longeard_18a,Longeard_20,Torrealba_19,Mau_20,Cerny_20,Chakrabarti_19,Gregory_20,Mutlu-Pakdil_19,Jenkins_21,Vasiliev_EDR3,Martinez-Garcia_21,Li_EDR3}. Triangles indicate works that used only stars with additional information on membership, usually from  spectroscopy, but also RRLyrae stars in some cases, as \citet{Simon_18}. HST measurements are from \citet{Piatek_03,Piatek_05,Piatek_06,Piatek_07,Pryor_15,Piatek_16,Sohn_13,Sohn_17}. Among them, those indicated by diamonds (pentagons) use background galaxies (QSOs) as references. The smaller error bars include only the random \textit{Gaia} error, the larger one also the systematic error when they are given as separated in the source. Note that we do not display the correlation between the PMs components here. The ellipses (when in the field of view of the plots) indicate the $\sigma=$100 km s$^{-1}$  prior (green) of \citet{McConnachie_DR2,McConnachie_EDR3}, and the escape velocity (grey) in the $1.6\times10^{12}$ M$_\odot$ Milky Way of \citet{Fritz_18}  centred on the expected reflex motion for the system. For galaxies at a distance $>$ 500~kpc, we only plot an ellipse (purple) corresponding to 200 km s$^{-1}$.}
         \label{fig:pms_lit}
   \end{figure*}

In this section, we compare our systemic PMs determinations with those in the literature. These were obtained with \textit{Gaia} DR2 data \citep{Helmi_18,Simon_18,Simon_20,Fritz_18,Fritz_19,Carlin_18,Massari_18,Kallivayalil_18,Pace_19,Pace_20,Fu_19,McConnachie_DR2,Longeard_18a,Longeard_20,Torrealba_19,Mau_20,Cerny_20,Chakrabarti_19,Gregory_20,Mutlu-Pakdil_19}, eGDR3 \citep{McConnachie_EDR3,Vasiliev_EDR3,Jenkins_21,Martinez-Garcia_21,Li_EDR3}, with HST \citep{Piatek_03,Piatek_05,Piatek_06,Piatek_07,Pryor_15,Piatek_16,Sohn_13,Sohn_17} and VLBI \citep{Brunthaler_05}. For the HST measurements, we ignore older determinations, when newer ones from the same group are available. We compare all measurements in Figs.~\ref{fig:pms_lit}, \ref{fig:pms_lit2}, \ref{fig:pms_lit3}, \ref{fig:pms_lit4}, \ref{fig:pms_lit5}. 

The agreement with \citet[][hereafter, MV20b]{McConnachie_EDR3} is in general very good, if not excellent, with the values being within 1 or 2-$\sigma$ at most, in each component (here we consider the largest of the two error-bars, since the methodology is very similar and the systematics should be directly comparable). There are however an handful of cases for which the $\mu_{\delta}$ component differs by more 2-$\sigma$: Antlia~II (3.3-$\sigma$), Reticulum~III (2.5-$\sigma$), Carina~III (2.3-$\sigma$), Segue~1  (2.5-$\sigma$) and in principle also for some of the brightest galaxies like Sextans when only the statistical error is used. Inspection of the spatial, CM and PM location of the probable members from our code does not reveal hints of specific issues with these galaxies; the differences are likely to be the results of the methodology applied, which for some systems turns out to have a more noticeable effect.

We note that there are cases in which the statistical  MV20b uncertainties are much smaller than those we determined, i.e. in each component separately they are between 10-30\% of ours for Bootes~IV, Leo~T, Cetus~III, and as small as 2-3\% for Indus~I, Virgo~I, Pegasus~III (there are also cases in which our uncertainties are smaller, but with a reduction of at most 60\%). Inspection of the output of our code corroborates the expectation of the large errors we find in these cases, given the small number of probable members with $P>0.5$ and their in general faint magnitudes; e.g. no stars with a probability of membership larger than 0.5 is found for Virgo~I, Pegasus~III, and only 2 for Cetus~III (see "Notes on individual galaxies" for more details). MV20b find only one likely member in Cetus~III, Pegasus~III and Virgo~I. Clearly the results for these systems need to be taken with a pinch of salt. 
The inclusion or not of spectroscopy in the determination does not seem to be the culprit of this difference, since in most cases we do not notice any significant reduction in the random error when we include the l.o.s. velocity likelihood term.

On the other hand, we suspect that the main reason for the difference might be the prior in systemic PM corresponding to a 100 km s$^{-1}$ velocity dispersion used by MV20b. This can be easily seen for the three most distant galaxies of their sample (Phoenix, Eridanus~II and Leo~T), where they give also the motions without this prior. For the others, if we model the ratio of ours and MV20b errors as the quadratic sum of one and the ratio of our statistical error over the 100 km s$^{-1}$ dispersion, then in median our error would be only 3\% larger, 
with excursion in both direction our error is between 53\% and 149\% 
of their scaled error, when we exclude Indus~I (324\%) and Virgo~I (201\%), for which we do not obtain a reliable measurement. This prior seems to be also mainly responsible for some PM differences, in cases where the absolute value of our PM is larger than in MV20b, like for Eridanus~III and Horologium~II.

Unlike MV20b, we do find a motion for Indus~II, with 6 stars with $P>0.5$ (1 with $P>0.95$); nonetheless, the distribution of the probable members on the various planes does not transmit confidence in the result. 

Typically our results compare well with the other works based on eGDR3. \citet{Li_EDR3} start with stars that have also spectroscopic observations and then add stars without spectroscopy but with astrometric properties compatible within 5-$\sigma$ to those of the spectroscopic sample.  Their errors (which includes systematics) are nearly always larger than our total errors (in median, about 1.18$\times$, 
clearly larger for many of the faintest systems). Due to their larger errors, the system with the most noticeable deviation is Bootes~I (2.3$\sigma$ in R.A. when we do not apply our correction for systematics). 

\citet{Martinez-Garcia_21} focus on 14 galaxies and also use a probabilistic approach, with quality cuts likely more conservative than those adopted here. Our statistical errors are in general smaller, about 0.7x their errors; the largest differences are found for the fainter and more diffuse galaxies, probably indicating that their method needs more stars to perform well. Their systematic errors are lower, since they also model small scales effects. They also apply a QSO-based zero-point correction on their systemic PMs. Our zero-point correction and that of \citet{Martinez-Garcia_21} differ by 0.009 mas yr$^{-1}$ on average, with the standard deviation in the ($\mu_{\alpha,*}$, $\mu_{\delta,*}$) zero-points for the galaxies in common being (0.012, 0.012) [mas yr$^{-1}$] for \citet{Martinez-Garcia_21} and (0.013, 0.008) [mas yr$^{-1}$] in our work. When comparing our motions before QSO correction, they agree usually within 2.0 $\sigma$, the exception is Reticulum~II which deviates by -2.9 $\sigma$
in R.A. (and 2.0 in Dec.) from ours and is also different from the other EDR3 determinations.  

Finally,  \citet{Jenkins_21} focus on Leo~IV, Leo~V and Bo\"{o}tes~I and use use only stars with spectroscopy, hence it is not surprising that their uncertainties on the systemic PMs are larger than ours. For Leo~IV and Leo~V the motions agree within 1 $\sigma$ and for Bo\"{o}tes~I within 2$\sigma$ in both dimensions.  

A comparison with \textit{Gaia} DR2 measurements tests mainly the performance of those, due to their larger astrometric errors, but that is still useful to perform. When comparing with \citet{Fritz_18} for 38 systems in common, the standard deviation of the distribution of differences in systemic PM normalised by the uncertainty\footnote{We use only the \citet{Fritz_18} error for normalisation, since the data from the two releases are not independent.} is 0.96/1.05 in the R.A. and Dec.
component, respectively, thus within expectations. At a closer look, it appears that the standard deviation for most of the sample would be smaller than $\sim$1 and it is inflated by a few cases with larger deviations (Segue~2, 3.8; Triangulum~II, 2.5 and Ursa~Major~I, 2.8 from the values in this work).
The accumulation at small deviation is probably understandable, since we might not be taking the correlation between the data sets well into account. 
 
We find good agreement with the preferred values by  \citet{Fritz_19}, but their sample of 4 UFDs is too small to tackle whether the small deviations found could be the result of chance. 

Regarding other estimates, we note that the sub-sample of those by \citet{Kallivayalil_18} for which spectroscopy was not used do not match our motion well. That is not necessarily surprising, since in those cases \citet{Kallivayalil_18} values were based on the assumption that the galaxy was a former satellites of the LMC, which is not the case for most, like Columba~I. It is more surprising for Phoenix~II which likely is a LMC satellite \citep[][but see also Sect.~\ref{sec:LMCsat}]{Fritz_19}. Probably the reason in this case is that the LMC model used in \citet{Kallivayalil_18} does not match reality sufficiently well, as it is likely not massive enough.

The agreement is slightly worse with \citet{Massari_18} with a standard deviation between ours and their results of 1.35 in R.A. and 1.06 in Dec. 
 This is mainly driven by Sagittarius~II, which deviates by 2.9/2.0 $\sigma$ from our value, despite the large errors; nonetheless, also other of \citet{Massari_18} determinations show deviations from other DR2 measurements. 

As for the galaxies found beyond the MW virial radius, 
we can compare with the GDR2 measurement of \citet{McConnachie_SOLO} for NGC~6822, WLM, Leo~A and IC~1613. Our statistical errors are between 27\% and 47\% of theirs, a clear improvement. The standard deviation between the values is 0.67/1.20 
in R.A./Dec. respectively in the expected range. 

Also for M33 our measurements agree reasonably well with the VLBI OH maser and HST ones by other two of \citet{Brunthaler_05,Marel_19}. 

It is interesting to compare \textit{Gaia} measurements to high precision PMs obtained in a completely independent way with other telescopes like HST and VLBI. The references used by HST are either QSOs or background galaxies. The accuracy of QSOs based measurements can suffer because of the small number of reference sources, since e.g. systematic errors cannot be well derived from the data. When compared with our errors, it seems that the uncertainties quoted in the literature are underestimated, as four out of five measurements have a deviation of at least 1.8 $\sigma$ (up to 3.5 $\sigma$) in one dimension.\footnote{In contrast to the previous comparisons with \textit{Gaia} measurement we apply here our QSO-based shifts, since these independent measurement are not affected by the \textit{Gaia} systematics.} 

In the cases with galaxies used as reference sources, 3 out of 4 works obtain deviations smaller than 1.6 $\sigma$ in both dimensions. The only exception is Sculptor (2.5-$\sigma$). 
Our QSO-based shift improves the comparison for Sculptor slightly, although less than the correction adopted by \citet{Martinez-Garcia_21}; nonetheless, also with this shift a difference remains with respect to the HST determination. Since \textit{Gaia} DR2 and EDR3 estimates agree with each other it seems unlikely that \textit{Gaia} systematics are the only reason for it.  Nevertheless overall HST PMs based on galaxies and \textit{Gaia} agree well, increasing the confidence in the precision and accuracy of both, see also the example of M31 \citep{Salomon_20,Marel_12a}.

\section{Orbit integration}\label{sec:orbits}

\subsection{Method} \label{sec:orb_method}

Using the PMs derived above with the distance modulus and the l.o.s. velocities from the literature listed in Tab.~\ref{tab:sample}, we integrated the orbits of each galaxy for which spectroscopic measurements are available in three MW potentials: in two of them, the MW is treated as an isolated system (hereafter {\it "isolated"} potentials), and we explore a mass for the MW DM halo that brackets the range of likely MW masses \citep{Boylan-Kolchin_13,Gibbons_14,Fritz_20,Wang_20m-review}; in the other potential (hereafter {\it "perturbed"} potential) a $8.8 \times 10^{11}$  M$_{\odot}$ MW is perturbed by a $1.5 \times 10^{11}$ M$_{\odot}$ LMC, as  published by \citet{Vasiliev_21}. The reason for including the latter case is that, although the mass of the LMC system is still subject to debate \citep[i.e. see][]{Wang19_Mstream}, recent observations, such as the rotational velocity of the LMC \citep{Marel_14}, some perturbation of the MW's disk \citep{Laporte_18}, the dynamic of the ATLAS, Tucana III, Orphan and Sagittarius streams \citep{Erkal_18,Erkal_19,Vasiliev_21,Li_21Atlas} and the dynamics of  distant halo stars \citep{Erkal_20}, are consistent with the idea of a massive LMC, i.e. with a mass of $1-2.5 \times 10^{11}$ M$_\odot$, perturbing significantly the gravitational potential of the MW \citep[see][]{Garavito-Camargo_19,Garavito-Camargo_20,Cunningham_20}; therefore, we wish to investigate the impact of a massive LMC on the past orbits of the dwarf galaxies of the MW and on the account of its possible satellites. 

The first isolated potential ("Light MW") is that published by \citet{Vasiliev_21} and composed of a spherical bulge of $1.2 \times 10^{10}$ M$_\odot$, an exponential disc of $5 \times 10^{10}$ M$_\odot$ and of a triaxial DM halo, with a total mass M$(<$R$_{vir})=8.8 \times 10^{11}$ M$_\odot$ within the virial radius of $251$ kpc, but unperturbed by the passage of LMC. The second isolated potential ("Heavy MW") is similar to the massive potential used by \citet{Fritz_18} and consists of a {\it MWPotential14} \citep{Bovy_15} with a more massive DM halo so that the system has a total mass of mass M$(<$r$_{\rm vir})=1.6 \times 10^{12}$ M$_\odot$ within the virial radius $r_{\rm vir} =307$ kpc . 
Orbits are integrated 6 Gyrs backward and forward, with a time step of 3 Myrs, through the {\sc Agama} package \citep{Vasiliev18}. The reason to integrate forward is that some of the most distant galaxies have not yet passed by pericenter, and therefore we have to integrate their orbit in the future to measure their orbital parameters. 

In the {\it perturbed} potential, published by \citet{Vasiliev_21}, the MW experiences the passage of a $M_{LMC}=1.5 \times 10^{11}$ M$_\odot$ LMC. In this model, the initial MW potential is as the {\it isolated} "Light MW" mentioned above, and the initial LMC is represented by a NFW halo with a scale radius of $r_s=10.84$ kpc and truncated at r$_{truc}=108.4$ kpc. In this case we integrate the orbits only backward; consider that before 5 Gyrs ago the model of \citet{Vasiliev_21} is stationary. 

In order to take into account the uncertainties on the different measurements, the systematic errors in the PMs, and the correlations between the PMs along the Right Ascension and the Declination, we integrated the orbits from 100 Monte Carlo realisations of the current position and velocity of the galaxies. For the galaxies having distance modulus uncertainties lower than 0.1 mag in Table~\ref{tab:sample}, we added in quadrature 0.1 mag, corresponding to typical systematic uncertainties between different methods of distances determinations. Additionally, we did not integrate the galaxies with heliocentric distances larger than 500 kpc since they are clearly not members of the MW system, and will require taking into account the LG in its globality \citep[see][]{McConnachie_SOLO}. 

The current (right-handed) Cartesian Galactoncentric coordinates were calculated with {\sc Astropy} \citep{astropy_2013, astropy_2018}, assuming that at the solar radius of $R_\odot=8.129$ kpc \citep{gravity_2018}, the circular velocity is of $V_{circ}(R_\odot)=229.0$ km\,s$^{-1}$ \citep{eilers_2019}, and with a Solar peculiar motion of (U$_\odot$, V$_\odot$, W$_\odot$)=(11.1, 12.24, 7.25) km.s$^{-1}$ \citep{schonrich_2010}. 

The orbital parameters, including their uncertainties, for the two {\it isolated} MW potentials are listed in Tab.~\ref{table_param}. The quantities listed in this table are the Galactocentric distance of the pericenter (Peri), of the apocenter (Apo), the eccentricity (ecc), the orbit period (T), the time since the last pericenter (T$_{last,peri}$), and the fraction of galaxies reaching their apocenter in the last/next 6 Gyr $\mathcal{F}_{apo}$. For each of these quantities (except $\mathcal{F}_{apo}$), the listed values correspond to the median of that parameter found for the 100 realisation of the orbits, and the uncertainties have been calculated from the 16$-th$ and 84$-th$ quantiles. When the majority of the orbits do not reach the apocenter within the time range of the integration, the uncertainties on the apocenter, eccentricity and period cannot be computed. Therefore, for those galaxies, we rather give the value of the 16$-th$ quantile. 

It is important to note here that the values given in this table are not free of biases. Indeed, the {\it observed} tangential velocity is known to be a biased estimator of the {\it real} tangential velocity of a given system \citep{Fritz_18,Marel_08}, where the former is inflated by the measurement uncertainties. Therefore, this bias also reverberates on the different derived parameters, like  apocenter or the pericenter. The bias on the tangential velocity can be understood with the following idealised experiment. Let us  assume that a galaxy is moving radially towards us with a {\it real} tangential velocity of exactly 0 km s$^{-1}$. In this case, the true pericenter is at 0 kpc. In reality, the uncertainties on the  systemic PM can only increase the {\it observed} tangential motion, this last one being positive by definition, increasing {\it de facto} the observed pericenter of the galaxy. 

In order to identify the galaxies least impacted by this bias, we follow the guidelines in the Appendix of \citet{Fritz_18}, where this effect was estimated with "backward" Monte-Carlo simulations. According to that study, assuming a typical velocity of $\sim$100 km s$^{-1}$ for galaxies in the MW system of satellites, one can expect the observed 3D/tangential velocity to be $<$1.5 the true one for measurements uncertainties $\lesssim$70 km $^{-1}$ on these quantities. For the rest of the analysis, we therefore mainly focus on systems with uncertainties on the total Galactocentric velocity $<$70 km $^{-1}$ (those with larger errors have their name in italics in Tab.~\ref{tab:table_param}).

\subsection{Results and discussion} \label{sec:orb_results}

\subsubsection{Effect of a massive LMC on the orbital properties}

\citet{Patel_20} examined the changes in the orbital properties of 13 UFDs and 5 classical dSphs introduced by the inclusion of the LMC (and SMC), for a range of LMC and MW masses, using GDR2 PMs. The authors found that the orbits of both classical dSphs and UFDs were noticeably affected by the inclusion of the Magellanic Clouds, in particular the LMC for the classical dSphs;  the inclusion of the SMC did not have an impact on the number of galaxies potentially classified as satellites of the LMC, but on the longevity of this association. 

Although, as we will see in Sect.~\ref{sec:LMCsat}, the large majority of the dwarf galaxies found around the MW that we examine were not bound to the LMC, also in this work we see that the presence of a massive LMC significantly perturbs the MW potential, modifying the orbits of most of the dwarfs. Figure~\ref{fig:apoperi} gives a view on how the pericenters and apocenters are impacted within the 68\% confidence interval.  In  Fig.~\ref{fig:orbits}, \ref{fig:orbits2}, \ref{fig:orbits3} we compare the orbital path of the dwarf galaxies found within 500$\,$kpc from the MW during the last 3 Gyr in the presence of the LMC ("perturbed") to the case of the isolated Light MW, which is equal to the former apart from the inclusion of the LMC and the corresponding response of the MW. This is one realisation coming from the integration of the observed systemic motions and l.o.s. velocities. For obvious reasons, the difference between the isolated and perturbed orbits become more important with lookback time since we used the same current dynamical parameters as a starting point. Thus by definition, the orbital shifting is of 0 kpc at t=0 Gyr. 

Interestingly, we can see that the orbits of the majority of the galaxies are affected by the presence of the LMC, regardless of their distance to the MW or to the LMC. The influence of the inclusion of a massive LMC can manifest itself in several ways, e.g. as a change in the pericentric and/or apocentric distance, which can become larger or smaller, as well as the timing of the passages, even for systems that are very distant from the LMC and away from its orbital plane. 

There are systems that are barely affected or affected in a minor way, e.g. Segue~1, Hydra~II, Leo~I as there are systems that are very strongly impacted, as e.g. 
Draco II, Sextans, Sculptor, but even NGC~6822. For example, Sculptor would have been infalling recently onto the MW in the perturbed potential rather than being compatible with having been a long term satellite. This is mainly either due to the proximity of the LMC or to the MW reflex motion to the gravitational wake caused by the LMC infall (also called collective response)  \citep{Garavito-Camargo_20, Petersen_20, Petersen_21,Vasiliev_21}. 

The perturbed potential explored here assumes a specific mass for the LMC and, of course, for the MW, but as discussed above, neither values are set in stone. That said, it is clear that if the LMC is indeed massive, the impact on the orbital properties of objects within and around the halo of the MW, and the conclusions one draw from them, can be significant. This is of course true also for individual stars in the MW (outer) stellar halo, a vast number of which will soon have 6D phase-space information thanks to large spectroscopic surveys (e.g. WHT/WEAVE, VISTA/4MOST, PSF etc.).

\subsubsection{Too-big-to-fail problem/Central DM halo densities} 

The determination of orbital parameters of MW satellite galaxies has also been used in the literature to examine aspects of the Too-big-to-fail (TBTF) problem and make considerations on the inner DM halo densities inferred. Recently, in their analysis of PHAT-II simulations,  \citet{Robles_21} found that at a given present-day maximum circular velocity, sub-haloes with small pericenters are more concentrated and have experienced a higher mass loss than those with a larger pericenter. Using GDR2 pericentric distances for the MW classical dSphs, they show that the allowed ranges for the maximum circular and peak velocities are both tightened than without the GDR2 information, with both quantities becoming smaller and going in the direction of exacerbating the TBTF problem. 

In comparison with the GDR2 pericentric distances in F18, used by \citet{Robles_21}, those we derive here for the similar isolated potentials are tighter and towards the upper range of what the GDR2 data were suggesting for Draco and Ursa Minor (Draco: "Light MW" eGDR3 $48-56$ kpc versus 31-58kpc in GDR2 and "Heavy MW" eGDR3  $34-42$ kpc versus 21-40 kpc in GDR2), which would go in the direction of slightly alleviating the issue. Interestingly, in the perturbed potential, Draco would have a 1-$\sigma$ range of pericentric distances of 81-122kpc, while Ursa~Minor would have 65-79kpc, which would push upwards the estimates of both the maximum and peak circular velocity.

Fornax is another interesting object, as dynamical modelling of the kinematic properties of its stellar component \citep[][]{WP_11, Amorisco_13, Pascale_18} as well as considerations and modelling of its system of globular clusters \citep[e.g.][]{Leung_20} suggest its DM halo to have a density core; this has sometimes been attributed to DM being heated up by stellar feedback \citep[e.g.][]{Read_19} and more recently to the possibility of significant mass loss due to tides \citep[][]{Genina_20}. The range of pericentric distances determined here (see Tab.~\ref{tab:table_param}) is rather similar to that of the GDR2 determinations in F18 for the isolated potentials, probably because in the case of Fornax the uncertainty in the distance measurement plays also a role \citep[][]{Borukhovetskaya_21}; also the values themselves are very similar. The 1-$\sigma$ range for pericenter in the perturbed potential  (66-124kpc) is in agreement with those given by the isolated potentials. These include orbital trajectories that can significantly reduce the peak circular velocity of the DM halo due to tidal mass loss and reconcile it with the kinematic properties measured at the half-light radius \citet{Genina_20, Borukhovetskaya_21}.

\subsubsection{Relation to the Milky Way}

\begin{figure*}
\centering
  \includegraphics[width=\textwidth]{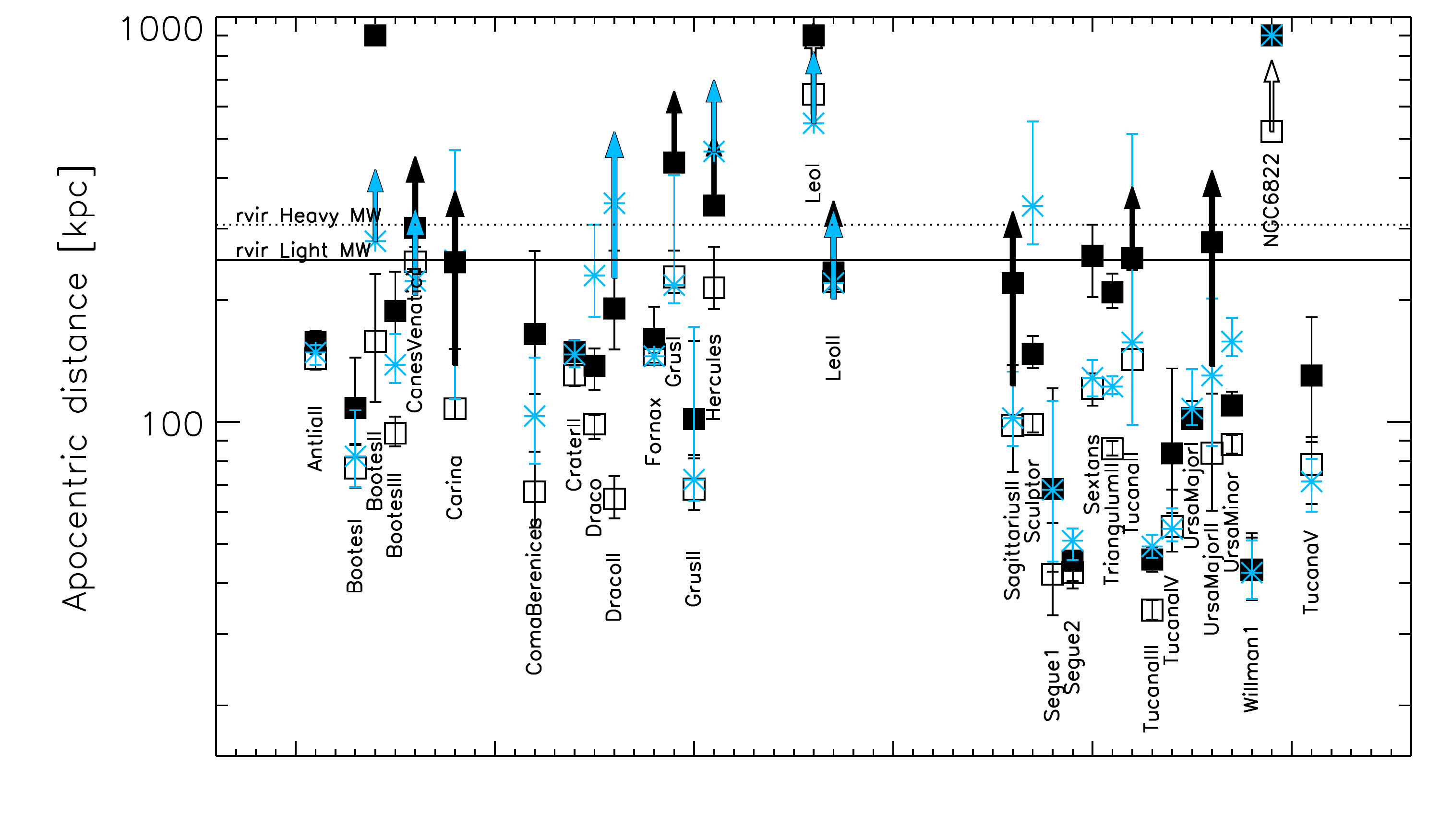}
  \includegraphics[width=\textwidth]{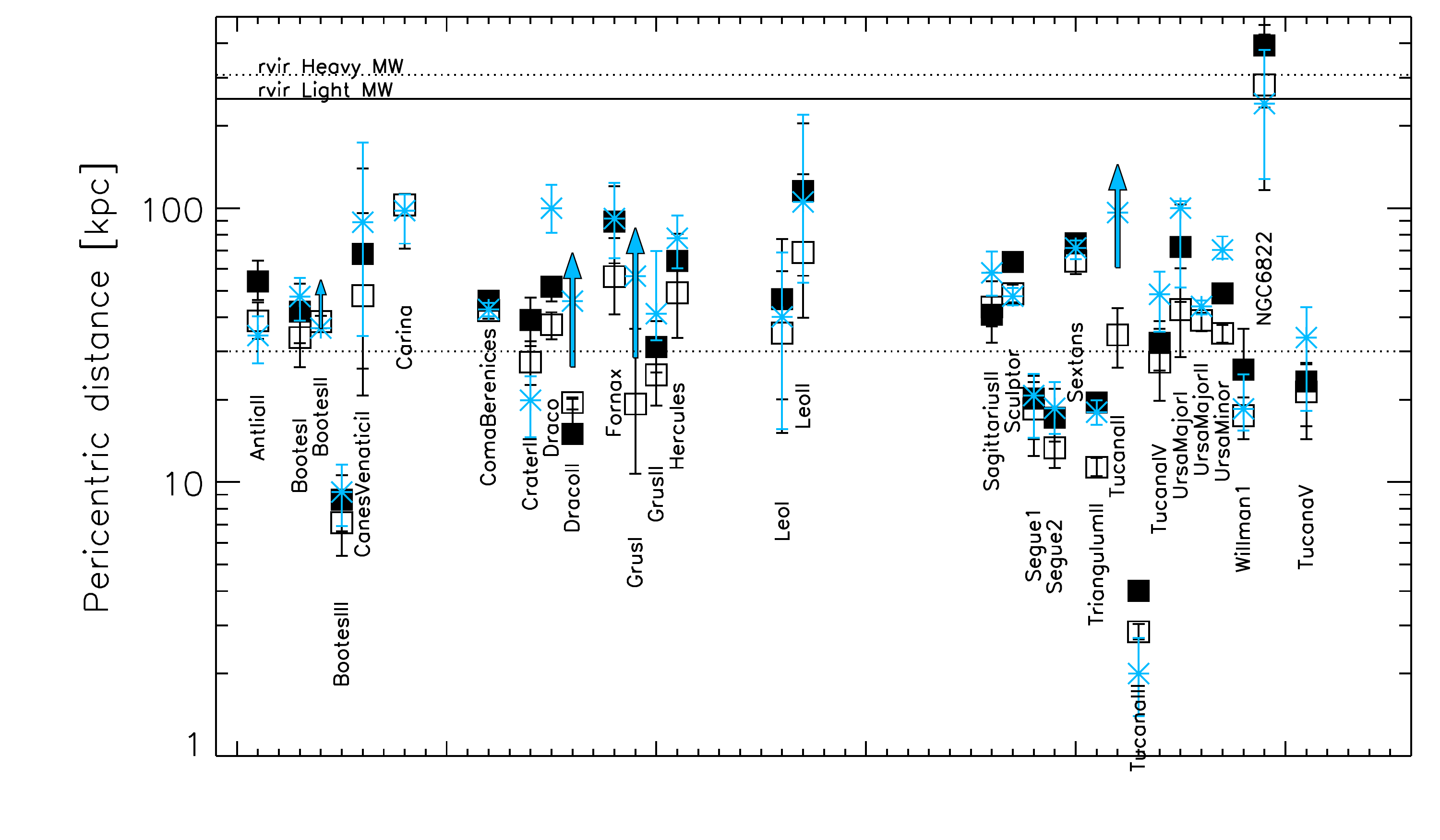}
   \caption{Apocentric (top) and pericentric (bottom) distance for the sample of likely MW satellite galaxies (excluding the high likely long-term LMC satellites of the LMC in Tab.~\ref{table_LMC_bound}) with error in 3D velocity less than 70 km s$^{-1}$. The filled and open squares show the results for the "light" and "heavy" isolated MW potentials and the light blue asterisks for the perturbed potentials. The arrows indicate those cases where either the median or the 84th percentiles were undefined. When not even the 16th percentile was defined, the symbols are placed at an apocentric distance = 900kpc. We only consider cases where the galaxy has already experienced a pericentric passage.} 
\label{fig:apoperi}
\end{figure*}

Figure~\ref{fig:apoperi} shows the apocentric (top) and pericentric (bottom) distances obtained in the three gravitational potentials for the systems with total Galactocentric velocity error $<$70 km $^{-1}$. Since here we wish to explore the relation to the MW, we exclude the systems that we find to be likely LMC satellites (which will be discussed in detail see below, in Sect.~\ref{sec:LMCsat}). 

Uncertainties in the MW gravitational potential do cause significant variations in the orbital parameters of some of the galaxies in the sample. Nonetheless, there are some considerations that we can make.

$\bullet$ {\it The system of Milky Way satellites} From the top panel, we can see that the 16th quantile of the distribution of apocenters never reaches within the MW virial radius for Leo~I and NGC~6822 within the time range of the orbit integration.
On the other hand, both galaxies do seem to have experienced one passage around the MW in the past; in particular when the perturbed and the "Heavy MW" potentials are considered for NGC~6822, this galaxy could have reached within $\sim$100kpc within the 68\% confidence interval, which supports the conclusions of \citet{teyssier12} based on a comparison of the Galactocentric distance and radial velocity of LG field galaxies with those of haloes in the Via Lactea~II simulations. 

If we perform no cut in the total Galactocentric velocity error, there are other galaxies which do not have an apocenter within the MW virial radius, i.e. Eridanus~II, Leo~T, Phoenix, Pisces~II.  However, the error in the total velocity is too large to draw meaningful conclusions without correcting for biases. Thus we cannot test yet the claim of \citet{teyssier12} that Leo~T and Phoenix are backsplash galaxies on the basis of the orbital trajectories\footnote{Note however that the radial velocities used by those authors for Phoenix is wrong, see \citet{Kacharov_17}}.  \citet{McConnachie_SOLO} use systemic PMs to understand which ones of the isolated galaxies studied in that work might have reached within 300kpc from the MW (or M31) but our uncertainties on Leo~T and Phoenix systemic motions are still too large to exclude this hypothesis in this way either. Nonetheless, the presence of young stars and HI gas in faint systems like Leo~T and Phoenix supports the hypothesis that they have never approached the MW before. Also, according to \citet{McConnachie_SOLO}, the possibility that Phoenix might have entered the MW virial radius is tiny and possible only if the MW mass is at the high end of the probable range. 

$\bullet$ {\it Fast-moving galaxies} Fast-moving galaxies are especially useful for placing   constraints on the MW mass  \citep{Boylan-Kolchin_13}. A clear example is that of Leo~I, whose large receding radial velocity and status as bound or unbound to the MW has caused several headaches for determinations of the MW mass since a long time \citep[e.g.][and references there in]{Wilkinson_99}. In our determination, Leo~I has a slightly smaller bulk PM value and error compared to the HST one  \citep{Sohn_13} used by \citet{Boylan-Kolchin_13}. This should cause a likely decrease in the MW mass but minor, since its Leo~I total velocity is dominated by the l.o.s. component. Also the impact of the LMC does not change the main conclusions on its orbital history. 

As for other fast-moving galaxies, most of those for which we can measure reliable PMs are likely to have come in with the LMC (see Sect.~\ref{sec:LMCsat}), as expected due its high velocity orbit, and are not of interest here. 

If we concentrate on objects not classified as likely LMC satellites and that are receding, to exclude those recently infalling: in the "perturbed" potential, Bootes~II would fall under this classification and has a large velocity compared to the escape velocity at its distance, as in \citet{Fritz_18}; CanesVenatici~I would do so in the "LightMW" but the inclusion of the LMC lowers the chances to have the apocenter beyond the MW virial radius for a 0.9$\times 10^{12}$ M$_{\odot}$ massive MW and makes the values get close to those of the "Heavy MW". The combined study of CanesVenatici~I, Draco~II and Hercules might turn out to be useful for considerations on the MW gravitational potential, because for the latter two objects, the likelihood of apocenters well outside of the MW virial radius has the opposite behaviour as for Canes Venatici~I, i.e. it increases with the inclusion of the effect of the LMC infall.

$\bullet$ {\it Tidal disturbances} Both Bootes~III and Tucana~III are known to be embedded in stellar streams \citep{Drlica-Wagner_15,Carlin_18}. 
Our analysis fully confirms the expectation that these features are the results of tidal disruption, since these two systems have pericenters in all the three potentials that bring them very close to the central regions of the MW, likely within 10 kpc or less; this is fully in line with the GDR2-based results \citep[see e.g.][]{Fritz_18, Simon_18, Carlin_18} and the eGDR3-ones by \citet{Li_EDR3} for Tucana~III. 

The stellar component of both Antlia~II and Crater~II has peculiar properties, with an extremely low surface brightness, large half-light radius and low l.o.s. velocity dispersion when compared to other MW satellites of similar stellar mass \citep{Torrealba_16a, Caldwell_17, Torrealba_19}. It has been postulated that also these galaxies have been "sculpted" by tidal disturbances by the MW \citep{Fattahi_17,Sanders_18,Torrealba_19}. According to the orbital parameters derived in this work, this hypothesis appears fairly robust for Crater~II, confirming the GDR2-based results, while for Antlia~II it appears more sound when considering the "perturbed" and "heavy MW" potentials, than in the "light MW"  one. The pericentric distances obtained for these two models are compatible with that explored by \citet{Torrealba_19} to study whether tidal effects onto a cored DM halo could explain the low surface brightness, large half-light radius and low l.o.s. velocity dispersion of Antlia~II. 

The spatial distribution of the high probability member stars returned by our method shows an elongation in the outer parts for Carina's stellar component (Fig.~\ref{fig:out_1}), compatible with what seen in previous studies, based on red giant branch stars observed spectroscopically \citep{Munoz_06} and deep wide-area photometry \citep{Battaglia_12, McMonigal_14}. Even if there are some intervening LMC stars in the Carina's line-of-sight, it is unlikely the feature is due to that, given that these would be included in our contamination model. Given the orbital parameters that we obtain, it appears very unlikely that this might be the result of a close interaction with the MW, nor with the LMC (see e.g. Fig.~\ref{fig:orbits_lmc_1}). The second last  pericenter in the perturbed potential, about 7 Gyr ago, might have brought Carina as close as 37kpc at a 1$\sigma$ level; but even if that would have been sufficient to strip its stellar component, it is highly unlikely that the elongation we see today is due to that, as tidal debris are not expected to be seen anymore after 15-20 crossing times \citep[e.g.][]{Penarrubia_09}, i.e., between 1 and 2 Gyr in this case. 

Within the 68\% confidence limit, Segue~1, Segue~2, Triangulum~II enter what can be potentially be seen as a dangerous zone, i.e. within 10-30~kpc from the MW centre in all the 3 potentials. We note that the clearly tidally disrupted Sagittarius (not included in this analysis) has had the most recent pericenter at about 16kpc and the second last at about 25~kpc, \citep{Vasiliev_21}, showing that tidal stripping can be efficient also at those Galactocentric distances.  

Depending on the potential and/or on the error-bars some galaxies might or might not have suffered strong tidal effects (e.g. CVen~I, Hercules, Willman~1, Tucana~V and several more).

\begin{figure}
\centering
  \includegraphics[width=\columnwidth]{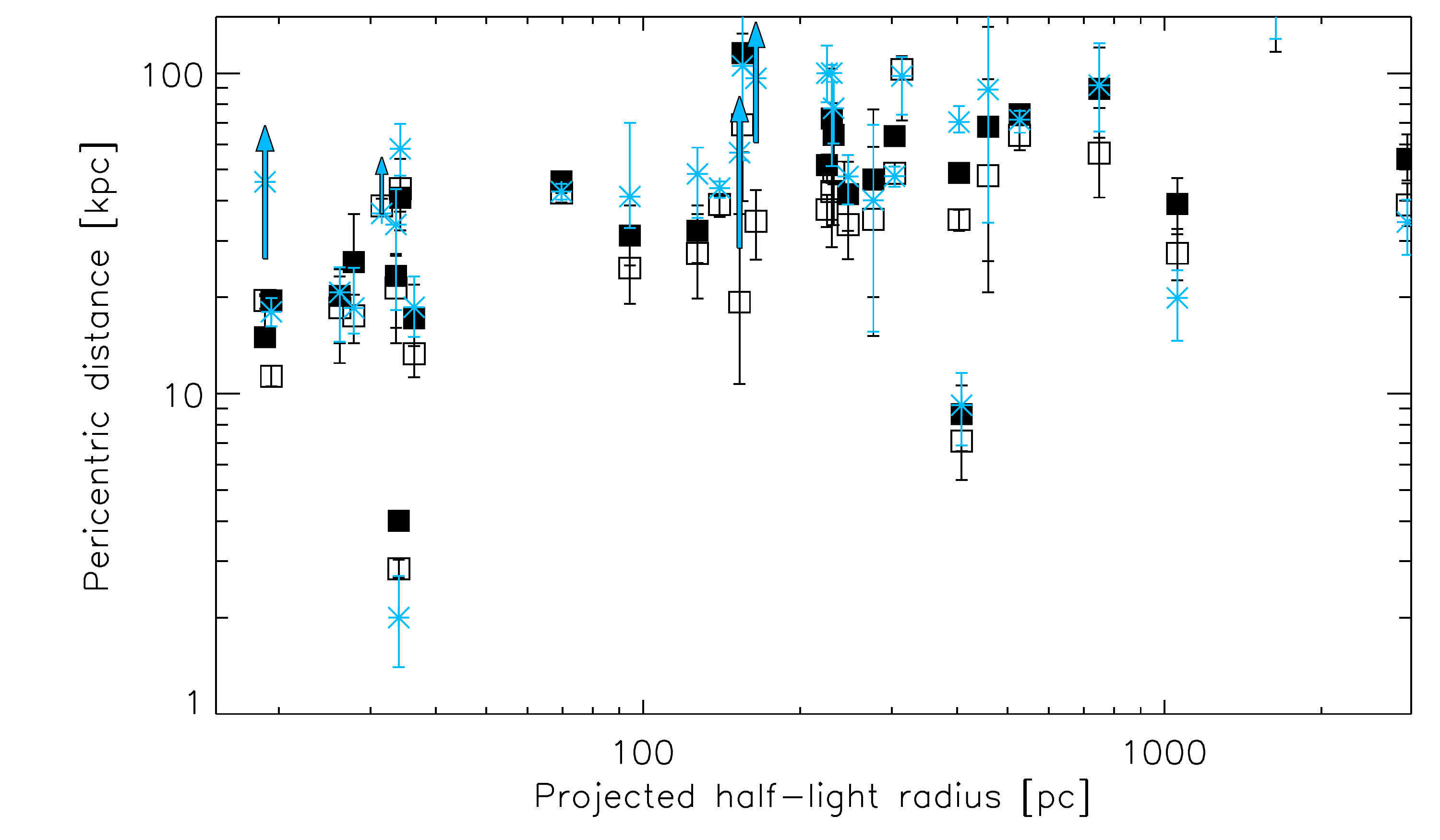}
   \caption{Pericentric distance compared to the projected semi-major axis half light radius for the sample of likely MW  satellite galaxies (excluding the highly likely long-term LMC satellites of the LMC in Tab.~\ref{table_LMC_bound}) with error in 3D velocity less than 70 km s$^{-1}$. The filled and open squares show the results for the "light" and "heavy" isolated MW potentials and the asterisks for perturbed potential. The arrows indicate those cases where either the median or the 84th percentiles were undefined. We only consider cases where the galaxy has already experienced a pericentric passage.} 
\label{fig:perisize}
\end{figure}

Especially in the perturbed potential, one can see hints that, excluding the galaxies with clearer tidal effects (streams or a very diffuse stellar component), the smallest galaxies in terms of half-light radius in general have the smallest pericenters, see Fig.~\ref{fig:perisize}. Some possible hypotheses are  that these systems had a fairly compact stellar structure at birth and have survived tides better than less compact systems,  or that this is a different manifestation of the type of features that tidal stripping might imprint on the galaxies it acts upon. A deeper investigation is deferred to the future.

We have investigated whether the ellipticity of the stellar component could be taken as a sign of strong tidal disturbances, but we find no clear trend between ellipticity and pericentric distances, neither in the sense of small pericenters having preferentially large ellipticities nor being preferentially round. 

$\bullet$ {\it Connection to star formation history (SFH)} The dwarf galaxies that inhabit the LG have long been known to exhibit a variety of SFHs \citep[e.g.][]{Mateo_98, Grebel_98, Skillman_05}, where this holds also when focusing only on those surrounding the MW. It is natural to ask whether a connection exists between e.g. the timing of strong enhancements of the star formation activity, or on the contrary, its quenching, and important times in the orbital history of these galaxies, e.g. infall into the MW halo or pericentric passages. 

The star formation histories of the vast majority of the dwarf galaxies that surround the MW halted 8-10 Gyr ago \citep[e.g.][]{Tolstoy_09, Brown_14, Gallart_15}. At that time, in a hierarchical formation framework, the DM halo of the MW was still growing rapidly; e.g. according to the formula in \citet{Wechsler_02}, around 8 Gyr ago, it would have assembled about half of its mass, while about 3 Gyr ago, already 80\% of it would have been in place (this excludes the infall of LMC-like systems, which are rare in the $\Lambda$CDM cosmogony). As it can be seen e.g. in \citet{Armstrong_21}, the trajectories  of MW satellites in a time evolving MW potential start deviating from those in a static potential around 3-4 Gyr ago, with the differences becoming more and more noticeable as a function of lookback time, as expected (see Fig.~4 in their article). Even though, the effect appears to be minor with respect to the mass and mass distribution of the MW, it is an additional source of uncertainty.  The time-variation of the potential is even stronger if one takes into account that around 8-10 Gyr ago, the MW accreated Gaia-Enceladus \citep[e.g.][]{Belokurov_18, Helmi_18_GE} and potentially there have been other subsequent events  \citep[e.g.][Sequoia and Kraken, respectively]{Myeong_19, Kruijssen_19}.

Therefore, since the gravitational potentials considered here do not include the growth of the MW DM halo as a function of time, we concentrate on those systems that have experienced star formation activity in the past few Gyrs and might have been linked to the MW: Leo~I, Fornax, Carina, NGC~6822. We refer the reader to works such as those by \citet{Fillingham_19} and \citet{Miyoshi_20} for an analysis of the connection between SFH and GDR2-based orbital trajectories of MW satellites via a comparison with simulations in the former and an analytical treatment of the growth of the MW halo in the latter. Even though both Phoenix and Leo~T host young stars, we exclude them since they are currently found beyond the MW virial radius and so far there is no evidence that they might have gone through a pericentric passage. As for Eridanus~II, recent SFH determinations seem to exclude the presence of young/intermediate-age stars \citep{Simon_21, Gallart_21}.   

The most detailed SFHs for Leo~I, Fornax and Carina were derived in \citet{Ruiz-Lara_21, Rusakov_21, deBoer_14}, respectively. The star formation activity of Leo~I saw the last episode of enhancement about $\sim$1 Gyr ago, after which it started decreasing till coming recently to a halt. The authors note that the timing is similar to that of the pericentric passage from HST PM measurements \citep{Sohn_13} and GDR2 ones \citep{Fritz_18,Helmi_18}, possibly indicating that the decrease and then stop of SFH was due to ram-pressure stripping of the gaseous component. Our eGDR3 based analysis confirms the timing of the last (and only) pericentric passage  for the 3 potentials explored (see Tab.~\ref{tab:table_param} and Fig.~\ref{fig:orbits2}). Compared to Fornax and Carina, Leo~I is highly likely to have come much closer to the MW center.

Within the time range considered here, \citet{Rusakov_21} detected intermittent episodes of enhanced star formation activity at $\sim$0.5, 1, 2 ago in Fornax. In our analysis, the timing of the last pericenter is at a lookback time of $-2.8_{-1.2}^{+0.8}$ Gyr (the negative sign means 2.8 Gyr ago) with an orbital period $4.4_{-0.8}^{+2.0}$ Gyr in the "Light MW" and $-1.5_{-0.4}^{+0.2}$ and an orbital period $2.5_{-0.3}^{+0.4}$ Gyr in the "Heavy MW". While it cannot be excluded that the burst occurring about 2 Gyr ago was due to a pericentric passage in either of the two isolated potentials, the orbital period would exclude that the more recent ones are due to the same cause. We examined the orbits determined for the perturbed potential and in this case, the timing of the last pericenter is similar to that of the "Heavy MW" case, with a period in the past exceeding 5 Gyr, leading to the same conclusions.

As for Carina, in the "Light MW" it has not passed pericenter yet, while in the "Heavy MW" the last pericenter occurred at a lookback time of $-1.0_{-0.3}^{+1.0}$ Gyr, with an orbital period of $2.6_{-0.5}^{+1.3}$ Gyr. The timing of the last pericenter passage in the perturbed potential is similar to that of the "Heavy MW", with the previous one occurring more than 7 Gyr ago. \citet{deBoer_14} find a long period of enhanced star formation activity, $4-6$ Gyrs ago, with a strong decrease about 1 Gyr ago, followed by an increase of SFR at the youngest ages probed, $0.25-0.5$ Gyr, till shut down. Also in this case, the decrease in star formation about 1 Gyr ago is potentially compatible with gas stripping.

Overall, a correlation can be found between some of the main features in the recent SFH of Leo~I, Fornax and Carina and their orbital histories (although correlation does not necessarily implies causality!); however, the same explanation seems not to be valid for the intermittent bursts of SFH seen in the past 1-2 Gyr in these galaxies, though those events are less strong than the older star forming events.

\citet{Fusco_14} determined the SFH of NGC~6822 in 6 fields at different distances from the centre, out to 4 kpc. Their shapes differ significantly from each other and there is no unambiguous signature of a drop at a common time. On the other hand, the timing of the pericentric passage has too large uncertainties to indicate anything conclusive. In any case, given the fairly large stellar mass of NGC~6822 in comparison to the other dwarfs, and the fact that it is likely to have kept in the outer regions of the MW halo, probably the expected effect of the pericentric passage on the gas content and distribution of this galaxy should be minor and more in the direction of an outside-in ram-pressure stripping and smooth reduction of the size of the region where the bulk of the star formation has occurred, rather than sharp features in the SFH.  

\subsubsection{LMC satellites} \label{sec:LMCsat}

\begin{figure*}
\centering
  \includegraphics[width=17.5cm]{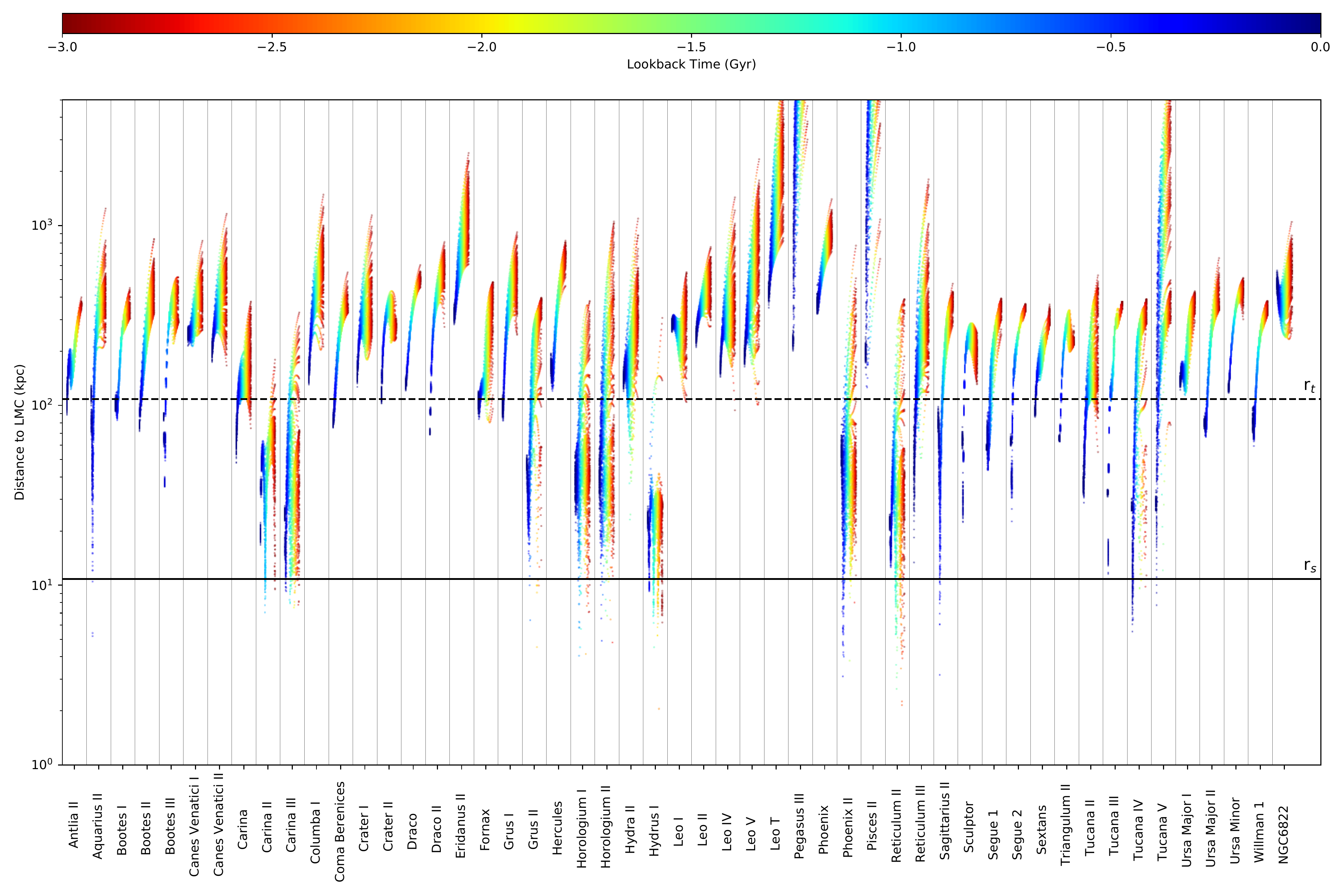}
   \caption{Distance from the LMC -as a function of lookback time for 100 Monte-Carlo realizations of the galaxies's orbit. The horizontal solid and dashed lines, labelled r$_s$ and r$_t$, correspond to the scale and truncation radius of the initial LMC potential, of 10.84 kpc and 108.4 kpc, respectively. } 
\label{dist_LMC}
\end{figure*}

Here we aim to explore which galaxies were likely part of the cortege of satellites that arrived with the LMC. For this, we used the 100 Monte-Carlo realisations of the orbits obtained in the {\it perturbed} MW potential to measure the time evolution of their relative distances to the LMC. As we can see from Fig.~\ref{dist_LMC}, the scatter increases with time. This is the consequence of the (un)accuracy of the different dynamical parameters measured for the galaxies (especially on the PMs, but also the distance), leading to a broad range of possible orbits for the galaxies with the least accurate measurements.

Although, the majority of the dwarf galaxies stay far from the LMC ($>$ 60 kpc) at every time step,  for 23 of them \footnote{Aquarius~II, B\"ootes~III, Canes~Venatici~II, Carina, Carina~II, Carina~III, Fornax, Grus~II, Horologium~I, Horologium~II, Hydra~II, Hydrus~I, Phoenix~II, Reticulum~II, Reticulum~III, Sagittarius~II, Sculptor, Segue~1, Segue~2, Tucana~II, Tucana~III, Tucana~IV, Tucana~V}, at least one of the 100 orbits went close enough ($<$ 60 kpc, corresponding to $\sim 6 r_s$) to potentially suggest a physical association. 

\begin{figure*}
\centering
  \includegraphics[width=17cm,clip,viewport=0 0 1025 350]{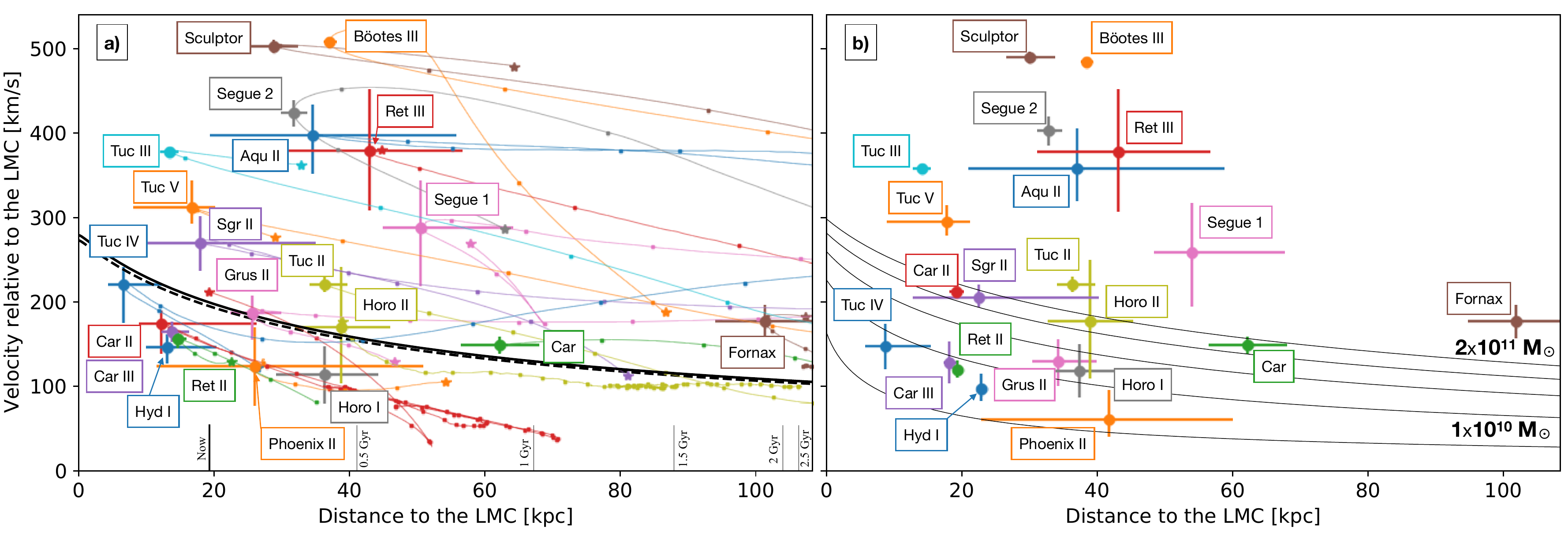}
   \caption{{\it Left panel:} Relative distance and velocity for the galaxies for which at least one of the 100 orbit realisations in the {\it perturbed} potential pass close to the LMC ($<$60 kpc). The black line represents the escape velocity of the initial a $1.5 \times 10^{11}$ M$_\odot$ LMC represented by a NFW profile with a scale radius $r_s=10.84$ kpc, while the dashed line shows the escape velocity of the current LMC. Each colored line shows the evolution of the relative distance and velocity of the galaxies with each point marking 100 Myr of evolution. For the galaxies that are not currently at their closest approach, the triangles show their current position on this diagram. The vertical lines at the bottom of the panel show the location of the Jacobi radius at different epochs. {\it Right panel:} Same as the left panel but assuming that the LMC follows the orbit of a point mass, and does not modify the MW potential. The black lines represent the escape velocity for a LMC with a mass of respectively $1, 5, 10, 15$ and $20 \times 10^{10}$ M$_\odot$ and a scale radius that respect the observational constraint following the requirement of \citet{Vasiliev_21}.}
\label{VescLMC}
\end{figure*}

\begin{figure*}
   \centering
   
   \includegraphics[width=0.23\textwidth]{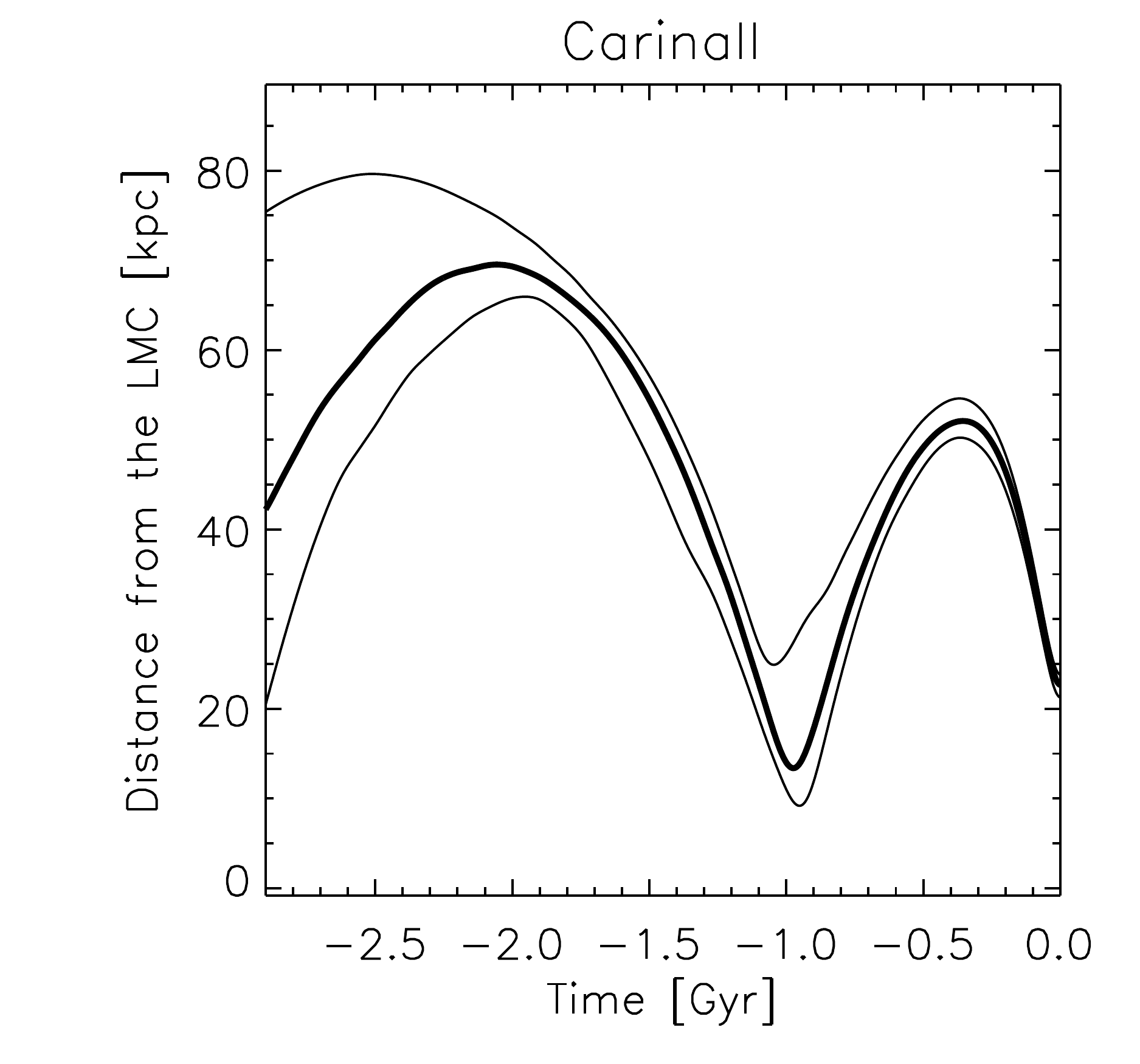}
  \includegraphics[width=0.23\textwidth]{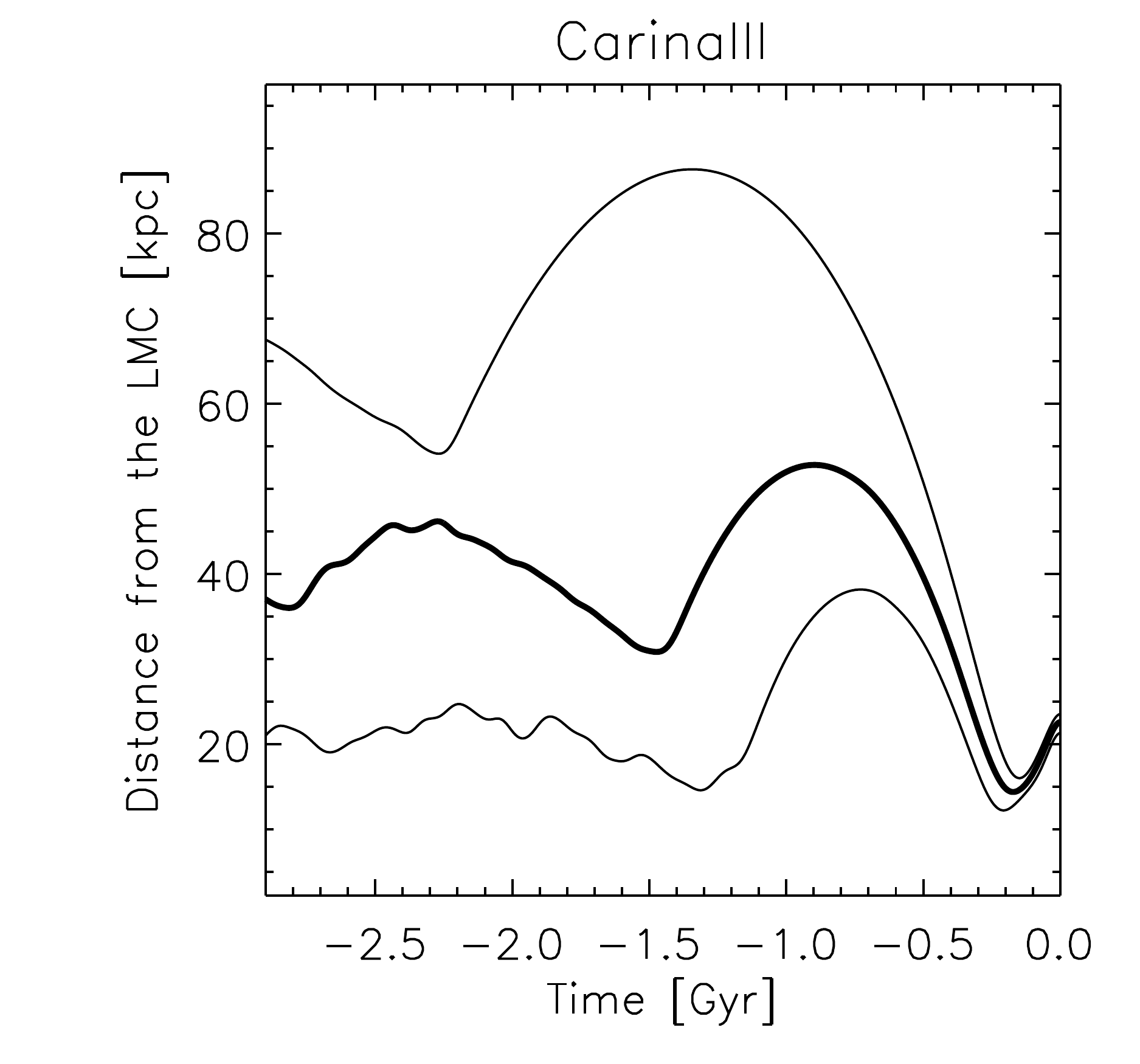}
  \includegraphics[width=0.23\textwidth]{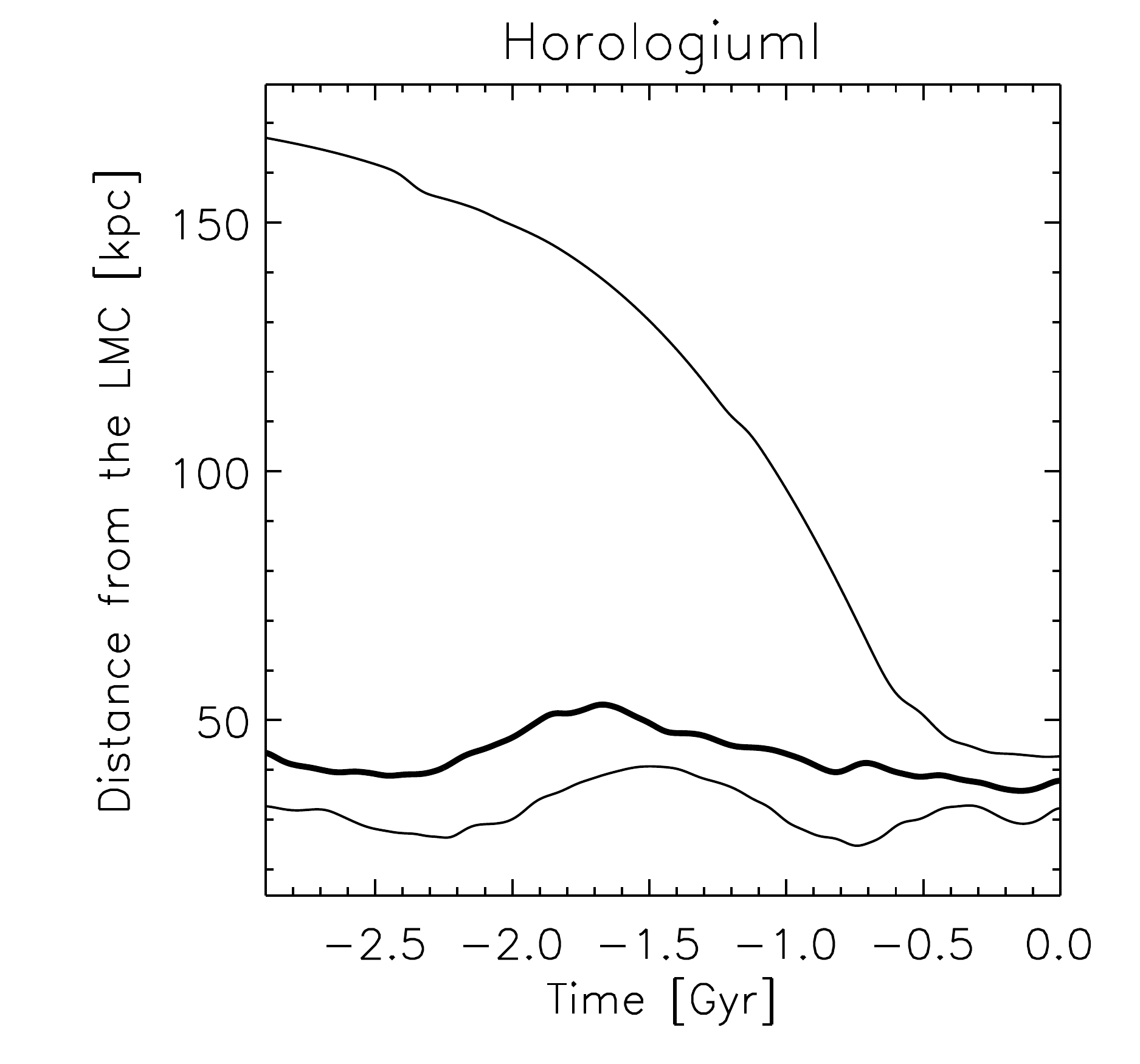} \\
 \includegraphics[width=0.23\textwidth]{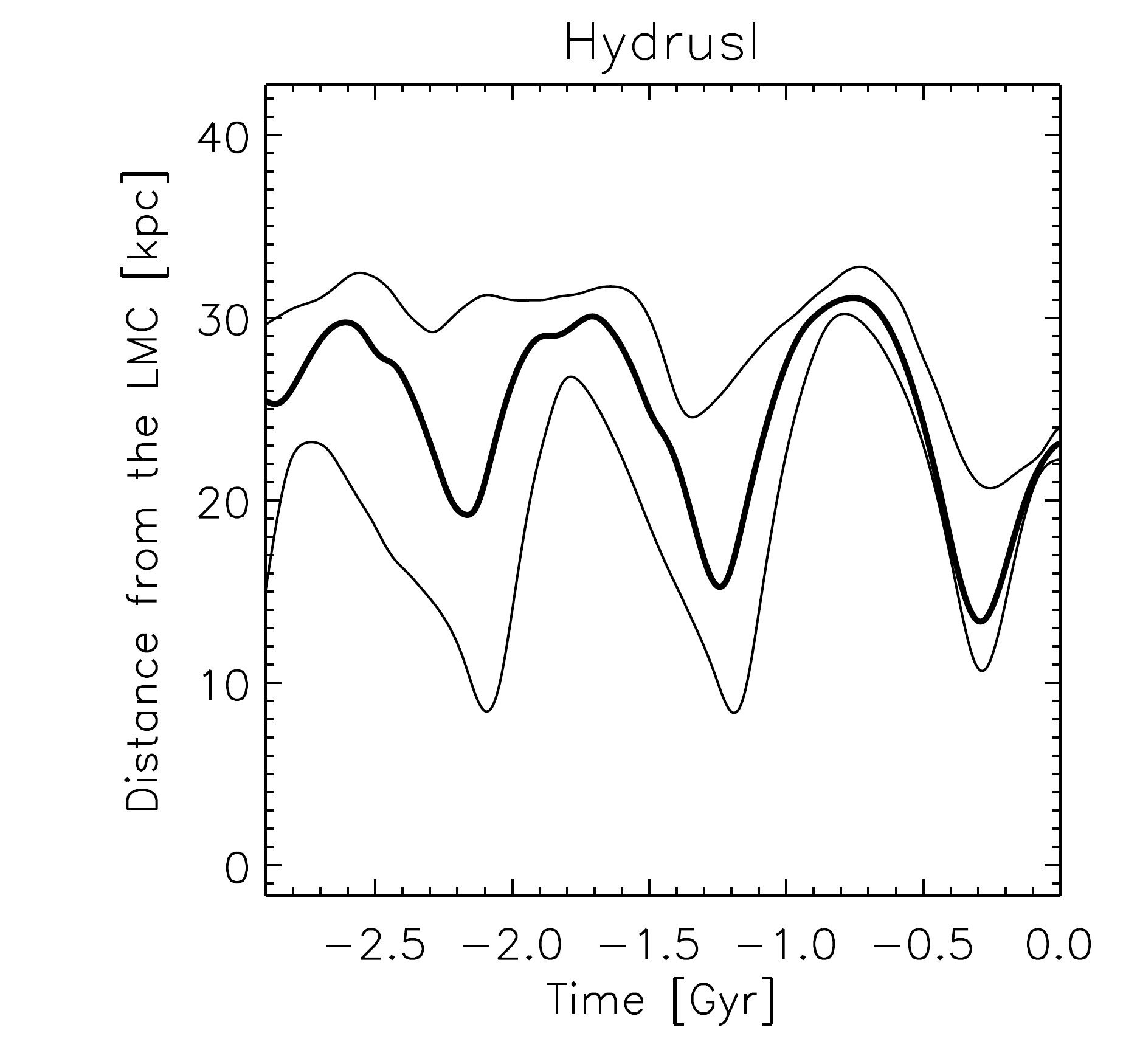}
 \includegraphics[width=0.23\textwidth]{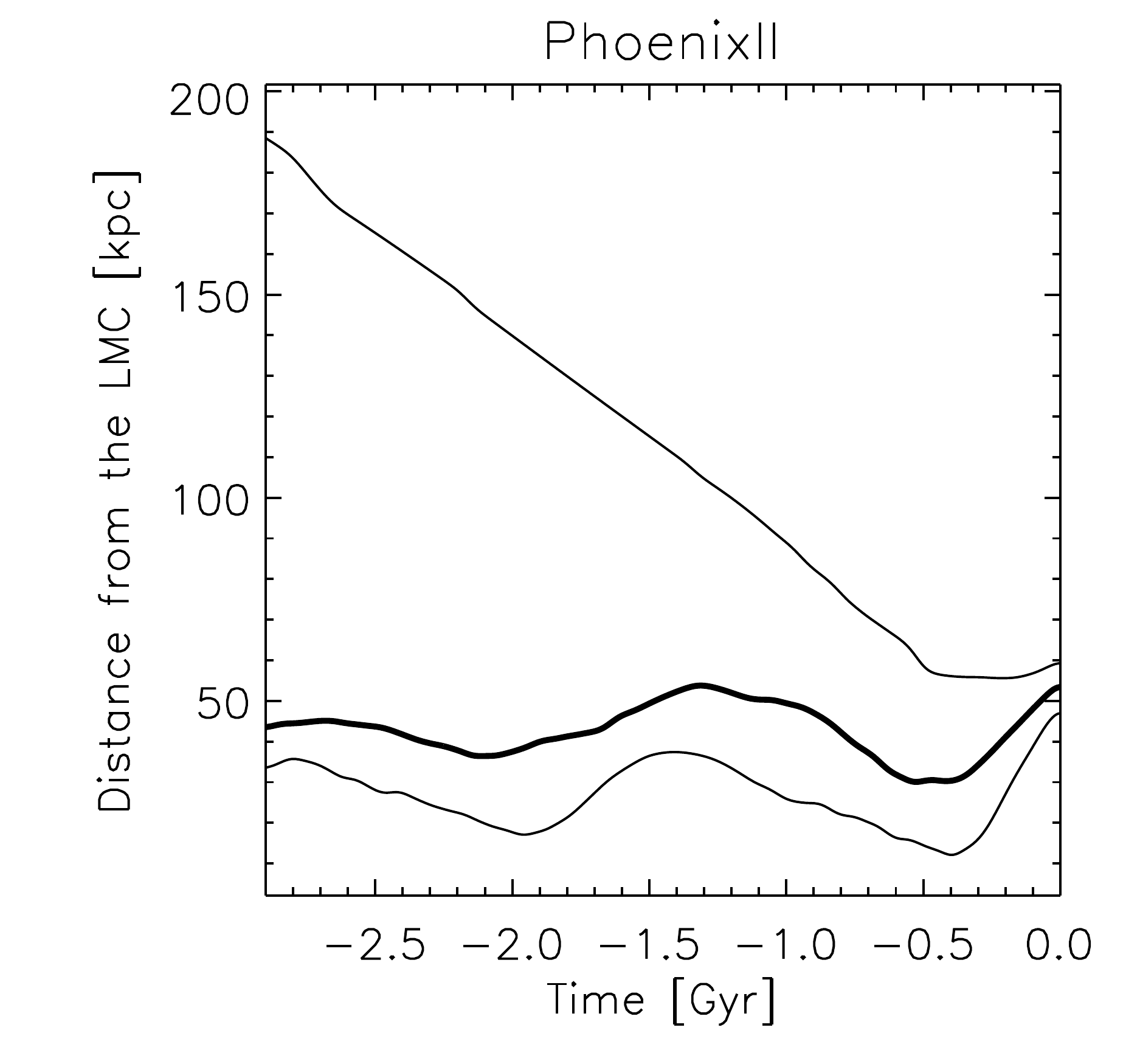}
 \includegraphics[width=0.23\textwidth]{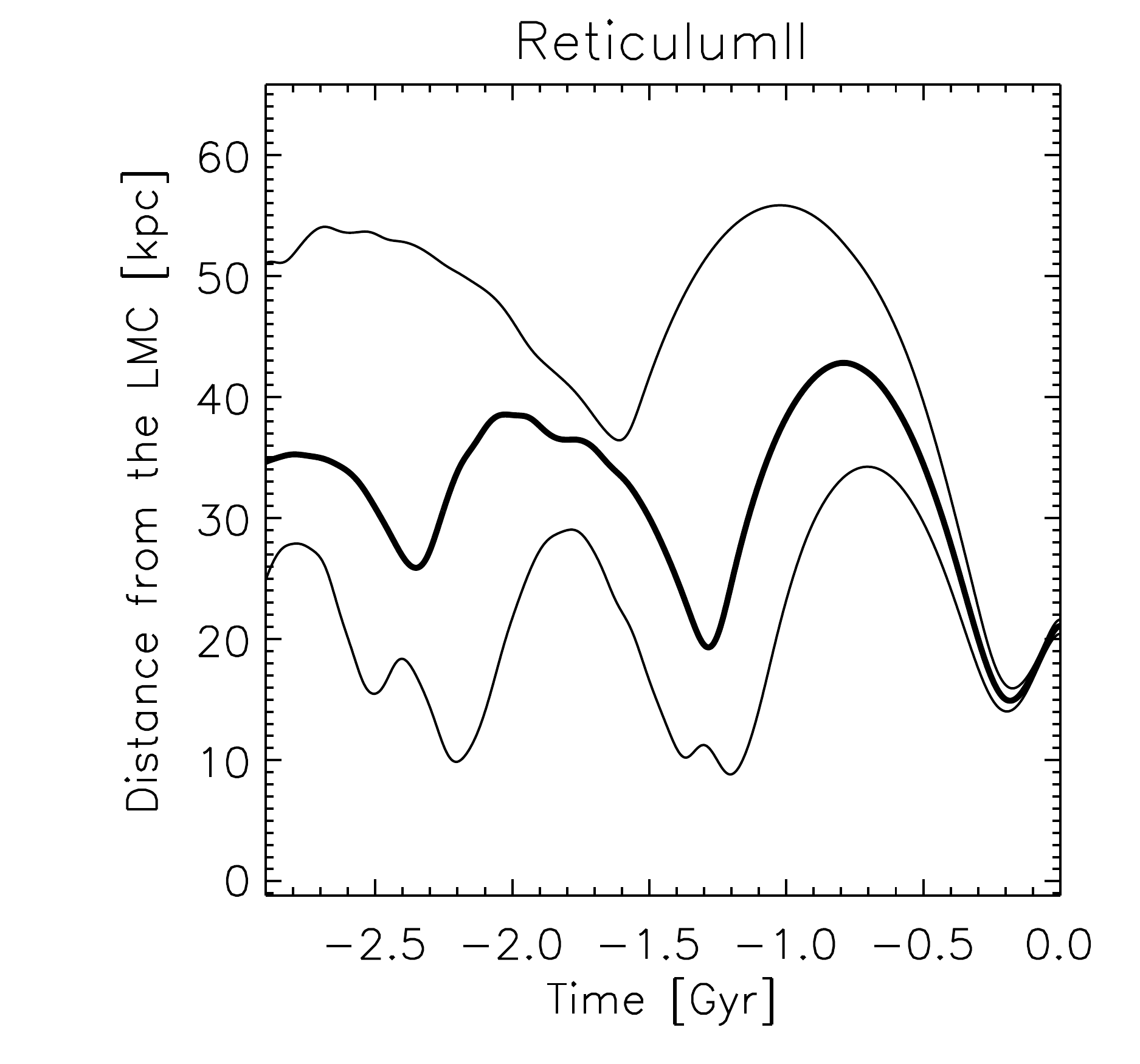}\\
 \includegraphics[width=0.23\textwidth]{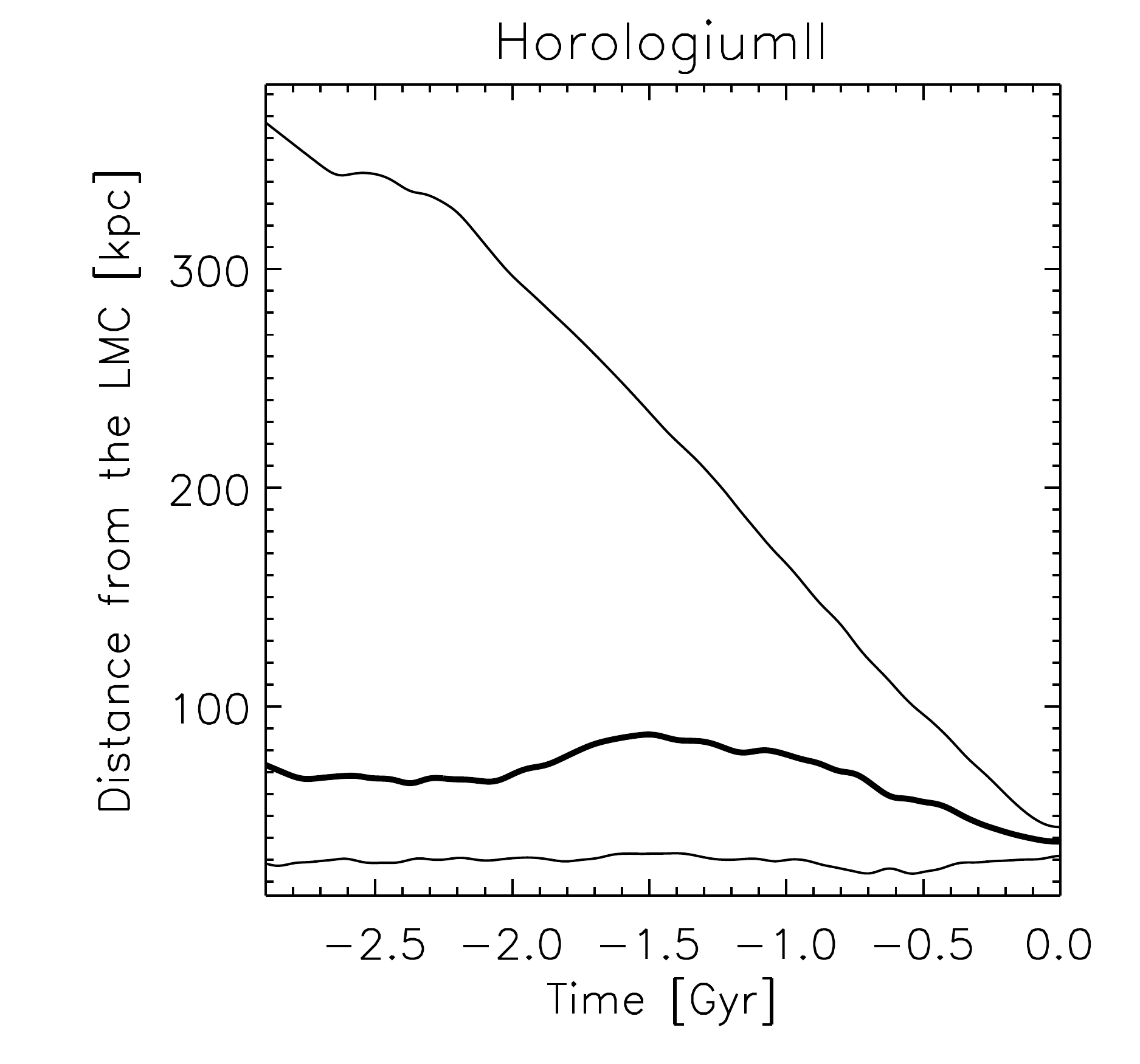} 
 \includegraphics[width=0.23\textwidth]{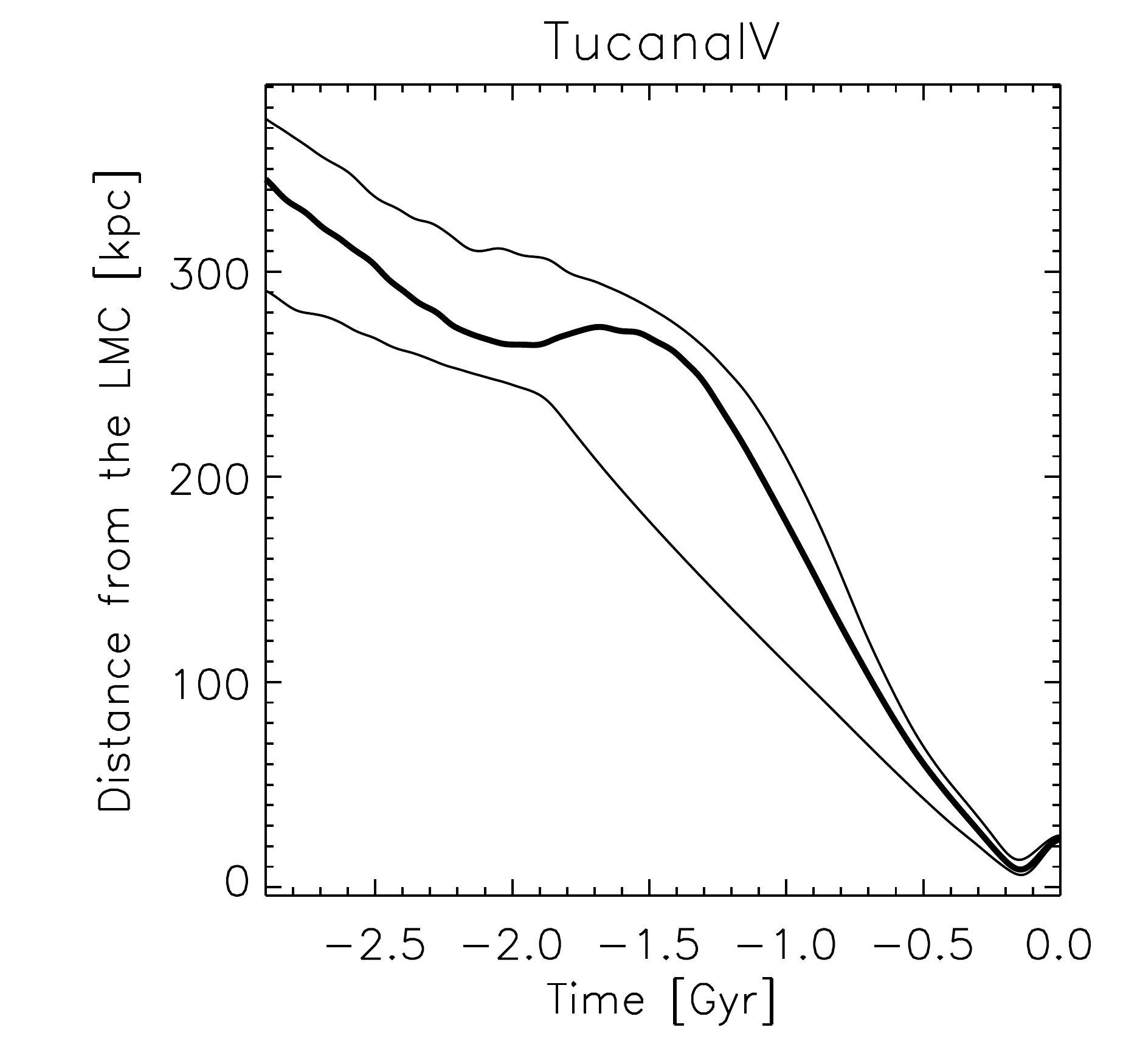}     
 \includegraphics[width=0.23\textwidth]{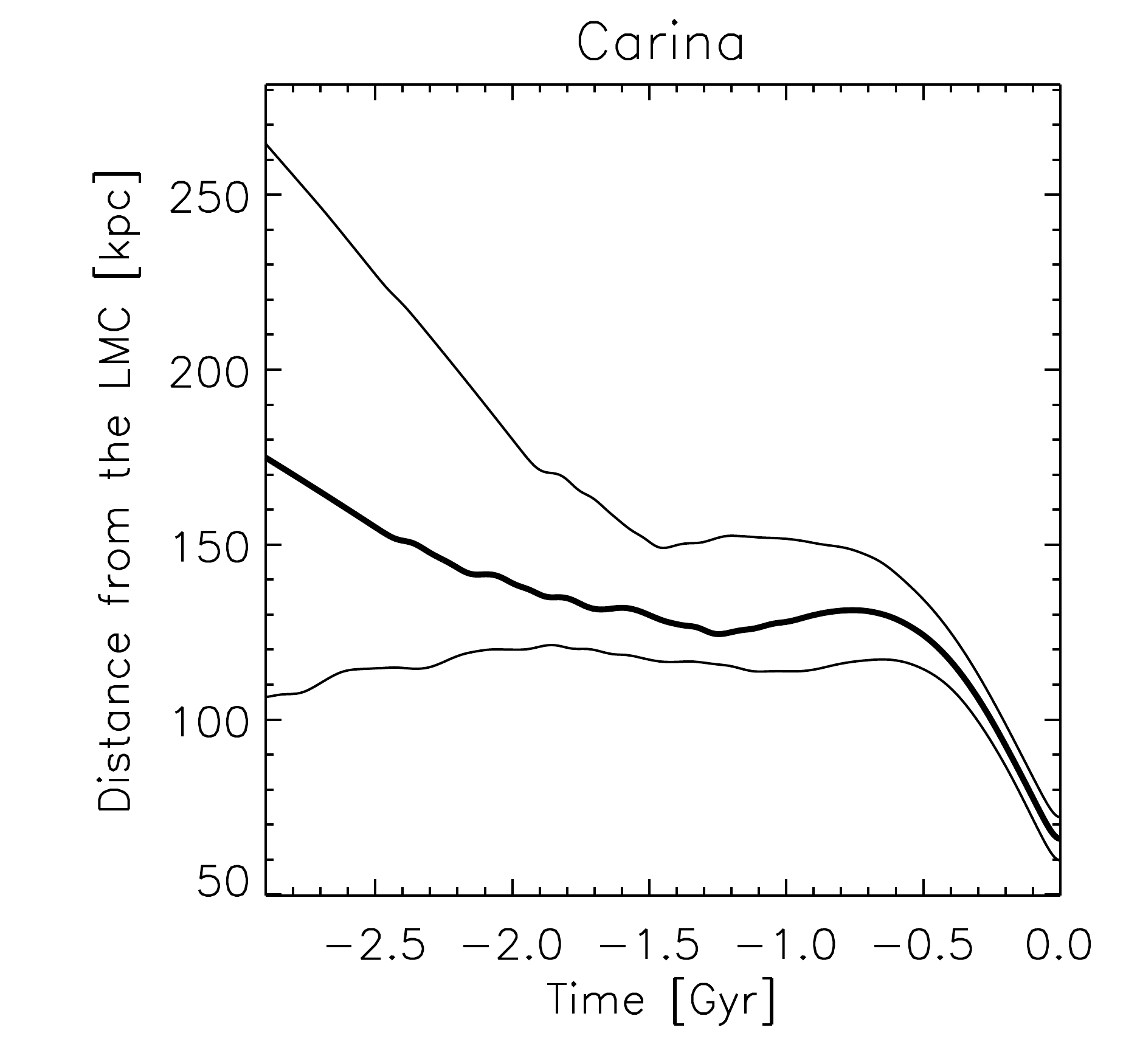}\\
  \includegraphics[width=0.23\textwidth]{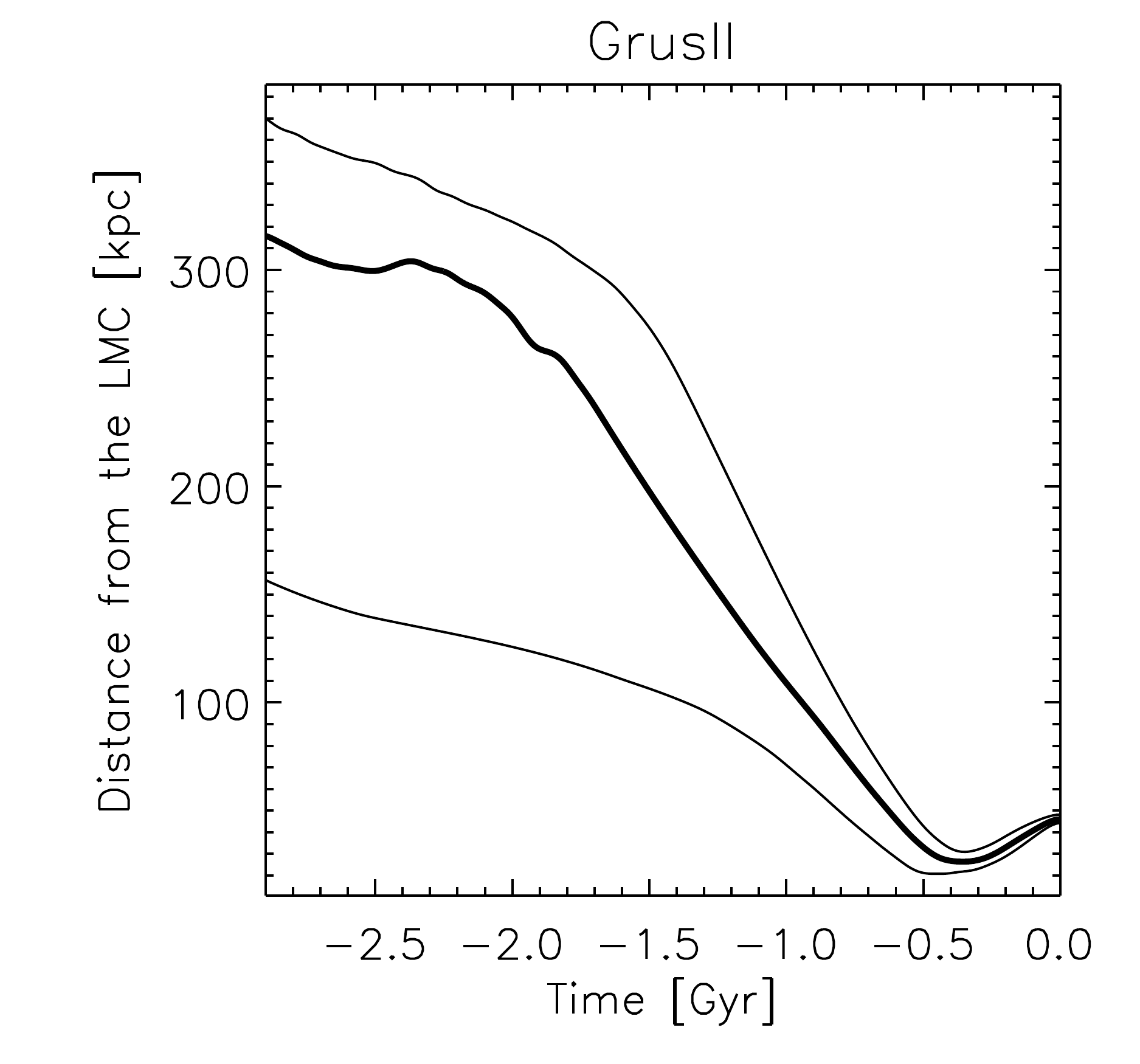}
      \caption{Distance from the LMC as a function of time for the last 3 Gyr. The thicker line gives the median of the 100 random realizations and the thinner lines the 16th and 84th percentiles. The first and second row from the top depict the likely long term satellites of the LMC; the third row the possible long term satellites of the LMC and the last row the recently captured satellites.}
         \label{fig:orbits_lmc_1}
   \end{figure*}

For those 23 galaxies, we increased the number of Monte-Carlo realisations of the orbits to 1000, so we can measure the fraction of orbits that have been linked to the LMC in the past.
The relative position and velocity with respect to the LMC at the moment of their closest approach is listed in Tab.~\ref{table_LMC_bound} and is shown in Fig.~\ref{VescLMC}. In this figure, we can see that the majority of the galaxies that pass close to the LMC are actually not linked to it, since their relative velocity is too high in comparison to the escape velocity of the LMC at any moment. On the other hand, 6 galaxies are clearly related to the LMC, with relative velocity at the moment of their closest approach significantly lower than the escape velocity: Carina~II, Carina~III, Hydrus~I, Reticulum~II, Phoenix~II and Horologium~I\footnote{The uncertainty on the total Galactocentric velocity of these 6 galaxies is $<$70 km s$^{-1}$.}.  A closer-look to the time evolution of the distance of these galaxies from the LMC is given in Fig.~\ref{fig:orbits_lmc_1} and will be provided as a movie\footnote{On http://research.iac.es/proyecto/GaiaDR3LocalGroup/pages/en/audiovisual-material.php}. 

Interestingly, none of those galaxies are currently bound to the Magellanic system. Indeed, assuming that the gravitational potential of the LMC inside the tidal radius is unchanged by the tidal stripping of the external DM halo at any time
\citep[but see][]{Errani_21} and that the mass of the MW is constant, one can measure the Jacobi radius (r$_J$) as a function of time $t$, such as:
\begin{equation}
\centering
    \mathrm{r}_J(t)= \mathrm{D_{LMC}}(t) \left( \frac{\mathrm{M_{LMC}}(< \mathrm{r}_{J})}{3 \mathrm{M_{MW}}(<\mathrm{D_{LMC}})} \right) \, ,
\end{equation}
where $\mathrm{D_{LMC}}$ is the Galactocentric distance of the LMC, $\mathrm{M_{LMC}}$ is the mass of the LMC inside the Jacobi radius and $\mathrm{M_{MW}}$ is the mass of the MW inside $\mathrm{D_{LMC}}$. With this formula, the Jacobi radius a t= 0, 0.5, 1.0, 1.5, 2.0, 2.5 Gyr ago is of r$_J$=19, 41, 67, 88, 104 and 106 kpc respectively. One can see that at present-day these 6 galaxies are found outside of the t=0 Jacobi radius. However, Carina~II, Carina~III, Hydrus~I, Reticulum~II and Phoenix~II were still bound to the LMC at the time of closest approach, as can also be gauged by the values listed in Tab.~\ref{table_LMC_bound}. This allows us to conclude that they are very likely related to the LMC.

For Horologium~I, the time of its closest approach to the LMC is more recent than the time when it escapes its gravitational attraction. However, over the last 5 Gyr, the distance of Horologium~I relative to the LMC oscillates between 35 and 55 kpc with a velocity systematically lower than the escape velocity. Thus, we conclude that Horologium~I used to be a satellite of the LMC prior to its escape. This is in agreement with the conclusions drawn by \citet{Erkal_20}. 

\citet{Patel_20} reached similar conclusions for Carina~II, Carina~ III and Hydrus~I, but for Reticulum~II and Phoenix~II they concluded that both have been recently captured by the LMC. It is likely that the reason for these different conclusions is due to the different method used. The method used in this study is relatively similar to \citet{Erkal_20} who measured the fraction of orbits energetically bound to the LMC, while in \citet{Patel_20}, they measured the orbits inside the radius of equi-density between the MW and the LMC. Moreover, the better accuracy on the systemic PMs derived in our study, which impacts significantly the probability of being related to the LMC, can also partly explain the different conclusions, as it is expected that more accurate and precise systemic PMs can deliver a stronger signal of association, when this is present, as noted by \citet{Patel_20}.

As for Horologium~II, despite having a median velocity relative to the LMC higher than the escape velocity, 56\% of its orbits are compatible with having been recently ($<$500 Myr) ejected from the LMC system. However, it seems that Horologium~II had a velocity relative to the LMC very close to the escape velocity, and this at any time. Thus, with the current precision on its systemic PM\footnote{We note that the uncertainty on its total velocity is $\sim$120 km s$^{-1}$, exceeding the quality cut of 70 km s$^{-1}$.}, which dominates the uncertainties on its past orbits, it is not possible to definitively conclude if Horologium~II is a former satellite of the LMC or if it has been interacting with it for a long period of time ($>$2 Gyr). A visual inspection of the different possible orbits seems to favour the first idea. 

On the contrary, Grus II, despite having a relative velocity at its closest approach similar to the escape velocity, has an orbit that clearly shows that the galaxy did not originate in the Magellanic system, but has just been interacting/was captured  with/by the LMC in the last 200 Myr.

Tucana~IV is also potentially parented by the LMC system. Despite its orbit reconstruction suggesting that it has been captured by the LMC about 500 Myr ago, its closest approach is at $6.57_{-2.21}^{+5.64s}$ kpc\footnote{We note that for such a small distance the positive bias likely increased the number.}, 200 Myr ago with a relative velocity to the LMC lower than the escape velocity at this radius. Moreover, it has to be noticed that at that distance, the orbit of Tucana~IV might have been highly perturbed by the SMC, which could have boosted its kinetic energy. Since, our model does not take into account the presence of the SMC, the orbit of Tucana~IV likely overestimates its past kinetic energy, especially more than 200 Myr ago. Thus it is very likely that this galaxy has always been bound to the LMC system.  

Although the majority of the possible orbits of Tucana II do not present any potential link with the LMC, 19\% of its orbits have a relative velocity lower than the escape velocity of the LMC for at least one time step, while the satellite was inside the tidal radius of the LMC at that time\footnote{Hereafter we refer to this kind of orbits as {\it "linked"} orbits.} However, even in that conditions, it is very unlikely that Tucana~II is related to the LMC. First because those "linked" orbits have an inclination of $\sim 47 \degree$ compared to the orbit of the LMC. Secondly, those orbits suggest that Tucana~II passed through the very outer region of the halo of the LMC more than 2.8 Gyr ago, and being influenced by it $\sim 4.5$ Gyr ago. Taking into account the simplistic potential model for both the MW and the LMC on which our model is based of, it is very unlikely that Tucana~II is parented to the LMC. 

As for the "classical" dSphs, 25\% of the potential orbits of Carina do pass through the external region of the LMC halo 2.7 Gyr ago, and for $\sim 70$\% of them interacted another time with the LMC $\sim 5-6$ Gyr (see online video). Moreover, contrary to Tucana~II, the orbital plane of the "linked" orbits of Carina are relatively close to the orbital plane of the LMC, with a typical angular separation between the two planes of $21 \degr$. Therefore, we cannot exclude with our study that Carina was orbiting in the external region of the LMC and has been ejected from it more than 5 Gyr, as suggested by \citet[]{Pardy_20}. However, our study tends to indicate that this scenario is unlikely, and a more accurate modelling of the MW-LMC accretion event, and/or better measurement of the current properties of Carina (especially of the distance) are required to definitively conclude something on the potential link between Carina and the Magellanic system. For Fornax the other "classical" dSphs that could be linked to the LMC, the fraction of "linked" orbits is even lower than for Carina (4\%) and majority of them just pass through the LMC halo 1.8 Gyr and do not show any clear common history with the LMC, at least in the last 6 Gyr, despite having an orbital plane relatively close to the LMC, with a typical separation of of $31 \degr$. Therefore, we concluded that it is improbable that Fornax was a part of the Magellanic system. 

It has to be noticed here that the fraction of "linked" orbits of Fornax and Carina that we found (0.04 and 0.25 respectively) is significantly different than the value found by \citet{Erkal_20} (0.128 and 0.004). This is the consequence of the difference in the systemic PMs that we measured above, compared to the values found with {\it Gaia} DR2 that they used for their work (see Figure \ref{fig:pms_lit}).

For the two others galaxies with non zero fraction of the orbits "linked" to the LMC, Hydra~II and Reticulum~III, these orbits are the consequence of the large uncertainties that remain on their systemic PM, which allow a very large range of potential orbits. However, even for the few "linked" orbits, they pass only once in the external halo of the LMC in the last 5 Gyr, indicating that they are not physically associated to it.

For all the other galaxies, we can unambiguously argue that they are satellite of the MW and never used to be satellites of the LMC.

\paragraph{Comparison with a non perturbing LMC} As said before, the mass of the Magellanic system is still heavily debated; for example based on hydro-dynamical simulations \citet{Wang19_Mstream} argued that a $10^{11}$ M$_\odot$ massive LMC cannot reproduce the Magellanic stream, neither the bridge between the LMC and the SMC. Their simulations favours a LMC with a mass of $1-2 \times 10^{10}$ M$_\odot$, which does not produce strong perturbations of the MW halo \citep[e.g. ][]{Law_10,Gomez_15}. Thus we decided to perform the analysis of which galaxies are/were linked to the LMC also by assuming different LMC masses, ranging from $1 \times 10^{10}$ M$_\odot$ to $2 \times 10^{11}$ M$_\odot$ and assuming that it does not perturb the halo of the MW. 

The result is shown on the right panel of Figure \ref{VescLMC}. One can see that the number of linked satellites is changing drastically depending of the mass of the LMC, with zero satellites for a LMC of $1 \times 10^{10}$, 3 (or 4 if Tucana~IV is included) for a $5 \times 10^{10}$. The figure also clearly shows the importance of taking into account the perturbations produced by a massive LMC, since by neglecting them will lead to a Carina~II not physically associated to the LMC even for a mass equal to $1.5 \times 10^{11}$ M$_{\odot}$, while we saw in the previous section that it is unambiguously linked to the LMC if the perturbations produced by a massive LMC are taking into account.

\paragraph{Estimation of the mass of the LMC}
We use the number of possible LMC long-term satellites in the context of the gravitational potential perturbed by the LMC infall to estimate the mass-ratio between the Magellanic system and the MW, as done by \citet{Fritz_19}. This assumes that the number of LMC/MW satellites is directly related to the mass-ratio between the two objects. Thus, the ratio of satellites between the LMC and the MW ($\mathcal{R}_{sat}$) is equal to:
$\mathcal{R}_{sat}=\frac{N_{LMC}}{N_{tot}-N_{LMC}}\, ,$
where $N_{LMC}$ is the number of LMC satellites, $N_{tot}$ is the total number of galaxies satellite, whether they are MW's or the LMC's, and $N_{MW} = N_{tot}-N_{LMC}$ the number of MW satellites. Despite being only a rough estimate, the mass so derived is still useful to verify the concordance with the mass of the LMC assumed used to find the number of LMC satellites.

Given that the exact number of $N_{LMC}$ and $N_{MW}$ is still subject to debate, mostly due to the uncertainties on systemic distances and PMs, we made 2 selections for both systems, a generous and a conservative one. For the LMC, the {\it generous} sample includes the 9 potential satellites listed in Tab.~\ref{table_LMC_bound} as "Highly likely parented" and "Potentially parented", plus the SMC \citep{Murai_80,Besla_12}; the {\it conservative} sample is composed of the 6 satellites "highly parented", plus the SMC. For the MW, we consider satellites those galaxies that have the $16^{th}$ quantile of their apocenter within the virial radius of the MW in any of the 3 potentials considered in Section~\ref{sec:orbits}, plus the Sgr dSph and the galaxies not assigned to the LMC; Sagittarius~II and Crater~I are not taken into account, since they are likely globular clusters \citep[e.g.][]{Laevens_14, Voggel_16, Longeard_21}. 
In the {\it generous} MW sample, we also add the galaxies for which spectroscopic measurements are not available. This leads to $N_{\rm tot}= 58$ ($N_{\rm MW}= 51-48$ for the {\it conservative} and {\it generous} LMC samples, respectively). 
The {\it conservative} MW sample is restricted to the galaxies with uncertainties on their total Galactocentric velocity $< 70$ km s$^{-1}$ (see Section~\ref{sec:orbits}), which obviously excludes those without spectroscopic measurements. This leads to $N_{\rm tot}= 37$ ($N_{\rm MW}= 27-30$ for the {\it conservative} and {\it generous} LMC samples, respectively). 
Doing the 4 possibles combinations that allow these 4 samples, we find a ratio of satellites between the LMC and the MW ranging from 0.14 to 0.37, with a mean of $\mathcal{R}_{sat}=0.24$, consistent with the values found by \citet[][]{Penarrubia_16} (0.2), \citet{Erkal_18} (0.13-0.19) and \citet{Fritz_19} ($0.18^{+0.09}_{-0.08}$). This translates into a mass of the LMC between $1.5-4.1 \times 10^{11}$ M$_{\odot}$ with a mean of $2.6 \times 10^{11}$ M$_{\odot}$ for a $1.1 \times 10^{12}$ M$_{\odot}$ MW \citep{Bland_16} and between $2.0 -5.5 \times 10^{11}$ M$_{\odot}$ with a mean of $3.6 \times 10^{11}$ M$_{\odot}$ for a $1.51 \times 10^{12}$ M$_{\odot}$ MW \citep{Fritz_20}. Although with this method the mass range for the LMC is broad, we can see that in all the case, it is consistent with a value of $\sim 1.5 \times 10^{11}$ M$_{\odot}$, but rejects the possibility of the LMC having a mass of $\sim 1-5 \times 10^{10}$ M$_{\odot}$. 

Assuming an extreme scenario,  with a number of LMC satellites as low as 2 (i.e. the SMC plus one other), as we found for a low-mass non-perturbing LMC, the ratio of satellites using the {\it conservative} and {\it generous} MW samples are of $\mathcal{R}_{sat}=$0.057 and 0.036 respectively. Assuming a MW with a mass of $1.1 \times 10^{12}$ M$_{\odot}$, this correspond to a mass of the LMC of $6.27 \times 10^{10}$ M$_{\odot}$ and $3.93 \times 10^{10}$ M$_{\odot}$, respectively. This shows a lower level of consistency between the results in the case of a low mass LMC, taking into account that the mass estimated here are those for the most extreme scenario considered. Indeed, if we use the higher mass for the MW found by \citet{Fritz_20}, the LMC mass estimated by its number of satellites and the mass assumed to find the number of its satellites will not be consistent anymore. However, it is important to stress that this is a ballpark estimate, since the method is based on is very simplistic. For example, it relies on the assumption that the number of luminous satellites surviving till present day is proportional to the mass of host haloes, something that is not necessarily true \citep[e.g.][for indications that LMC-like haloes destroy less satellites than more massive haloes]{Jahn_19}. What we can say here is that the observations tend to favour a massive LMC of $1.5-2.5 \times 10^{11}$ M$_{\odot}$ rather than a $1-2 \times 10^{10}$ M$_{\odot}$ LMC.  

\begin{table*}
\centering
\footnotesize
\caption{List of galaxies reaching within 60 kpc from the LMC center for at least one out of the 100 initial Monte-Carlo realisations of their orbit in the {\it perturbed} potential. The table lists the median time of the closest approach (t$_{ca}$), the median relative distance to the LMC at that time (R$_{ca}$), the median relative velocity (V$_{ca}$) the fraction of orbit linked to the LMC for at least one time step (see text) ($\mathcal{F}_{link}$), and the median time of escape for those linked orbits (T$_{esc}$). It have to be notice that this escape time refer to the most recent moment when the galaxy escape the LMC according to the criteria we defined in the text, and do not take into account galaxy that escape the LMC and have been recaptured aftermath.}\label{table_LMC_bound}
\begin{tabular}{l|l|l|l|l|l}
\hline
\hline
  \multicolumn{1}{c|}{Galaxy} &
  \multicolumn{1}{c|}{t$_{ca}$ [Gyr]} &
  \multicolumn{1}{c|}{R$_{ca}$ [kpc]} &
  \multicolumn{1}{c|}{V$_{ca}$ [km.s$^{-1}$]} &
  \multicolumn{1}{c|}{$\mathcal{F}_{link}$} &
  \multicolumn{1}{c}{T$_{esc}$ [Gyr]} \\
\hline
  \multicolumn{6}{c}{} \\
  \multicolumn{6}{c}{{\bf Highly parented to the LMC}} \\
\hline
Carina II& -0.97 & $12.18_{-3.33}^{+14.02}$ & $174.12_{-35.65}^{+15.77}$ & 1.00 & -0.58\\ 
 
Carina III& -0.18 & $13.74_{-1.54}^{+2.60}$ & $164.54_{-12.12}^{+11.59}$ & 0.99 & -0.07\\
Horologium I& -0.12 & $36.35_{-7.20}^{+7.94}$ & $113.90_{-34.31}^{+33.44}$ & 0.84 & -0.46\\ 
Hydrus I& -0.30 & $13.09_{-3.11}^{+7.14}$ & $146.25_{-18.54}^{+14.48}$ & 1.00 & -0.14\\ 
Phoenix II& -0.43 & $25.99_{-14.49}^{+24.83}$ & $123.57_{-47.04}^{+45.76}$ & 0.90 & -0.31\\
Reticulum II& -0.19 & $14.69_{-0.85}^{+1.21}$ & $155.57_{-9.30}^{+7.01}$ & 1.00 & -0.08\\
\hline
  \multicolumn{6}{c}{} \\
  \multicolumn{6}{c}{{\bf Potentially parented to the LMC}} \\
\hline
Horologium II& -0.04 & $38.81_{-6.89}^{+7.11}$ & $170.06_{-65.89}^{+71.39}$ & 0.56 & -0.43\\
Tucana IV& -0.15 & $6.57_{-2.21}^{+5.64}$ & $220.88_{-45.68}^{+19.99}$ & 0.94 & -0.06\\ 
Carina& 0.00 & $62.27_{-5.81}^{+5.67}$ & $148.31_{-8.86}^{+10.24}$ & 0.25 & -2.77\\ 
\hline
  \multicolumn{6}{c}{} \\
  \multicolumn{6}{c}{{\bf Recently captured (<1 Gyr) by the LMC}} \\
\hline
Grus II& -0.34 & $25.62_{-3.33}^{+4.23}$ & $186.06_{-30.59}^{+22.15}$ & 0.57 & -0.29\\ 
\hline
  \multicolumn{6}{c}{} \\
  \multicolumn{6}{c}{{\bf Satellites of the MW}} \\
\hline
Aquarius II& -0.29 & $34.43_{-15.08}^{+21.26}$ & $396.99_{-43.99}^{+36.47}$ & 0.00 & $-$\\ 
Bootes III& -0.20 & $37.07_{-0.89}^{+1.04}$ & $507.41_{-6.35}^{+3.62}$ & 0.00 & $-$\\ 
Canes Venatici II& 0.00 & $196.13_{-8.30}^{+8.50}$ & $380.03_{-56.31}^{+59.44}$ & 0.00 & $-$\\ 
Fornax & -0.13 & $101.33_{-7.26}^{+7.11}$ & $177.90_{-20.00}^{+19.05}$ & 0.04 & -1.81\\ 
Hydra II & 0.00 & $137.36_{-7.51}^{+8.02}$ & $160.19_{-30.45}^{+54.30}$ & 0.04 & -2.02\\ 
Reticulum III& -0.03 & $42.96_{-11.89}^{+13.76}$ & $378.09_{-69.13}^{+73.07}$ & 0.02 & -1.43\\ 
Sagittarius II& -0.35 & $18.08_{-8.55}^{+16.82}$ & $269.88_{-33.21}^{+31.83}$ & 0.00 & $-$\\ 
Sculptor& -0.11 & $28.88_{-3.44}^{+3.62}$ & $503.05_{-7.97}^{+7.38}$ & 0.00 & $-$\\ 
Segue 1& -0.31 & $50.61_{-5.43}^{+13.29}$ & $288.02_{-68.34}^{+55.05}$ & 0.00 & $-$\\ 
Segue 2& -0.15 & $31.82_{-1.86}^{+1.89}$ & $423.72_{-15.11}^{+15.66}$ & 0.00 & $-$\\ 
Tucana II& 0.00 & $36.40_{-2.31}^{+3.26}$ & $220.16_{-8.60}^{+9.17}$ & 0.19 & -2.84\\ 
Tucana III& -0.08 & $13.47_{-1.32}^{+1.31}$ & $377.66_{-3.60}^{+3.44}$ & 0.00 & $-$\\
Tucana V& -0.08 & $16.69_{-8.57}^{+3.44}$ & $310.79_{-18.06}^{+32.45}$ & 0.00 & $-$\\ 
\hline\hline
\end{tabular}
\end{table*}

\section{Further applications} \label{sec:applications}

The existence of systemic PMs and catalogues of member stars with astrometric properties opens a wealth of possibilities for the study of the internal and orbital properties of LG galaxies, which goes beyond what can be addressed in one article. Below we make a (non-exhaustive) list of the  applications that could make use either of the systemic PM measurements derived or of the list of probable member stars and associated quantities provided by this work:

\begin{itemize}
    \item It is natural to expect that in the future LG dwarf galaxies will continue being the subject of intensive spectroscopic follow-up for the acquisition of large samples of individual stars with l.o.s. velocities and stellar atmospheric parameters (and chemical abundances). The large field-of-view and multiplex power of instruments like DESI, WEAVE, 4MOST, MOONS, PFS, MSE, and the collective power of the telescopes they are/will be mounted on, makes them particularly suitable for a comprehensive study of these systems. Lists of probable members allow to enhance the success rate of such observations by reducing the amount of contamination, as well as to assign priorities. 
    
    \item The 3D motions of MW satellites, independently on their nature as galaxies or stellar clusters, can be used for determinations of the MW mass, either on their own \citep[for recent works see e.g.][]{Callingham_19, Fritz_20, Li20_MW-mass} or as a useful addition to samples of other halo tracers, since their predominantly probe the outer parts of the MW gravitational potential and their tangential motions are known with a much higher precision than similarly distant individual MW halo stars. These 3D motions can also be used to determine the velocity anisotropy of the MW system of satellites \citep[see][for such determinations based on GDR2]{Riley_19, Fritz_20}, typically an important ingredient for mass modeling and interesting to compare to that of other objects/tracers and to the properties of satellite systems in cosmologically simulated MW-like haloes.
    
    \item The eGDR3 3D motions of MW satellite galaxies can be used to update limits on the density of the MW hot gas corona that would be required to ram-pressure strip them of their gaseous component \citep[see][for a GDR2-based analysis]{Putnam_21}. Since the common assumption is that ram-pressure stripping is most effective at pericenter, it would be interesting to perform this type of analysis both taking into account the growth of the MW DM halo (and possibly of its hot gas corona), since the SFHs of the great majority of MW satellite galaxies would suggest that they had lost their gaseous component already 8-10 Gyr ago, and combining it with models including the recent infall of a massive LMC, for those dwarf galaxies showing star formation activity in the last couple of Gyrs.  
    
    \item \citet{Marel_19} presented the first {\it Gaia}-based study of the dynamics of the M31-M33 system, resolved the PM rotation of both galaxies, and argued that, thanks for the complete view of these rotating galaxies, the {\it Gaia} DR2 PMs determinations allow for an independent assessment of possible biases of the systemic PM measurements based on small field-of-views. The authors found that the motions of M31 and M33 support the hypothesis in which M33 is on its first infall onto M31. In our determinations of M33 motion, the statistical errors are a factor 3-4 smaller in the GDR2 based value by \citet{Marel_19}; but systematic errors still dominate. Nonetheless, these new measurements can be used to revisit the above issues, in conjunction with eGDR3 measurements of M31 motion \citep[e.g.][]{Salomon_20}.
    
    \item For those systems that enjoy both a determination of the systemic PM and l.o.s. velocity, phase-space information can be used to look for associations among dwarf galaxies, and/or with globular clusters and streams. This will be the subject of a future work. This kinematic information can also be used to determine the orbital poles to further investigates the main plane of satellites around the MW, the vast polar structure of satellites (VPOS) \citep{Pawlowski_13}. The eGDR3 determinations  will be particularly useful for the most distant satellites, for which the PMs in DR2 were often not precise enough for a good determination of membership to the VPOS \citep{Fritz_18,Fritz_19}. Further it is interesting to investigate how many of the members of the VPOS were once satellites of the LMC. 
\end{itemize}

\section{Summary and conclusions} \label{sec:conclusions}

In this work we have jointly analyzed the spatial  distribution and the distributions onto the colour-magnitude and PM planes of  individual {\it Gaia} eDR3 sources with full astrometric solutions in the direction of 74 Local Group dwarf galaxies to determine systemic PMs of these systems. The sample includes 14 galaxies outside of the virial radius of the MW, out to $\sim$1.4 Mpc. 

Our method is largely based on that by \citet{McConnachie_DR2} and \citet{McConnachie_SOLO}, to which we have introduced some modifications, aimed at a more realistic treatment of the information on the colour-magnitude diagram of the dwarf galaxies.

We are able to determine systemic PMs for 72 systems when the analysis makes no use of complementary spectroscopic information, and for 73 of them when we make use of such additional data. Overall, we consider certainly reliable the measurements for 66 of them, including all those for the galaxies outside of the MW virial radius. 

The output of our analysis, including the list of members/non-members, the plots showing their  distribution on the observables used for the maximum likelihood analysis and the posterior distribution function of the systemic PMs will be made available after publication of the article. 

In general our results compares very well to those in the literature based on eGDR3 data; however, we notice that our measurement uncertainties are larger than those by \citet{McConnachie_EDR3} for a few systems; the main explanation for this difference is probably the prior on the velocity dispersion of the MW halo used by the other work and in a few cases differences in the quality cuts applied to the eGDR3 data. 

We use eGDR3 astrometry for QSOs in the line-of-sight to the dwarf galaxies to calculate the effect of {\it Gaia} systematics on the  
systemic PMs and their uncertainties. These corrections, as well as uncertainties on the distance module, are taken into account for the determination of the 3D velocities used for the reconstruction of the orbital trajectories of the galaxies around the MW (for systems out to the distance of NGC~6822).  

In order to tackle the effect of the MW mass onto the orbital history and parameters, we integrate the orbits in two MW static potentials, with a mass between 0.9 and 1.6$\times 10^{12}$ M$_{\odot}$. 
In addition, we complement the analysis by integrating the orbits also in a MW potential perturbed by the infall of a massive LMC, for which we use the model by \citet{Vasiliev_21}. In this way, also the reflex motion imprinted onto the objects found in the outskirts of the MW halo is factored in. It should be pointed out that the errors on the transverse and 3D velocities are still very large for several systems, and this can cause biases on the determined orbital parameters. In order to limit the impact of such biases, our considerations based on the results of the orbit integration analysis mainly concern galaxies with uncertainties in the observed 3D velocities $<$70 km s$^{-1}$, which we expect to be inflated with respect to the true 3D velocity of a factor $\lesssim 0.5$.   

The inclusion of a massive LMC, and the response of the MW, is found to modify the orbits of the majority of the MW satellites, regardless of their distance from the MW or the LMC, in a variety of ways, e.g. increasing/decreasing the pericentric, apocentric distance and the timing of these crucial events. Significant differences are also seen, as it would be expected, in the results from the two isolated potentials.

In general though, orbit integration of the eGDR3-based systemic PMs and literature l.o.s. velocities lead to the following conclusions in the three gravitational potentials used:

\begin{itemize}

\item Leo~I and NGC~6822 seem not to be currently bound to the MW, although both, including NGC~6822 are likely to have entered within its virial radius once in the past. 

\item Bootes~III and Tucana~III are very likely to have reached within 10~kpc from the MW center, fully confirming the expectations that the streams in which they are embedded are the result of tidal disruption.

\item The orbital properties of Crater~II confirm those from GDR2-based systemic PMs and are in line with those explored by models that explain its very low surface brightness, large half-light radius and low l.o.s. velocity dispersion in the context of strong tidal disturbances from the MW.

\item There are hints that the half-light radius of systems that are likely to have reached within 30~kpc from the MW center is smaller than those whose orbits kept them in more external regions.  

\item Concerning the SFH of MW satellites hosting young stars, i.e. Leo~I, Fornax and Carina, a correlation can be found between some of the main features seen in their recent SFH, such as bursts or a clear decrease in the activity, the same explanation does not fit all the main features observed. 
\end{itemize}

We also carried out an analysis aimed at identifying which ones of the galaxies surrounding the MW might have been/are physically linked to the LMC. We first identify what galaxies have at least an orbit that brings them within $\sim$60~kpc from the LMC, and then look at their total velocity and position as compared to the escape velocity curve of the LMC as a function of time. In the hypothesis that the gravitational potential and time evolution of the MW + LMC system is well represented by the \citet{Vasiliev_21} model, we find 6 systems that are highly likely to have been satellites of the LMC (Carina~II, Carina~III, Horologium~I, Hydrus~I, Phoenix~II, Reticulum~II), 3 that might have been potentially associated as satellites (Horologium~II, Tucana~IV, Carina) and one that seems to have been recently captured (Grus~II). On the other hand, we find it unlikely that Fornax was associated to the LMC. 

Exploring some generous and conservative estimates in the assignment of satellites to the LMC or the MW, we find that the ratio between two ranges between 0.14 and 0.37. 
A simple rescaling of the number of satellite galaxies with the DM halo mass would suggest the DM halo of the LMC being in the range 1.5-4.1$\times 10^{11}$ M$_{\odot}$ for a 1.1$\times 10^{12}$ M$_{\odot}$ MW mass. It should be however pointed out that the number of (and which) galaxies are classified as LMC satellites would change when applying the same methodology using a MW potential not perturbed by the presence of the LMC and allowing for a smaller LMC DM halo mass. It would be interesting to explore further which combinations of the LMC+MW gravitational potential lead to a number of LMC satellites consistent with the expectations of cosmological theories for LMC-like haloes. 

The significantly more accurate and precise eGDR3 astrometry has allowed to us to expand and improve our view of the dynamical properties of galaxies in the LG and its immediate surroundings with respect to GDR2, and even though this has been only a skimming of the potential applications of these measurements, we can only look impatiently forward to the fourth data release!

\begin{acknowledgements}
The authors acknowledge financial support through the grant (AEI/FEDER, UE) AYA2017-89076-P, as well as by the Ministerio de Ciencia, Innovación y Universidades (MCIU), through the State Budget and by the Consejería de Economía, Industria, Comercio y Conocimiento of the Canary Islands Autonomous Community, through the Regional Budget. GT acknowledges support from the Agencia Estatal de Investigaci\'on of the Ministerio de Ciencia e Innovaci\'on under grant 
FJC2018-037323-I. The authors are thankful to Santi Cassisi for kindly providing the simulated Basti-IAC CMD in eGDR3 passbands. 
This work has made use of data from the European Space Agency (ESA) mission
{\it Gaia} (\url{https://www.cosmos.esa.int/gaia}), processed by the {\it Gaia}
Data Processing and Analysis Consortium (DPAC,
\url{https://www.cosmos.esa.int/web/gaia/dpac/consortium}). Funding for the DPAC
has been provided by national institutions, in particular the institutions
participating in the {\it Gaia} Multilateral Agreement.
\end{acknowledgements}

%
%
\bibliographystyle{aa} 
\bibliography{eGDR3_ppm_arxiv} 

\begin{appendix}

\section{Comments on individual galaxies} \label{sec:individual}

In the following, unless said otherwise, the comments refer to the results of the "baseline" analysis, i.e. with no spectroscopic information for the whole sample, apart from Pisces~II and Tucana~V. The comments referring to the number of member stars returned by the routine are limited to the trickiest cases, i.e. those in the low statistics regime, $\lesssim$10 members.


\subsection{Antlia~II}

For this galaxy the correction for photometric completeness in the determination of the CMD likelihood term has the most noticeable effect.  
Without this correction our motion is more similar to those by \citet{Li_EDR3,McConnachie_EDR3}.

\subsection{Aquarius~II} 

There is significant scatter in the PM plane for the $P>0.5$ stars; on the other hand, the distribution on the colour-magnitude plane and on the sky seems reasonable.  
We noticed that for this system the cut  ipd\_gof\_harmonic\_amplitude $<0.2$ excludes several sources whose spatial, photometric and astrometric properties are perfectly compatible with those of the sources classified as probable members. This is probably the reason for the significantly larger uncertainties in the systemic PM in this work with respect to  \citet{McConnachie_EDR3}.

\subsection{Bootes~III}

The system is clearly detected in the distribution of $P>0.5$ stars in the three observables (PMs, CMD< location on the sky). This strongly argues in favour of the actual existence of the system. The resulting spatial distribution is clearly lopsided with respect to the East-West axis, probably a result of tidal disruption, given the very small pericentric distance (7-9 kpc) found in both the potentials explored in this work (see Tab.~\ref{tab:table_param}) and the work by \citet{Carlin_18} based on {\it Gaia} DR2. Our PM determination is in excellent agreement with that by those authors and with the predictions for the retrograde orbit of the Styx stream. 

\subsection{Bootes~IV}
Our routine returns only 5 members with P$>0.5$ and all very faint. The distribution in PM has 3 stars clumping at $\mu_\delta$ around $\sim$2mas yr$^{-1}$ and the other 2 stars are found at at 0 mas yr$^{-1}$ and -2 mas yr$^{-1}$. The error-bars in this component of the  systemic PM are such that within 2$\sigma$ the clump at ~2 mas yr$^{-1}$ would be included. 

It is one of the objects with the most elongated stellar structure, with an ellipticity $\sim$0.6.

\subsection{Canes Venatici~I}
There seems to be a possible elongation on the N-W side in the spatial distribution of probable members. The pericentric distance could reach 20-30 kpc within the 1-$\sigma$ uncertainties in both isolated potentials explored; while the results from the perturbed potential tend to suggest larger pericentric distances within the 1-$\sigma$ confidence interval, we find that 5 out of the 100 realizations have pericenters below 20kpc, hence the results from the perturbed potential do not necessarily go against the possibility of Canes Venatici~I having reached quite inward into the MW halo. 

\citet{Matus-Carillo_20} model Canes Venatici~I as a DM free object and look for orbits that match several of its photometric and kinematic properties, including its ellipticity, half-light radius, position angle, velocity dispersion. Both their PM predictions and the orbital parameters are in line with our values. 

\subsection{Carina}

The spatial distribution of the high probability member stars returned by our routine shows an elongation in the outer parts (Fig.~\ref{fig:out_1}), compatible with what seen in previous studies of the spatial distribution of Carina stars, based on red giant branch stars observed spectroscopically \citep{Munoz_06} and deep wide-area photometry \citep{Battaglia_12, McMonigal_14}. Even if there are some intervening LMC stars in the Carina's line-of-sight, it is unlikely the feature is due to that, given that these would be included in our contamination model. Given the orbital parameters that we obtain, it since very unlikely this might be the result of a close interaction with the MW, nor with the LMC (see e.g. Fig.~\ref{fig:orbits_lmc_1}) as suggested by \citet{Fritz_18}). 

It is possible that this galaxy was linked to the LMC.

\subsection{Carina~II}

Highly likely to have been part of the cohort of LMC satellites. 

\subsection{Carina~III}

Highly likely to have been part of the cohort of LMC satellites.

\subsection{Cetus~II}

It is possible that there is quite some amount of residual contamination among the stars with high probability of membership: about half are found beyond 3x the half-light radius and they display a large scatter in PM. 

\subsection{Cetus~III}
The PDF of systemic PM has extended wings, but of low amplitude.
It is one of the objects with the most elongated stellar structure, with an ellipticity $\sim$0.75. 

\subsection{Columba~I}

The \citet{Fritz_19} determination of Columba~I systemic PM is well compatible with that obtained in this work; other GDR2-based systemic PMs agree less well with our values (while there is a good agreement between the various eGDR3-based measurements).

\subsection{Crater~I}

This object is likely to be a stellar cluster \citep{Kirby_15,Weisz_16,Voggel_16}. The distribution of probable members is rather sparse on the PM plane. However, there is a 2-$\sigma$ agreement between our value of systemic PM in the $\alpha,*$ component and excellent agreement in the $\delta$ component with the determination by \citet{Vasiliev_EDR3}.

\subsection{Delve~1}

 It is argued to be a faint halo cluster due to its compactness \citep{Mau_20} but it lacks spectroscopic information that could confirm/validate this classification.
The distribution of $P>$0.5 stars is quite sparse in PM but well clumped in space. Comparing the PDF and the distribution of members on PM plane, one might be led to think that the error-bars are underestimated. Our systemic motion is incompatible with the DR2 motion of \citet{Mau_20}.

\subsection{Draco~II}

The distribution of probable member stars is reasonably well clumped in all properties. The spatial distribution seems asymmetric, although we have not verified whether this is statistically significant. \citet{Longeard_18a} find hints of tidal extension along the major axis. 
Since this system galaxy is currently at 24 kpc from the MW, it can be presently experiencing significant tidal forces,  independent of its past orbit.

\subsection{DESJ0225+0304}

Problematic PDF, with strong lopsideness and/or very extended wings of high amplitude.
It is one of the objects with the most elongated stellar structure, with an ellipticity $\sim$0.6.

\subsection{Eridanus~II}

Our proper motion PM is in better agreement with a previous entry into the MW halo than the PM in \citet{McConnachie_SOLO}. However, the error of about 200 km s$^{-1}$ in transverse velocity is still so large that forward Monte Carlo simulations do not lead to useful results. 

\subsection{Eridanus~III}

 It is argued to be a globular cluster due to its compactness \citep{Conn_18a} but it lacks spectroscopic information that could confirm/validate this classification. Its PM suggests a total velocity at the edge of what expected for the escape speed in the "Heavy MW" potential at Eridanus~III distance. Smaller errors are needed to understand whether Eridanus~III is unbound. 

\subsection{Grus~II} 

Recently captured by the LMC.

\subsection{Hercules}

This is one of the UFDs for which several features possibly attributable to tidal effects have been detected, see references within the review article by \citet{Simon_19}.
It is one of the objects with the most elongated stellar structure, with an ellipticity $=$0.7.

\subsection{Horologium~I}

Highly likely to have been part of the cohort of LMC satellites.

\subsection{Horologium~II}

It is one of the objects with the most elongated stellar structure, with an ellipticity $\sim$0.7. 
Sparse distribution of members on the PM plane. All of them are outside the 1 rh ellipse. Compared to \citet{Fritz_19}, the uncertainties are only slightly reduced  and the value of the new systemic PM is between the two GDR2-based options. This system remains still tricky. There is also a relevant difference between our and \citet{McConnachie_EDR3} systemic PM, which might be partly caused by their prior. 

The bulk line of sight velocity of Horologium~II is likely the most uncertain of all the systems in the sample, since its identification in \citet{Fritz_19} is based on only 3 potential members stars. In our analysis the faintest of these 3 stars is not used because of the quality cuts applied to the \textit{Gaia} data. 
Without the use of spectroscopy, the faintest stars, closer to the centre, has a probability of membership of 99.6\% member, and the brighter one, located further out, has 5.5\%. Thus, one of the stars classified as spectroscopic members is clearly a certain member, which makes the use of its line-of-sight velocity  trustworthy. The other star has a non-negligible likelihood despite its large distance from the centre, because it belongs to the now more visible PM peak of Horologium~II. When we use also spectroscopy the probabilities of membership increase to 99.9\% for the fainter star and 69.2\% for the brighter one. 
Our PM changes by 
0.57/0.15 $\sigma$ when spectroscopy is used and the error decreases by 19\%.

It is possible that this galaxy was linked to the LMC.

\subsection{Hydrus~I}

Highly likely to have been part of the cohort of LMC satellites.
There seems to be a secondary clump on the PM plane, around (1.7, -1) [mas yr$^{-1}$].

\subsection{IC1613}

Our systemic motion, combined it to its small error bars, moves it away from the region of values that make a passage within 300kpc from M31 likely according to \citet{McConnachie_SOLO}. Thus IC1613 likely evolved in isolation.

\subsection{Indus~I/Kim~2}

This system is likely a globular cluster \citep{Kim_15b}. 
Our routine returns two members with P$>0.5$, of a range of magnitudes, but quite offset from the centre. Double peaked PDF. 

This is the system for which the uncertainty in the systemic PM increases the most compared \citet{McConnachie_EDR3}, independently on their 100 km s$^{-1}$ prior. It is possible that we lose members with our conservative quality cuts.

\subsection{Indus~II}

Only one P$>0.95$ member, at very high PM. PDF with extended wings, but of low amplitude. According to \citet{Cantu_20}, based on a deep photometric study, Indus~II is a false-positive. Our analysis does not lead to a clean detection of the system either. The fact that our systemic PM leads to the object being clearly unbound from the MW does not increase our confidence in its existence.

\subsection{Leo~A}

While our error bars are a factor 3-4 smaller than of \citet{McConnachie_SOLO}, they are still clearly too large to restrict its orbits relative to the MW or M31.

\subsection{Leo~I}

It has experienced only one passage around the MW and its currently on its way out. 

\subsection{Leo~V}
Detections of over-densities, members at large radii and possibly a l.o.s. velocity gradient \citep[e.g.][]{Sand_12, Collins_17} have been interpreted as possible signs of tidal disturbance from the MW. \citet{Mutlu-Pakdil_19} do not confirm those signs, but do find members at large distances; \citet{Jenkins_21} find a weak velocity gradient, with only a 2$\,\sigma$ significance. 

The member stars identified by our routine are at most within 3 half-light radii, or just beyond. While the pericentric distances in Tab.~\ref{tab:table_param} are rather well constrained and do not suggest Leo~V coming close enough to the MW to experience tidal disruption, the error in the transverse velocity is about 130 km s$^{-1}$ per dimension, hence the current determination of orbital parameters is likely to be biased and might have benefited from backward Monte-Carlo simulations. 

\subsection{Leo~T}

The routine returns 8 members with P$>0.5$ and all very faint.

Our statistical PM errors improve compared to \citet{McConnachie_SOLO}, but they are still too large to exclude a backsplash origin for Leo~T (see their Fig.~4).  The error in the transverse velocity is extremely large, $\sim$660 km s$^{-1}$, hence the current determination of orbital parameters cannot be considered robust. Nonetheless, it is unlikely that such a faint galaxy might have hold on to its gas if entering the MW halo.

\subsection{M33}

The systemic PM of M~33 seems to be somewhat sensitive to the spatial region used for the selection of the stars to be analyzed. If we were to use a region within a semi-major axis radius of 0.2\degree, the $\mu_{\alpha,*}$ component would remain practically unchanged, whilst the $\mu_{\delta}$ component would decrease to $\sim -0.009 \pm 0.005$ mas yr$^{-1}$. However, it should be noticed that this change does not appear significant, since it is of the size of the systematic error related to a scale length of 0.6\degree, and this systematic error would increase when considering a smaller spatial region.

\subsection{NGC3109}

Despite the large distance of 1.46 Mpc,  the transverse velocity error of $\sim$220 km s$^{-1}$ is already of the size of the velocity dispersion between the isolated systems, thus possibly already useful for scientific applications and certainly will be so in the fourth {\it Gaia} data release.

\subsection{NGC6822}
Our statistical PM errors are about half of those by \citet{McConnachie_SOLO}. 

According to our orbit integration, it is possible that NGC~6822 passed within the virial radius of the MW. If we compare our PMs to the predictions in \citet{McConnachie_SOLO}, we confirm that a passage within the virial radius of M31 is to be excluded.

\subsection{Pegasus~III}

No stars with probability of membership larger than 0.5. Problematic PDF, with strong lopsideness and/or very extended wings of high amplitude.

\subsection{Phoenix}

The spatial distribution of probable members has a cross-like shape. Wide-area photometric studies showed that a disc-like structure tilted of 90\degree with respect to the main body is visible in young ($<$1 Gyr old) stars and absent in stars $>$5 Gyr old \citep{Battaglia_12}. We are probably seeing traces of this feature in our sample of members, which by construction should be RGB stars, unless of some young main-sequence stars scattered on the RGB by photometric errors. 

The error in the transverse velocity of Phoenix is about 80 km s$^{-1}$, hence the determination of the orbital parameters is likely to be biased. 
When comparing our systemic PM to the range of values that would allow a passage within the virial radius of the MW \citet{McConnachie_SOLO}, the uncertainties, while smaller than in that study, are still such that a backsplash origin cannot be excluded if the MW is more massive than 1.3$\times 10^{12}$ M$_{\odot}$ (with a MW DM halo mass of 1.3$\times 10^{12}$ M$_{\odot}$ there are no orbits that lead to a passage within the MW virial radius). 

\subsection{Phoenix~II}

The systemic PM of \citet{Fritz_19} is in excellent agreement with ours; in this work the statistical uncertainties are reduced of a factor of 2 with respect to that GDR2 based analysis. 

Highly likely to have been part of the cohort of LMC satellites.

\subsection{PiscesII}

No stars with probability of membership larger than 0.5 an flat PDF of the systemic PM if the spectroscopy is not taken into account. When the spectroscopic information is included, two stars with P$>0.95$ are found and the PDF becomes clearly peaked.  

\subsection{Reticulum~II}

It is one of the objects with the most elongated stellar structure, with an ellipticity $\sim$0.6. 
Possibly lopsided spatial distribution of member stars. 

Highly likely to have been part of the cohort of LMC satellites.

\subsection{Reticulum~III}
The distribution of $P>0.5$ stars on the sky, PM and CM-plane does not appear overly convincing. 
The different EDR3 estimates error bars overlap only partly, overall the error is still large in km s$^{-1}$.
The determination by \citet{Fritz_19} is just compatible with ours, given the large error-bars in both cases. 
While with \citet{Pace_19} systemic PM, Reticulum~III would be unbound to the MW even for a massive (1.6$\times$10$^{12}$ M$_{\odot}$) MW DM halo,  the other measurements in the literature, and ours, suggest it is bound  (see Fig.~\ref{fig:pms_lit4}). See also the work by \citet{Li_EDR3} for the quantification of the probability to be bound in several MW potentials.

The l.o.s determination in \citet{Fritz_19} was uncertain, as based on 3 stars classified as probable spectroscopic members. Our analysis finds that without the inclusion of spectroscopic information, the star named  ret3\_2\_70 in F19 has a probability of membership of only 0.2\%. This only increases to 2.4\% when including the spectroscopic information, thus star ret3\_2\_70 is likely not a member. The other two stars have probabilities of membership of 96\% and 67\% (brighter star) when not including  spectroscopic information and therefore they are likely members. 
The brighter spectroscopic member star causes also a relatively large change between the PM determinations with and without spectroscopic information of 0.08/0.72 $\sigma$ and an error reduction of 0.29 $\sigma$ when spectroscopy is used.

\subsection{Sagittarius~II}

Likely a globular cluster according to \citep{Longeard_21}.

\subsection{Segue~1}

This system displays possible extra-tidal features, as summarised by \citet{Simon_19}. 

\subsection{Segue~2}

Outlier in the mass-metallicity relation and argued to have become an ultra-faint through tidal stripping of a dwarf galaxy with a much larger stellar mass by \citet{Kirby_13_segue2}. 

\subsection{Sextans~B}
The uncertainty in the transverse velocity for this very distant galaxy exceeds 1000 km s$^{-1}$. For a 10 years extended \textit{Gaia} mission, we expect a much improved uncertainty,  at least a factor 6.6 due to the scaling of PMs with time \citep{Lindegren_20a}.

\subsection{Tucana~III}

Clearly embedded in a tidal stream \citep{Drlica-Wagner_15}, likely originating from a dwarf galaxy \citep{Li_18b,Marshall_19}.

\subsection{Tucana~IV}

According to our analysis, it is possible that this galaxy was linked to the LMC.

\subsection{Tucana~V}

Our routine returns three members with P$>0.5$, of various magnitudes, but the brightest ones are quite offset from the centre. Problematic PDF, with strong lopsideness and/or very extended wings of high amplitude when not including the spectroscopic information. The situation improves very significantly when including the spectroscopic information. 

\subsection{UGC~4879}
Two of the handful of members have G-mag around 18.5. The absolute magnitude of these stars, if belonging to UGC~4879, would be about $\sim -7$, too bright even for OB stars \citep[e.g.][]{Wegner_00}. These might be contaminants or unresolved/partly resolved clusters or H~II regions. 

\subsection{Ursa Major~I}

It is one of the objects with the most elongated stellar structure, with an ellipticity $\sim$0.6.

\subsection{Ursa Major~II}
There is a clear overdensity in the PM plane, not associated to Ursa Major II, that seems to show up as a brighter HB in the CMD of non-member stars at G$\sim$17.
According to \citet{Munoz_10,Munoz_18}, its radial surface density profile and morphology suggest that the object has been tidally destroyed. 

\subsection{Ursa Minor}
The spatial distribution of members in the outskirts seems rounder than the assumed value of 0.55 for the global ellipticity of Ursa Minor's stellar component. This can be a manifestation of \citet{Pace_20} finding that the metal-poor stars have a more extended and rounder spatial distribution than the metal-rich stars (with ellipticities of $0.33_{-0.09}^{+0.12}$ for the former and $0.75 \pm 0.03$ for the latter).

\subsection{Virgo~1}
No systemic PM determination. No stars with probability of membership larger than 0.5 and flat PDF.  
It is one of the objects with the most elongated stellar structure, with an ellipticity $\sim$0.6. 

\subsection{WLM}

Our smaller error bars and slightly different systemic motion relatively to that in \citet{McConnachie_SOLO} makes it less likely that in the past it reached within the virial radius of M31, although the errors are still too large to be certain.

\section{Tables} 

\onecolumn
\begin{landscape}
\tiny
\begin{longtable}[c]{l|r|r|r|r|r|r|r|r|r|r|r|l}
\caption{Sample of systems analyzed in this work, together with their main global properties. These are the coordinates of the optical center (cols.2 \& 3), the distance modulus (4), half-light radius along the projected major axis (5), ellipticity, defined as 1 - $b/a$, with $b$ and $a$ being the projected minor and major axes of the stellar component (6), position angle, measured from North to East (7), heliocentric systemic l.o.s. velocity (8) and velocity dispersion (9), mean stellar metallicity (10), type (11), dispersion of the stars metallicity distribution function (12). Col.~13 lists the corresponding references, whose numeric code corresponds to: (1) \citet{Torrealba_19}; (2) \citet{Torrealba_16a} ; (3) \citet{Okamoto_12}; (4) \citet{Dallora_06}; (5) \citet{Roderick_16}; (6) \citet{Koposov_11};(7) \citet{Norris_10}; (8) \citet{Walsh_08}; (9) \citet{Koch_09}; (10) \citet{Ji_16}; (11) \citet{Grillmair_09}; (12) \citet{Correnti_09}; (13) \citet{Carlin_09}; 
(14) \citet{Carlin_18}; (15) \citet{Homma_19}; (16) \citet{Munoz_18}; (17) \citet{Kuehn_08}; (18) \citet{simon_07}; (19) \citet{Kirby_13_massmet}; (20) \citet{Greco_08}; (21) \citet{Karczmarek_15}; (22) \citet{Walker_09}; (23) \citet{Fabrizio_12}; (24) \cite{Torrealba_18}; (25) \citet{Li_18a}; (26) \citet{Ji_20}; (27) \citet{Mau_20}; (28) \citet{Drlica-Wagner_15}; (29) \citet{Conn_18a}; (30) \citet{Homma_18}; (31) \citet{Carlin_17}; (32) \citet{Fritz_19}; (33) \citet{Musella_09}; (34) \citet{Belokurov_14}; (35) \citet{Weisz_16}; (36) \citet{Kirby_15}; (37) \citet{Vivas_20}; (38) \cite{Torrealba_16b}; (39) \citet{Caldwell_17}; (40) \citet{Cerny_21}; (41) \citet{Luque_17}; (42) \citet{Muraveva_20}; (43) \citet{Spencer_18}; (44) \citet{Longeard_18a}; (45) \citet{Crnojevic_16}; (46) \citet{Li_17}; (47) \citet{Conn_18b}; (48) \citet{Battaglia_06}; (49) \citet{Rizzi_07}; (50) \citet{Cantu_20}; (51) \citet{Martinez-Vazquez_19}; (52) \citet{Walker_16}; (53) \citet{Koposov_15a}; (54) \citet{Koposov_15b}; (55) \citet{Kim_15a}; (56) \citet{Vivas_16}; (57) \citet{Koposov_18}; (58) \citet{Kim_15b}; (59) \citet{Stetson_14}; (60) \citet{Mateo_08}; (61) \citet{Gullieuszik_08}; (62) \citet{Spencer_17}; (63) \citet{Moretti_09}; (64) \citet{Jenkins_21}; (65) \citet{Mutlu-Pakdil_19}; (66) \citet{Clementini_12}; (67) \citet{Kim_16}; (68) \citet{McConnachie_12}; (69) \citet{Holtzman_00}; (70) \citet{Battaglia_12}; (71) \citet{Kacharov_17}; (72) \citet{Mutlu-Pakdil_18}; (73) \citet{Simon_20}; (74) \citet{Garling_18}; (75) \citet{Drlica-Wagner_16}; (76) \citet{Belokurov_10}; (77) \citet{Longeard_21}; (78) \citet{Martinez-Vazquez_15}; (79) \citet{Battaglia_08};(80) \citet{Belokurov_07}; (81) \citet{Simon_11}; (82) \citet{Frebel_14}; (83) \citet{Boettcher_13}; (84) \citet{Kirby_13_segue2}; (85) \citet{Cicuendez_18}; (86) \citet{Vivas_19}; (87) \citet{Battaglia_11}; (88) \citet{Kirby_17a}; (89) \citet{Chiti_18}; (90) \citet{Li_18b}; (91) \citet{Garofalo_13}; (92) \citet{Dallora_12}; (93) \citet{Bellazzini_02}; (94) \citet{Homma_16}; (95) \citet{Willman_06}; (96) \citet{Willman_11}; (97) \citet{Gerbrandt_15}; (98) \citet{Kim_09}; (99) \citet{Higgs_21}; (100) \citet{Kirby_17b}; (101) \citet{Bernard_10}; (102) Taibi et al. in prep.; 
(103) \citet{Lee_95}; (104) \citet{Cook_99}; (105) \citet{Clementini_03}; (106) \citet{Kirby_14};  (107) \citet{Gallagher_98}; (108) \citet{Leaman_09}; (109) \citet{Momany_02}; (110) \citet{Bellazzini_11}; (111) \citet{Bellazzini_14}; (112) \citet{Dolphin_03}; (113) \citet{Kim_02}; (114) \citet{Kam_17}; (115) \citet{Conn_12}; (116) \citet{Crnojevic_14}; (117) \citet{Geha_10}; (118) \citet{Ho_15}; (119) \citet{McConnachie_05}; (120) \citet{Geha_06}; (121) \citet{Longeard_20}. Notes a: According to \citet{Koposov_11}, the losvd of Bootes~I is best described by
2 Gaussians; the value of the velocity dispersion in this table corresponds to the
colder gaussian, which includes 70\% of the stars. b: No half-light radius is given; the value of 30arcmin is taken from visual inspection of Fig.10 of Grillmair 2009; c: No uncertainty given on the ellipticity; we assume 0.2; d: internal velocity dispersion is not completely resolved when they exclude binaries and dependent on the
prior; e) updated online compilation. }\label{tab:sample}\\
\hline
\hline
\multicolumn{1}{c|}{Galaxy} &
\multicolumn{1}{c|}{RA} &
\multicolumn{1}{c|}{DEC} &
\multicolumn{1}{c|}{dm} &
\multicolumn{1}{c|}{Rh} &
\multicolumn{1}{c|}{ell} &
\multicolumn{1}{c|}{PA} &
\multicolumn{1}{c|}{<v$_{los}$>} &
\multicolumn{1}{c|}{$\sigma_{\rm los}$} &
\multicolumn{1}{c|}{<[Fe/H]>} &
\multicolumn{1}{c|}{type} &
\multicolumn{1}{c|}{$\sigma_{\rm [Fe/H]}$} &
\multicolumn{1}{c}{References} \\

\multicolumn{1}{c|}{} &
\multicolumn{1}{c|}{[degrees]} &
\multicolumn{1}{c|}{[degrees]} &
\multicolumn{1}{c|}{} &
\multicolumn{1}{c|}{[']} &
\multicolumn{1}{c|}{} &
\multicolumn{1}{c|}{[degrees]} &
\multicolumn{1}{c|}{ [km s$^{-1}$]} &
\multicolumn{1}{c|}{[km s$^{-1}$]} &
\multicolumn{1}{c|}{[dex]} &
\multicolumn{1}{c|}{} &
\multicolumn{1}{c|}{[dex]} &
\multicolumn{1}{c}{} \\
\hline
\endfirsthead
\multicolumn{13}{c}%
{{\bfseries \tablename\ \thetable{} -- continued from previous page}} \\
\hline
\multicolumn{1}{c|}{Galaxy} &
\multicolumn{1}{c|}{RA} &
\multicolumn{1}{c|}{DEC} &
\multicolumn{1}{c|}{dm} &
\multicolumn{1}{c|}{Rh} &
\multicolumn{1}{c|}{ell} &
\multicolumn{1}{c|}{PA} &
\multicolumn{1}{c|}{<v$_{los}$>} &
\multicolumn{1}{c|}{$\sigma_{\rm los}$} &
\multicolumn{1}{c|}{<[Fe/H]>} &
\multicolumn{1}{c|}{type} &
\multicolumn{1}{c|}{$\sigma_{\rm [Fe/H]}$} &
\multicolumn{1}{c}{References} \\
\multicolumn{1}{c|}{} &
\multicolumn{1}{c|}{[degrees]} &
\multicolumn{1}{c|}{[degrees]} &
\multicolumn{1}{c|}{} &
\multicolumn{1}{c|}{[']} &
\multicolumn{1}{c|}{} &
\multicolumn{1}{c|}{[degrees]} &
\multicolumn{1}{c|}{ [km s$^{-1}$]} &
\multicolumn{1}{c|}{[km s$^{-1}$]} &
\multicolumn{1}{c|}{[dex]} &
\multicolumn{1}{c|}{} &
\multicolumn{1}{c|}{[dex]} &
\multicolumn{1}{c}{} \\
\hline
\endhead
\hline 
\endfoot
\hline \hline
\endlastfoot
            AntliaII &  143.88670 &  -36.76730 &   20.60$_{-0.11}^{+0.11}$ &   76.20$_{-7.20}^{+7.20}$ & 0.38$_{-0.08}^{+0.08}$ &        156.0$_{-6.0}^{+6.0}$ &        290.70$_{-0.50}^{+0.50}$ &       5.71$_{-1.08}^{+1.08}$ &   -1.36$_{-0.04}^{+0.04}$ &        UFD & 0.57$_{-0.03}^{+0.03}$ &                        1;1;1;1;1;1;1;1;1 \\
          AquariusII &  338.48130 &   -9.32740 &   20.16$_{-0.07}^{+0.07}$ &    5.10$_{-0.80}^{+0.80}$ & 0.39$_{-0.09}^{+0.09}$ &        121.0$_{-9.0}^{+9.0}$ &        -71.10$_{-2.50}^{+2.50}$ &       5.40$_{-0.90}^{+3.40}$ &   -2.30$_{-0.50}^{+0.50}$ &        UFD &  &                        2;2;2;2;2;2;2;2;- \\
             BootesI &  210.02250 &   14.50060 &   19.11$_{-0.08}^{+0.08}$ &   12.80$_{-0.70}^{+0.70}$ & 0.26$_{-0.01}^{+0.01}$ &         13.8$_{-3.9}^{+3.9}$ &        101.80$_{-0.70}^{+0.70}$ &       2.40$_{-0.50}^{+0.90}$ &   -2.55$_{-0.11}^{+0.11}$ &        UFD & 0.37$_{-0.08}^{+0.08}$ &                       3;4;3;5;5;6;6a;7;7 \\
            BootesII &  209.51792 &   12.85250 &   18.10$_{-0.06}^{+0.06}$ &    2.60$_{-0.80}^{+0.80}$ & 0.34$_{-0.15}^{+0.15}$ &      -28.5$_{-57.0}^{+57.0}$ &       -117.00$_{-5.20}^{+5.20}$ &      10.50$_{-7.40}^{+7.40}$ &   -2.79$_{-0.10}^{+0.06}$ &        UFD & $<$0.35 &                      8;8;8;8;8;9;9;10;10 \\
           BootesIII &  209.30000 &   26.80000 &   18.35$_{-0.10}^{+0.10}$ & 30.00$_{-15.00}^{+15.00}$ & 0.50$_{-0.20}^{+0.20}$ &       90.0$_{-20.0}^{+20.0}$ &        197.50$_{-3.80}^{+3.80}$ &      14.00$_{-3.20}^{+3.20}$ &   -2.10$_{-0.20}^{+0.20}$ &        UFD & 0.55$_{-0.19}^{+0.19}$ &             11;11;11b;12c;11;13;13;13;14 \\
            BootesIV &  233.68920 &   43.72610 &   21.60$_{-0.20}^{+0.20}$ &    7.60$_{-0.80}^{+0.80}$ & 0.64$_{-0.05}^{+0.05}$ &          3.0$_{-4.0}^{+4.0}$ &                              &                          &                        &        UFD &  &                   15;15;15;15;15;-;-;-;- \\
      CanesVenaticiI &  202.00910 &   33.55210 &   21.62$_{-0.05}^{+0.05}$ &    7.48$_{-0.20}^{+0.20}$ & 0.45$_{-0.02}^{+0.02}$ &         80.0$_{-2.0}^{+2.0}$ &         30.90$_{-0.60}^{+0.60}$ &       7.60$_{-0.40}^{+0.40}$ &   -1.91$_{-0.01}^{+0.01}$ &        UFD & 0.44 &               16;17;16;16;16;18;18;19;19 \\
     CanesVenaticiII &  194.29270 &   34.32260 &   21.02$_{-0.06}^{+0.06}$ &    1.43$_{-0.24}^{+0.24}$ & 0.41$_{-0.13}^{+0.13}$ &        9.0$_{-13.0}^{+13.0}$ &       -128.90$_{-1.20}^{+1.20}$ &       4.60$_{-1.00}^{+1.00}$ &   -2.12$_{-0.05}^{+0.05}$ &        UFD & 0.59 &               16;20;16;16;16;18;18;19;19 \\
              Carina &  100.40650 &  -50.95930 &   20.12$_{-0.12}^{+0.12}$ &   10.20$_{-0.10}^{+0.10}$ & 0.37$_{-0.01}^{+0.01}$ &         60.0$_{-1.0}^{+1.0}$ &        222.90$_{-0.10}^{+0.10}$ &       6.60$_{-1.20}^{+1.20}$ &   -1.72$_{-0.01}^{+0.01}$ &       dSph & 0.24 &               16;21;16;16;16;22;22;23;23 \\
            CarinaII &  114.10670 &  -57.99910 &   17.79$_{-0.05}^{+0.05}$ &    8.69$_{-0.75}^{+0.75}$ & 0.34$_{-0.07}^{+0.07}$ &        170.0$_{-9.0}^{+9.0}$ &        477.20$_{-1.20}^{+1.20}$ &       3.40$_{-0.80}^{+1.20}$ &   -2.44$_{-0.09}^{+0.09}$ &        UFD & 0.22$_{-0.07}^{+0.10}$ &               24;24;24;24;24;25;25;25;25 \\
           CarinaIII &  114.62980 &  -57.89970 &   17.22$_{-0.10}^{+0.10}$ &    3.75$_{-1.00}^{+1.00}$ & 0.55$_{-0.18}^{+0.18}$ &      150.0$_{-14.0}^{+14.0}$ &        284.60$_{-3.10}^{+3.40}$ &       5.60$_{-2.10}^{+4.30}$ &   -1.80$_{-0.20}^{+0.20}$ &        UFD & 1.15$_{-0.58}^{+0.58}$ &               24;24;24;24;24;25;25;24;26 \\
          CentaurusI &  189.58500 &  -40.90200 &   20.33$_{-0.10}^{+0.10}$ &    2.90$_{-0.40}^{+0.50}$ & 0.40$_{-0.10}^{+0.10}$ &       20.0$_{-11.0}^{+11.0}$ &                             &                            & -1.80                  &        UFD &  &                  27;27;27;27;27;-;27;-;- \\
             CetusII &   19.47000 &  -17.42000 &   17.38$_{-0.20}^{+0.20}$ &    1.90$_{-0.50}^{+1.00}$ & 0.00$_{-0.00}^{+0.40}$ &                           &                             &                            &   -1.28$_{-0.07}^{+0.07}$ &        UFD &  &                  28;28;28;28;28;-;-;29;- \\
            CetusIII &   31.33100 &   -4.27000 &   22.00$_{-0.10}^{+0.20}$ &    1.23$_{-0.19}^{+0.42}$ & 0.76$_{-0.08}^{+0.06}$ &        101.0$_{-6.0}^{+5.0}$ &                             &                            &                        &        UFD &  &                   30;30;30;30;30;-;-;-;- \\
            ColumbaI &   82.85696 &  -28.04253 &   21.31$_{-0.11}^{+0.11}$ &    2.20$_{-0.20}^{+0.20}$ & 0.30$_{-0.10}^{+0.10}$ &         24.0$_{-9.0}^{+9.0}$ &        153.70$_{-4.80}^{+5.00}$ &      0.00$_{-0.00}^{+16.10}$ &   -2.37$_{-0.34}^{+0.35}$ &        UFD & 0.71$_{-0.24}^{+0.49}$ &               31;31;31;31;31;32;32;32;32 \\
       ComaBerenices &  186.74580 &   23.90690 &   18.13$_{-0.08}^{+0.08}$ &    5.67$_{-0.32}^{+0.32}$ & 0.37$_{-0.05}^{+0.05}$ &        -58.0$_{-4.0}^{+4.0}$ &         98.10$_{-0.90}^{+0.90}$ &       4.60$_{-0.80}^{+0.80}$ &   -2.25$_{-0.05}^{+0.05}$ &        UFD & 0.43 &               16;33;16;16;16;18;18;19;19 \\
             CraterI/Laevens~1 &  174.06600 &  -10.87780 &   20.81$_{-0.05}^{+0.05}$ &    0.43$_{-0.01}^{+0.01}$ & 0.00$_{-0.00}^{+0.05}$ &                           &        149.30$_{-1.20}^{+1.20}$ &       0.00$_{-0.00}^{+3.90}$ &   -1.68$_{-0.05}^{+0.05}$ &        UFD & 0.00$_{-0.00}^{+0.40}$ &               34;35;35;35;35;36;36;36;36 \\
            CraterII &  177.32800 &  -18.41800 &   20.33$_{-0.07}^{+0.07}$ &   31.20$_{-2.50}^{+2.50}$ & 0.12$_{-0.02}^{+0.02}$ &        135.0$_{-4.0}^{+4.0}$ &         87.50$_{-0.40}^{+0.40}$ &       2.70$_{-0.30}^{+0.30}$ &   -1.98$_{-0.10}^{+0.10}$ &        UFD & 0.22$_{-0.03}^{+0.04}$ &               37;37;38;37;37;39;39;39;39 \\
              Delve1 &  247.72500 &   -0.97200 &   16.39$_{-0.10}^{+0.10}$ &    0.97$_{-0.17}^{+0.24}$ & 0.20$_{-0.20}^{+0.10}$ &       21.0$_{-30.0}^{+26.0}$ &                               &                          & -1.50                  &        UFD &  &                  27;27;27;27;27;-;-;27;- \\
              Delve2 &   28.77200 &  -68.25300 &   19.26$_{-0.10}^{+0.10}$ &    1.04$_{-0.15}^{+0.19}$ & 0.03$_{-0.03}^{+0.15}$ &       74.0$_{-40.0}^{+84.0}$ &                               &                          &   -2.00$_{-0.50}^{+0.20}$ &        UFD &  &                  40;40;40;40;40;-;40;-;- \\
       DESJ0225+0304 &   36.42670 &    3.06950 &   16.88$_{-0.05}^{+0.06}$ &    2.68$_{-0.70}^{+1.33}$ & 0.61$_{-0.23}^{+0.14}$ &       31.3$_{-13.4}^{+11.5}$ &                               &                          &   -1.26$_{-0.03}^{+0.03}$ &        UFD &  &                   41;41;41;41;41;-;-;-;- \\
               Draco &  260.06840 &   57.91850 &   19.53$_{-0.07}^{+0.07}$ &    9.61$_{-0.10}^{+0.10}$ & 0.30$_{-0.01}^{+0.01}$ &         87.0$_{-1.0}^{+1.0}$ &       -292.30$_{-0.40}^{+0.40}$ &       9.00$_{-0.30}^{+0.30}$ &   -1.98$_{-0.01}^{+0.01}$ &       dSph & 0.35$_{-0.02}^{+0.02}$ &               16;42;16;16;16;43;43;19;19 \\
             DracoII &  238.17400 &   64.57900 &   16.67$_{-0.05}^{+0.05}$ &    3.00$_{-0.50}^{+0.70}$ & 0.23$_{-0.15}^{+0.15}$ &       76.0$_{-32.0}^{+22.0}$ &       -342.50$_{-1.20}^{+1.10}$ &       0.00$_{-0.00}^{+5.90}$ &   -2.70$_{-0.10}^{+0.10}$ &        UFD & $<$0.24 &               44;44;44;44;44;44;44;44;44 \\
          EridanusII &   56.08375 &  -43.53380 &   22.80$_{-0.10}^{+0.10}$ &    2.31$_{-0.12}^{+0.12}$ & 0.48$_{-0.04}^{+0.04}$ &         72.6$_{-3.3}^{+3.3}$ &         75.60$_{-3.30}^{+3.30}$ &       6.90$_{-0.90}^{+1.20}$ &   -2.38$_{-0.13}^{+0.13}$ &        UFD & 0.47$_{-0.09}^{+0.12}$ &               45;45;45;45;45;46;46;46;46 \\
         EridanusIII &   35.68970 &  -52.28370 &   19.80$_{-0.04}^{+0.04}$ &    0.32$_{-0.03}^{+0.04}$ & 0.44$_{-0.03}^{+0.02}$ &        109.0$_{-5.0}^{+5.0}$ &                              &                           &   -2.40$_{-0.12}^{+0.19}$ &        UFD &  &                  47;47;47;47;47;-;-;47;- \\
              Fornax &   39.96667 &  -34.51361 &   20.72$_{-0.04}^{+0.04}$ &   18.50$_{-0.10}^{+0.10}$ & 0.30$_{-0.01}^{+0.01}$ &         46.8$_{-1.6}^{+1.6}$ &         54.10$_{-0.50}^{+0.50}$ &      11.40$_{-0.40}^{+0.40}$ &   -1.04$_{-0.01}^{+0.01}$ &       dSph & 0.33 &               48;49;48;48;48;48;48;19;19 \\
               GrusI &  344.16600 &  -50.16800 &   20.51$_{-0.10}^{+0.10}$ &    4.16$_{-0.74}^{+0.54}$ & 0.44$_{-0.10}^{+0.08}$ &        153.0$_{-7.0}^{+8.0}$ &       -140.50$_{-1.60}^{+2.40}$ &       0.00$_{-0.00}^{+9.80}$ &   -1.88$_{-0.03}^{+0.09}$ &        UFD & 0.00$_{-0.00}^{+0.90}$ &               50;51;50;50;50;52;52;50;50 \\
              GrusII &  331.02500 &  -46.44200 &   18.70$_{-0.08}^{+0.08}$ &    5.90$_{-0.50}^{+0.50}$ & 0.00$_{-0.00}^{+0.21}$ &                           &       -110.00$_{-0.50}^{+0.50}$ &       0.00$_{-0.00}^{+2.00}$ &   -2.51$_{-0.11}^{+0.11}$ &        UFD & 0.00$_{-0.00}^{+0.45}$ &               73;73;73;73;73;73;73;73;73 \\
            Hercules &  247.77220 &   12.78520 &   20.68$_{-0.17}^{+0.17}$ &    5.83$_{-0.65}^{+0.65}$ & 0.70$_{-0.03}^{+0.03}$ &        -74.0$_{-2.0}^{+2.0}$ &         45.00$_{-1.10}^{+1.10}$ &       5.10$_{-0.90}^{+0.90}$ &   -2.39$_{-0.04}^{+0.04}$ &        UFD & 0.51 &               16;74;16;16;16;18;18;19;19 \\
         HorologiumI &   43.88130 &  -54.11600 &   19.50$_{-0.20}^{+0.20}$ &    1.71$_{-0.37}^{+0.37}$ & 0.32$_{-0.13}^{+0.13}$ &       53.0$_{-27.0}^{+27.0}$ &        112.80$_{-2.60}^{+2.50}$ &       4.90$_{-0.90}^{+2.80}$ &   -2.76$_{-0.10}^{+0.10}$ &        UFD & 0.17$_{-0.03}^{+0.20}$ &               16;53;16;16;16;54;54;54;54 \\
        HorologiumII &   49.10770 &  -50.04860 &   19.46$_{-0.20}^{+0.20}$ &    2.17$_{-0.59}^{+0.59}$ & 0.71$_{-0.17}^{+0.17}$ &      137.0$_{-12.0}^{+12.0}$ &      168.70$_{-12.60}^{+12.90}$ &      0.00$_{-0.00}^{+54.60}$ &   -1.87$_{-0.50}^{+0.36}$ &        UFD & 0.00$_{-0.00}^{+1.93}$ &               16;55;16;16;16;32;32;32;32 \\
             HydraII &  185.42510 &  -31.98600 &   20.89$_{-0.12}^{+0.12}$ &    1.65$_{-0.39}^{+0.39}$ & 0.25$_{-0.16}^{+0.16}$ &       13.0$_{-28.0}^{+28.0}$ &        303.10$_{-1.40}^{+1.40}$ &       0.00$_{-0.00}^{+3.60}$ &   -2.02$_{-0.08}^{+0.08}$ &        UFD & 0.40$_{-0.26}^{+0.48}$ &               16;56;16;16;16;36;36;36;36 \\
             HydrusI &   37.38920 &  -79.30890 &   17.20$_{-0.04}^{+0.04}$ &    7.42$_{-0.54}^{+0.62}$ & 0.21$_{-0.07}^{+0.15}$ &       97.0$_{-14.0}^{+14.0}$ &         80.40$_{-0.60}^{+0.60}$ &       2.69$_{-0.43}^{+0.51}$ &   -2.52$_{-0.09}^{+0.09}$ &        UFD & 0.41$_{-0.08}^{+0.08}$ &               57;57;57;57;57;57;57;57;57 \\
              IndusI/Kim2 &  317.20821 &  -51.16350 &   20.10$_{-0.10}^{+0.10}$ &    0.42$_{-0.02}^{+0.02}$ & 0.12$_{-0.10}^{+0.10}$ &         35.0$_{-5.0}^{+5.0}$ &                              &                          &                        &        UFD &  &                   58;58;58;58;58;-;-;-;- \\
             IndusII &  309.72000 &  -46.16000 &   21.65$_{-0.16}^{+0.16}$ &    2.90$_{-1.00}^{+1.10}$ & 0.00$_{-0.00}^{+0.40}$ &                           &                              &                           &                        &        UFD &  &                   28;28;28;28;28;-;-;-;- \\
                LeoI &  152.11460 &   12.30590 &   22.15$_{-0.10}^{+0.10}$ &    3.53$_{-0.03}^{+0.03}$ & 0.31$_{-0.01}^{+0.01}$ &         78.0$_{-1.0}^{+1.0}$ &        282.90$_{-0.50}^{+0.50}$ &       9.20$_{-0.40}^{+0.40}$ &   -1.45$_{-0.01}^{+0.01}$ &       dSph & 0.32 &               16;59;16;16;16;60;60;19;19 \\
               LeoII &  168.36270 &   22.15290 &   21.68$_{-0.11}^{+0.11}$ &    2.46$_{-0.03}^{+0.03}$ & 0.07$_{-0.02}^{+0.02}$ &         40.0$_{-9.0}^{+9.0}$ &         78.30$_{-0.60}^{+0.60}$ &       7.40$_{-0.40}^{+0.40}$ &   -1.63$_{-0.01}^{+0.01}$ &       dSph & 0.40 &               16;61;16;16;16;62;62;19;19 \\
               LeoIV &  173.24050 &   -0.54530 &   20.94$_{-0.07}^{+0.07}$ &    2.61$_{-0.32}^{+0.32}$ & 0.19$_{-0.09}^{+0.09}$ &      -28.0$_{-30.0}^{+30.0}$ &        131.40$_{-1.10}^{+1.20}$ &       3.60$_{-1.10}^{+1.00}$ &   -2.47$_{-0.14}^{+0.14}$ &        UFD & 0.42$_{-0.09}^{+0.12}$ &               16;63;16;16;16;64;64;64;64 \\
                LeoV &  172.78570 &    2.21940 &   21.25$_{-0.08}^{+0.08}$ &    1.05$_{-0.39}^{+0.39}$ & 0.45$_{-0.18}^{+0.18}$ &      -64.0$_{-33.0}^{+33.0}$ &       173.00$_{--0.80}^{+1.00}$ &   0.00                    &   -2.28$_{-0.16}^{+0.15}$ &        UFD & 0.34$_{-0.10}^{+0.17}$ &              16;65;16;16;16;64;64d;64;64 \\
                LeoT &  143.72920 &   17.04820 &   23.06$_{-0.15}^{+0.15}$ &    1.27$_{-0.13}^{+0.13}$ & 0.24$_{-0.09}^{+0.09}$ &     -104.0$_{-20.0}^{+20.0}$ &         38.10$_{-2.00}^{+2.00}$ &       7.50$_{-1.60}^{+1.60}$ &   -1.74$_{-0.04}^{+0.04}$ &        UFD & 0.54 &               16;66;16;16;16;18;18;19;19 \\
          PegasusIII &  336.10200 &    5.40500 &   21.66$_{-0.12}^{+0.12}$ &    0.85$_{-0.22}^{+0.22}$ & 0.38$_{-0.38}^{+0.22}$ &      114.0$_{-17.0}^{+19.0}$ &       -222.90$_{-2.60}^{+2.60}$ &       5.40$_{-2.50}^{+3.00}$ &   -2.55$_{-0.15}^{+0.15}$ &        UFD &  &               67;67;67;67;67;67;67;67;67 \\
             Phoenix &   27.77625 &  -44.44472 &   23.06$_{-0.12}^{+0.12}$ &    2.30$_{-0.07}^{+0.07}$ & 0.30$_{-0.03}^{+0.03}$ &          8.0$_{-4.0}^{+4.0}$ &        -21.20$_{-1.00}^{+1.00}$ &       9.30$_{-0.70}^{+0.70}$ &   -1.49$_{-0.04}^{+0.04}$ &        UFD & 0.51$_{-0.04}^{+0.04}$ &              68e;69;70;70;70;71;71;71;71 \\
           PhoenixII &  354.99279 &  -54.40495 &   19.60$_{-0.20}^{+0.20}$ &    1.50$_{-0.30}^{+0.30}$ & 0.40$_{-0.10}^{+0.10}$ &      156.0$_{-13.0}^{+13.0}$ &         32.40$_{-3.80}^{+3.70}$ &      11.00$_{-5.30}^{+9.40}$ &   -2.51$_{-0.17}^{+0.19}$ &        UFD & 0.33$_{-0.16}^{+0.29}$ &               72;72;72;72;72;32;32;32;32 \\
             PictorI &   70.94900 &  -50.28540 &   20.30$_{-0.20}^{+0.20}$ &    0.89$_{-0.36}^{+0.36}$ & 0.57$_{-0.19}^{+0.19}$ &       69.0$_{-21.0}^{+21.0}$ &                              &                           &                        &        UFD &  &                   16;53;16;16;16;-;-;-;- \\
            PictorII &  101.18000 &  -59.89700 &   18.30$_{-0.15}^{+0.12}$ &    3.80$_{-1.00}^{+1.50}$ & 0.13$_{-0.13}^{+0.22}$ &       14.0$_{-66.0}^{+60.0}$ &                              &                           &   -1.80$_{-0.30}^{+0.30}$ &        UFD &  &                   75;75;75;75;75;-;-;-;- \\             
            PiscesII &  344.63450 &    5.95260 &   21.31$_{-0.17}^{+0.17}$ &    1.18$_{-0.20}^{+0.20}$ & 0.39$_{-0.10}^{+0.10}$ &       98.0$_{-13.0}^{+13.0}$ &       -226.50$_{-2.70}^{+2.70}$ &       5.40$_{-2.40}^{+3.60}$ &   -2.45$_{-0.07}^{+0.07}$ &        UFD & 0.48$_{-0.29}^{+0.70}$ &               16;76;16;16;16;36;36;36;36 \\
         ReticulumII &   53.94929 &  -54.04661 &   17.50$_{-0.10}^{+0.10}$ &    6.30$_{-0.40}^{+0.40}$ & 0.60$_{-0.10}^{+0.10}$ &         68.0$_{-2.0}^{+2.0}$ &         64.70$_{-0.80}^{+1.30}$ &       3.22$_{-0.49}^{+1.64}$ &   -2.46$_{-0.10}^{+0.09}$ &        UFD & 0.29$_{-0.05}^{+0.13}$ &               72;72;72;72;72;54;54;54;54 \\
        ReticulumIII &   56.36000 &  -60.45000 &   19.81$_{-0.31}^{+0.31}$ &    2.40$_{-0.80}^{+0.90}$ & 0.00$_{-0.00}^{+0.40}$ &                           &        274.20$_{-7.40}^{+7.50}$ &      0.00$_{-0.00}^{+31.20}$ &   -2.81$_{-0.29}^{+0.29}$ &        UFD & 0.35$_{-0.09}^{+0.21}$ &               75;75;75;75;75;32;32;32;32 \\
       SagittariusII/Laevens5 &  298.16471 &  -22.06505 &   19.20$_{-0.20}^{+0.20}$ &    1.70$_{-0.50}^{+0.50}$ & $<$0.12 &      103.0$_{-17.0}^{+28.0}$ &       -177.20$_{-0.60}^{+0.50}$ &       1.70$_{-0.50}^{+0.50}$ &   -2.23$_{-0.07}^{+0.07}$ &        UFD & 0.20 &            72;72;121;121;121;77;77;77;77 \\
            Sculptor &   15.01830 &  -33.71860 &   19.62$_{-0.04}^{+0.04}$ &   12.43$_{-0.18}^{+0.18}$ & 0.36$_{-0.01}^{+0.01}$ &         92.0$_{-1.0}^{+1.0}$ &        110.60$_{-0.50}^{+0.50}$ &      10.10$_{-0.30}^{+0.30}$ &   -1.68$_{-0.01}^{+0.01}$ &       dSph & 0.46 &                  16;78;16;16;16;79;19;19 \\
              Segue1 &  151.75040 &   16.07560 &   16.80$_{-0.20}^{+0.20}$ &    3.93$_{-0.42}^{+0.42}$ & 0.32$_{-0.13}^{+0.13}$ &       75.0$_{-18.0}^{+18.0}$ &        208.50$_{-0.90}^{+0.90}$ &       3.70$_{-1.10}^{+1.40}$ &   -2.71$_{-0.39}^{+0.45}$ &        UFD & 0.95$_{-0.26}^{+0.42}$ &               16;80;16;16;16;81;81;82;82 \\
              Segue2 &   34.82260 &   20.16240 &   17.68$_{-0.17}^{+0.17}$ &    3.64$_{-0.29}^{+0.29}$ & 0.21$_{-0.07}^{+0.07}$ &      166.0$_{-16.0}^{+16.0}$ &        -40.20$_{-0.90}^{+0.90}$ &       0.00$_{-0.00}^{+2.60}$ &   -2.22$_{-0.13}^{+0.13}$ &        UFD & 0.43 &               16;83;16;16;16;84;84;84;84 \\
             Sextans &  153.26800 &   -1.61800 &   19.64$_{-0.01}^{+0.01}$ &   21.40$_{-0.60}^{+0.70}$ & 0.27$_{-0.03}^{+0.03}$ &         52.0$_{-3.0}^{+3.0}$ &        226.00$_{-0.60}^{+0.60}$ &       8.40$_{-0.40}^{+0.40}$ &   -1.90$_{-0.01}^{+0.01}$ &       dSph & 0.60 &               85;86;85;85;85;87;87;87;87 \\
        TriangulumII &   33.32520 &   36.17020 &   17.27$_{-0.11}^{+0.11}$ &    2.34$_{-0.58}^{+0.58}$ & 0.48$_{-0.17}^{+0.17}$ &       28.0$_{-19.0}^{+19.0}$ &       -381.70$_{-1.10}^{+1.10}$ &       0.00$_{-0.00}^{+4.20}$ &   -2.24$_{-0.05}^{+0.05}$ &        UFD & 0.53$_{-0.12}^{+0.38}$ &               16;31;16;16;16;88;88;88;88 \\
            TucanaII &  342.97960 &  -58.56890 &   18.80$_{-0.20}^{+0.20}$ &    9.83$_{-1.11}^{+1.66}$ & 0.39$_{-0.20}^{+0.10}$ &      107.0$_{-18.0}^{+18.0}$ &       -129.10$_{-3.50}^{+3.50}$ &       8.60$_{-2.70}^{+4.40}$ &   -2.23$_{-0.12}^{+0.18}$ &        UFD & 0.29$_{-0.12}^{+0.15}$ &                  53;53;53;53;52;52;89;89 \\
           TucanaIII &  359.10750 &  -59.58332 &   16.80$_{-0.10}^{+0.10}$ &    5.10$_{-1.20}^{+1.20}$ & 0.20$_{-0.10}^{+0.10}$ &       25.0$_{-38.0}^{+38.0}$ &       -101.20$_{-0.50}^{+0.50}$ &       0.90$_{-0.50}^{+0.60}$ &   -2.42$_{-0.08}^{+0.07}$ &        UFD & $<$0.19 &               72;72;72;72;72;90;90;90;90 \\
            TucanaIV &    0.71700 &  -60.83000 &   18.36$_{-0.19}^{+0.18}$ &    9.30$_{-0.90}^{+1.40}$ & 0.39$_{-0.10}^{+0.07}$ &         27.0$_{-8.0}^{+9.0}$ &         15.90$_{-1.70}^{+1.80}$ &       4.30$_{-1.00}^{+1.70}$ &   -2.49$_{-0.16}^{+0.15}$ &        UFD & 0.00$_{-0.00}^{+0.64}$ &               73;73;73;73;73;73;73;73;73 \\
             TucanaV &  354.34700 &  -63.26600 &   18.70$_{-0.34}^{+0.11}$ &    2.10$_{-0.40}^{+0.60}$ & 0.51$_{-0.20}^{+0.10}$ &       29.0$_{-11.0}^{+11.0}$ &        -36.20$_{-2.20}^{+2.50}$ &       0.00$_{-0.00}^{+7.40}$ &   -2.17$_{-0.23}^{+0.23}$ &        UFD &  &               73;73;73;73;73;73;73;73;73 \\
          UrsaMajorI &  158.77060 &   51.94790 &   19.94$_{-0.13}^{+0.13}$ &    8.13$_{-0.31}^{+0.31}$ & 0.59$_{-0.03}^{+0.03}$ &         67.0$_{-2.0}^{+2.0}$ &        -55.30$_{-1.40}^{+1.40}$ &       7.60$_{-1.00}^{+1.00}$ &   -2.10$_{-0.03}^{+0.03}$ &        UFD & 0.65 &               16;91;16;16;16;18;18;19;19 \\
         UrsaMajorII &  132.87260 &   63.13350 &   17.70$_{-0.16}^{+0.16}$ &   13.90$_{-0.40}^{+0.40}$ & 0.55$_{-0.03}^{+0.03}$ &        -76.0$_{-2.0}^{+2.0}$ &       -116.50$_{-1.90}^{+1.90}$ &       6.70$_{-1.40}^{+1.40}$ &   -2.18$_{-0.05}^{+0.05}$ &        UFD & 0.66 &               16;92;16;16;16;18;18;19;19 \\
           UrsaMinor &  227.24200 &   67.22210 &   19.41$_{-0.12}^{+0.12}$ &   18.20$_{-0.10}^{+0.10}$ & 0.55$_{-0.01}^{+0.01}$ &         50.0$_{-1.0}^{+1.0}$ &       -246.90$_{-0.40}^{+0.40}$ &       8.00$_{-0.30}^{+0.30}$ &   -2.13$_{-0.01}^{+0.01}$ &       dSph & 0.43 &               16;93;16;16;16;43;43;19;19 \\
              VirgoI &  180.03800 &    0.68100 &   19.80$_{-0.10}^{+0.20}$ &    1.76$_{-0.40}^{+0.49}$ & 0.59$_{-0.14}^{+0.12}$ &        62.0$_{-13.0}^{+8.0}$ &                              &                           &                        &        UFD &  &                   94;94;94;94;94;-;-;-;- \\
            Willman1 &  162.34360 &   51.05010 &   17.90$_{-0.40}^{+0.40}$ &    2.52$_{-0.21}^{+0.21}$ & 0.47$_{-0.06}^{+0.06}$ &         74.0$_{-4.0}^{+4.0}$ &        -14.10$_{-1.00}^{+1.00}$ &       4.00$_{-0.80}^{+0.80}$ & -2.10                  &        UFD &  &                16;95;16;16;16;96;96;96;- \\
                LeoA &  149.86042 &   30.74639 &   24.28$_{-0.05}^{+0.05}$ &    2.30$_{-0.90}^{+0.90}$ & 0.42$_{-0.05}^{+0.05}$ &        116.0$_{-6.0}^{+6.0}$ &         26.20$_{-0.90}^{+1.00}$ &       9.00$_{-0.60}^{+0.80}$ &   -1.70$_{-0.10}^{+0.10}$ &  late-type & 0.42 &            68;99;99;99;99;100;100;100;19 \\
              IC1613 &   16.19917 &    2.11778 &   24.40$_{-0.01}^{+0.01}$ &    7.57$_{-0.05}^{+0.05}$ & 0.20$_{-0.05}^{+0.05}$ &         90.0$_{-1.0}^{+1.0}$ &       -234.00$_{-0.60}^{+0.60}$ &      11.30$_{-0.50}^{+0.50}$ &   -1.10$_{-0.10}^{+0.10}$ &  late-type & 0.28 &          68;101;99;99;99;102;102;102;102 \\
             NGC6822 &  296.23583 &  -14.78917 &   23.36$_{-0.17}^{+0.17}$ &   11.95$_{-0.70}^{+0.70}$ & 0.28$_{-0.15}^{+0.15}$ &       67.0$_{-15.0}^{+15.0}$ &        -54.50$_{-1.70}^{+1.70}$ &      23.20$_{-1.20}^{+1.20}$ &   -1.05$_{-0.10}^{+0.10}$ &  late-type & 0.49 &            68;105;99;99;99;106;106;19;19 \\
            Peg-dIrr &  352.15125 &   14.74306 &   24.40$_{-0.20}^{+0.20}$ &    3.81$_{-0.05}^{+0.05}$ & 0.56$_{-0.05}^{+0.05}$ &        126.3$_{-0.3}^{+0.3}$ &       -179.50$_{-1.50}^{+1.50}$ &      12.30$_{-1.10}^{+1.20}$ &   -1.39$_{-0.01}^{+0.01}$ &  late-type & 0.56 &            68;107;99;99;99;106;106;19;19 \\
                 WLM &    0.49250 &  -15.46083 &   24.85$_{-0.08}^{+0.08}$ &    4.10$_{-0.13}^{+0.13}$ & 0.54$_{-0.06}^{+0.06}$ &        177.0$_{-0.5}^{+0.5}$ &       -130.00$_{-1.00}^{+1.00}$ &      17.50$_{-2.00}^{+2.00}$ &   -1.14$_{-0.04}^{+0.04}$ &  late-type & 0.39 &           68;68;99;99;99;108;108;108;108 \\
             Sg-dIrr &  292.49583 &  -17.68083 &   25.14$_{-0.18}^{+0.18}$ &    1.43$_{-0.08}^{+0.08}$ & 0.56$_{-0.18}^{+0.18}$ &         86.9$_{-3.4}^{+3.4}$ &        -78.40$_{-1.60}^{+1.60}$ &       9.40$_{-1.10}^{+1.50}$ &   -1.90$_{-0.10}^{+0.13}$ &  late-type &  &           68e;109;99;99;99;100;100;100;- \\
             UGC4879 &  139.00917 &   52.84000 &   25.61$_{-0.13}^{+0.13}$ &    1.13$_{-0.10}^{+0.10}$ & 0.43$_{-0.06}^{+0.06}$ &         81.2$_{-6.5}^{+6.5}$ &        -29.20$_{-1.60}^{+1.60}$ &       9.60$_{-1.20}^{+1.30}$ &   -1.43$_{-0.02}^{+0.02}$ &  late-type & 0.52 &           68e;110;99;99;99;106;106;19;19 \\
             NGC3109 &  150.77875 &  -26.15972 &   25.57$_{-0.08}^{+0.08}$ &    4.30$_{-0.10}^{+0.10}$ & 0.82$_{-0.01}^{+0.01}$ &         91.5$_{-1.0}^{+1.0}$ &        403.00$_{-2.00}^{+2.00}$ &                           &   -1.84$_{-0.20}^{+0.20}$ &  late-type &  &          68e;68e;68e;68e;68e;68e;-;68e;- \\
            SextansA &  152.75333 &   -4.69278 &   25.77$_{-0.12}^{+0.12}$ &    1.80                & 0.00                &    0.0                    &        324.00$_{-1.00}^{+1.00}$ &                           &   -1.45$_{-0.00}^{+0.00}$ &  late-type &  &           68e;111;111;111;111;68;-;112;- \\
                 M33 &   23.46208 &   30.66028 &   24.80$_{-0.05}^{+0.05}$ &                        & 0.41                &   23.0                    &       -180.00$_{-3.00}^{+3.00}$ &                           &                        &     spiral &  &                68e;113;-;68;68;114;-;-;- \\
            SextansB &  150.00042 &    5.33222 &   25.80$_{-0.12}^{+0.12}$ &    1.90                & 0.31                &       95.0$_{-15.0}^{+15.0}$ &        304.00$_{-1.00}^{+1.00}$ &                           & -1.60                  &  late-type &  &           68e;111;111;111;111;68;-;111;- \\
\hline
\end{longtable}
\end{landscape}

\onecolumn
\begin{longtable}[c]{l|rrrr|rrrr}
\caption{Systemic PMs (determined using the completeness correction for systems treated with the synthetic CMD, and without prior from spectroscopy). We highlight in red and orange the determinations we do not trust and those that might be particularly uncertainty, respectively (see main text). For Pisces~II and Tucana~V we list also the determination from the run that included the spectroscopic information. The columns are: (1) galaxy name; (2,3) systemic PM in the $\alpha,*$ and $\delta$ components, respectively; (4) fraction of stars in the galaxy under consideration; (5) the correlation coefficient; (6) Number of stellar objects analyzed; (7,8,9) Number of stars with probability of membership to the galaxy $>$0.5, 0.8 and 0.95, respectively. The zero-points and additional uncertainties from {\it Gaia} eDR3 systematics are kept separate from this table, and listed in \ref{tab:systematics}.} 
\label{tab:sysmotions} \\
\hline
\hline
\multicolumn{1}{l|}{Galaxy} &
\multicolumn{1}{r}{$\mu_{\rm \alpha,*, sys}$} &
\multicolumn{1}{r}{$\mu_{\rm \delta, sys}$} &
\multicolumn{1}{r}{$f_{\rm gal}$} &
\multicolumn{1}{r|}{C} &
\multicolumn{1}{r}{N$_i$} &
\multicolumn{1}{r}{P$(>0.5)$} &
\multicolumn{1}{r}{P$(>0.8)$} &
\multicolumn{1}{r}{P$(>0.95)$} \\

\multicolumn{1}{l|}{ } &
\multicolumn{1}{r}{[mas\,yr$^{-1}$] } &
\multicolumn{1}{r}{[mas\,yr$^{-1}$]} &
\multicolumn{1}{r}{} &
\multicolumn{1}{r|}{} &
\multicolumn{1}{r}{} &
\multicolumn{1}{r}{} &
\multicolumn{1}{r}{} &
\multicolumn{1}{r}{} \\
\hline
\endfirsthead
\caption{Continued.}\\
\hline
\multicolumn{1}{c|}{Galaxy} &
\multicolumn{1}{c}{$\mu_{\rm \alpha,*, sys}$} &
\multicolumn{1}{c}{$\mu_{\rm \delta, sys}$} &
\multicolumn{1}{c}{$f_{\rm gal}$} &
\multicolumn{1}{c|}{C} &
\multicolumn{1}{c}{N$_i$} &
\multicolumn{1}{c}{P$(>0.5)$} &
\multicolumn{1}{c}{P$(>0.8)$} &
\multicolumn{1}{c}{P$(>0.95)$} \\

\multicolumn{1}{c|}{ } &
\multicolumn{1}{c}{[mas\,yr$^{-1}$] } &
\multicolumn{1}{c}{[mas\,yr$^{-1}$]} &
\multicolumn{1}{c}{} &
\multicolumn{1}{c|}{} &
\multicolumn{1}{c}{} &
\multicolumn{1}{c}{} &
\multicolumn{1}{c}{} &
\multicolumn{1}{c}{} \\
\hline
\endhead
\hline 
\endfoot
\hline \hline
\endlastfoot
	AntliaII & $-0.10_{-0.01}^{+0.01}$ & $0.09_{-0.01}^{+0.01}$ & $0.0047_{-0.0001}^{+0.0001}$ & $0.20$ & $401473$ & $1321$ & $690$ & $255$\\
	\textcolor{orange}{AquariusII} & $-0.03_{-0.19}^{+0.23}$ & $-0.47_{-0.16}^{+0.15}$ & $0.04_{-0.01}^{+0.01}$ & $-0.22$ & $262$ & $8$ & $5$ & $4$\\
	BootesI & $-0.39_{-0.02}^{+0.02}$ & $-1.06_{-0.01}^{+0.01}$ & $0.13_{-0.01}^{+0.01}$ & $-0.14$ & $1658$ & $204$ & $164$ & $128$\\
	BootesII & $-2.46_{-0.07}^{+0.07}$ & $-0.45_{-0.06}^{+0.05}$ & $0.22_{-0.04}^{+0.04}$ & $-0.17$ & $105$ & $22$ & $20$ & $17$\\
	BootesIII & $-1.16_{-0.02}^{+0.02}$ & $-0.88_{-0.01}^{+0.01}$ & $0.020_{-0.002}^{+0.002}$ & $0.33$ & $6687$ & $97$ & $60$ & $44$\\
	\textcolor{orange}{BootesIV} & $0.03_{-0.38}^{+0.41}$ & $0.40_{-0.76}^{+0.92}$ & $0.01_{-0.01}^{+0.01}$ & $0.12$ & $680$ & $5$ & $4$ & $3$\\
	CanesVenaticiI & $-0.10_{-0.03}^{+0.03}$ & $-0.12_{-0.02}^{+0.02}$ & $0.29_{-0.02}^{+0.02}$ & $0.18$ & $511$ & $152$ & $140$ & $129$\\
	CanesVenaticiII & $-0.12_{-0.11}^{+0.10}$ & $-0.29_{-0.08}^{+0.08}$ & $0.25_{-0.05}^{+0.05}$ & $0.33$ & $65$ & $16$ & $15$ & $15$\\
	Carina & $0.53_{-0.01}^{+0.01}$ & $0.13_{-0.01}^{+0.01}$ & $0.044_{-0.001}^{+0.001}$ & $-0.11$ & $53804$ & $2358$ & $2120$ & $1732$\\
	CarinaII & $1.88_{-0.01}^{+0.01}$ & $0.13_{-0.02}^{+0.01}$ & $0.008_{-0.001}^{+0.001}$ & $-0.00$ & $9827$ & $68$ & $46$ & $28$\\
	CarinaIII & $3.10_{-0.04}^{+0.03}$ & $1.39_{-0.04}^{+0.04}$ & $0.008_{-0.002}^{+0.002}$ & $0.01$ & $1984$ & $13$ & $11$ & $9$\\
	CentaurusI & $-0.13_{-0.05}^{+0.05}$ & $-0.21_{-0.04}^{+0.04}$ & $0.04_{-0.01}^{+0.01}$ & $-0.01$ & $842$ & $33$ & $28$ & $16$\\
	CetusII & $2.85_{-0.05}^{+0.05}$ & $0.47_{-0.06}^{+0.06}$ & $0.13_{-0.04}^{+0.04}$ & $0.13$ & $86$ & $10$ & $9$ & $7$\\
	\textcolor{orange}{CetusIII} & $0.47_{-0.81}^{+0.85}$ & $-0.56_{-0.87}^{+0.94}$ & $0.03_{-0.02}^{+0.03}$ & $0.29$ & $75$ & $2$ & $1$ & $1$\\
	ColumbaI & $0.17_{-0.07}^{+0.07}$ & $-0.41_{-0.08}^{+0.08}$ & $0.03_{-0.01}^{+0.01}$ & $-0.10$ & $397$ & $8$ & $8$ & $5$\\
	ComaBerenices & $0.43_{-0.02}^{+0.02}$ & $-1.72_{-0.02}^{+0.02}$ & $0.17_{-0.03}^{+0.03}$ & $-0.29$ & $257$ & $45$ & $35$ & $26$\\
	CraterI & $-0.04_{-0.12}^{+0.12}$ & $-0.12_{-0.10}^{+0.10}$ & $0.09_{-0.02}^{+0.02}$ & $-0.23$ & $160$ & $13$ & $13$ & $12$\\
	CraterII & $-0.07_{-0.02}^{+0.02}$ & $-0.11_{-0.01}^{+0.01}$ & $0.027_{-0.001}^{+0.001}$ & $-0.05$ & $22266$ & $536$ & $366$ & $199$\\
	\textcolor{orange}{Delve1} & $0.04_{-0.07}^{+0.07}$ & $-1.54_{-0.05}^{+0.05}$ & $0.01_{-0.01}^{+0.01}$ & $-0.13$ & $763$ & $9$ & $8$ & $7$\\
	Delve2 & $0.92_{-0.11}^{+0.12}$ & $-0.97_{-0.08}^{+0.09}$ & $0.04_{-0.01}^{+0.01}$ & $0.23$ & $215$ & $8$ & $7$ & $6$\\
	\textcolor{red}{DESJ0225+0304} & $0.80_{-1.40}^{+1.80}$ & $-0.40_{-2.70}^{+2.88}$ & $0.03_{-0.02}^{+0.03}$ & $0.19$ & $77$ & $2$ & $2$ & $0$\\
	Draco & $0.04_{-0.01}^{+0.01}$ & $-0.19_{-0.01}^{+0.01}$ & $0.082_{-0.002}^{+0.002}$ & $0.23$ & $20861$ & $1727$ & $1636$ & $1490$\\
	DracoII & $1.12_{-0.09}^{+0.09}$ & $0.91_{-0.10}^{+0.10}$ & $0.16_{-0.03}^{+0.03}$ & $-0.02$ & $178$ & $29$ & $25$ & $21$\\
	EridanusII & $0.15_{-0.11}^{+0.11}$ & $0.03_{-0.13}^{+0.14}$ & $0.16_{-0.03}^{+0.04}$ & $-0.15$ & $119$ & $19$ & $16$ & $16$\\
	EridanusIII & $1.39_{-0.13}^{+0.13}$ & $-0.64_{-0.14}^{+0.14}$ & $0.04_{-0.02}^{+0.03}$ & $-0.28$ & $82$ & $3$ & $3$ & $3$\\
	Fornax & $0.381_{-0.001}^{+0.001}$ & $-0.358_{-0.002}^{+0.002}$ & $0.797_{-0.002}^{+0.002}$ & $-0.43$ & $30261$ & $24172$ & $23963$ & $23606$\\
	GrusI & $0.07_{-0.05}^{+0.05}$ & $-0.27_{-0.07}^{+0.08}$ & $0.07_{-0.02}^{+0.02}$ & $0.04$ & $196$ & $13$ & $9$ & $8$\\
	GrusII & $0.39_{-0.04}^{+0.03}$ & $-1.51_{-0.04}^{+0.04}$ & $0.07_{-0.01}^{+0.01}$ & $0.30$ & $657$ & $39$ & $26$ & $14$\\
	Hercules & $-0.04_{-0.05}^{+0.04}$ & $-0.34_{-0.04}^{+0.04}$ & $0.03_{-0.01}^{+0.01}$ & $0.55$ & $1536$ & $46$ & $36$ & $28$\\
	HorologiumI & $0.85_{-0.03}^{+0.04}$ & $-0.60_{-0.04}^{+0.04}$ & $0.16_{-0.03}^{+0.04}$ & $0.03$ & $116$ & $18$ & $16$ & $15$\\
	HorologiumII & $0.98_{-0.19}^{+0.18}$ & $-0.84_{-0.23}^{+0.21}$ & $0.06_{-0.02}^{+0.03}$ & $0.04$ & $89$ & $4$ & $4$ & $3$\\
	HydraII & $-0.37_{-0.14}^{+0.14}$ & $-0.03_{-0.10}^{+0.10}$ & $0.05_{-0.01}^{+0.01}$ & $0.04$ & $445$ & $21$ & $15$ & $12$\\
	HydrusI & $3.79_{-0.01}^{+0.01}$ & $-1.50_{-0.01}^{+0.01}$ & $0.07_{-0.01}^{+0.01}$ & $0.01$ & $1973$ & $125$ & $95$ & $61$\\
	\textcolor{red}{IndusI} & $0.78_{-1.84}^{+5.12}$ & $-0.85_{-1.62}^{+6.96}$ & $0.004_{-0.003}^{+0.007}$ & $0.16$ & $278$ & $2$ & $1$ & $1$\\
	\textcolor{orange}{IndusII} & $4.58_{-0.63}^{+0.36}$ & $-1.29_{-0.53}^{+0.34}$ & $0.02_{-0.01}^{+0.01}$ & $-0.13$ & $413$ & $6$ & $4$ & $1$\\
	LeoI & $-0.06_{-0.01}^{+0.01}$ & $-0.12_{-0.01}^{+0.01}$ & $0.20_{-0.01}^{+0.01}$ & $-0.46$ & $6775$ & $1342$ & $1331$ & $1310$\\
	LeoII & $-0.11_{-0.03}^{+0.03}$ & $-0.14_{-0.03}^{+0.03}$ & $0.08_{-0.00}^{+0.00}$ & $-0.30$ & $4260$ & $338$ & $336$ & $330$\\
	LeoIV & $-0.03_{-0.14}^{+0.14}$ & $-0.28_{-0.12}^{+0.11}$ & $0.08_{-0.02}^{+0.03}$ & $-0.19$ & $111$ & $8$ & $8$ & $6$\\
	LeoV & $0.10_{-0.21}^{+0.21}$ & $-0.41_{-0.15}^{+0.15}$ & $0.09_{-0.03}^{+0.03}$ & $-0.13$ & $95$ & $8$ & $8$ & $7$\\
	LeoT & $0.23_{-0.37}^{+0.36}$ & $-0.12_{-0.22}^{+0.22}$ & $0.07_{-0.02}^{+0.03}$ & $-0.34$ & $112$ & $8$ & $7$ & $7$\\
	\textcolor{red}{PegasusIII} & $2.13_{-4.53}^{+5.28}$ & $1.81_{-4.22}^{+5.37}$ & $0.01_{-0.01}^{+0.01}$ & $0.03$ & $139$ & $0$ & $0$ & $0$\\
	Phoenix & $0.08_{-0.03}^{+0.03}$ & $-0.06_{-0.04}^{+0.04}$ & $0.74_{-0.03}^{+0.03}$ & $-0.14$ & $283$ & $209$ & $207$ & $202$\\
	PhoenixII & $0.50_{-0.05}^{+0.05}$ & $-1.20_{-0.06}^{+0.06}$ & $0.11_{-0.03}^{+0.03}$ & $-0.45$ & $104$ & $11$ & $11$ & $9$\\
	PictorI & $0.15_{-0.08}^{+0.08}$ & $0.08_{-0.12}^{+0.12}$ & $0.05_{-0.02}^{+0.02}$ & $-0.20$ & $171$ & $9$ & $7$ & $6$\\
	PictorII & $1.15_{-0.06}^{+0.06}$ & $1.14_{-0.05}^{+0.06}$ & $0.02_{-0.01}^{+0.01}$ & $-0.05$ & $877$ & $13$ & $10$ & $4$\\
	\textcolor{red}{PiscesII} & $2.64_{-5.08}^{+4.93}$ & $2.50_{-5.09}^{+5.00}$ & $0.01_{-0.01}^{+0.01}$ & $0.02$ & $119$ & $0$ & $0$ & $0$\\
	PiscesII (spec, preferred) & $0.71_{-0.41}^{0.43}$ &  $-0.61_{-0.24}^{0.30}$ & $0.02_{-0.01}^{0.02}$ & 0.242 &  119  &  2 &    2 &  2  \\
	ReticulumII & $2.37_{-0.02}^{+0.02}$ & $-1.35_{-0.02}^{+0.02}$ & $0.14_{-0.02}^{+0.01}$ & $-0.13$ & $540$ & $75$ & $70$ & $53$\\
	\textcolor{orange}{ReticulumIII} & $0.31_{-0.14}^{+0.14}$ & $-0.61_{-0.30}^{+0.22}$ & $0.05_{-0.02}^{+0.02}$ & $0.07$ & $149$ & $7$ & $3$ & $3$\\
	SagittariusII & $-0.77_{-0.04}^{+0.04}$ & $-0.91_{-0.02}^{+0.02}$ & $0.04_{-0.01}^{+0.01}$ & $-0.02$ & $1767$ & $71$ & $65$ & $52$\\
	Sculptor & $0.099_{-0.002}^{+0.002}$ & $-0.159_{-0.002}^{+0.002}$ & $0.612_{-0.005}^{+0.005}$ & $-0.40$ & $11134$ & $6832$ & $6750$ & $6576$\\
	Segue1 & $-2.06_{-0.05}^{+0.05}$ & $-3.42_{-0.04}^{+0.05}$ & $0.13_{-0.03}^{+0.03}$ & $-0.44$ & $195$ & $25$ & $17$ & $9$\\
	Segue2 & $1.43_{-0.05}^{+0.06}$ & $-0.31_{-0.05}^{+0.04}$ & $0.12_{-0.02}^{+0.03}$ & $0.26$ & $187$ & $24$ & $20$ & $14$\\
	Sextans & $-0.40_{-0.01}^{+0.01}$ & $0.02_{-0.01}^{+0.01}$ & $0.098_{-0.002}^{+0.003}$ & $-0.37$ & $15315$ & $1504$ & $1362$ & $1148$\\
	TriangulumII & $0.58_{-0.05}^{+0.05}$ & $0.08_{-0.06}^{+0.06}$ & $0.03_{-0.01}^{+0.01}$ & $0.33$ & $581$ & $17$ & $13$ & $8$\\
	TucanaII & $0.91_{-0.02}^{+0.02}$ & $-1.27_{-0.03}^{+0.03}$ & $0.03_{-0.01}^{+0.01}$ & $-0.24$ & $1670$ & $39$ & $28$ & $17$\\
	TucanaIII & $-0.08_{-0.02}^{+0.02}$ & $-1.62_{-0.02}^{+0.02}$ & $0.12_{-0.02}^{+0.02}$ & $-0.24$ & $405$ & $51$ & $33$ & $19$\\
	TucanaIV & $0.56_{-0.04}^{+0.05}$ & $-1.69_{-0.05}^{+0.05}$ & $0.014_{-0.004}^{+0.004}$ & $-0.08$ & $1287$ & $14$ & $7$ & $6$\\
	\textcolor{red}{TucanaV} & $-0.13_{-0.20}^{+3.79}$ & $-1.15_{-0.13}^{+5.07}$ & $0.02_{-0.02}^{+0.02}$ & $0.34$ & $150$ & $3$ & $3$ & $2$\\
	Tucana~V (spec, preferred) & 
	$-0.14_{-0.05}^{+0.04}$ & $-1.18_{-0.06}^{+0.05}$ & $0.03_{-0.01}^{+0.02}$ & $0.22$ & $150$ & $4$ & $3$ & $3$ \\
	UrsaMajorI & $-0.40_{-0.03}^{+0.03}$ & $-0.63_{-0.04}^{+0.04}$ & $0.11_{-0.01}^{+0.01}$ & $-0.03$ & $476$ & $54$ & $50$ & $36$\\
	UrsaMajorII & $1.73_{-0.02}^{+0.02}$ & $-1.90_{-0.02}^{+0.02}$ & $0.020_{-0.003}^{+0.003}$ & $0.09$ & $3368$ & $65$ & $42$ & $22$\\
	UrsaMinor & $-0.124_{-0.004}^{+0.004}$ & $0.071_{-0.005}^{+0.005}$ & $0.151_{-0.003}^{+0.003}$ & $-0.08$ & $13921$ & $2116$ & $2009$ & $1811$\\
	\textcolor{red}{VirgoI} & $2.06_{-4.75}^{+5.54}$ & $2.34_{-5.03}^{+5.16}$ & $0.01_{-0.01}^{+0.01}$ & $-0.01$ & $100$ & $0$ & $0$ & $0$\\
	Willman1 & $0.28_{-0.07}^{+0.06}$ & $-1.11_{-0.07}^{+0.08}$ & $0.14_{-0.04}^{+0.04}$ & $-0.13$ & $87$ & $11$ & $11$ & $9$\\
	LeoA & $-0.06_{-0.09}^{+0.09}$ & $-0.06_{-0.08}^{+0.09}$ & $0.61_{-0.06}^{+0.06}$ & $-0.23$ & $69$ & $43$ & $41$ & $35$\\
	IC1613 & $0.04_{-0.02}^{+0.02}$ & $0.01_{-0.01}^{+0.01}$ & $0.55_{-0.02}^{+0.02}$ & $0.48$ & $788$ & $441$ & $418$ & $363$\\
    NGC6822 & $-0.06_{-0.01}^{+0.01}$ & $-0.07_{-0.01}^{+0.01}$ & $0.126_{-0.004}^{+0.004}$ & $0.32$ & $8611$ & $1073$ & $863$ & $730$\\
	Peg-dIrr & $0.15_{-0.14}^{+0.13}$ & $0.07_{-0.11}^{+0.12}$ & $0.14_{-0.05}^{+0.06}$ & $0.12$ & $51$ & $7$ & $5$ & $4$\\
	WLM & $0.09_{-0.03}^{+0.03}$ & $-0.07_{-0.02}^{+0.02}$ & $0.64_{-0.04}^{+0.04}$ & $0.20$ & $206$ & $134$ & $133$ & $117$\\
	Sg-dIrr & $0.11_{-0.18}^{+0.19}$ & $-0.37_{-0.17}^{+0.17}$ & $0.07_{-0.02}^{+0.02}$ & $0.23$ & $156$ & $10$ & $9$ & $9$\\
	UGC4879 & $-0.00_{-0.11}^{+0.11}$ & $-0.04_{-0.09}^{+0.09}$ & $0.66_{-0.15}^{+0.13}$ & $0.12$ & $11$ & $7$ & $7$ & $6$\\
	NGC3109 & $-0.04_{-0.03}^{+0.03}$ & $-0.01_{-0.03}^{+0.03}$ & $0.47_{-0.02}^{+0.02}$ & $0.03$ & $625$ & $301$ & $276$ & $227$\\
	SextansA & $-0.15_{-0.04}^{+0.05}$ & $-0.03_{-0.05}^{+0.04}$ & $0.72_{-0.04}^{+0.04}$ & $-0.20$ & $149$ & $109$ & $107$ & $87$\\
    M33 & $0.062_{-0.004}^{+0.004}$ & $0.011_{-0.003}^{+0.003}$ & $0.628_{-0.005}^{+0.005}$ & $0.06$ & $11240$ & $7172$ & $5949$ & $4721$\\
	SextansB & $-0.29_{-0.16}^{+0.16}$ & $-0.28_{-0.17}^{+0.17}$ & $0.38_{-0.08}^{+0.08}$ & $-0.51$ & $38$ & $15$ & $13$ & $12$\\    
	\hline
\end{longtable}

\begin{table*}
\caption{Zero-points and additional uncertainties from {\it Gaia} eDR3 systematics: $\mu_{\rm \alpha,*, QSO}$ (Col.~1) and $\mu_{\rm \delta, QSO}$ (Col.~2) are the weighted average of the PMs of QSO within 7deg from the galaxy's center, and "error" (Col.~3) is the additional uncertainty per proper motion component. Col.~4 indicates whether the uncertainties in the distance modulus (dm) or in the systemic PM (PM) are the dominant source of error; for the latter we consider both statistical and systematic uncertainties; "stat" or "syst" indicate whether statistical or systematic errors dominate PM uncertainties. For Crater~II in one dimension the statistical error is larger than the systematic one and viceversa. 
} \label{tab:systematics}
\tiny
\begin{tabular}{l|r|r|r|c}
\hline
\hline
  \multicolumn{1}{c|}{Galaxy} &
  \multicolumn{1}{c|}{$\mu_{\rm \alpha,*, QSO}$} &
  \multicolumn{1}{c|}{$\mu_{\rm \delta, QSO}$ } &
  \multicolumn{1}{c|}{error} &
  \multicolumn{1}{c}{Dominant source} \\
  \multicolumn{1}{c|}{ } &
  \multicolumn{1}{c|}{[mas yr$^{-1}$]} &
  \multicolumn{1}{c|}{[mas yr$^{-1}$]} &
  \multicolumn{1}{c|}{[mas yr$^{-1}$]} &
  \multicolumn{1}{c}{} \\
\hline
  AntliaII & 0.0040 & 0.0010 & 0.013 & PM \\
  AquariusII & -0.015 & -0.01 & 0.021 & PM-stat \\
  BootesI & -0.021 & 0.011 & 0.019 & dm \\
  BootesII & -0.017 & 0.011 & 0.021 & dm \\
  BootesIII & -0.023 & 0.01 & 0.016 & dm \\
  BootesIV & -0.0070 & -0.0070 & 0.02 & PM-stat \\
  CanesVenaticiI & 0.0050 & 0.0010 & 0.02 & PM-stat \\
  CanesVenaticiII & 0.015 & -0.0090 & 0.022 & PM-stat \\
  Carina & -0.01 & 0.0040 & 0.019 & dm \\
  CarinaII & 0.0040 & -0.0070 & 0.02 & dm \\
  CarinaIII & 0.0010 & -0.0060 & 0.022 & dm \\
  CentaurusI & 0.0010 & -0.0060 & 0.021 & PM-stat\\
  CetusII & -0.033 & -0.0090 & 0.021 & dm \\
  CetusIII & -0.0070 & -0.02 & 0.022 & PM-stat\\
  ColumbaI & -0.0010 & -0.014 & 0.021 & PM-stat \\
  ComaBerenices & 0.0010 & -0.017 & 0.02 & dm \\
  CraterI & 0.012 & -0.0030 & 0.022 & PM-stat \\
  CraterII & 0.026 & -0.0030 & 0.016 & PM-stat/syst \\
  Delve1 & 0.0010 & -0.013 & 0.022 & dm \\
  Delve2 & -0.0020 & -0.0040 & 0.022 & PM-stat \\
  DESJ0225+0304 & -0.017 & -0.025 & 0.022 & PM-stat\\
  Draco & 0.0070 & 0.0010 & 0.019 & PM-syst \\
  DracoII & 0.0010 & 0.0010 & 0.021 & PM-stat \\
  EridanusII & -0.0030 & -0.0040 & 0.021 & PM-stat \\
  EridanusIII & -0.0080 & -0.0060 & 0.022 & PM-stat \\
  Fornax & -0.0040 & -0.0070 & 0.017 & dm \\
  GrusI & -0.0040 & 0.0040 & 0.021 & PM-stat \\
  GrusII & -0.0010 & 0.012 & 0.02 & dm \\
  Hercules & 0.02 & -0.01 & 0.021 & PM-stat \\
  HorologiumI & -0.0 & -0.0080 & 0.022 & dm \\
  HorologiumII & 0.0020 & -0.0080 & 0.022 & PM-stat\\
  HydraII & 0.0090 & 0.01 & 0.022 & PM-stat\\
  HydrusI & 0.0030 & -0.0 & 0.02 & dm \\
  IndusI & 0.0020 & 0.018 & 0.022 & PM-stat \\
  IndusII & -0.0040 & 0.0030 & 0.021 & PM-stat\\
  LeoI & -0.0020 & 0.0050 & 0.021 & PM-syst\\
  LeoII & 0.01 & -0.014 & 0.021 & PM-stat\\
  LeoIV & -0.0010 & -0.021 & 0.022 & PM-stat\\
  LeoV & -0.0 & -0.021 & 0.022 & PM-stat\\
  LeoT & 0.0050 & 0.0080 & 0.022 & PM-stat\\
  PegasusIII & -0.021 & -0.0080 & 0.023 & PM-stat\\
  Phoenix & -0.01 & -0.0080 & 0.021 & PM-stat\\
  PhoenixII & -0.013 & 0.0020 & 0.022 & dm \\
   PictorI & -0.0040 & 0.0040 & 0.022 & PM-stat\\
    PictorII & 0.0010 & -0.0050 & 0.021 & dm \\
  PiscesII & -0.012 & -0.0020 & 0.022 & PM-stat\\
  ReticulumII & 0.0020 & -0.0050 & 0.02 & dm \\
  ReticulumIII & 0.0080 & -0.0030 & 0.021 & PM-stat\\
  SagittariusII & -0.0 & -0.0 & 0.022 & dm \\
  Sculptor & -0.0010 & -0.012 & 0.019 & PM-syst\\
  Segue1 & 0.0010 & 0.013 & 0.021 & dm \\
  Segue2 & 0.0010 & -0.014 & 0.022 & dm \\
  Sextans & -0.025 & 0.011 & 0.018 & PM/dm-syst\\
  TriangulumII & 0.016 & -0.021 & 0.022 & PM-syst\\
  TucanaII & -0.013 & 0.0040 & 0.019 & dm \\
  TucanaIII & -0.011 & 0.0050 & 0.02 & dm \\
  TucanaIV & -0.013 & 0.0050 & 0.019 & dm \\
  TucanaV & -0.01 & 0.01 & 0.021 & dm\\
  UrsaMajorI & 0.0050 & -0.0060 & 0.02 & PM/dm-stat \\
  UrsaMajorII & 0.0090 & -0.0070 & 0.019 & dm \\
  UrsaMinor & -0.0010 & 0.0050 & 0.018 & PM-syst\\
  VirgoI & -0.0060 & -0.0070 & 0.022 & PM-stat\\
  Willman1 & 0.0030 & -0.0070 & 0.021 & dm \\
  LeoA & 0.016 & 0.0050 & 0.021 & PM-stat\\
  IC1613 & -0.0030 & -0.0030 & 0.02 & PM-syst\\
  NGC6822 & -0.0080 & -0.016 & 0.02 & PM-syst\\
  Peg-dIrr & -0.0 & -0.0010 & 0.021 & PM-stat\\
  WLM & -0.0030 & 0.0030 & 0.021 & PM-stat\\
  Sg-dIrr & -0.0010 & 0.0020 & 0.023 & PM-stat\\
  UGC4879 & 0.016 & -0.012 & 0.022 & PM-stat\\
  NGC3109 & 0.0040 & -0.012 & 0.022 & PM-stat\\
  SextansA & -0.021 & 0.0060 & 0.022 & PM-stat\\
  M33 & 0.015 & 0.003 &  0.019 & PM-syst\\
  SextansB & -0.012 & 0.011 & 0.022 & PM-stat\\
\hline
\hline
\end{tabular}
\end{table*}

\onecolumn
\begin{landscape}
\tiny
\begin{longtable}[c]{l|c|c|c|c|c|c||c|c|c|c|c|c}
\caption{Orbital parameters for the case of the 2 isolated MW potentials as described in Section \ref{sec:orbits}): "Light MW" (light) and "Heavy MW" (heavy). For both of these models, are given: the pericenter (peri), apocenter (apo), eccentricity (ecc), periode (T), time since last pericenter (T$_{last,peri}$) and fraction of orbit reaching there apocenter in the last 8 Gyr ($\mathcal{F}_{apo}$). The values correspond to the median of these parameters calculated fro 100 Monte-Carlo realisations, and the uncertainties correspond to the $16^{th}$ and $84^{th}$ quantiles. When the majority of there orbits do not reach the apocenter, the uncertainties on the apocenter, eccentricity and period can not be computed. Thus, for those galaxies, we rather give the values of the $16^{th}$ quantile. These values are marked by a $\star$.  Galaxies with name in italics indicate those for which the uncertainty on the total Galactocentric velocity is $>$70 km s$^{-1}$ (see Sect.~\ref{sec:orbits}).
The systemic PM from the run with the spectroscopic information were used for Pisces~II and Tucana~V. Pegasus~III is omitted due to the lack of a trustworthy systemic PM.
}\label{tab:table_param}\\
\hline
\hline
\multicolumn{1}{c|}{Galaxy} &
\multicolumn{1}{c|}{Peri(light)} &
\multicolumn{1}{c|}{Apo(light)} &
\multicolumn{1}{c|}{ecc(light)} &
\multicolumn{1}{c|}{T(light)} &
\multicolumn{1}{c|}{T$_{last,peri}$(light)} &
\multicolumn{1}{c||}{$\mathcal{F}_{apo}$(light)} &
\multicolumn{1}{c|}{Peri(heavy)}&
\multicolumn{1}{c|}{Apo(heavy)} &
\multicolumn{1}{c|}{ecc(heavy)} &
\multicolumn{1}{c|}{T(heavy)} &
\multicolumn{1}{c|}{T$_{last,peri}$(heavy)} &
\multicolumn{1}{c}{$\mathcal{F}_{apo}$(heavy)} \\ 
\multicolumn{1}{c|}{} &
\multicolumn{1}{c|}{[kpc]} &
\multicolumn{1}{c|}{[kpc]} &
\multicolumn{1}{c|}{} &
\multicolumn{1}{c|}{[Gyr]} &
\multicolumn{1}{c|}{[Gyr]} &
\multicolumn{1}{c||}{} &
\multicolumn{1}{c|}{[kpc]} &
\multicolumn{1}{c|}{[kpc]} &
\multicolumn{1}{c|}{} &
\multicolumn{1}{c|}{[Gyr]} &
\multicolumn{1}{c|}{[Gyr]} &
\multicolumn{1}{c}{} \\ 
\hline
Antlia II        & $ 54.02_{-7.77}^{+10.34} $ & $ 157.70_{-10.29}^{+10.59} $  & $ 0.66_{-0.04}^{+0.04} $  & $ 3.62_{-0.44}^{+0.44} $ & $ -0.99_{-0.05}^{+0.07} $  & $ 1.00 $ & $ 38.80_{-5.45}^{+6.46} $ & $ 143.37_{-8.57}^{+6.68} $  & $ 0.73_{-0.04}^{+0.03} $  & $ 2.20_{-0.19}^{+0.17} $ & $ -0.77_{-0.05}^{+0.05} $ & $ 1.00 $ \\ 
{\it \textcolor{orange}{Aquarius II}}     & $ 74.82_{-50.36}^{+25.95} $ & $ 123.67_{-15.22}^{+178.71} $  & $ 0.62_{-0.26}^{+0.30} $  & $ 3.06_{-1.04}^{+6.42} $ & $ -0.69_{-0.20}^{+0.50} $  & $ 0.86 $ & $ 48.64_{-30.16}^{+44.82} $ & $ 110.46_{-7.04}^{+17.71} $  & $ 0.59_{-0.26}^{+0.23} $  & $ 1.88_{-0.49}^{+1.08} $ & $ -0.64_{-0.11}^{+0.12} $ & $ 0.99 $ \\ 
Bootes I         & $ 41.93_{-9.73}^{+11.03} $ & $ 108.29_{-19.87}^{+36.07} $  & $ 0.62_{-0.01}^{+0.03} $  & $ 2.25_{-0.49}^{+1.11} $ & $ -0.30_{-0.02}^{+0.04} $  & $ 1.00 $ & $ 33.70_{-7.45}^{+8.78} $ & $ 77.00_{-8.13}^{+10.72} $  & $ 0.56_{-0.05}^{+0.06} $  & $ 1.18_{-0.18}^{+0.25} $ & $ -0.30_{-0.01}^{+0.02} $ & $ 1.00 $ \\ 
Bootes II        & $ 38.74_{-1.55}^{+2.03} $ & $ - $  & $ - $  & $ - $ & $ - $  & $ 0.11 $ & $ 38.52_{-2.10}^{+2.02} $ & $ 158.38_{-46.47}^{+73.30} $  & $ 0.75_{-0.09}^{+0.08} $  & $ 2.44_{-0.73}^{+1.25} $ & $ -2.14_{-1.18}^{+0.80} $ & $ 0.98 $ \\ 
Bootes III       & $ 8.63_{-2.01}^{+1.97} $ & $ 188.00_{-23.01}^{+47.29} $  & $ 0.96_{-0.01}^{+0.00} $  & $ 3.43_{-0.54}^{+1.14} $ & $ -0.14_{-0.01}^{+0.01} $  & $ 1.00 $ & $ 7.11_{-1.73}^{+2.01} $ & $ 93.83_{-6.79}^{+9.31} $  & $ 0.92_{-0.01}^{+0.02} $  & $ 1.13_{-0.10}^{+0.14} $ & $ -0.13_{-0.01}^{+0.01} $ & $ 1.00 $ \\ 
Canes Venatici I  & $ 68.09_{-42.17}^{+71.49} $ & $ 301.35_{-28.15}^{-} $  & $ 0.85_{-0.12}^{-} $  & $ 7.95_{-2.01}^{-} $ & $ -1.39_{-0.09}^{+0.14} $  & $ 0.75 $ & $ 47.90_{-27.20}^{+47.98} $ & $ 248.47_{-12.70}^{+21.86} $  & $ 0.80_{-0.18}^{+0.11} $  & $ 4.04_{-0.41}^{+1.39} $ & $ -1.17_{-0.10}^{+0.08} $ & $ 1.00 $ \\ 
{\it Canes Venatici II} & $ 49.44_{-32.83}^{+60.85} $ & $ 243.36_{-41.88}^{-} $  & $ 0.88_{-0.12}^{-} $  & $ 5.52_{-1.45}^{-} $ & $ -3.20_{-1.53}^{-} $  & $ 0.80 $ & $ 36.20_{-22.02}^{+48.82} $ & $ 195.03_{-15.80}^{+33.87} $  & $ 0.82_{-0.16}^{+0.11} $  & $ 2.99_{-0.50}^{+1.40} $ & $ -2.08_{-0.93}^{+0.33} $ & $ 0.97 $ \\ 
Carina          & $ 106.66_{-5.41}^{+6.43} $ & $ 247.96_{-109.45}^{-} $  & $ 0.56_{-0.29}^{-} $  & $ 7.10_{-2.92}^{-} $ & $ -0.00_{-0.01}^{+0.00} $  & $ 0.71 $ & $ 102.77_{-31.65}^{+10.10} $ & $ 107.87_{-6.01}^{+43.66} $  & $ 0.19_{-0.14}^{+0.14} $  & $ 2.58_{-0.53}^{+1.28} $ & $ -1.02_{-0.32}^{+1.01} $ & $ 1.00 $ \\ 
Carina II        & $ 28.24_{-2.24}^{+2.65} $ & $ - $  & $ - $  & $ - $ & $ -0.07_{-0.00}^{+0.00} $  & $ 0.01 $ & $ 26.94_{-2.35}^{+2.73} $ & $ 162.27_{-35.10}^{+54.01} $  & $ 0.83_{-0.03}^{+0.03} $  & $ 2.35_{-0.57}^{+1.00} $ & $ -0.08_{-0.00}^{+0.00} $ & $ 1.00 $ \\ 
Carina III       & $ 28.69_{-1.33}^{+1.24} $ & $ - $  & $ - $  & $ - $ & $ -0.01_{-0.00}^{+0.00} $  & $ 0.05 $ & $ 28.65_{-1.36}^{+1.24} $ & $ 132.69_{-35.28}^{+59.16} $  & $ 0.78_{-0.07}^{+0.06} $  & $ 1.91_{-0.52}^{+0.96} $ & $ -0.01_{-0.00}^{+0.00} $ & $ 0.99 $ \\ 
{\it Columba I}        & $ 183.32_{-10.13}^{+7.72} $ & $ - $  & $ - $  & $ - $ & $ - $  & $ 0.11 $ & $ 179.93_{-28.97}^{+9.48} $ & $  \geq 206.30^{\star} $  & $  \geq 0.38^{\star} $  & $  \geq 5.23^{\star} $ & $ \geq -2.17^{\star} $ & $ 0.45 $ \\ 
Coma Berenices   & $ 45.96_{-3.64}^{+1.86} $ & $ 164.92_{-47.69}^{+99.67} $  & $ 0.72_{-0.08}^{+0.11} $  & $ 3.48_{-1.10}^{+2.66} $ & $ -0.03_{-0.01}^{+0.01} $  & $ 0.91 $ & $ 42.32_{-2.85}^{+2.64} $ & $ 67.33_{-12.46}^{+17.15} $  & $ 0.37_{-0.09}^{+0.10} $  & $ 1.16_{-0.19}^{+0.27} $ & $ -0.06_{-0.02}^{+0.02} $ & $ 1.00 $ \\ 
{\it Crater I}         & $ 99.64_{-72.58}^{+45.39} $ & $ 151.64_{-10.49}^{-} $  & $ 0.63_{-0.36}^{-} $  & $ 4.20_{-1.39}^{-} $ & $ -1.49_{-0.85}^{-} $  & $ 0.79 $ & $ 59.01_{-39.75}^{+79.12} $ & $ 147.87_{-9.33}^{+14.43} $  & $ 0.65_{-0.35}^{+0.24} $  & $ 2.48_{-0.47}^{+1.97} $ & $ -1.21_{-0.55}^{+0.23} $ & $ 0.95 $ \\ 
Crater II        & $ 39.08_{-7.65}^{+8.03} $ & $ 148.72_{-10.41}^{+10.93} $  & $ 0.73_{-0.04}^{+0.05} $  & $ 3.06_{-0.24}^{+0.38} $ & $ -2.27_{-0.38}^{+0.20} $  & $ 1.00 $ & $ 27.42_{-4.83}^{+5.20} $ & $ 130.54_{-7.74}^{+6.60} $  & $ 0.79_{-0.03}^{+0.04} $  & $ 1.86_{-0.14}^{+0.14} $ & $ -1.28_{-0.11}^{+0.11} $ & $ 1.00 $ \\ 
Draco           & $ 51.68_{-6.06}^{+4.07} $ & $ 137.59_{-17.40}^{+14.19} $  & $ 0.63_{-0.01}^{+0.01} $  & $ 3.06_{-0.49}^{+0.35} $ & $ -2.67_{-0.38}^{+0.47} $  & $ 1.00 $ & $ 37.56_{-4.40}^{+4.16} $ & $ 98.87_{-8.06}^{+5.03} $  & $ 0.62_{-0.02}^{+0.03} $  & $ 1.53_{-0.17}^{+0.11} $ & $ -1.15_{-0.09}^{+0.15} $ & $ 1.00 $ \\ 
Draco II         & $ 15.00_{-0.60}^{+5.14} $ & $ 190.69_{-39.41}^{+74.09} $  & $ 0.92_{-0.05}^{+0.02} $  & $ 3.65_{-0.89}^{+1.87} $ & $ -3.15_{-1.35}^{+0.96} $  & $ 0.94 $ & $ 19.52_{-1.09}^{+0.89} $ & $ 64.60_{-6.78}^{+8.92} $  & $ 0.70_{-0.02}^{+0.03} $  & $ 0.87_{-0.10}^{+0.12} $ & $ -0.82_{-0.12}^{+0.08} $ & $ 1.00 $ \\ 
{\it Eridanus II}      & $ 351.10_{-37.21}^{+24.87} $ & $ - $  & $ - $  & $ - $ & $ - $  & $ 0.00 $ & $ 348.21_{-62.42}^{+25.84} $ & $ - $  & $ - $  & $ - $ & $ - $ & $ 0.09 $ \\ 
Fornax          & $ 89.38_{-26.44}^{+30.76} $ & $ 160.41_{-12.52}^{+32.15} $  & $ 0.46_{-0.07}^{+0.12} $  & $ 4.36_{-0.77}^{+1.95} $ & $ -2.81_{-1.20}^{+0.79} $  & $ 0.98 $ & $ 56.25_{-15.32}^{+21.67} $ & $ 146.87_{-6.87}^{+7.72} $  & $ 0.61_{-0.13}^{+0.09} $  & $ 2.50_{-0.27}^{+0.41} $ & $ -1.53_{-0.38}^{+0.19} $ & $ 1.00 $ \\ 
Grus I           & $ 25.19_{-11.46}^{+20.99} $ & $  \geq 436.98^{\star} $  & $  \geq 0.97^{\star} $  & $  \geq 11.02^{\star} $ & $ - $  & $ 0.22 $ & $ 19.29_{-8.58}^{+17.00} $ & $ 228.23_{-19.85}^{+36.73} $  & $ 0.91_{-0.05}^{+0.04} $  & $ 3.31_{-0.30}^{+0.91} $ & $ -2.87_{-0.87}^{+0.29} $ & $ 1.00 $ \\ 
Grus II          & $ 31.14_{-6.03}^{+7.55} $ & $ 101.74_{-20.41}^{+56.89} $  & $ 0.70_{-0.01}^{+0.05} $  & $ 1.94_{-0.47}^{+1.30} $ & $ -1.76_{-1.21}^{+0.48} $  & $ 0.99 $ & $ 24.65_{-5.62}^{+8.54} $ & $ 68.38_{-7.76}^{+14.46} $  & $ 0.63_{-0.04}^{+0.05} $  & $ 0.97_{-0.15}^{+0.29} $ & $ -0.76_{-0.31}^{+0.14} $ & $ 1.00 $ \\ 
Hercules        & $ 64.22_{-19.19}^{+16.39} $ & $ \geq 342.16^{\star} $  & $  \geq 0.87^{\star} $  & $  \geq 8.40^{ \star} $ & $ -0.56_{-0.05}^{+0.07} $  & $ 0.33 $ & $ 49.16_{-15.56}^{+16.25} $ & $ 214.43_{-24.30}^{+56.66} $  & $ 0.77_{-0.03}^{+0.05} $  & $ 3.49_{-0.59}^{+1.60} $ & $ -0.53_{-0.05}^{+0.05} $ & $ 1.00 $ \\ 
Horologium I     & $ 74.15_{-13.02}^{+8.96} $ & $ 138.81_{-53.92}^{+190.26} $  & $ 0.47_{-0.14}^{+0.29} $  & $ 3.56_{-1.43}^{+6.11} $ & $ -2.24_{-1.46}^{-} $  & $ 0.87 $ & $ 61.23_{-21.47}^{+19.06} $ & $ 83.73_{-8.20}^{+28.75} $  & $ 0.31_{-0.07}^{+0.18} $  & $ 1.62_{-0.38}^{+0.69} $ & $ -1.09_{-1.00}^{+0.35} $ & $ 1.00 $ \\ 
{\it Horologium II}    & $ 78.90_{-7.59}^{+7.60} $ & $ \geq 303.60^{\star} $  & $  \geq 0.71^{\star} $  & $  \geq 8.33^{\star} $ & $ -0.02_{-0.05}^{+0.02} $  & $ 0.18 $ & $ 78.81_{-10.46}^{+6.98} $ & $ 253.48_{-140.76}^{-} $  & $ 0.65_{-0.25}^{-} $  & $ 4.66_{-2.35}^{-} $ & $ -0.03_{-0.50}^{+0.02} $ & $ 0.70 $ \\ 
{\it Hydra II}         & $ 104.24_{-57.40}^{+30.62} $ & $  \geq 324.70^{\star} $  & $  \geq 0.88^{\star} $  & $  \geq 8.09^{\star} $ & $ -0.51_{-0.23}^{+0.25} $  & $ 0.23 $ & $ 86.15_{-50.31}^{+43.74} $ & $ 306.08_{-101.61}^{-} $  & $ 0.81_{-0.09}^{-} $  & $ 6.12_{-2.93}^{-} $ & $ -0.57_{-0.09}^{+0.27} $ & $ 0.71 $ \\ 
Hydrus I         & $ 25.58_{-1.89}^{+1.22} $ & $  \geq 228.88^{\star} $  & $  \geq 0.90^{\star} $  & $  \geq 4.64^{\star} $ & $ \geq -2.98^{\star} $  & $ 0.31 $ & $ 25.53_{-1.66}^{+1.24} $ & $ 102.91_{-35.47}^{+43.88} $  & $ 0.75_{-0.10}^{+0.07} $  & $ 1.44_{-0.49}^{+0.66} $ & $ -1.43_{-0.66}^{+0.50} $ & $ 1.00 $ \\ 
Leo I            & $ 46.56_{-26.54}^{+30.50} $ & $ - $  & $ - $  & $ - $ & $ -1.20_{-0.06}^{+0.05} $  & $ 0.00 $ & $ 34.97_{-19.84}^{+23.94} $ & $ 643.59_{-69.57}^{-} $  & $ 0.98_{-0.05}^{-} $  & $ 12.62_{-2.10}^{-} $ & $ -1.06_{-0.05}^{+0.05} $ & $ 0.57 $ \\ 
Leo II           & $ 115.55_{-58.87}^{+88.35} $ & $ 233.99_{-14.60}^{-} $  & $ 0.64_{-0.20}^{-} $  & $ 6.86_{-1.45}^{-} $ & $ -2.29_{-0.35}^{+1.32} $  & $ 0.77 $ & $ 69.01_{-29.16}^{+63.85} $ & $ 222.17_{-9.48}^{+13.47} $  & $ 0.68_{-0.28}^{+0.13} $  & $ 3.99_{-0.47}^{+1.25} $ & $ -1.67_{-0.26}^{+0.16} $ & $ 1.00 $ \\ 
{\it Leo IV}           & $ 143.17_{-107.12}^{+16.37} $ & $ 189.69_{-39.56}^{-} $  & $ 0.89_{-0.46}^{-} $  & $ 7.17_{-4.02}^{-} $ & $ -0.02_{-1.77}^{-} $  & $ 0.60 $ & $ 93.48_{-68.86}^{+63.83} $ & $ 159.36_{-12.30}^{+240.81} $  & $ 0.67_{-0.44}^{+0.29} $  & $ 3.07_{-0.90}^{+5.94} $ & $ -1.16_{-0.75}^{-} $ & $ 0.86 $ \\ 
{\it Leo V}            & $ 171.65_{-15.33}^{+12.00} $ & $ - $  & $ - $  & $ - $ & $ -0.09_{-0.39}^{+0.06} $  & $ 0.12 $ & $ 170.22_{-23.25}^{+12.38} $ & $  \geq 246.13^{\star} $  & $  \geq 0.50^{\star} $  & $  \geq 5.81^{\star} $ & $ -0.11_{-0.80}^{+0.08} $ & $ 0.31 $ \\ 
{\it Leo T}            & $ 409.73_{-22.91}^{+37.36} $ & $ - $  & $ - $  & $ - $ & $ - $  & $ 0.00 $ & $ 409.12_{-22.60}^{+37.52} $ & $ - $  & $ - $  & $ - $ & $ - $ & $ 0.02 $ \\ 
{\it Phoenix}         & $ 212.77_{-134.79}^{+119.84} $ & $ - $  & $ - $  & $ - $ & $ - $  & $ 0.00 $ & $ 150.95_{-94.95}^{+143.40} $ & $  \geq 667.84^{\star} $  & $ 0.93^{\star} $  & $  \geq 14.04^{\star} $ & $ - $ & $ 0.23 $ \\ 
Phoenix II       & $ 80.09_{-8.40}^{+5.70} $ & $ \geq 421.49^{\star} $  & $  \geq 0.83^{\star} $  & $  \geq 11.96^{\star} $ & $ - $  & $ 0.16 $ & $ 79.55_{-10.09}^{+6.01} $ & $ 247.37_{-131.03}^{-} $  & $ 0.68_{-0.28}^{-} $  & $ 4.59_{-2.38}^{-} $ & $ -1.55_{-2.46}^{-} $ & $ 0.80 $ \\ 
{\it Pisces II}        & $ 179.16_{-15.52}^{+15.73} $ & $ - $  & $ - $  & $ - $ & $ - $  & $ 0.00 $ & $ 178.93_{-15.31}^{+15.41} $ & $ - $  & $ - $  & $ - $ & $ - $ & $ 0.00 $ \\ 
Reticulum II     & $ 28.29_{-2.87}^{+2.12} $ & $ 82.36_{-21.96}^{+36.16} $  & $ 0.65_{-0.07}^{+0.09} $  & $ 1.57_{-0.43}^{+0.71} $ & $ -1.52_{-0.71}^{+0.47} $  & $ 1.00 $ & $ 25.46_{-4.09}^{+3.20} $ & $ 47.18_{-6.28}^{+9.57} $  & $ 0.48_{-0.01}^{+0.02} $  & $ 0.72_{-0.11}^{+0.15} $ & $ -0.62_{-0.16}^{+0.14} $ & $ 1.00 $ \\ 
{\it \textcolor{orange}{Reticulum III}}   & $ 81.90_{-34.84}^{+13.23} $ & $  \geq 138.43^{\star} $  & $  \geq 0.67^{\star} $  & $  \geq 3.03^{\star} $ & $ -0.20_{-0.23}^{+0.13} $  & $ 0.41 $ & $ 76.81_{-41.24}^{+15.12} $ & $ 165.70_{-63.35}^{-} $  & $ 0.63_{-0.13}^{-} $  & $ 3.08_{-1.51}^{-} $ & $ -0.29_{-0.15}^{+0.21} $ & $ 0.77 $ \\
Sagittarius II   & $ 40.85_{-3.87}^{+17.25} $ & $ 220.27_{-97.14}^{-} $  & $ 0.81_{-0.11}^{-} $  & $ 4.92_{-2.46}^{-} $ & $ -1.93_{-1.88}^{-} $  & $ 0.72 $ & $ 43.59_{-11.32}^{+10.51} $ & $ 98.26_{-22.84}^{+40.29} $  & $ 0.58_{-0.02}^{+0.04} $  & $ 1.60_{-0.45}^{+0.74} $ & $ -1.36_{-0.77}^{+0.45} $ & $ 1.00 $ \\ 
Sculptor        & $ 63.65_{-3.09}^{+4.47} $ & $ 147.35_{-11.59}^{+15.79} $  & $ 0.57_{-0.01}^{+0.02} $  & $ 3.46_{-0.32}^{+0.46} $ & $ -0.38_{-0.05}^{+0.06} $  & $ 1.00 $ & $ 48.74_{-3.51}^{+3.81} $ & $ 98.69_{-4.37}^{+6.17} $  & $ 0.50_{-0.03}^{+0.03} $  & $ 1.66_{-0.08}^{+0.13} $ & $ -0.45_{-0.03}^{+0.04} $ & $ 1.00 $ \\ 
Segue 1          & $ 20.18_{-5.83}^{+4.29} $ & $ 68.39_{-25.70}^{+52.65} $  & $ 0.71_{-0.04}^{+0.09} $  & $ 1.21_{-0.49}^{+1.03} $ & $ -0.09_{-0.02}^{+0.02} $  & $ 0.99 $ & $ 18.51_{-6.06}^{+4.72} $ & $ 42.10_{-8.77}^{+14.18} $  & $ 0.58_{-0.02}^{+0.06} $  & $ 0.59_{-0.15}^{+0.22} $ & $ -0.09_{-0.01}^{+0.01} $ & $ 1.00 $ \\ 
Segue 2          & $ 17.18_{-3.10}^{+4.71} $ & $ 45.51_{-4.93}^{+2.72} $  & $ 0.61_{-0.08}^{+0.04} $  & $ 0.76_{-0.10}^{+0.11} $ & $ -0.26_{-0.02}^{+0.02} $  & $ 1.00 $ & $ 13.33_{-2.07}^{+3.02} $ & $ 42.49_{-3.67}^{+2.76} $  & $ 0.68_{-0.05}^{+0.03} $  & $ 0.55_{-0.06}^{+0.05} $ & $ -0.21_{-0.02}^{+0.02} $ & $ 1.00 $ \\ 
Sextans         & $ 74.45_{-5.68}^{+4.38} $ & $ 257.24_{-53.90}^{+49.89} $  & $ 0.71_{-0.05}^{+0.04} $  & $ 6.62_{-1.69}^{+1.72} $ & $ -0.28_{-0.02}^{+0.02} $  & $ 0.96 $ & $ 64.03_{-6.48}^{+4.97} $ & $ 121.11_{-11.43}^{+10.84} $  & $ 0.48_{-0.00}^{+0.01} $  & $ 2.21_{-0.27}^{+0.24} $ & $ -0.40_{-0.02}^{+0.03} $ & $ 1.00 $ \\ 
Triangulum II    & $ 19.53_{-0.90}^{+0.81} $ & $ 208.95_{-18.02}^{+23.72} $  & $ 0.91_{-0.00}^{+0.01} $  & $ 4.03_{-0.48}^{+0.59} $ & $ -4.01_{-0.61}^{+0.50} $  & $ 1.00 $ & $ 11.33_{-0.80}^{+0.90} $ & $ 85.93_{-3.24}^{+3.82} $  & $ 0.87_{-0.01}^{+0.01} $  & $ 1.06_{-0.04}^{+0.06} $ & $ -0.97_{-0.06}^{+0.05} $ & $ 1.00 $ \\ 
\hline
\end{longtable}

\begin{longtable}[c]{l|c|c|c|c|c|c||c|c|c|c|c|c}
\caption{Continued.}\label{table_param}\\
\hline
\multicolumn{1}{c|}{Galaxy} &
\multicolumn{1}{c|}{Peri(light)} &
\multicolumn{1}{c|}{Apo(light)} &
\multicolumn{1}{c|}{ecc(light)} &
\multicolumn{1}{c|}{T(light)} &
\multicolumn{1}{c|}{T$_{last,peri}$(light)} &
\multicolumn{1}{c||}{$\mathcal{F}_{apo}$(light)} &
\multicolumn{1}{c|}{Peri(heavy)}&
\multicolumn{1}{c|}{Apo(heavy)} &
\multicolumn{1}{c|}{ecc(heavy)} &
\multicolumn{1}{c|}{T(heavy)} &
\multicolumn{1}{c|}{T$_{last,peri}$(heavy)} &
\multicolumn{1}{c}{$\mathcal{F}_{apo}$(heavy)} \\ 
\multicolumn{1}{c|}{} &
\multicolumn{1}{c|}{[kpc]} &
\multicolumn{1}{c|}{[kpc]} &
\multicolumn{1}{c|}{} &
\multicolumn{1}{c|}{[Gyr]} &
\multicolumn{1}{c|}{[Gyr]} &
\multicolumn{1}{c||}{} &
\multicolumn{1}{c|}{[kpc]} &
\multicolumn{1}{c|}{[kpc]} &
\multicolumn{1}{c|}{} &
\multicolumn{1}{c|}{[Gyr]} &
\multicolumn{1}{c|}{[Gyr]} &
\multicolumn{1}{c}{} \\ 
\hline
Tucana II        & $ 38.24_{-5.87}^{+7.71} $ & $  \geq 253.78^{\star} $  & $  \geq 0.87^{\star} $  & $  \geq 5.48^{\star} $ & $  \geq -3.23^{\star} $  & $ 0.27 $ & $ 34.49_{-8.30}^{+8.74} $ & $ 142.65_{-44.09}^{+94.58} $  & $ 0.76_{-0.03}^{+0.05} $  & $ 2.14_{-0.75}^{+1.70} $ & $ -1.91_{-1.04}^{+0.72} $ & $ 0.95 $ \\ 
Tucana III       & $ 4.01_{-0.19}^{+0.16} $ & $ 45.81_{-3.03}^{+3.25} $  & $ 0.91_{-0.01}^{+0.00} $  & $ 0.64_{-0.05}^{+0.04} $ & $ -0.58_{-0.05}^{+0.05} $  & $ 1.00 $ & $ 2.84_{-0.17}^{+0.19} $ & $ 34.35_{-1.85}^{+2.04} $  & $ 0.92_{-0.01}^{+0.00} $  & $ 0.37_{-0.02}^{+0.02} $ & $ -0.31_{-0.02}^{+0.02} $ & $ 1.00 $ \\ 
Tucana IV        & $ 32.23_{-6.67}^{+6.44} $ & $ 83.72_{-24.14}^{+51.93} $  & $ 0.63_{-0.05}^{+0.09} $  & $ 1.65_{-0.52}^{+1.12} $ & $ -1.41_{-0.80}^{+0.64} $  & $ 0.95 $ & $ 27.40_{-7.59}^{+8.94} $ & $ 55.25_{-7.39}^{+12.92} $  & $ 0.52_{-0.05}^{+0.09} $  & $ 0.83_{-0.16}^{+0.26} $ & $ -0.64_{-0.28}^{+0.16} $ & $ 1.00 $ \\ 
Tucana V         & $ 23.30_{-7.28}^{+3.59} $ & $ 130.25_{-41.10}^{+50.98} $  & $ 0.82_{-0.02}^{+0.05} $  & $ 2.40_{-0.91}^{+1.20} $ & $ -2.22_{-1.20}^{+0.84} $  & $ 1.00 $ & $ 21.36_{-6.99}^{+5.91} $ & $ 78.58_{-15.79}^{+13.30} $  & $ 0.73_{-0.03}^{+0.05} $  & $ 1.07_{-0.27}^{+0.23} $ & $ -0.89_{-0.23}^{+0.24} $ & $ 1.00 $ \\
Ursa Major I      & $ 72.22_{-26.81}^{+31.00} $ & $ 102.34_{-5.00}^{+10.61} $  & $ 0.32_{-0.22}^{+0.25} $  & $ 2.52_{-0.42}^{+1.50} $ & $ -1.16_{-0.33}^{+0.71} $  & $ 0.98 $ & $ 42.68_{-14.05}^{+17.72} $ & $ 101.93_{-4.60}^{+4.67} $  & $ 0.58_{-0.17}^{+0.12} $  & $ 1.63_{-0.20}^{+0.29} $ & $ -0.81_{-0.12}^{+0.10} $ & $ 1.00 $ \\ 
Ursa Major II     & $ 39.60_{-2.70}^{+2.28} $ & $ 277.93_{-140.47}^{-} $  & $ 0.86_{-0.13}^{-} $  & $ 6.35_{-3.64}^{-} $ & $  \geq -3.79^{\star} $  & $ 0.64 $ & $ 38.93_{-3.33}^{+2.54} $ & $ 83.79_{-23.28}^{+33.72} $  & $ 0.54_{-0.13}^{+0.11} $  & $ 1.33_{-0.34}^{+0.52} $ & $ -1.27_{-0.55}^{+0.35} $ & $ 1.00 $ \\ 
Ursa Minor       & $ 48.85_{-3.28}^{+3.44} $ & $ 109.94_{-5.40}^{+9.03} $  & $ 0.56_{-0.01}^{+0.01} $  & $ 2.47_{-0.17}^{+0.14} $ & $ -2.03_{-0.22}^{+0.15} $  & $ 1.00 $ & $ 34.92_{-2.69}^{+2.56} $ & $ 88.07_{-4.37}^{+4.87} $  & $ 0.61_{-0.02}^{+0.02} $  & $ 1.34_{-0.07}^{+0.10} $ & $ -0.93_{-0.08}^{+0.05} $ & $ 1.00 $ \\ 
Willman 1        & $ 25.74_{-5.40}^{+10.57} $ & $ 43.26_{-6.85}^{+9.89} $  & $ 0.38_{-0.10}^{+0.10} $  & $ 0.89_{-0.20}^{+0.35} $ & $ -0.38_{-0.08}^{+0.08} $  & $ 1.00 $ & $ 17.47_{-3.12}^{+6.67} $ & $ 43.14_{-6.58}^{+8.71} $  & $ 0.58_{-0.06}^{+0.06} $  & $ 0.59_{-0.10}^{+0.16} $ & $ -0.29_{-0.07}^{+0.05} $ & $ 1.00 $ \\ 
NGC 6822         & $ 393.15_{-160.02}^{+72.84} $ & $ - $  & $ - $  & $ - $ & $ -1.98_{-2.46}^{+1.27} $  & $ 0.01 $ & $ 282.00_{-165.56}^{+149.05} $ & $  \geq 521.16^{\star} $  & $  \geq 0.58^{\star} $  & $  \geq 12.91^{\star} $ & $ -3.33_{-0.59}^{+2.05} $ & $ 0.37 $ \\ 
\hline
\hline
\end{longtable}

\normalsize

\end{landscape}
\twocolumn

\section{Plots on tests and validation}
Here we include plots showing the systemic PMs of the galaxies in the sample derived in different ways.

\begin{figure*}
   \centering
  \includegraphics[width=\textwidth]{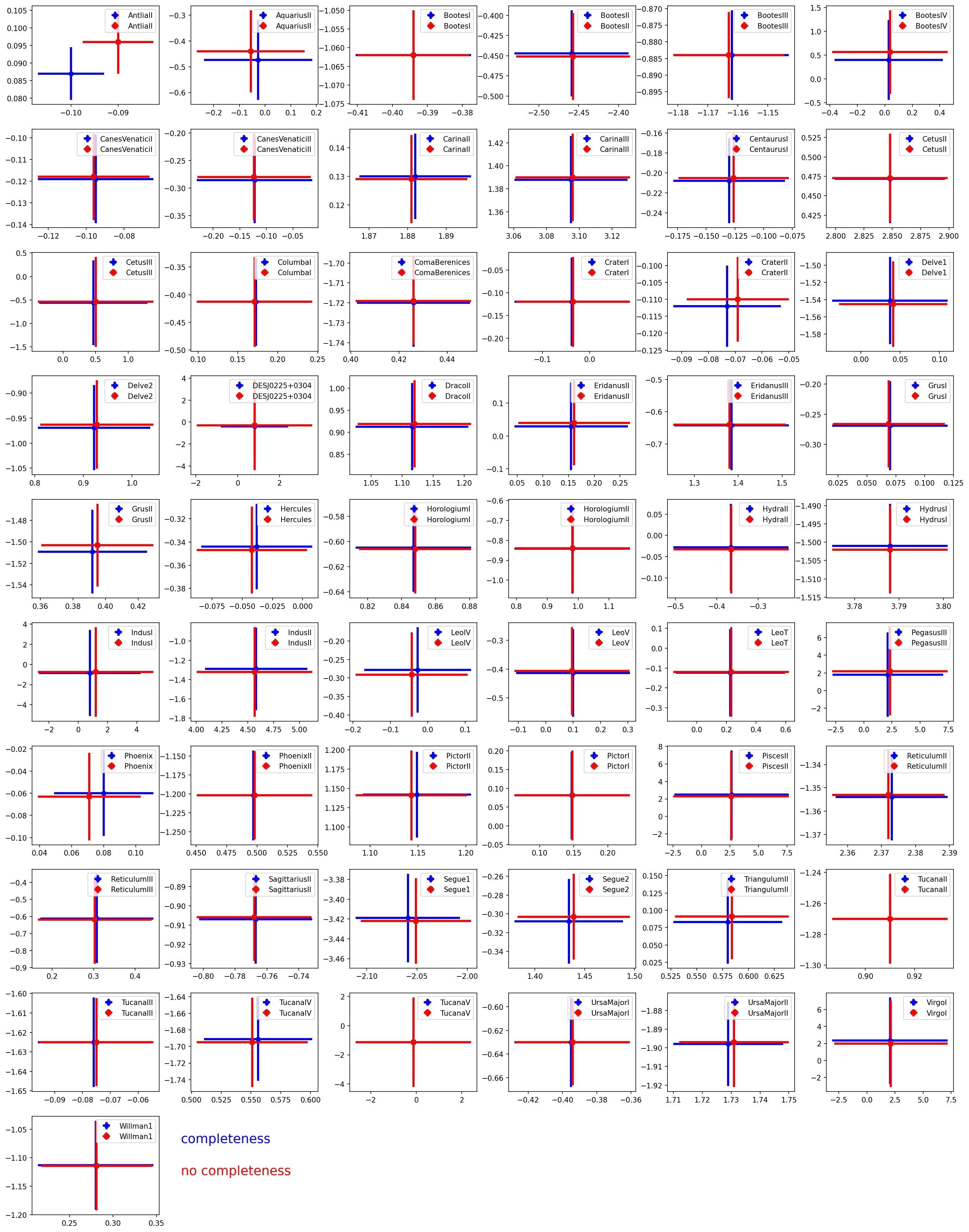}
      \caption{Comparison of systemic PMs determined with (blue) and without (red) the correction for the photometric completeness of eGDR3 data, for the systems whose CMD probability distribution was calculated using a synthetic CMD (see Tab.~\ref{tab:method}), in the run not including spectroscopic information.  The x-axis and y-axis show the $\mu_{\alpha,*}$ and $\mu_{\delta}$ component, respectively.
              }
         \label{fig:compl}
   \end{figure*}

   \begin{figure*}
   \centering
  \includegraphics[width=\textwidth]{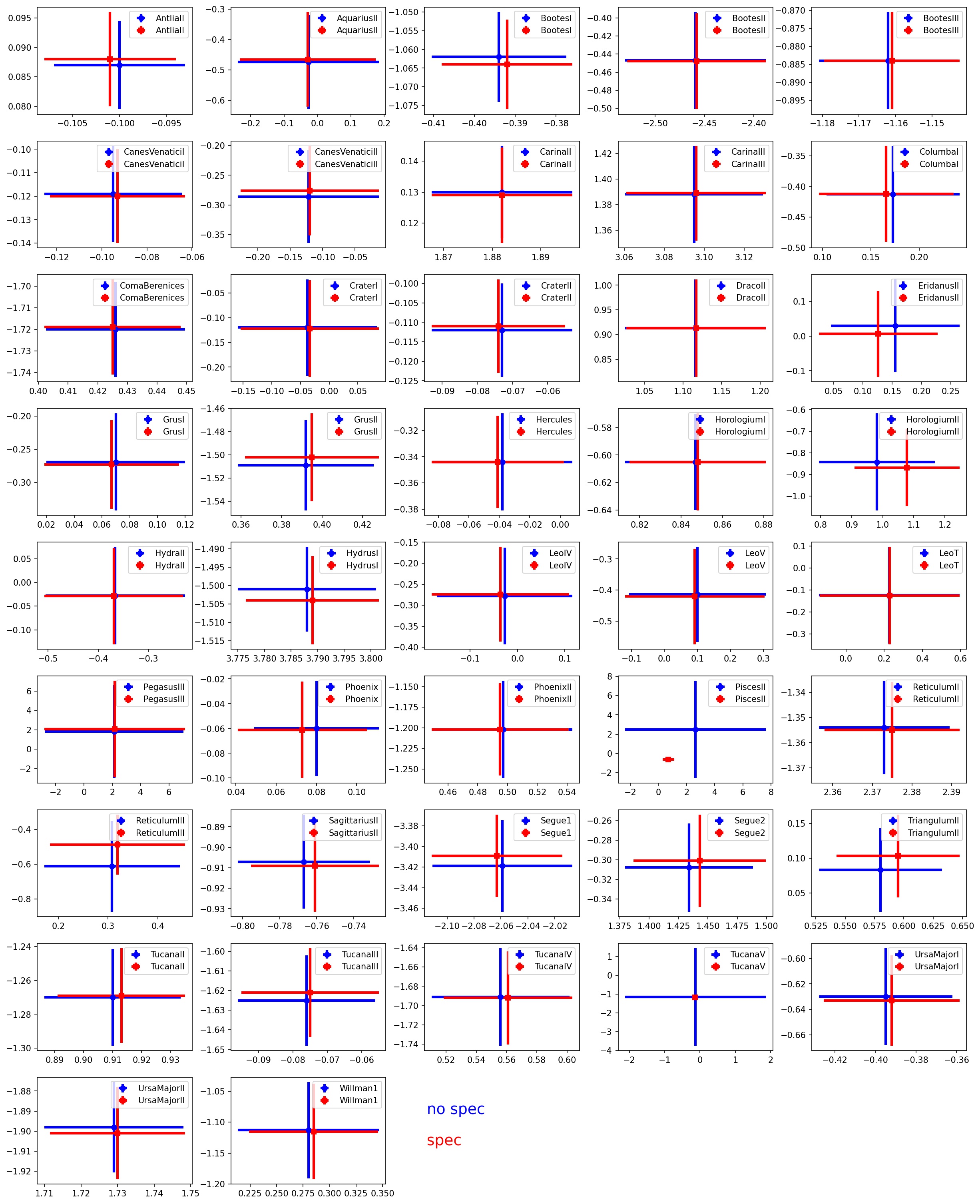}
      \caption{Comparison of systemic PMs determined with (red) and without (blue) the likelihood term for line-of-sight velocities. The x-axis and y-axis show the $\mu_{\alpha,*}$ and $\mu_{\delta}$ component, respectively. The sources of the spectroscopic works are those used in \citet[][see references to the original studies therein]{Fritz_18}, \citet{Kirby_13_massmet},  \citet{Kirby_15},  \citet{Carlin_18}, \citet{Torrealba_19},  \citet{Simon_20}, \citet{Longeard_21}.
}
         \label{fig:spec}
   \end{figure*}

             \begin{figure*}
   \centering
   \includegraphics[width=0.30\textwidth,angle=0]{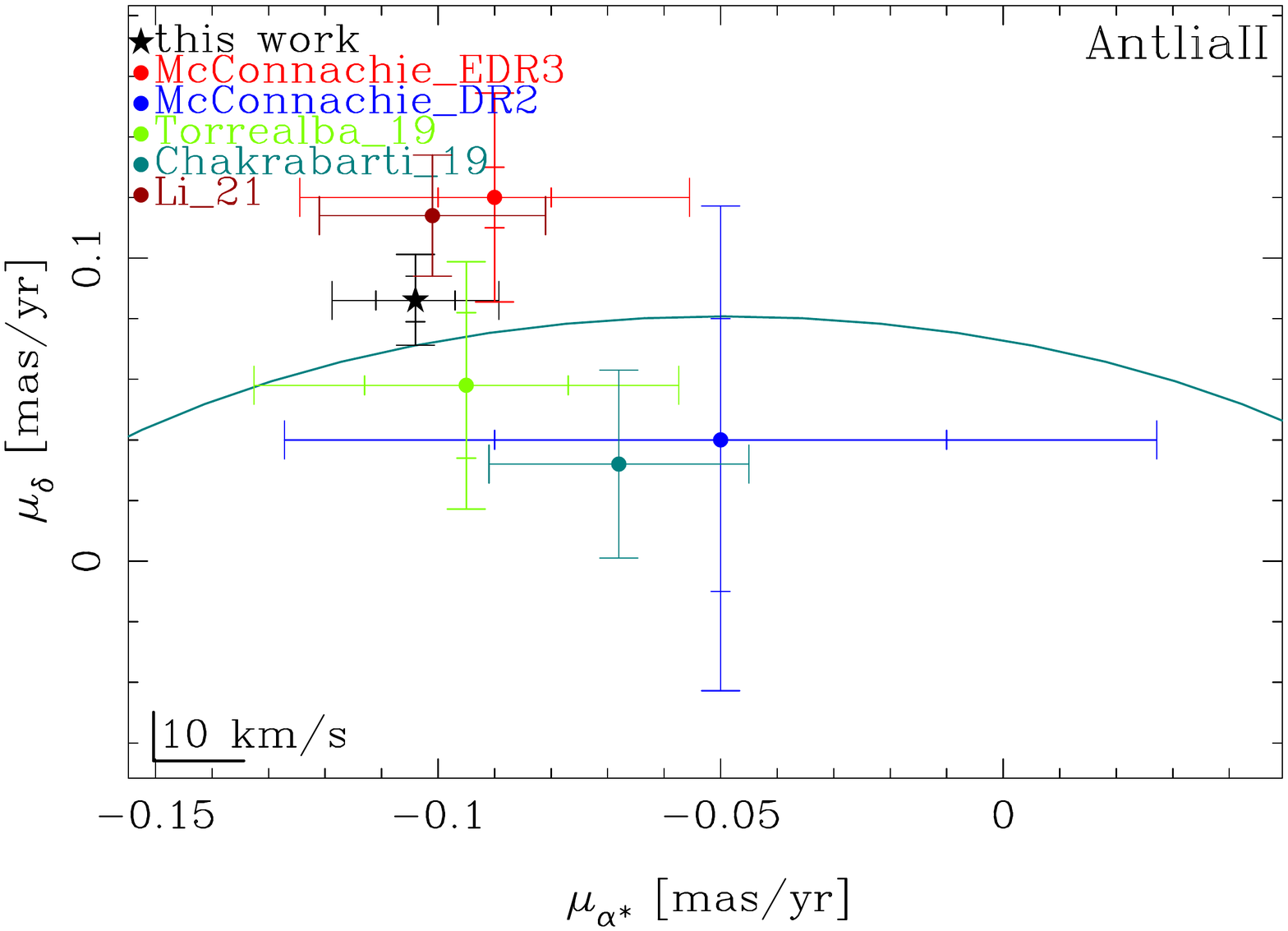}
      \includegraphics[width=0.30\textwidth,angle=0]{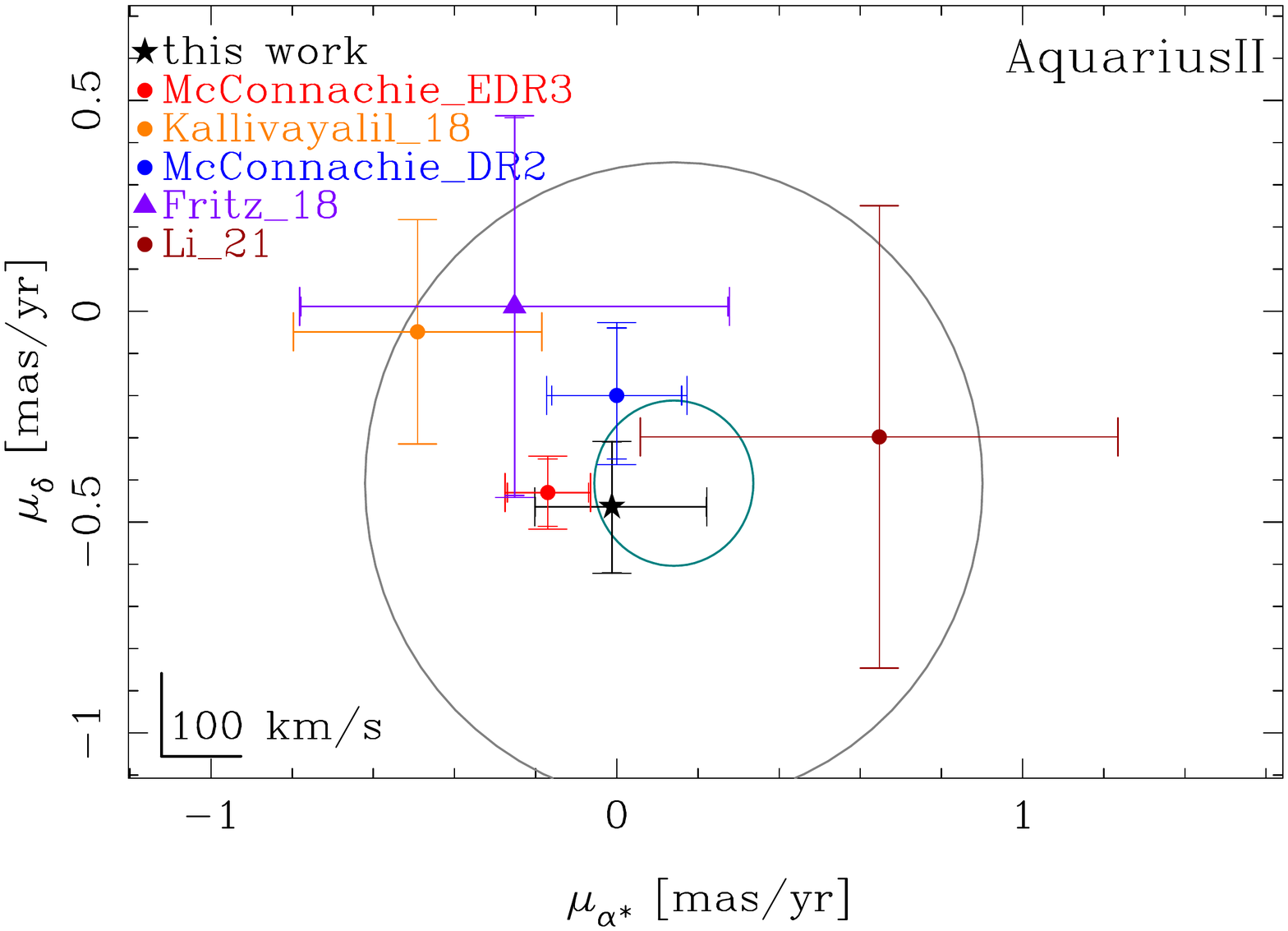}  
            \includegraphics[width=0.30\textwidth,angle=0]{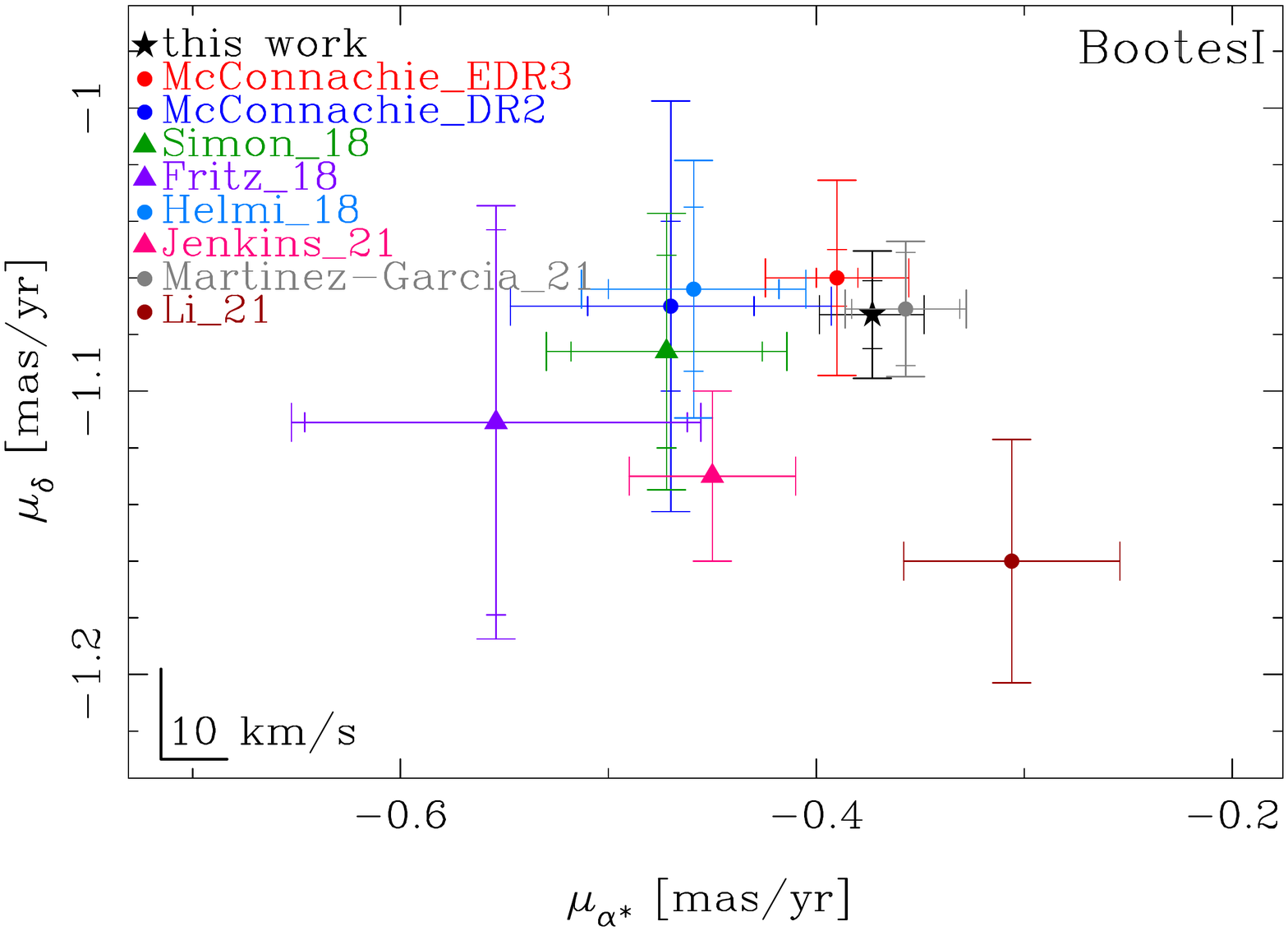}
\includegraphics[width=0.30\textwidth,angle=0]{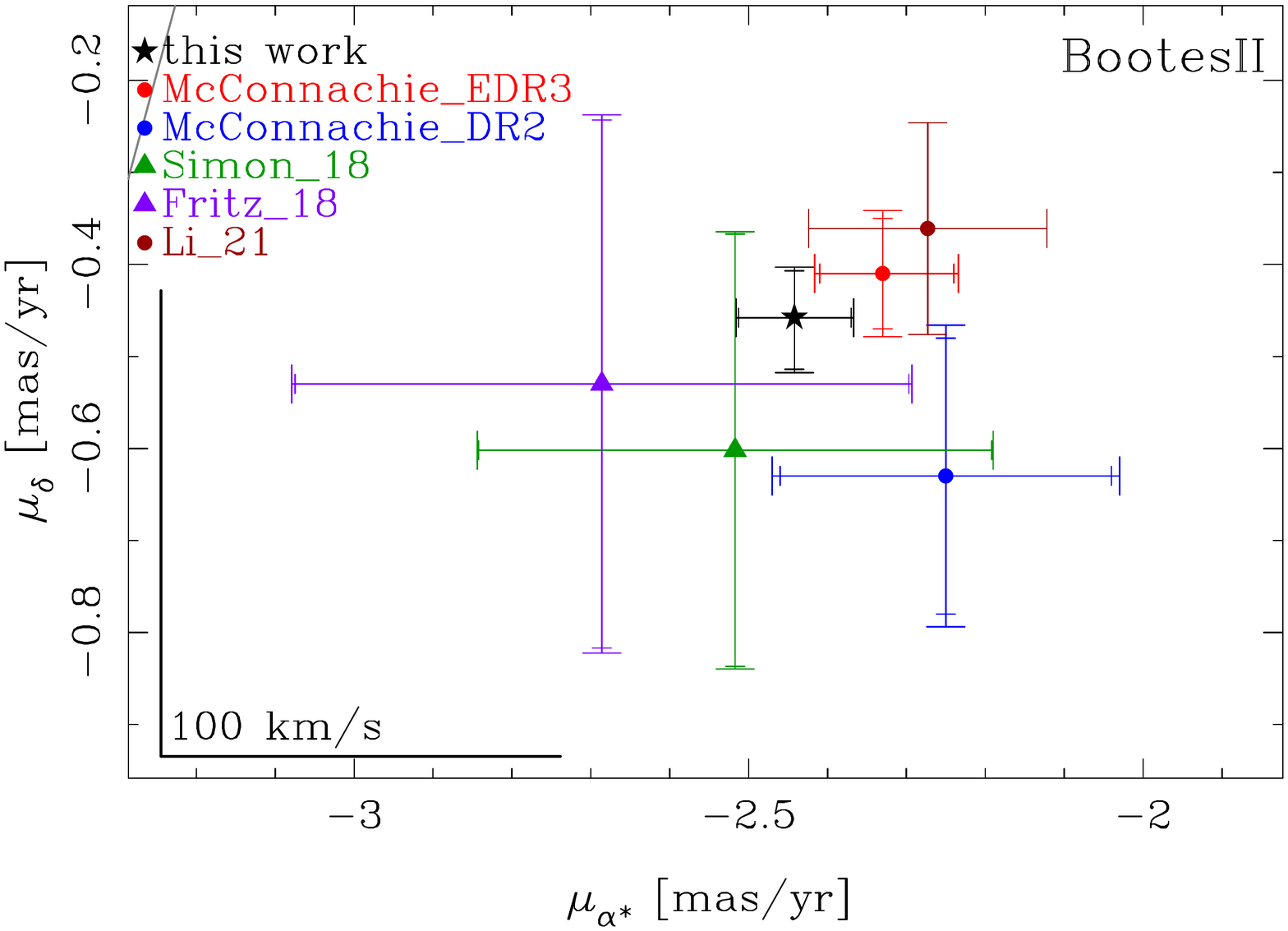}
\includegraphics[width=0.30\textwidth,angle=0]{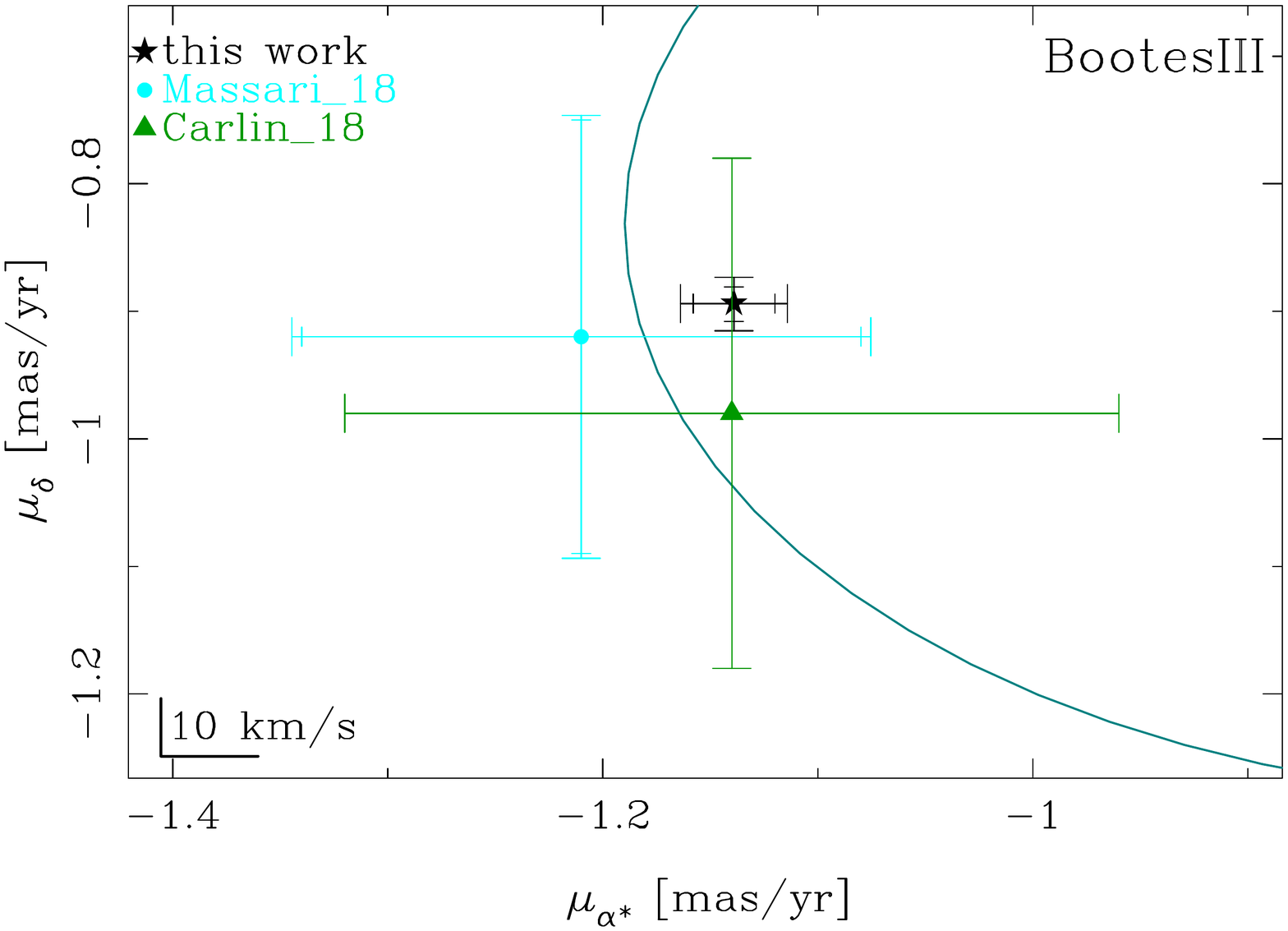}
\includegraphics[width=0.30\textwidth,angle=0]{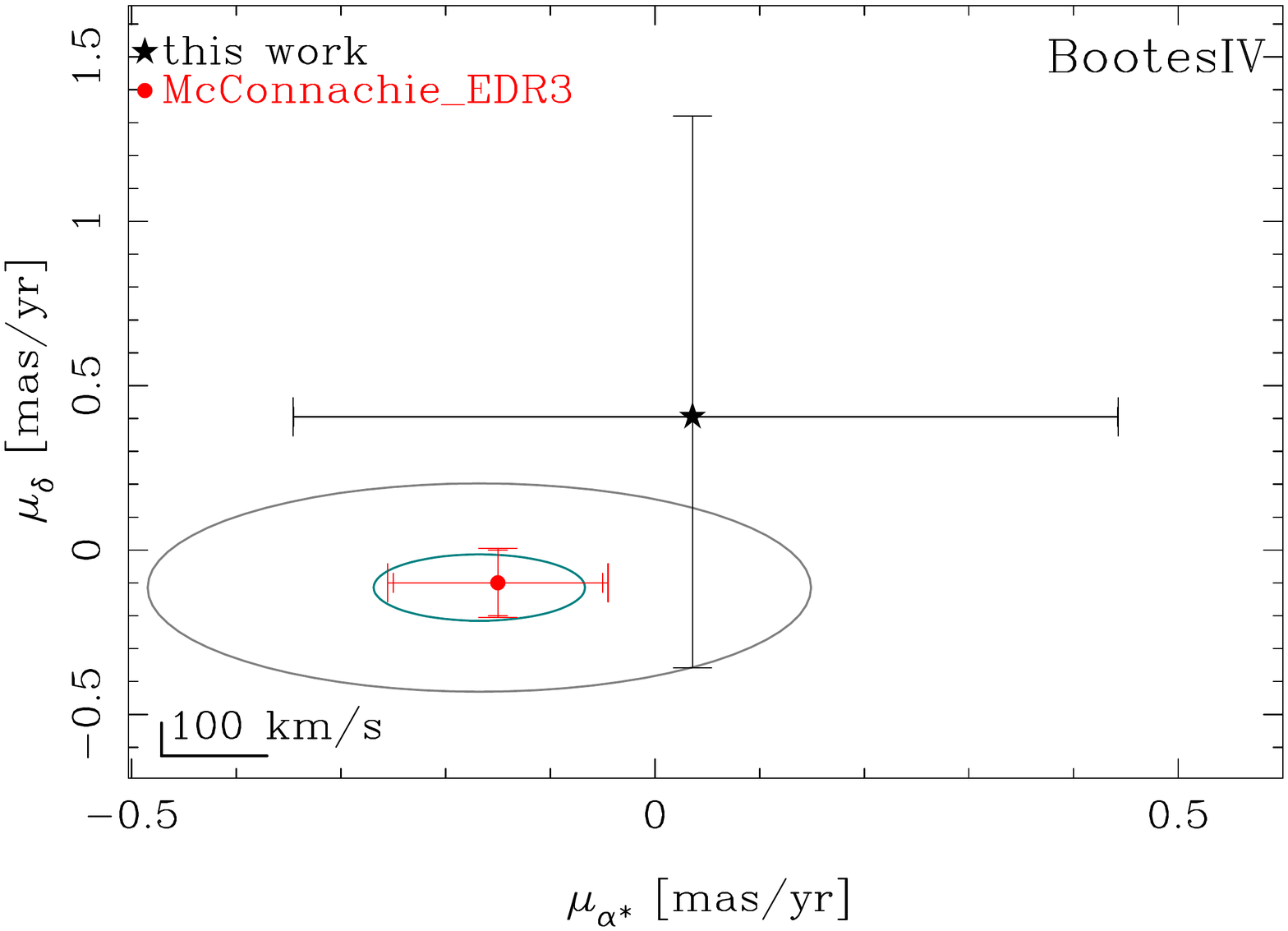}
\includegraphics[width=0.30\textwidth,angle=0]{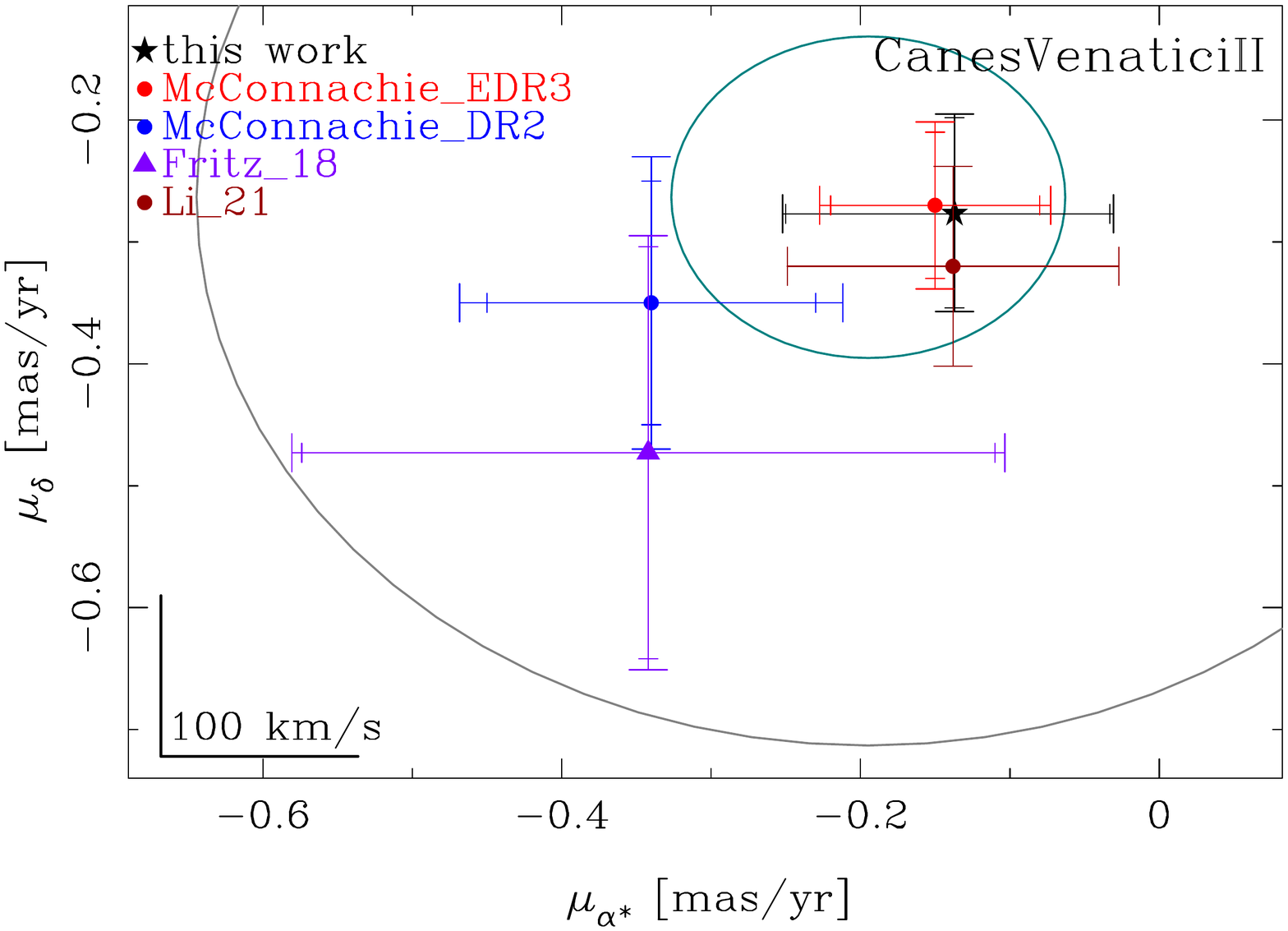}
\includegraphics[width=0.30\textwidth,angle=0]{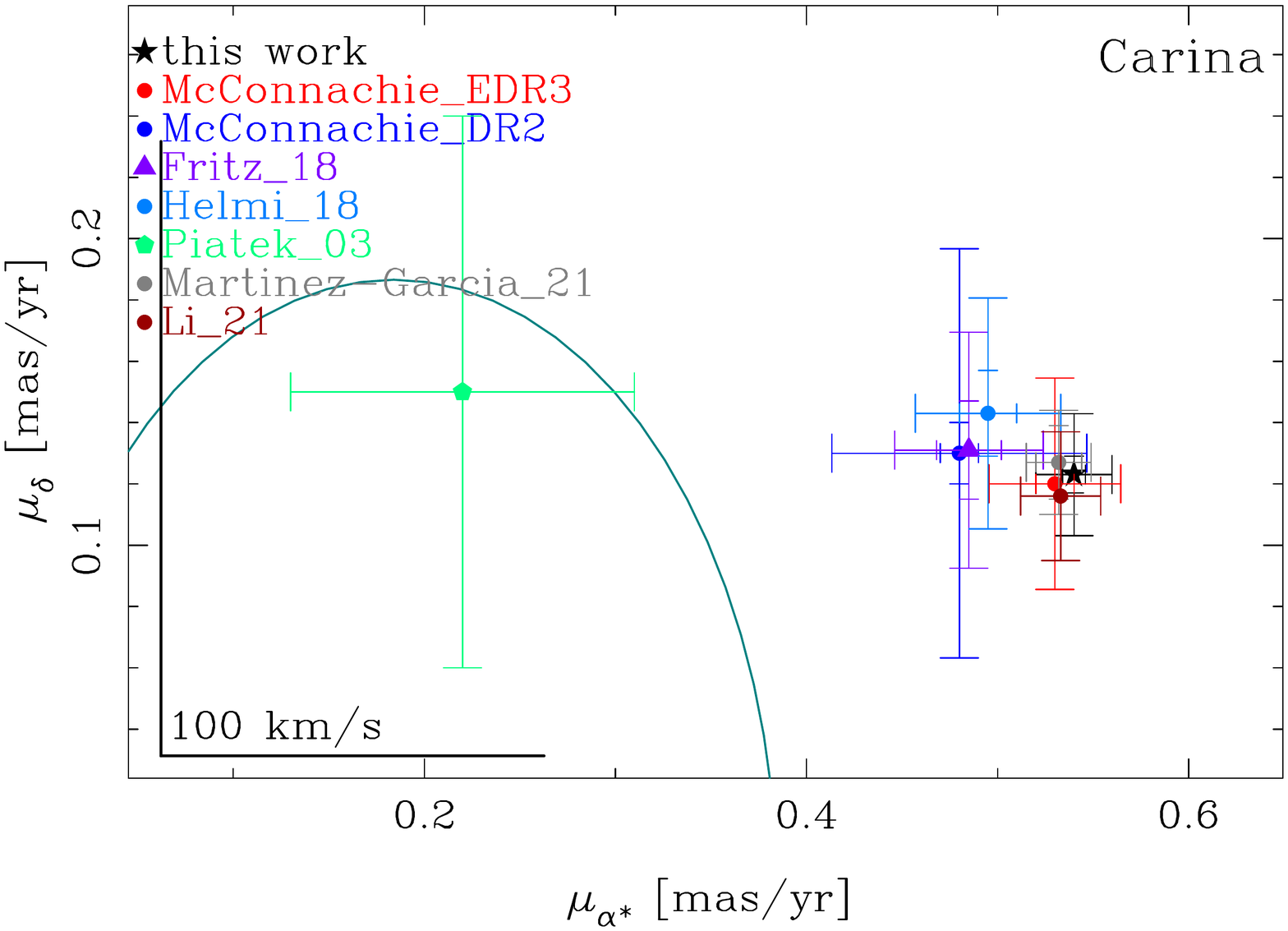}
\includegraphics[width=0.30\textwidth,angle=0]{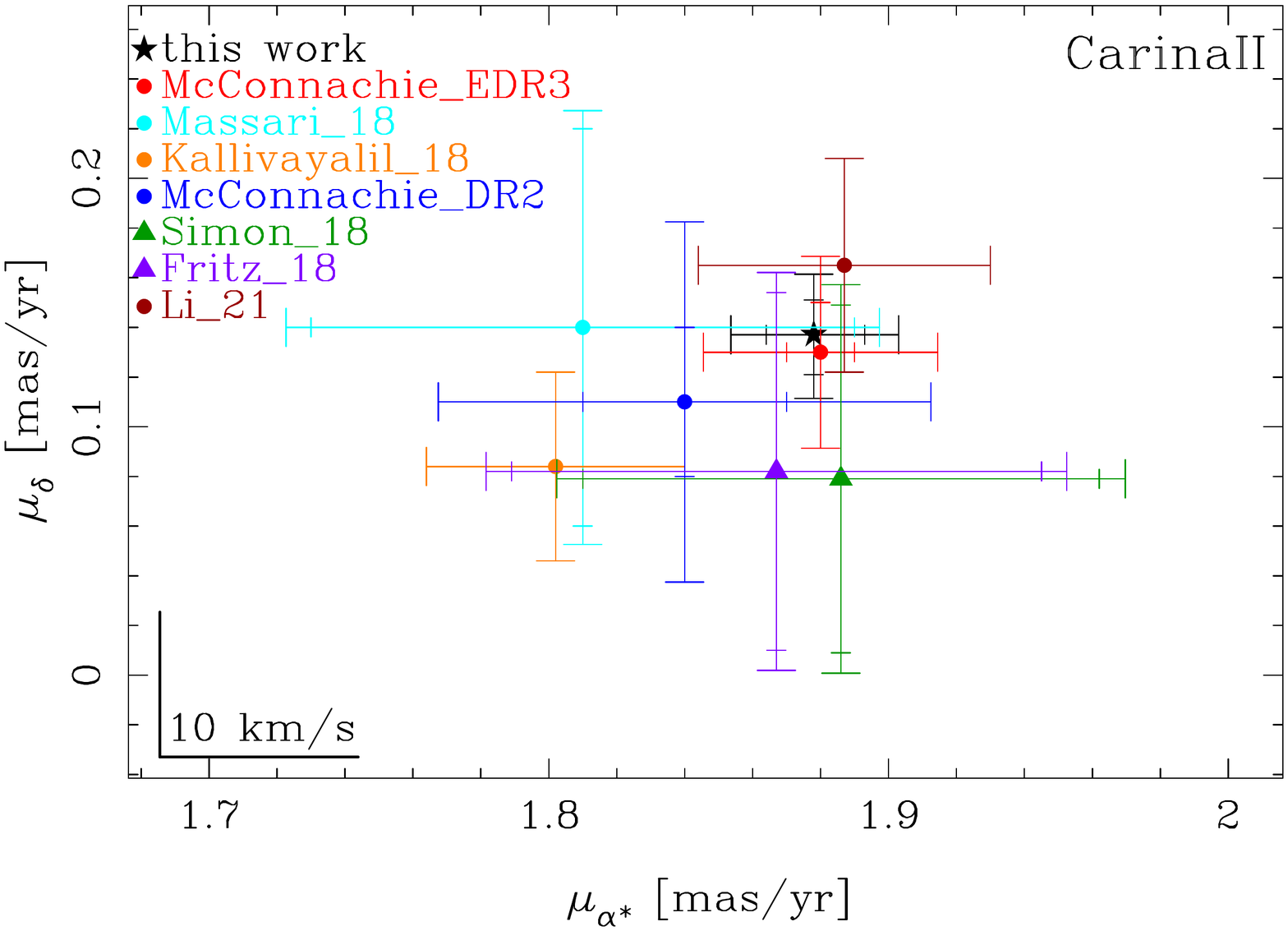}
\includegraphics[width=0.30\textwidth,angle=0]{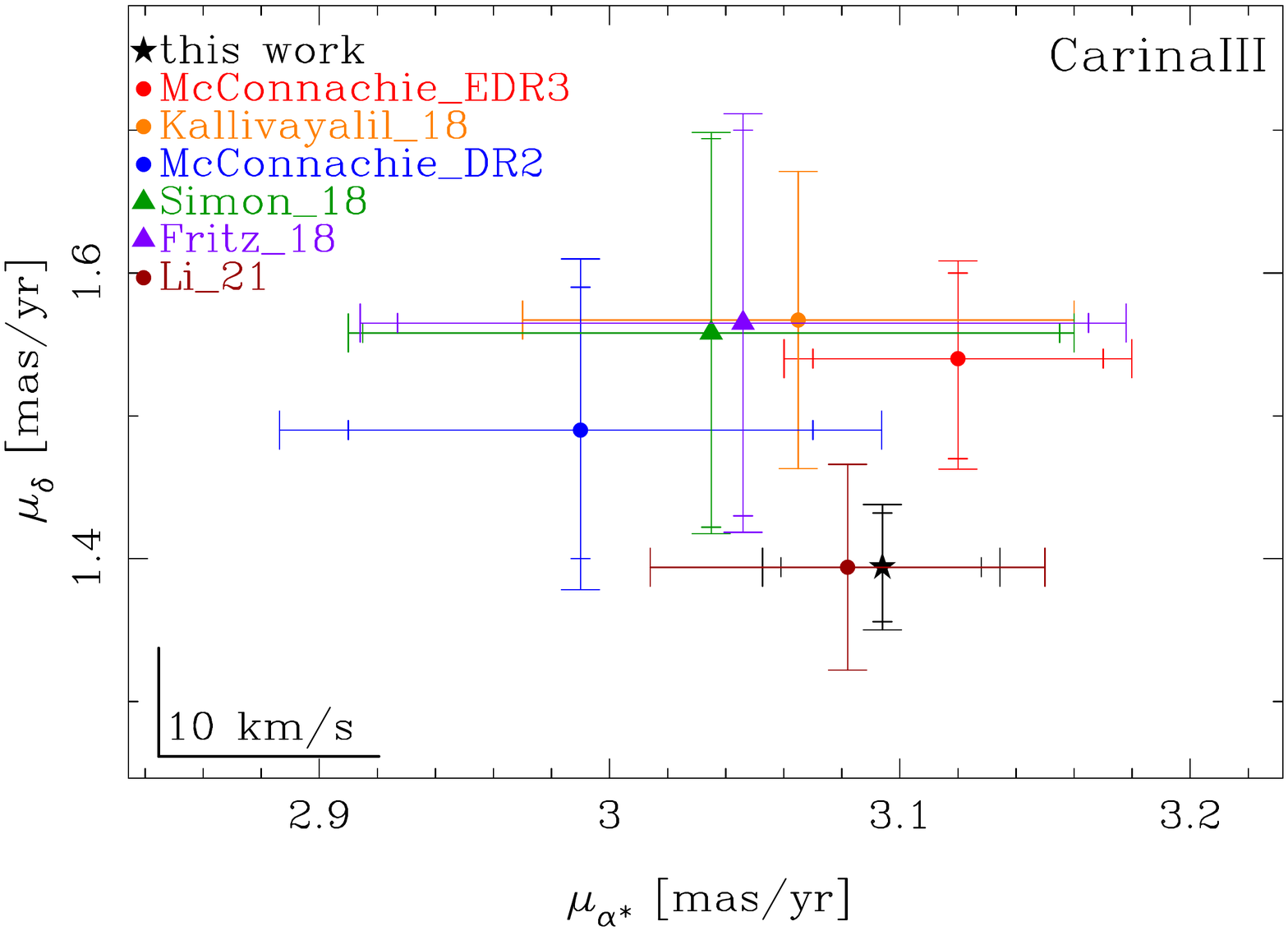}
\includegraphics[width=0.30\textwidth,angle=0]{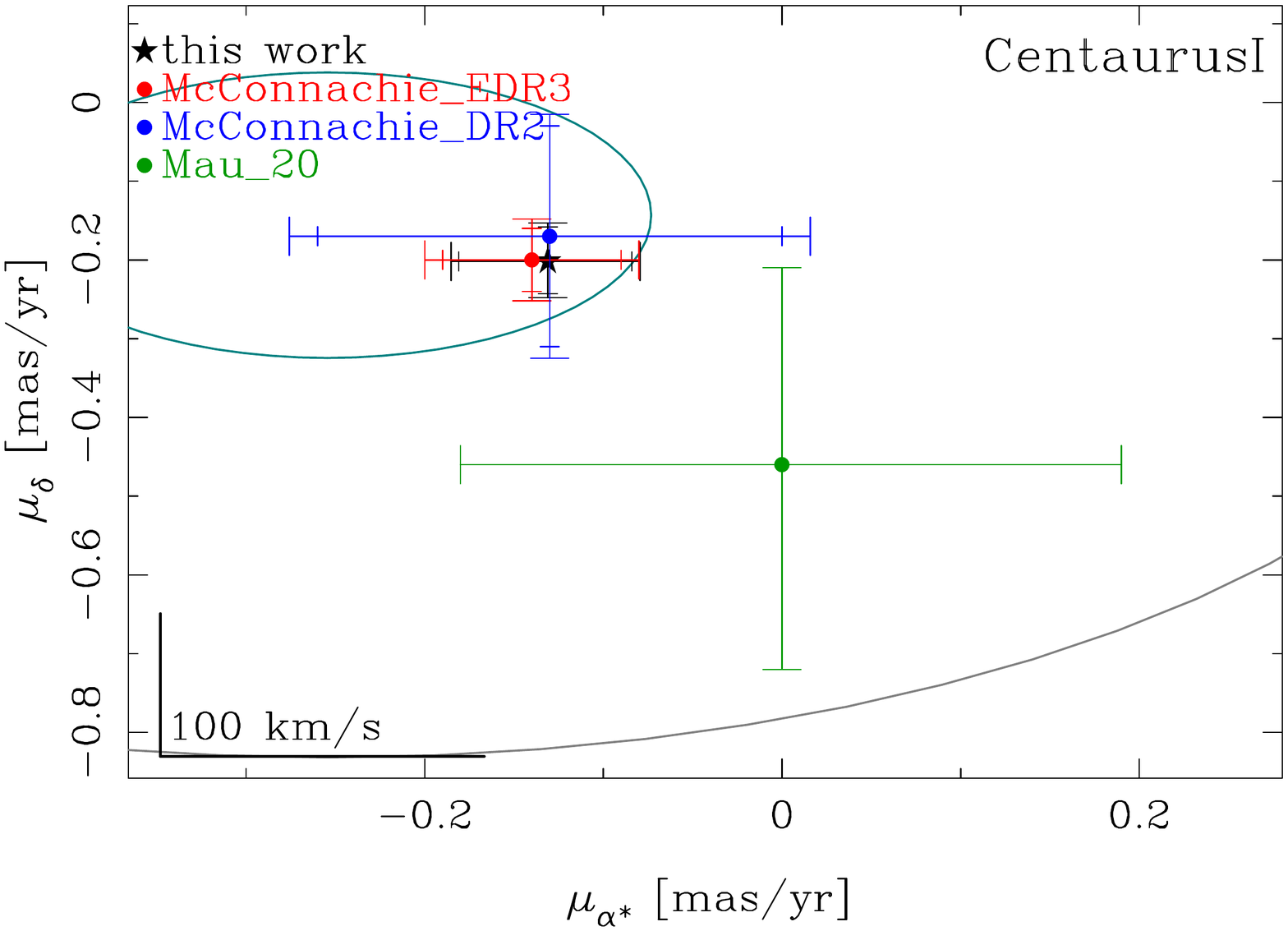}
\includegraphics[width=0.30\textwidth,angle=0]{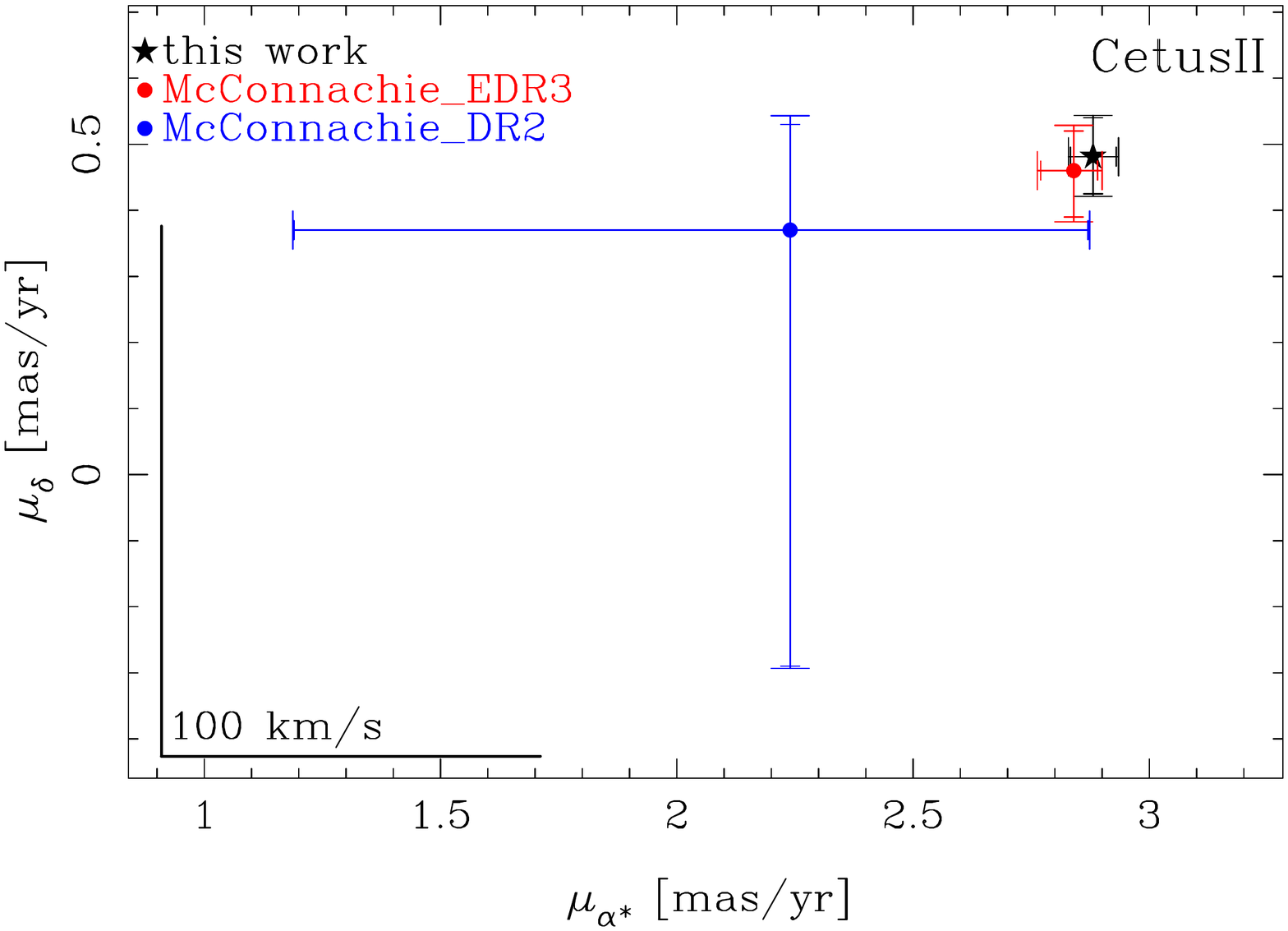}
\includegraphics[width=0.30\textwidth,angle=0]{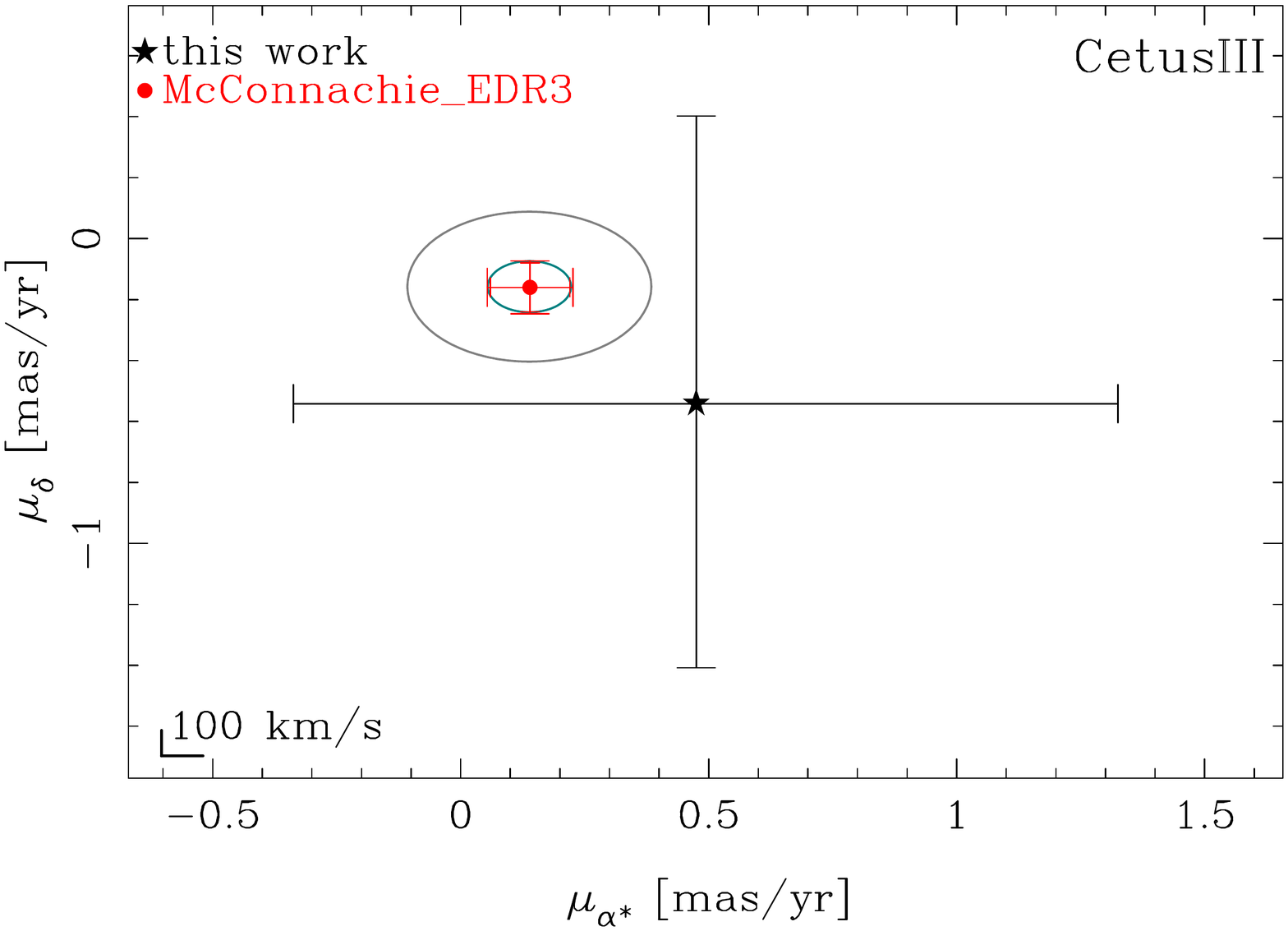}
\includegraphics[width=0.30\textwidth,angle=0]{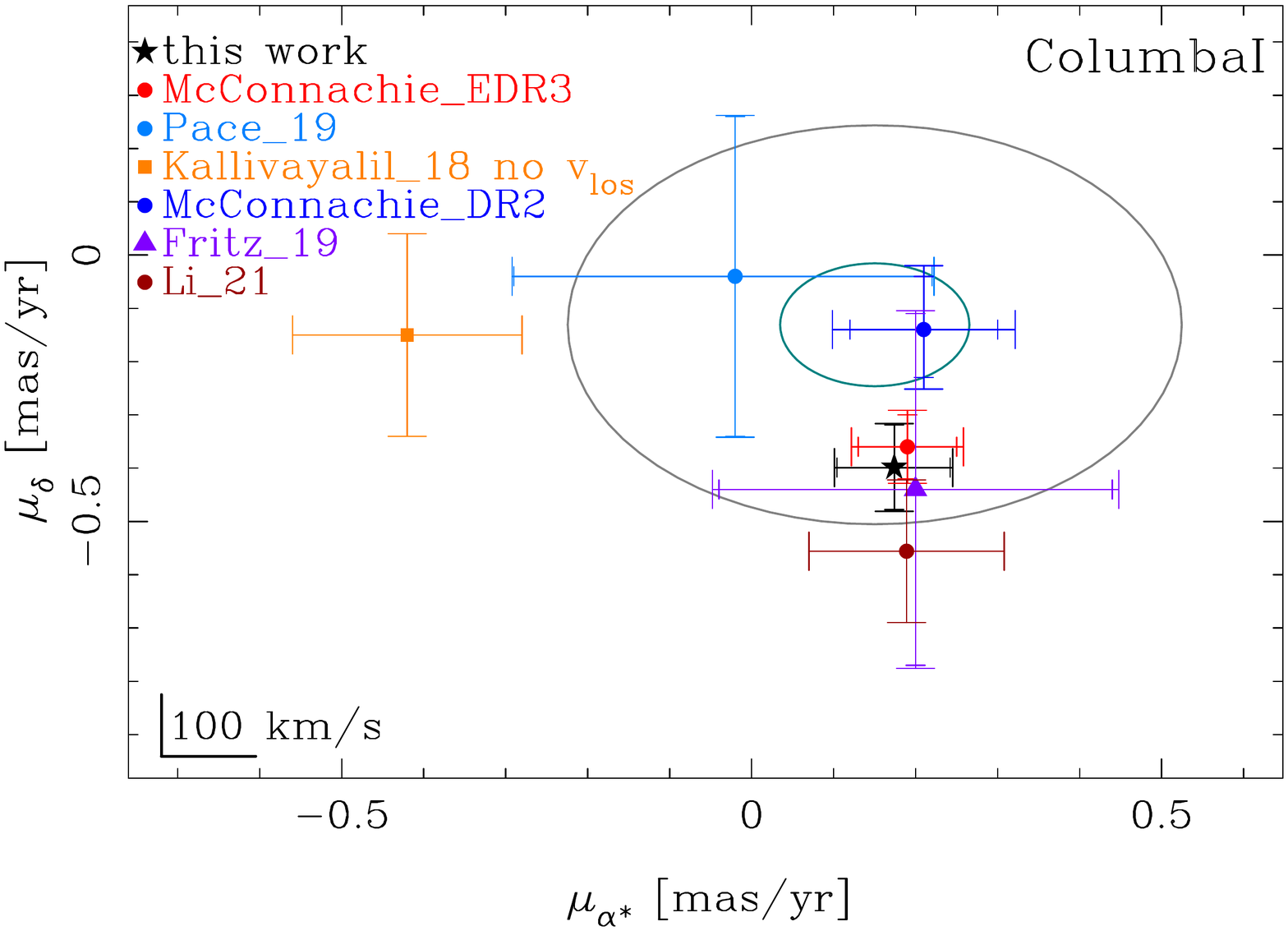}
\includegraphics[width=0.30\textwidth,angle=0]{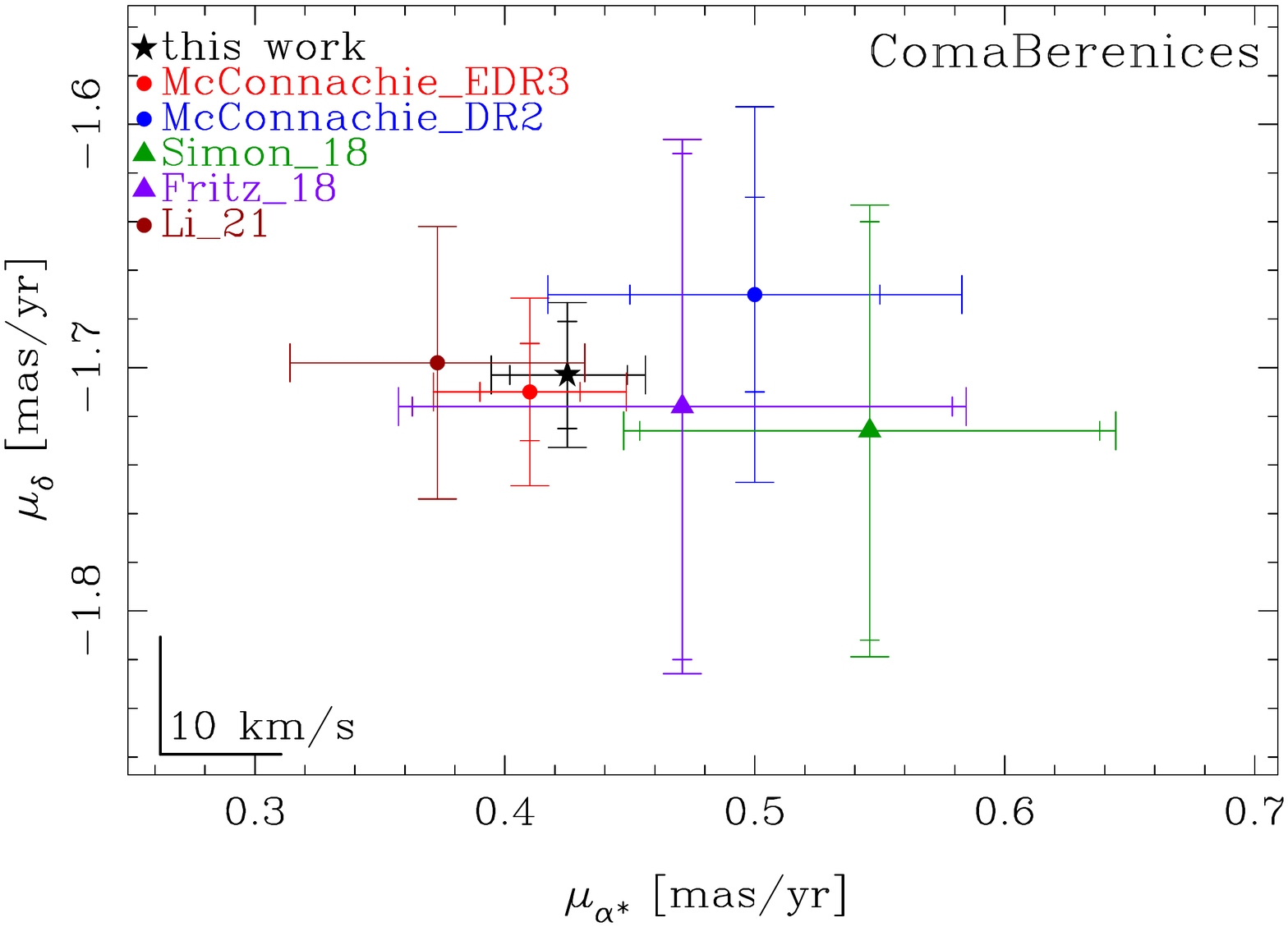}
      \caption{As Figure~\ref{fig:pms_lit}.}
         \label{fig:pms_lit2}
   \end{figure*}   
   
       \begin{figure*}
   \centering
   \includegraphics[width=0.30\textwidth,angle=0]{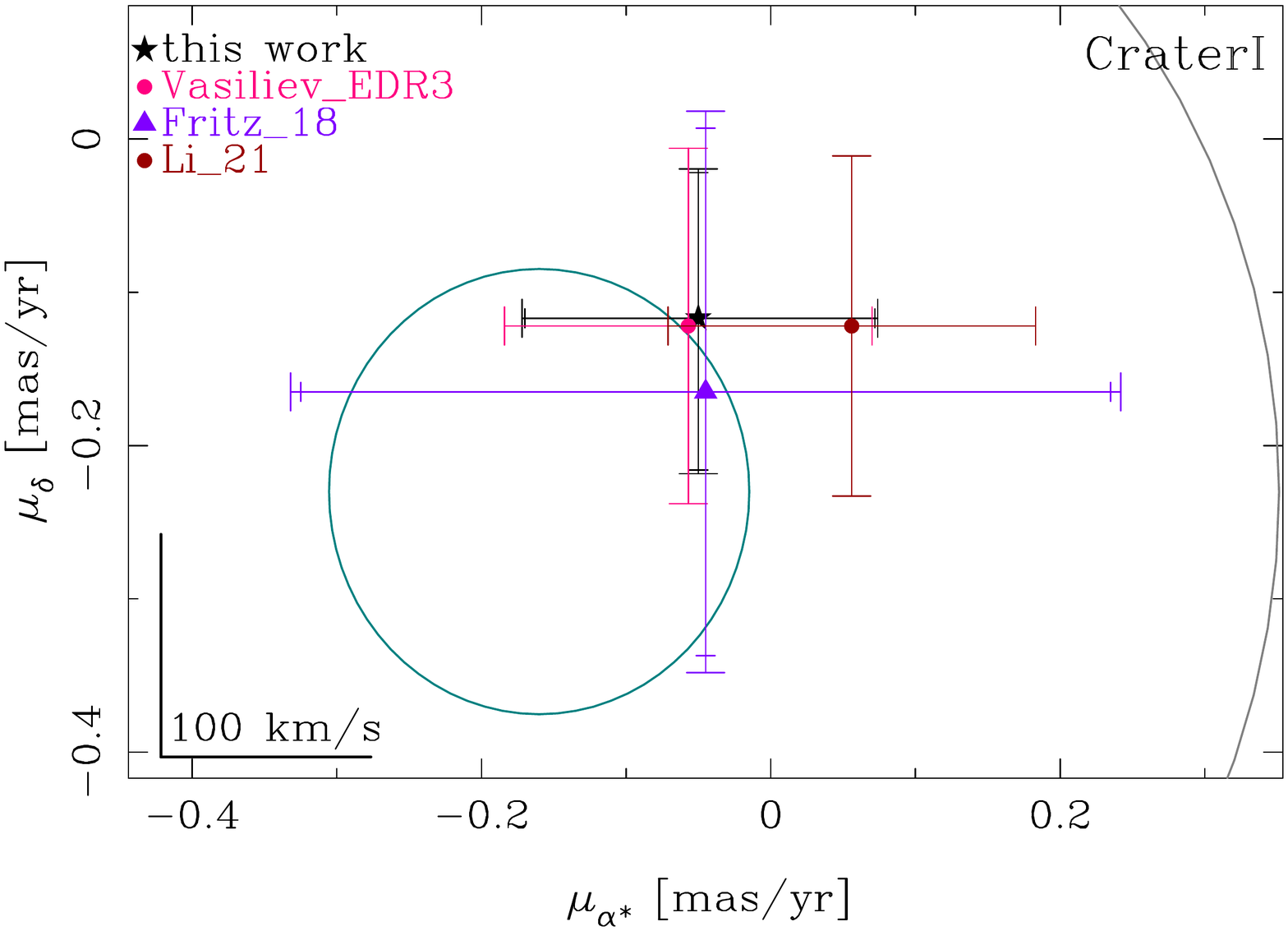}
\includegraphics[width=0.30\textwidth,angle=0]{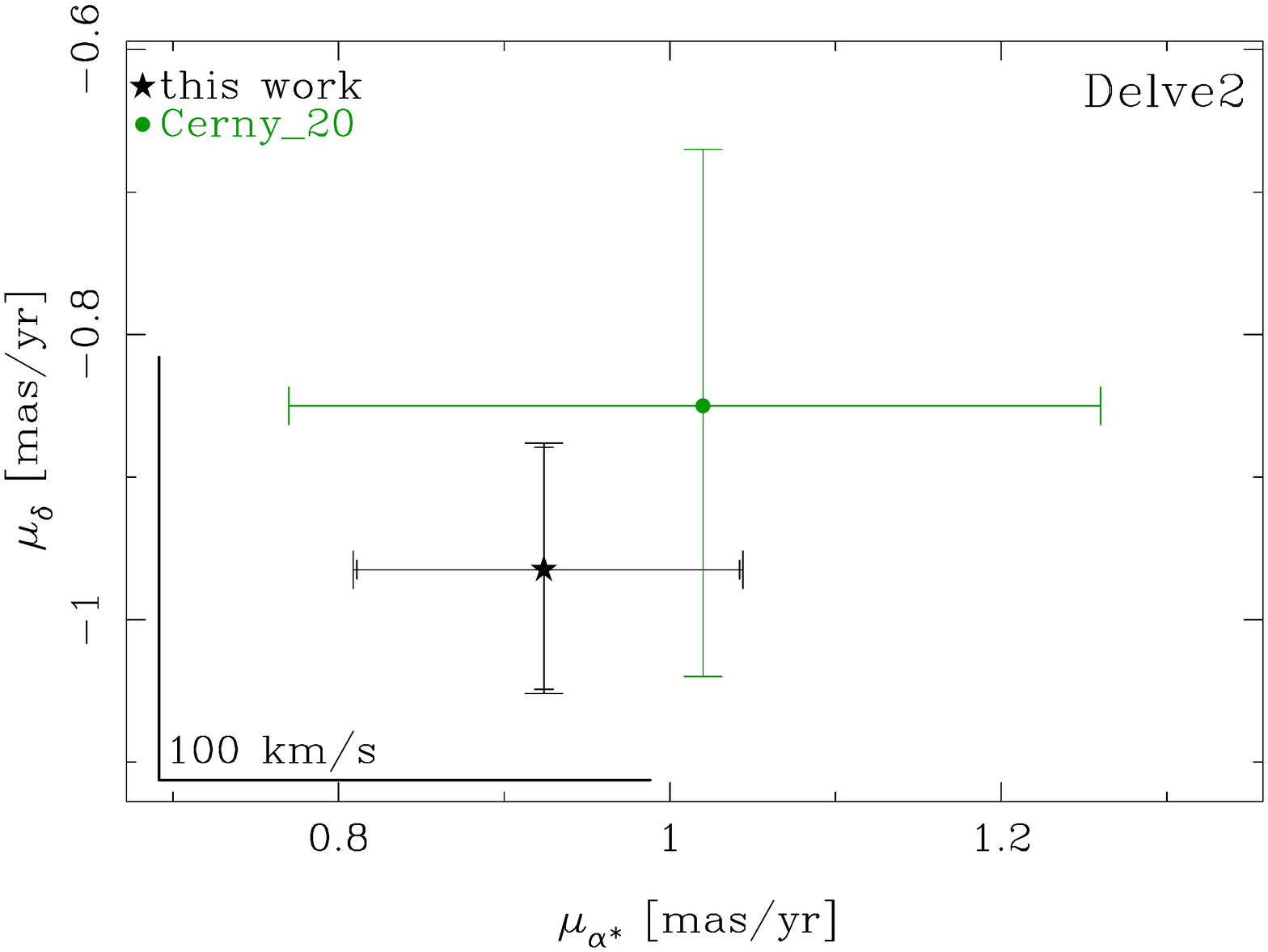}
\includegraphics[width=0.30\textwidth,angle=0]{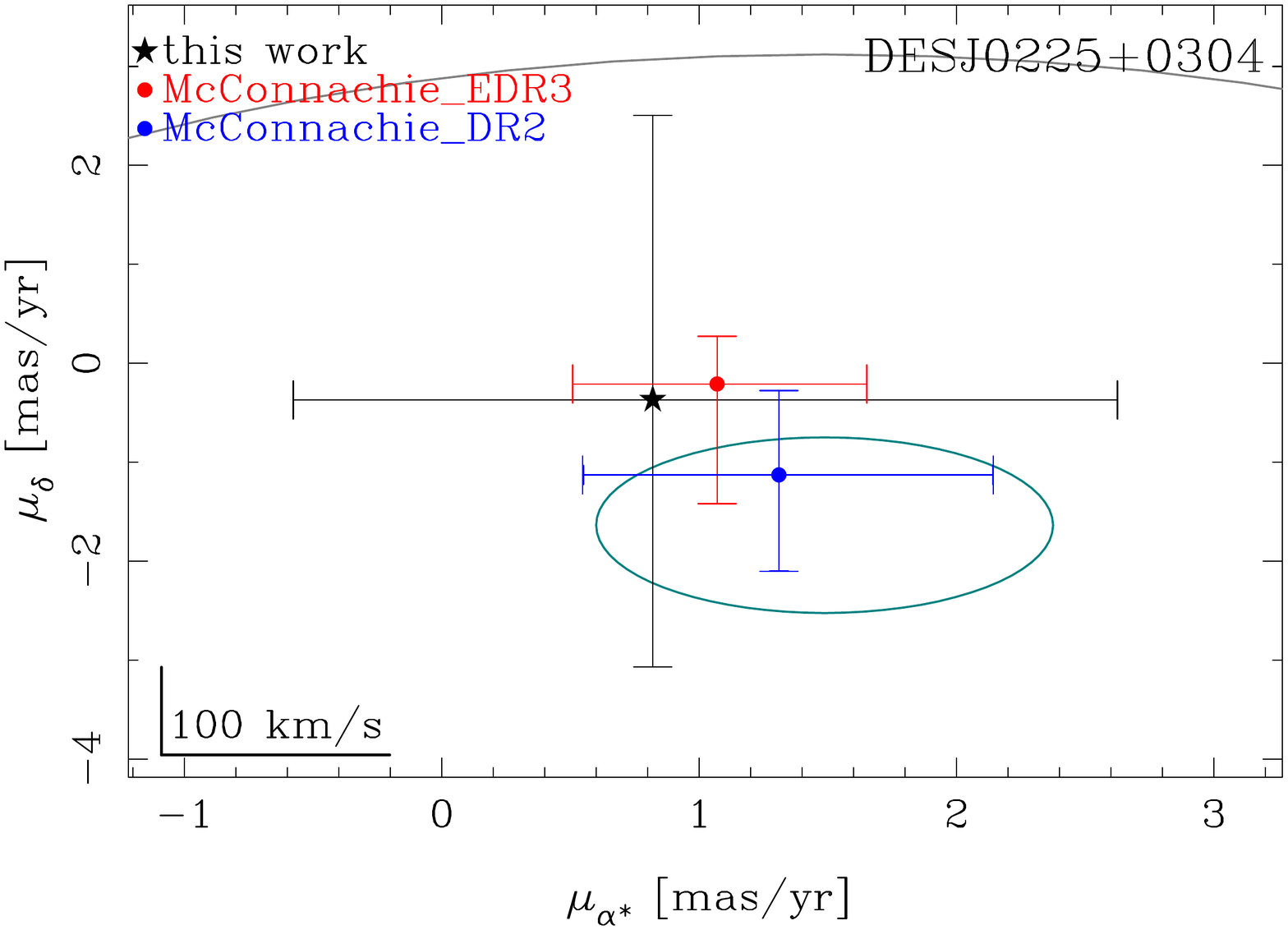}
\includegraphics[width=0.30\textwidth,angle=0]{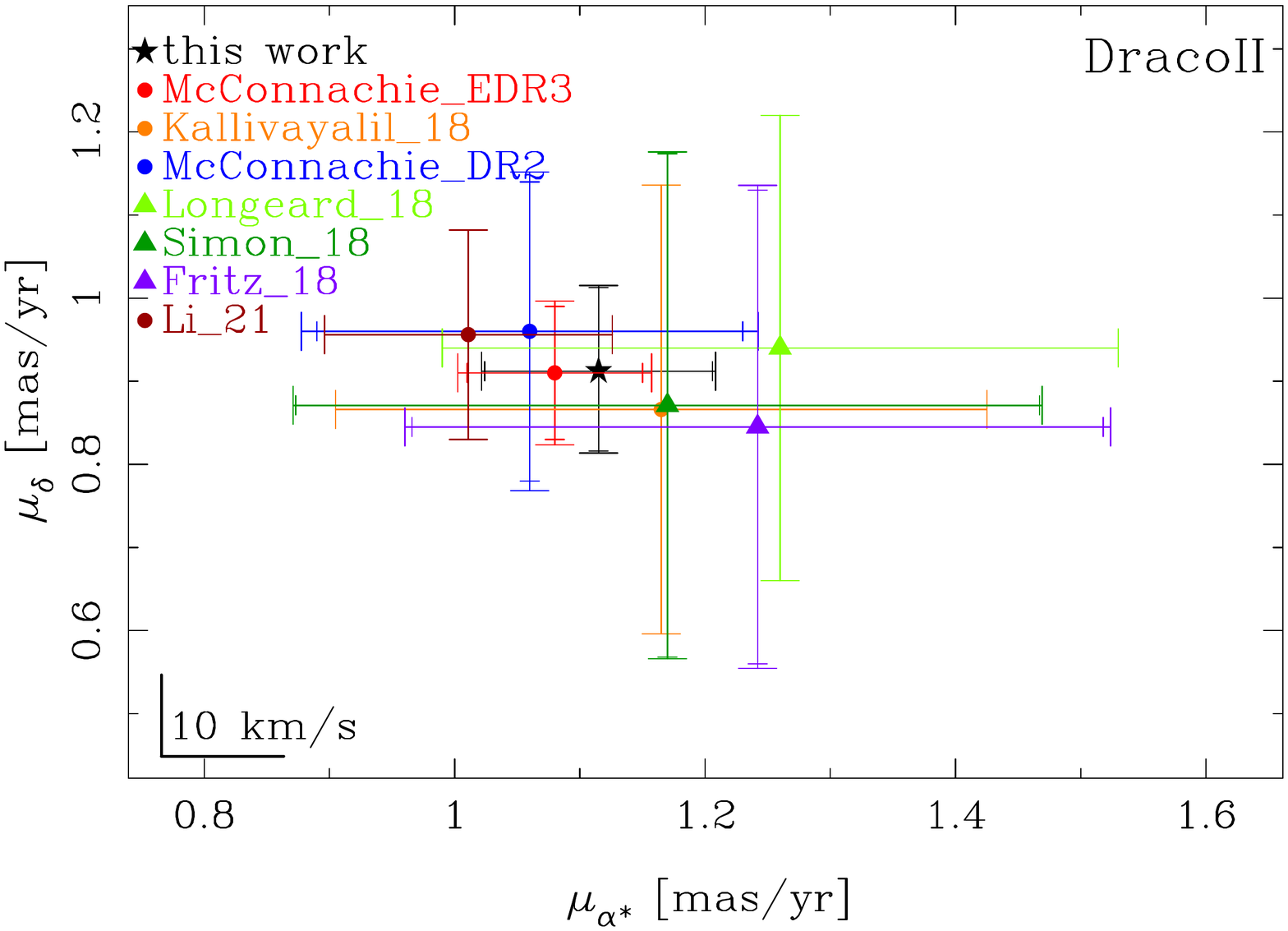}
\includegraphics[width=0.30\textwidth,angle=0]{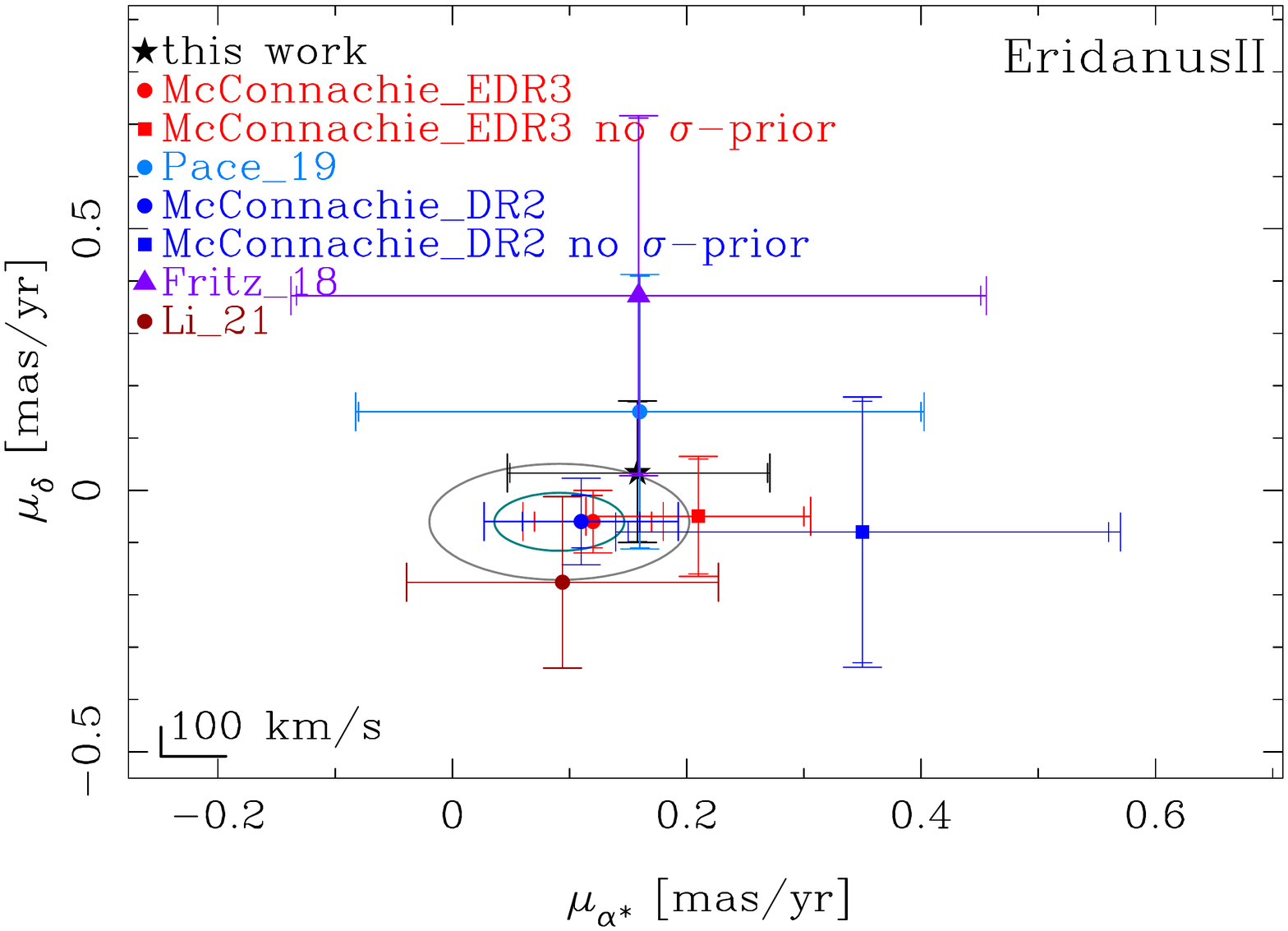}
\includegraphics[width=0.30\textwidth,angle=0]{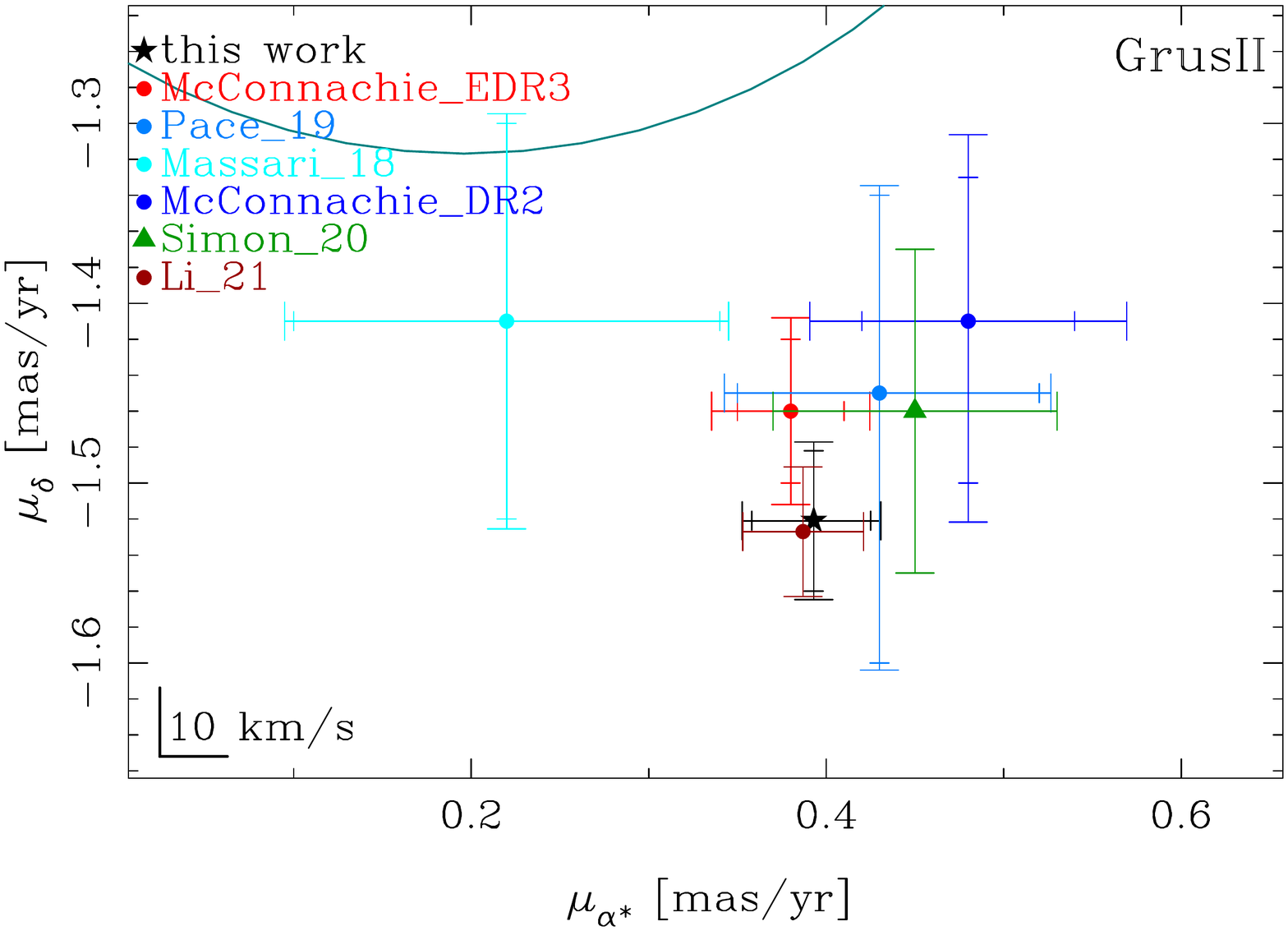}
\includegraphics[width=0.30\textwidth,angle=0]{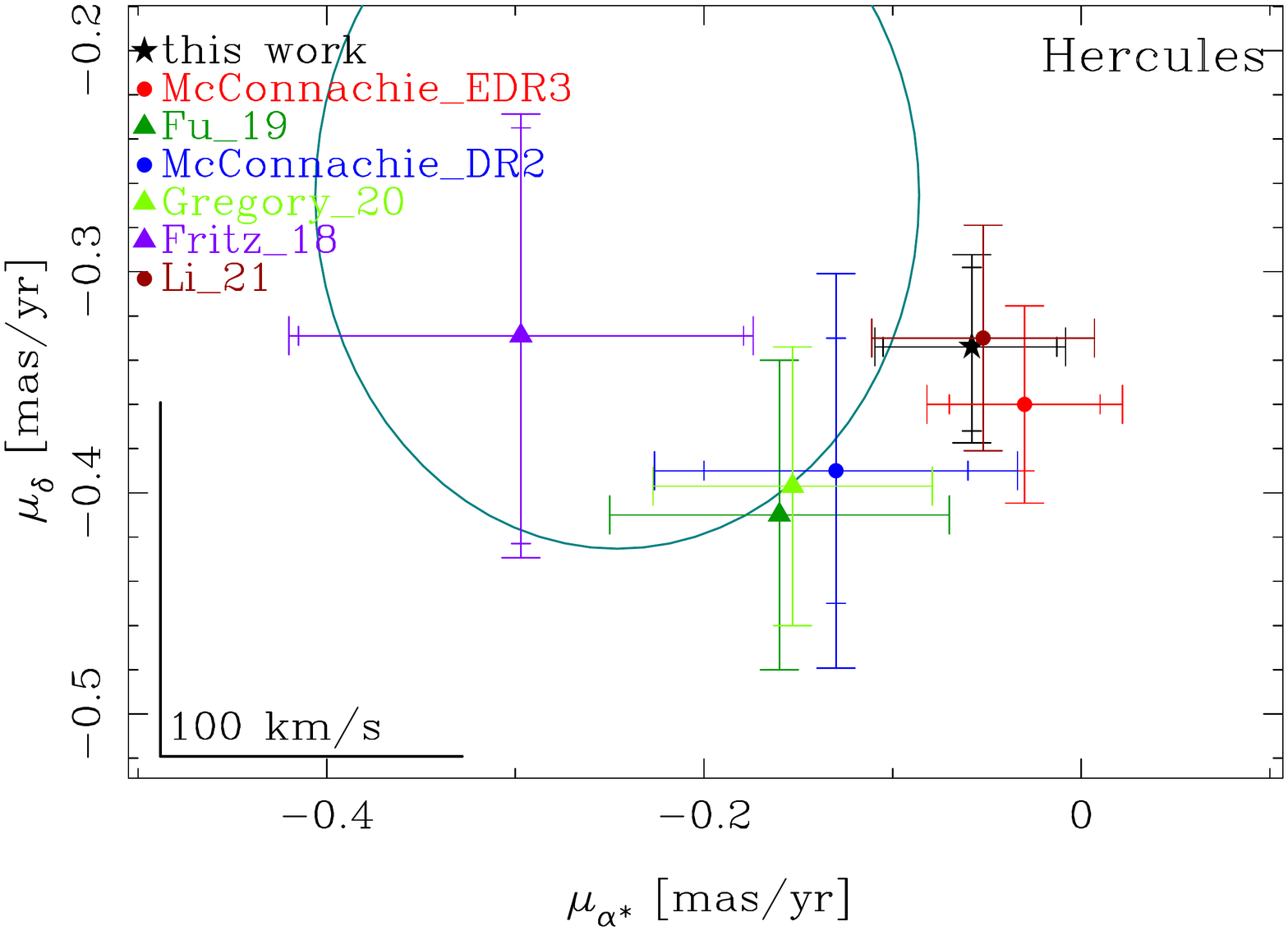}
\includegraphics[width=0.30\textwidth,angle=0]{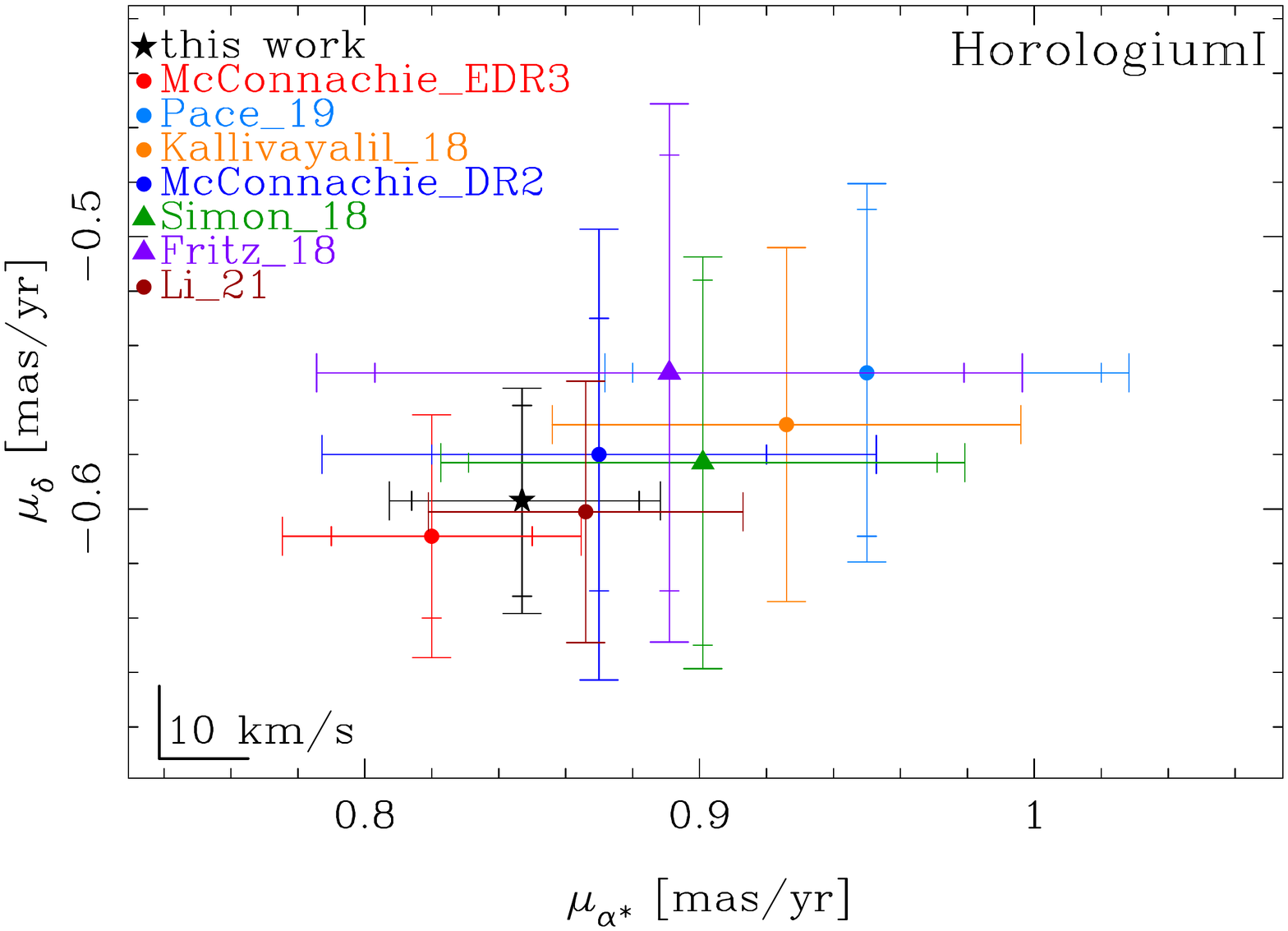}
\includegraphics[width=0.30\textwidth,angle=0]{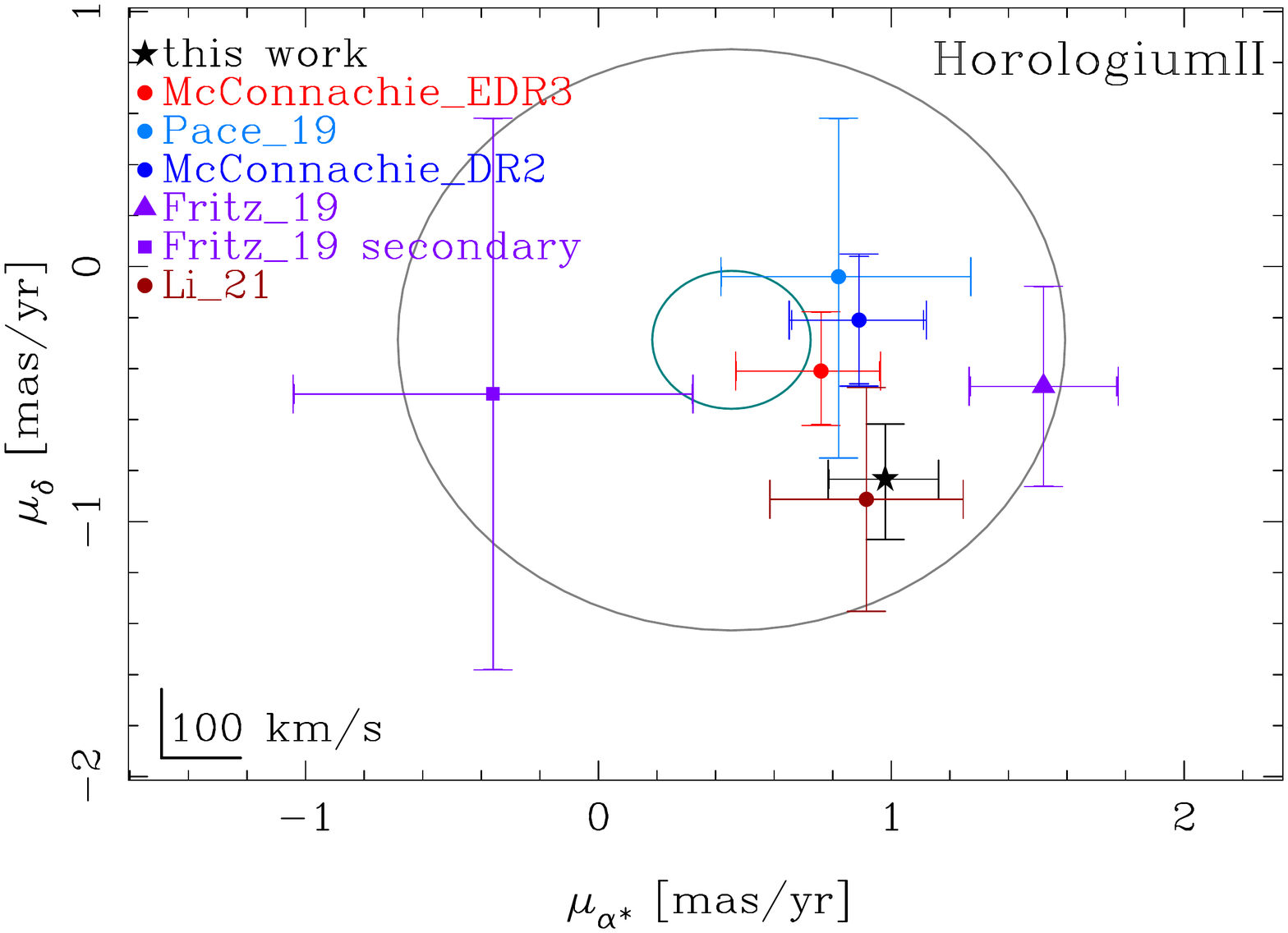}
   \includegraphics[width=0.30\textwidth,angle=0]{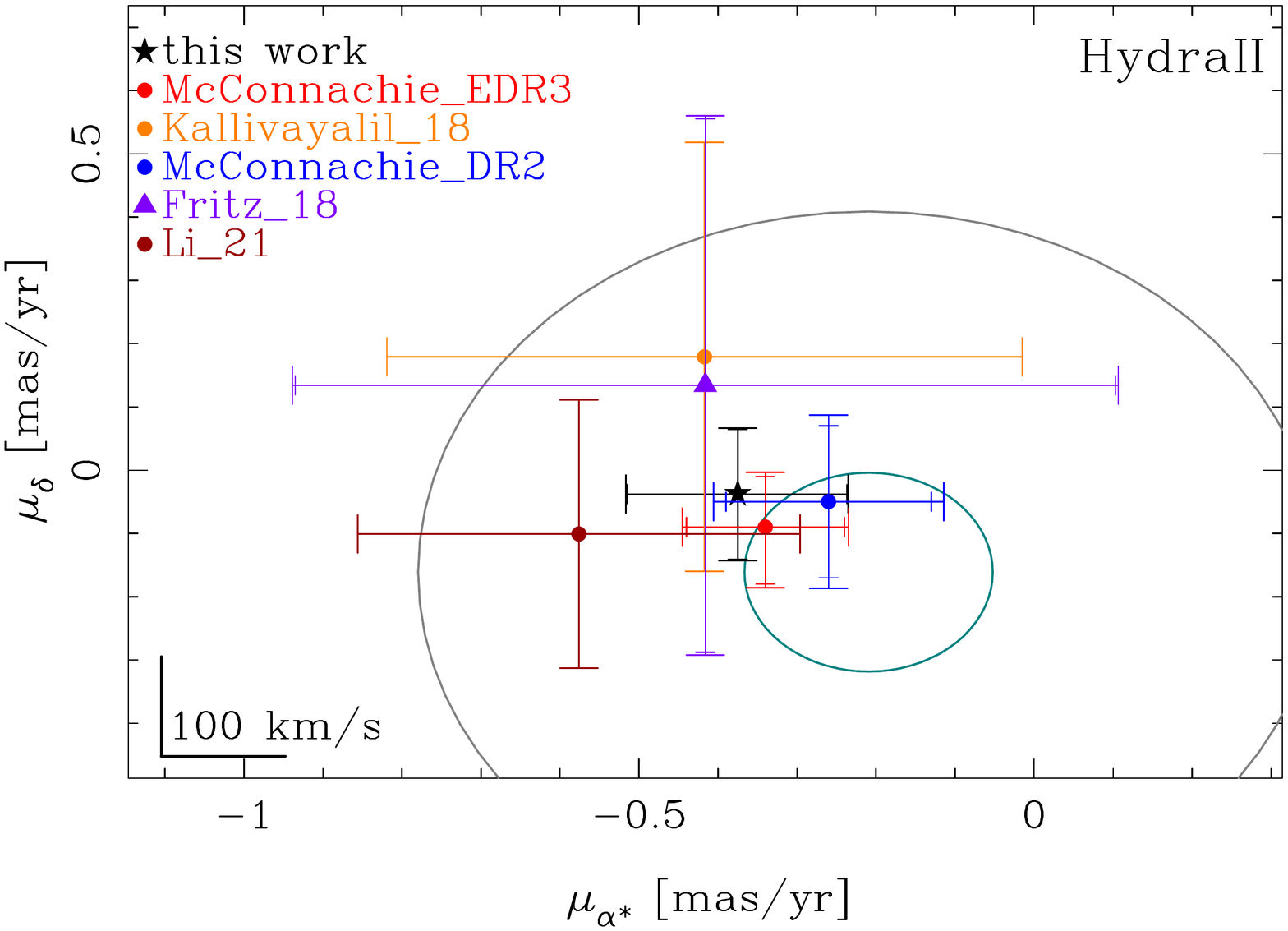}
      \includegraphics[width=0.30\textwidth,angle=0]{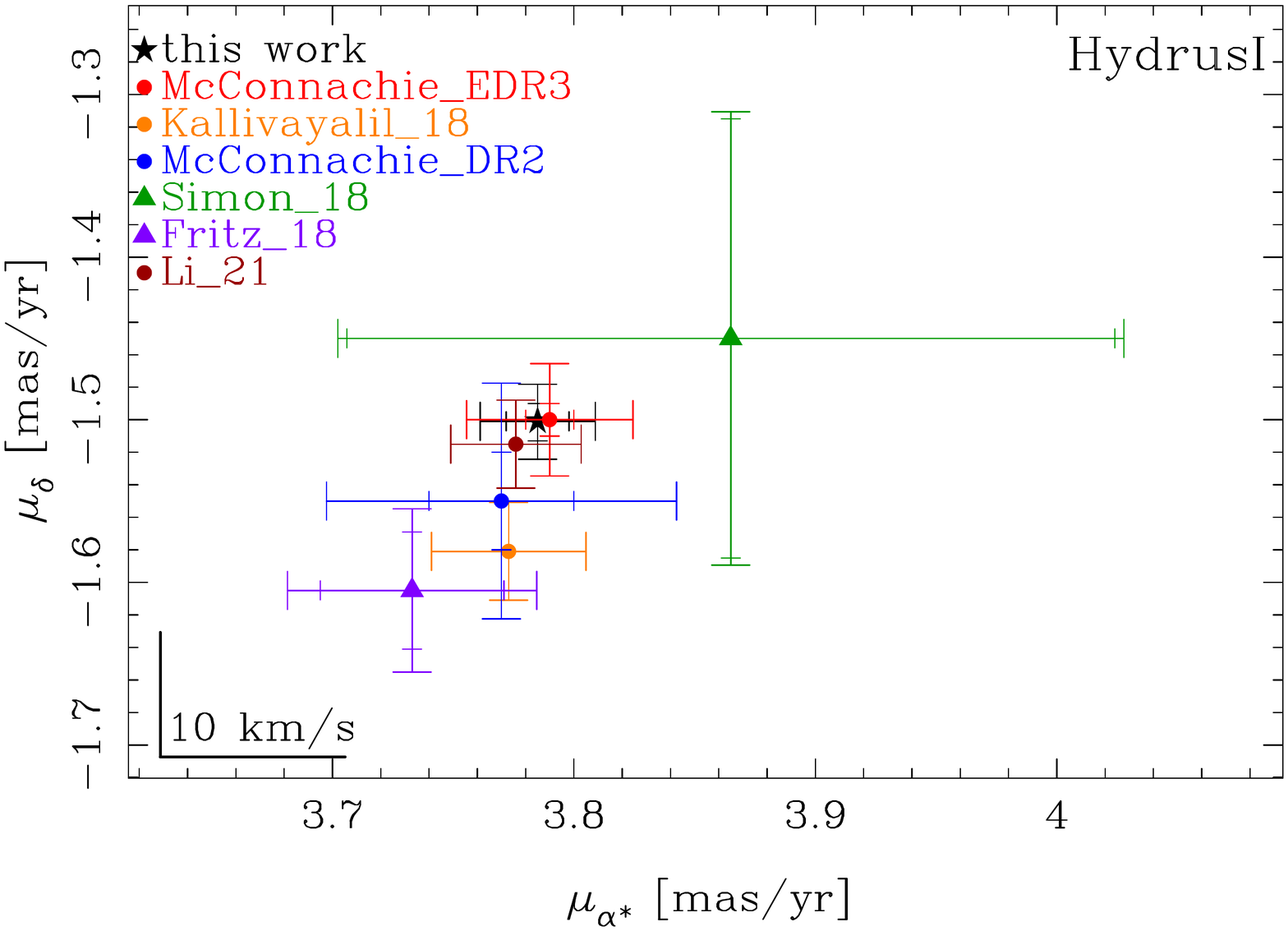} 
      \includegraphics[width=0.30\textwidth,angle=0]{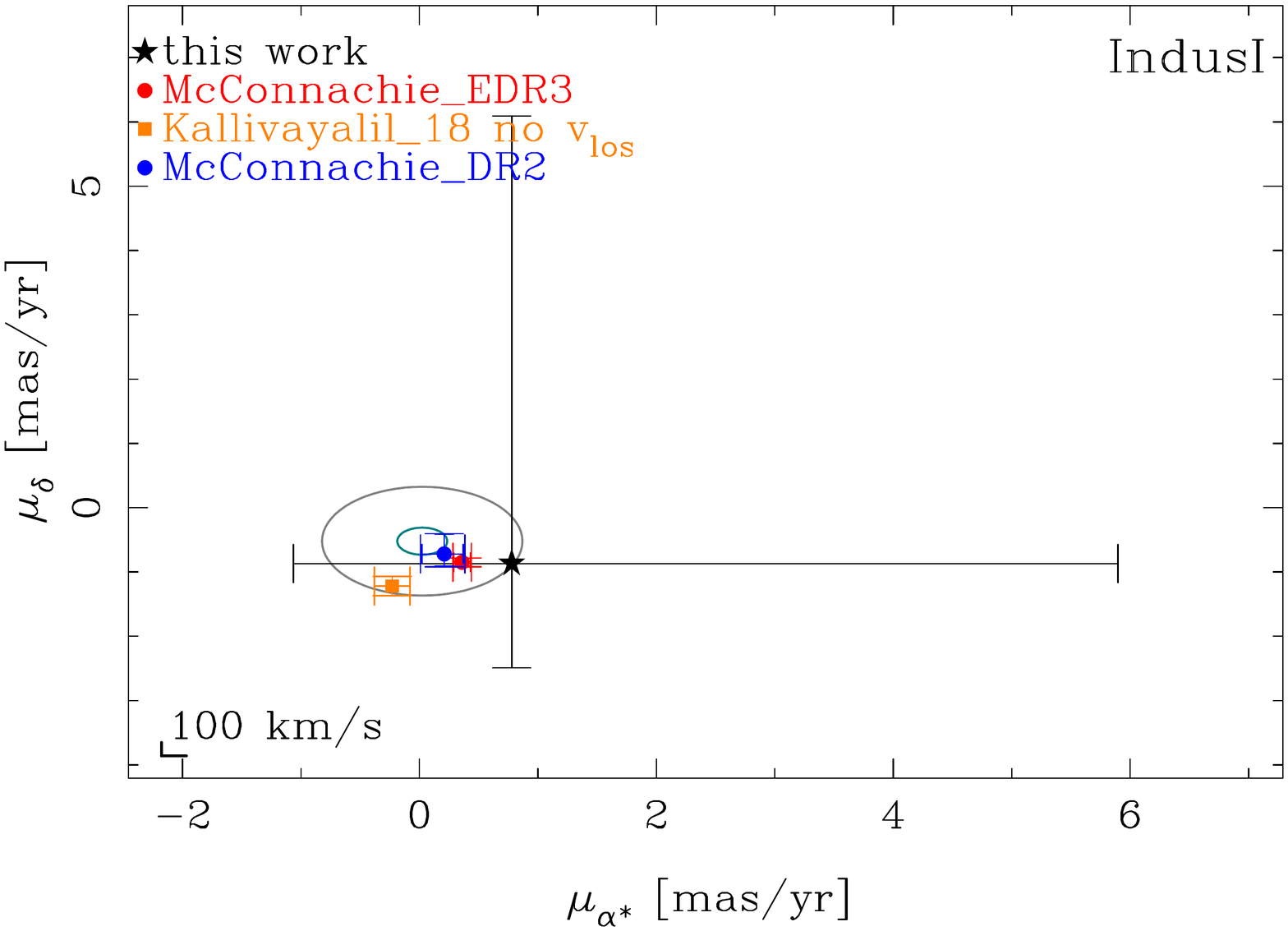}
      \includegraphics[width=0.30\textwidth,angle=0]{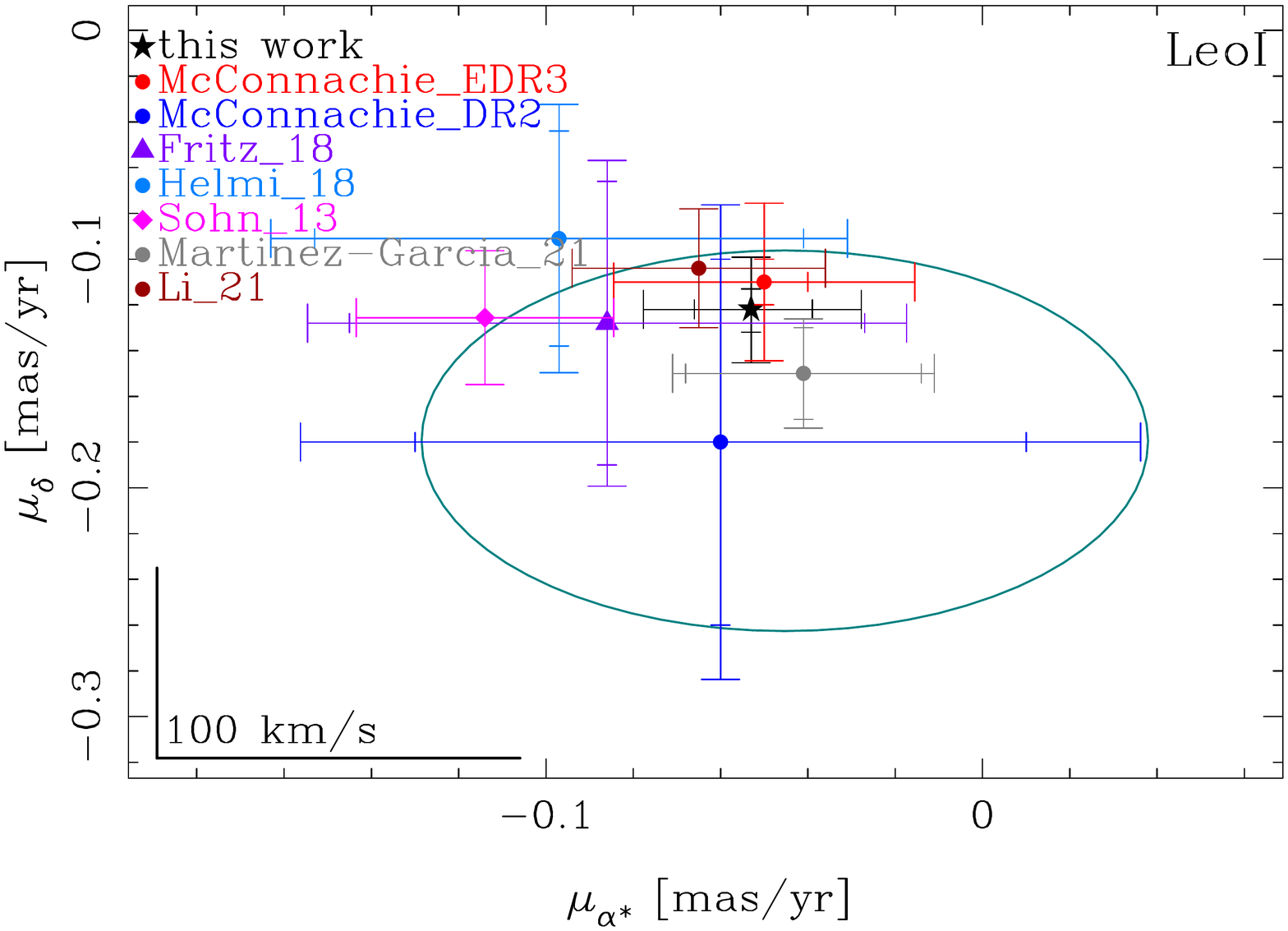}
\includegraphics[width=0.30\textwidth,angle=0]{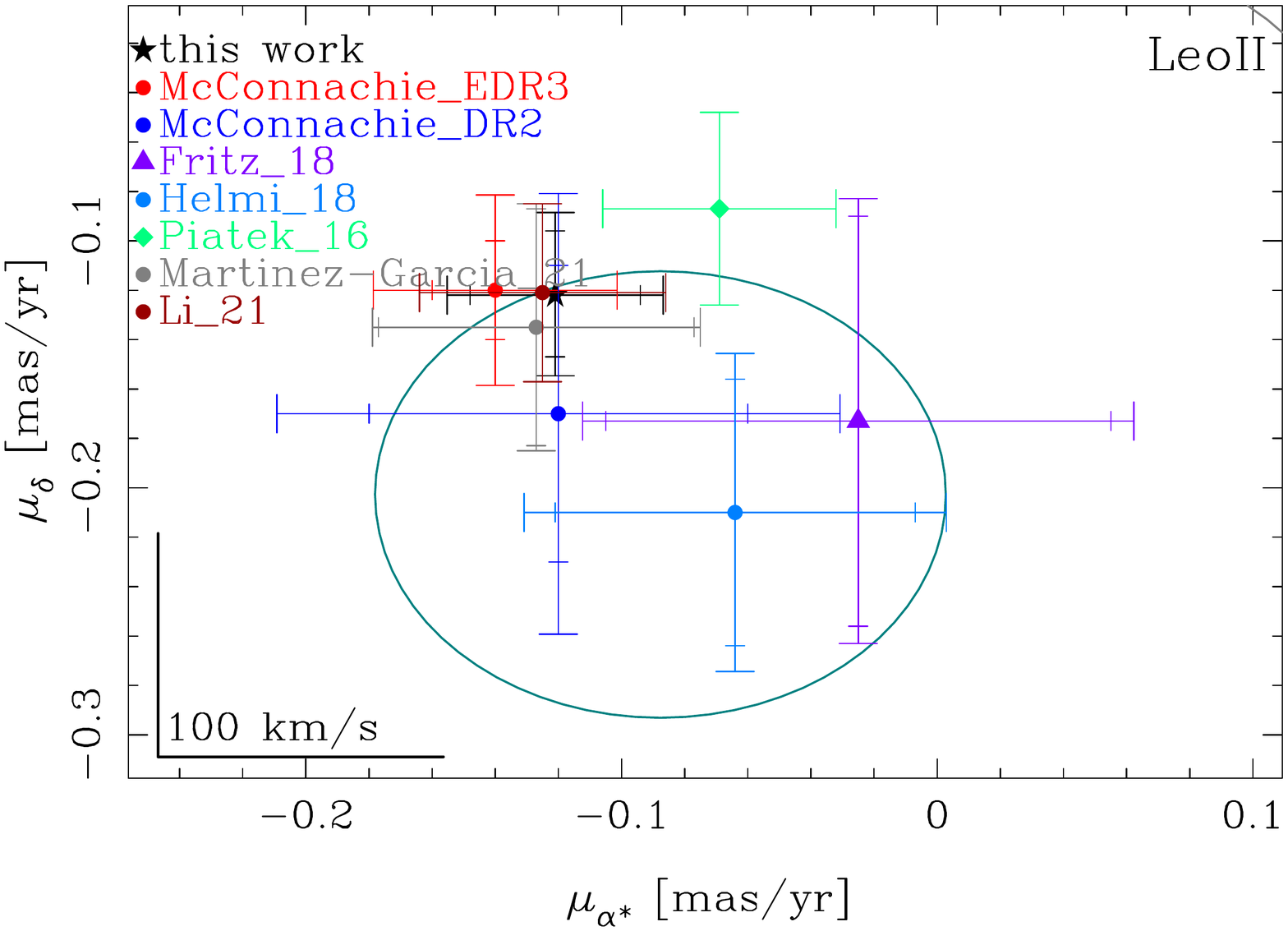}
\includegraphics[width=0.30\textwidth,angle=0]{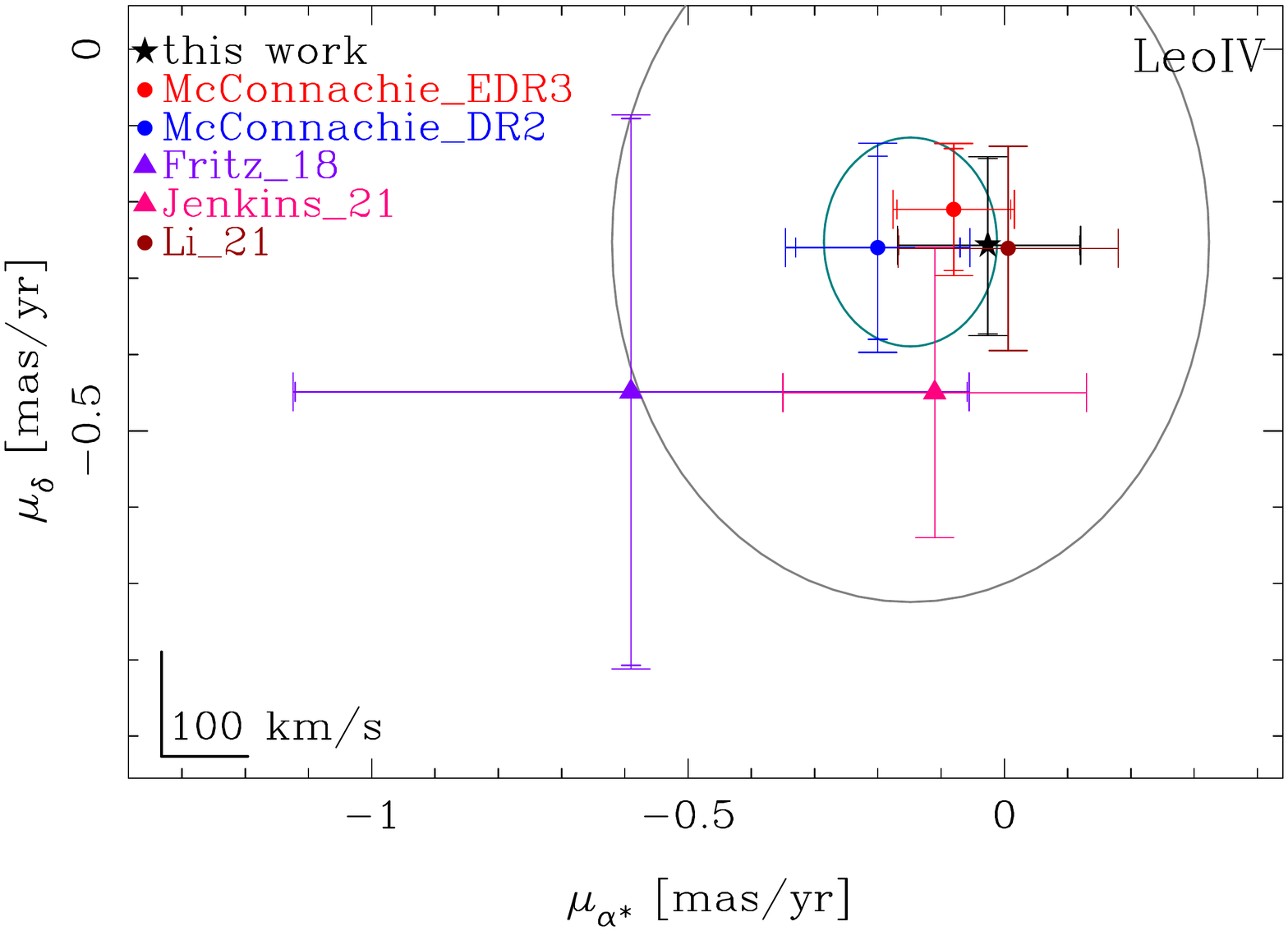}
      \caption{As Figure~\ref{fig:pms_lit}.}
         \label{fig:pms_lit3}
   \end{figure*}

       \begin{figure*}
   \centering
\includegraphics[width=0.30\textwidth,angle=0]{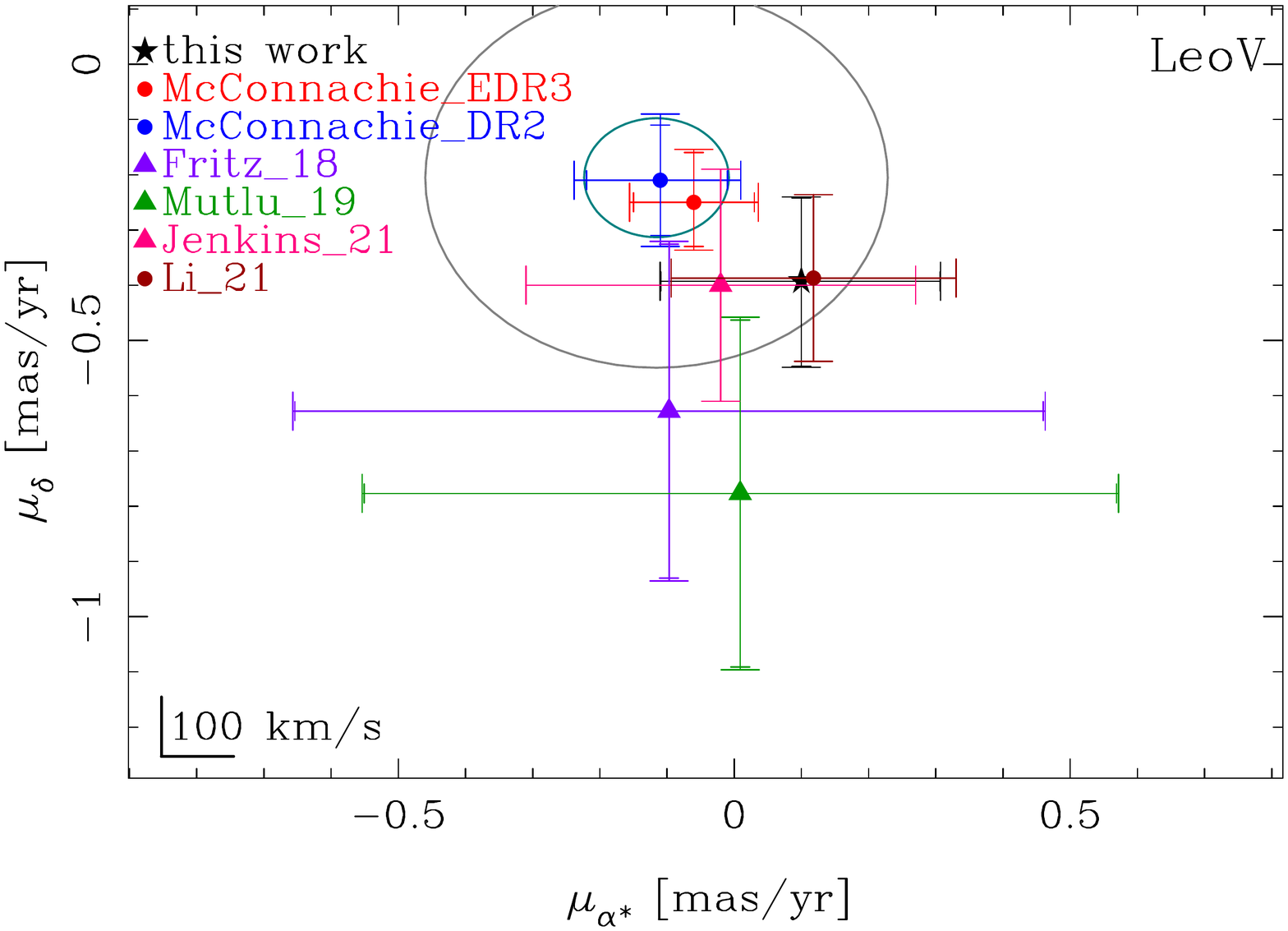}
\includegraphics[width=0.30\textwidth,angle=0]{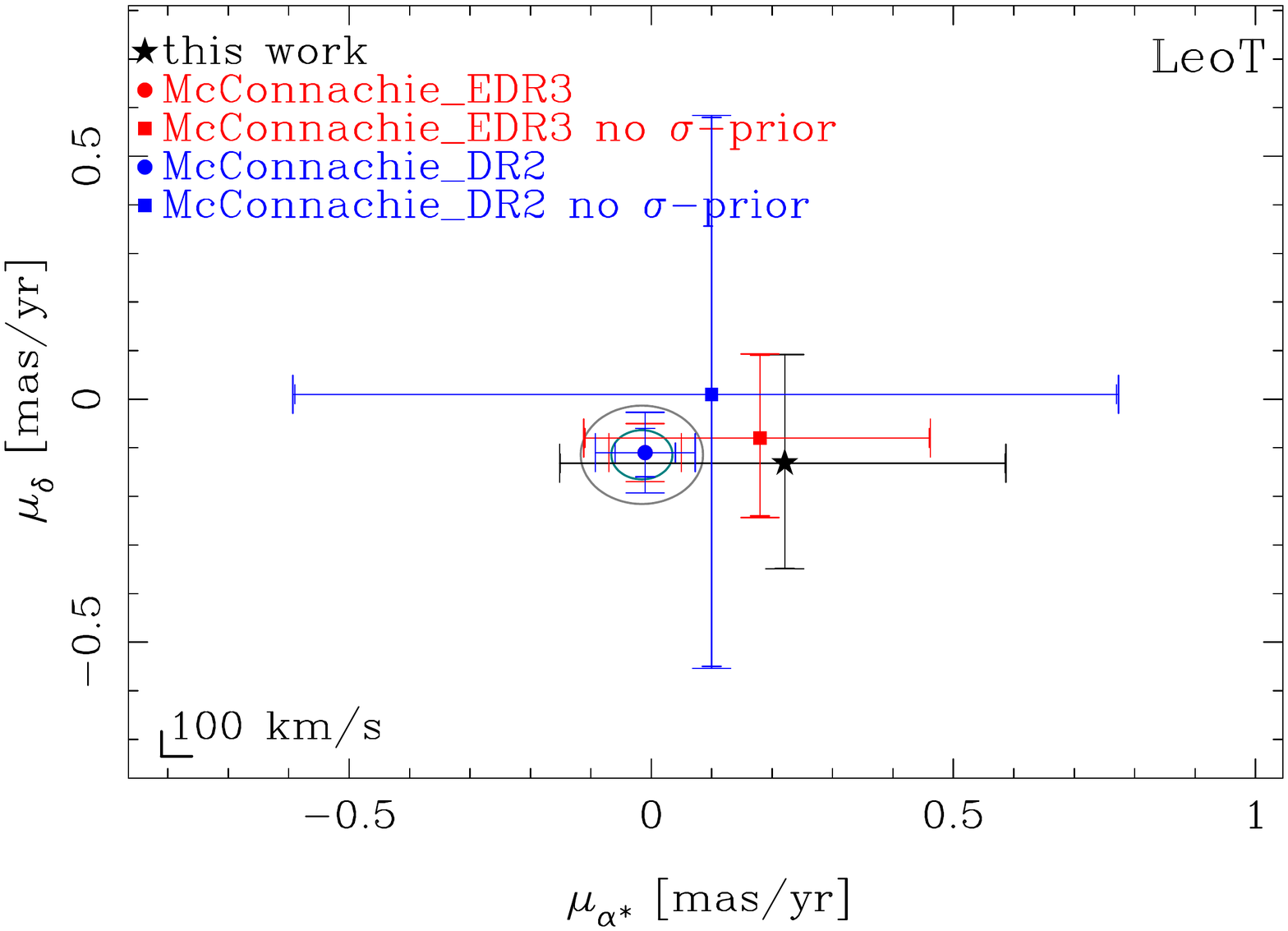}
\includegraphics[width=0.30\textwidth,angle=0]{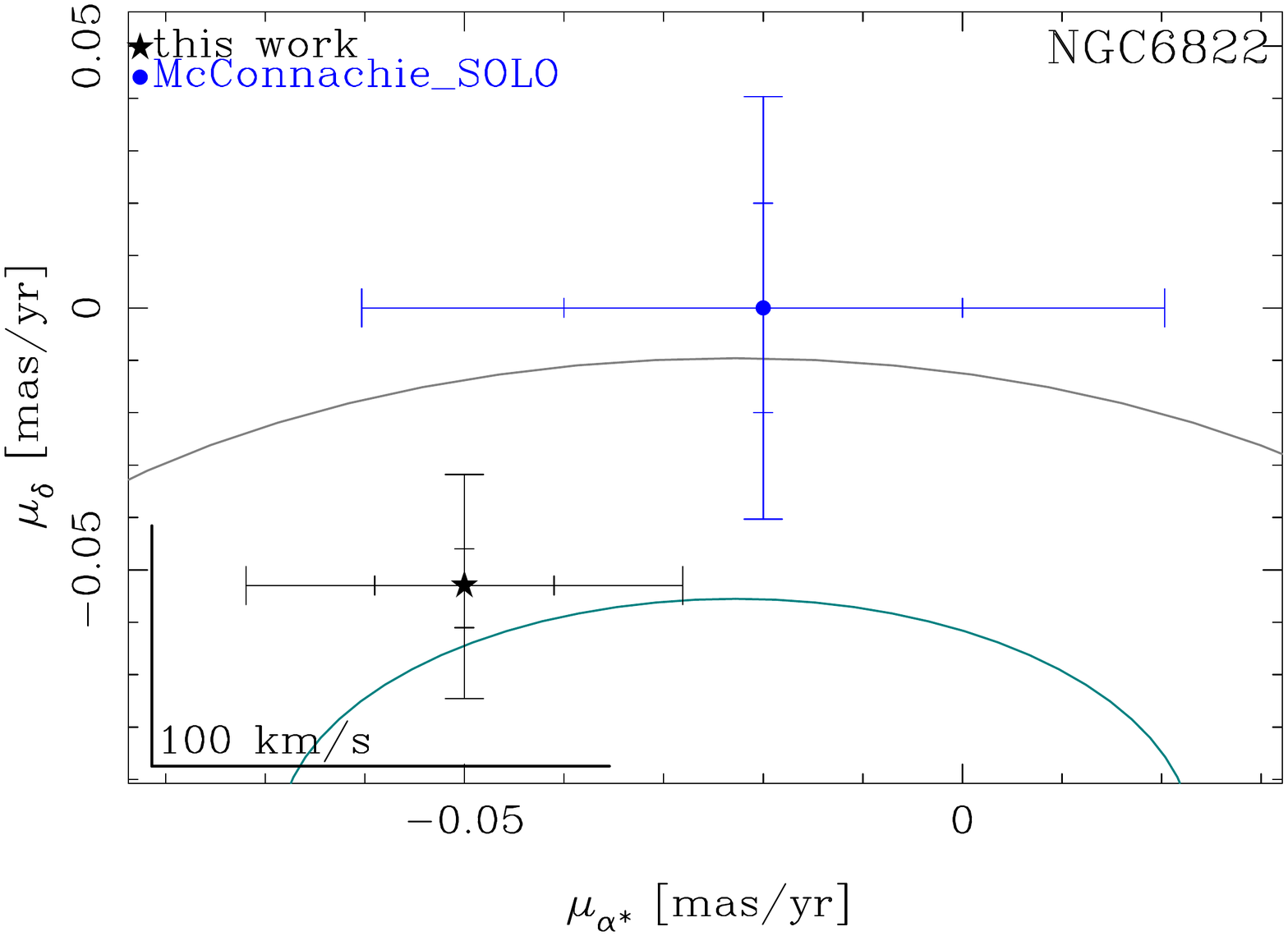}
\includegraphics[width=0.30\textwidth,angle=0]{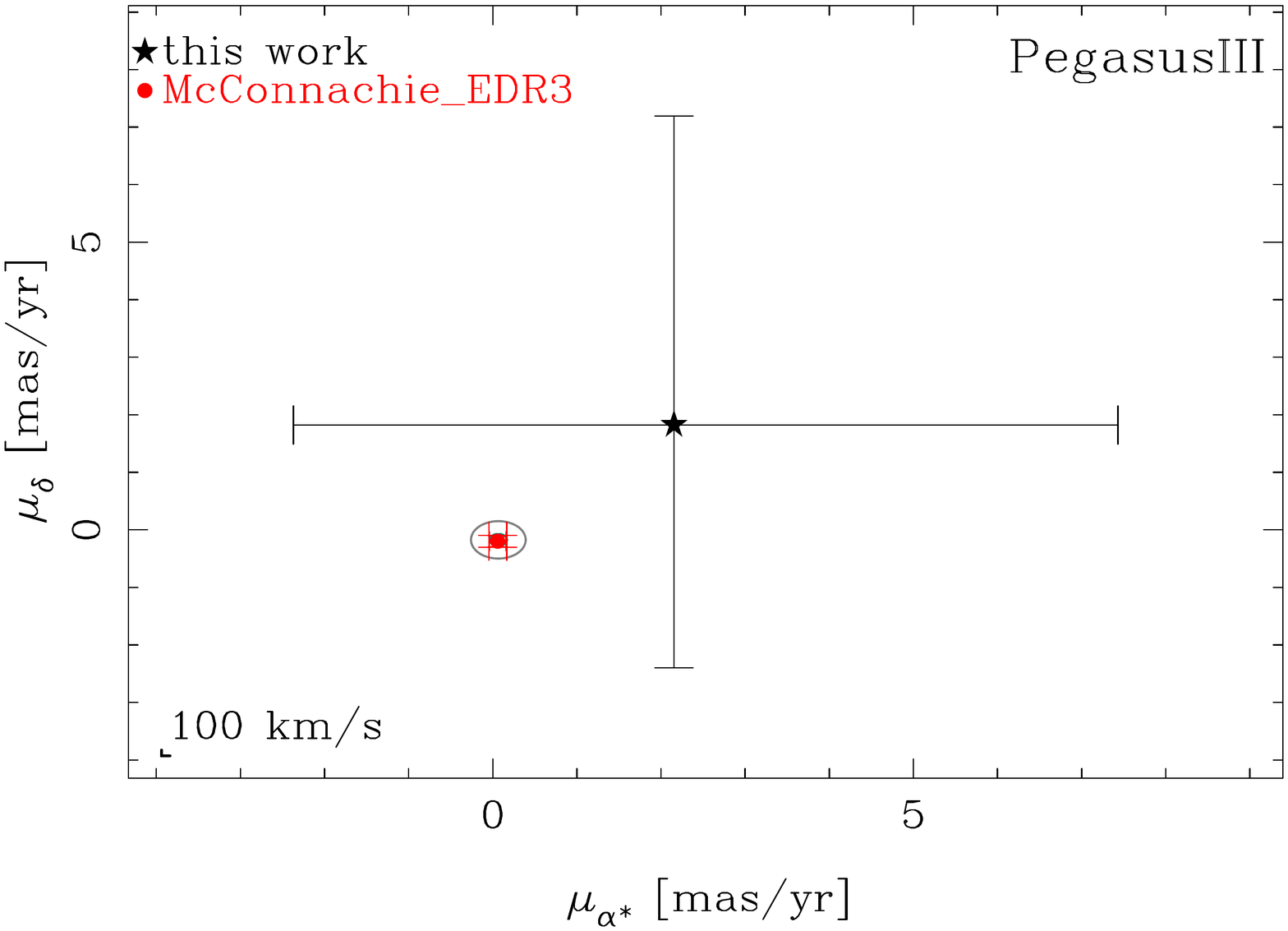}
\includegraphics[width=0.30\textwidth,angle=0]{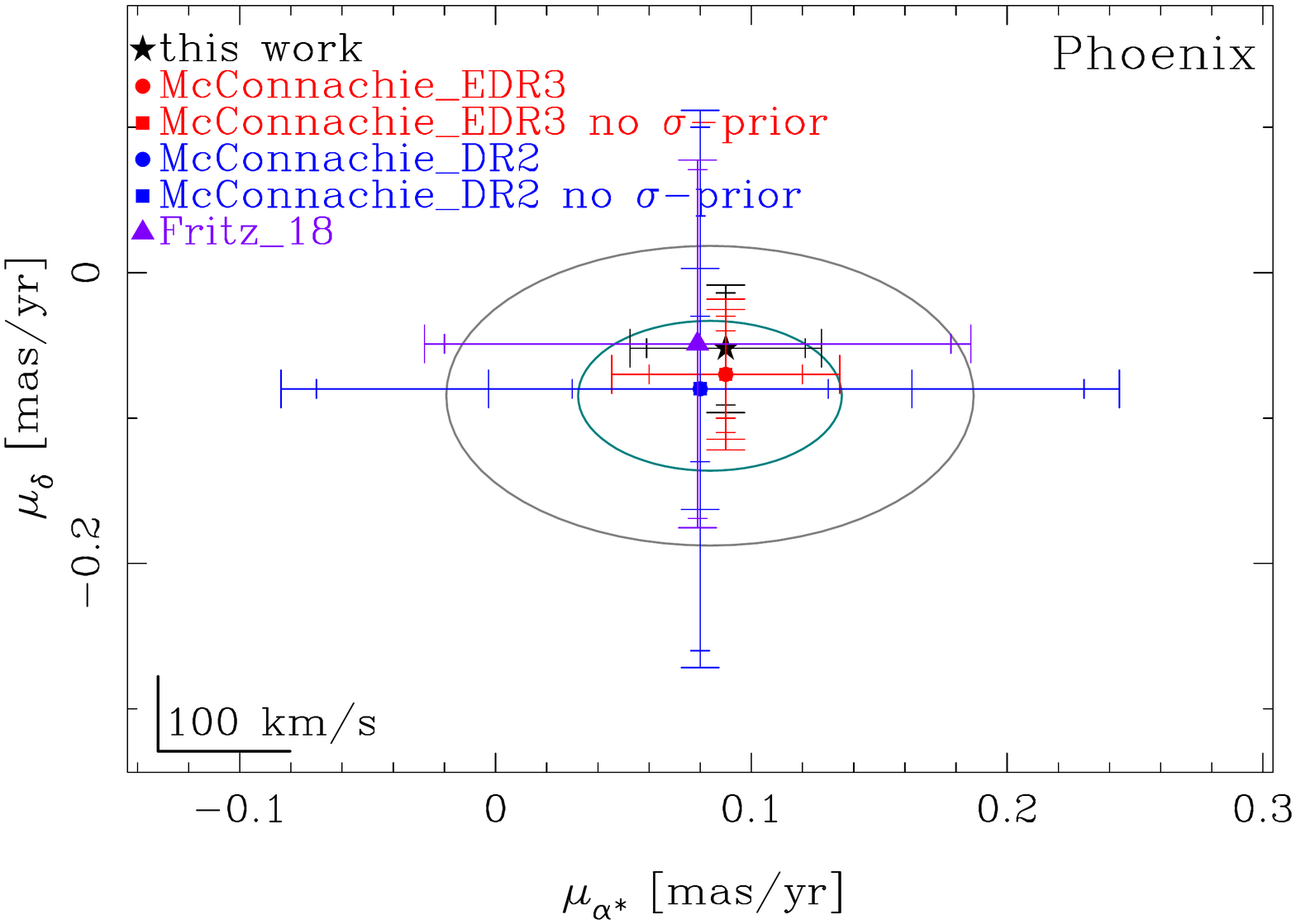}
   \includegraphics[width=0.30\textwidth,angle=0]{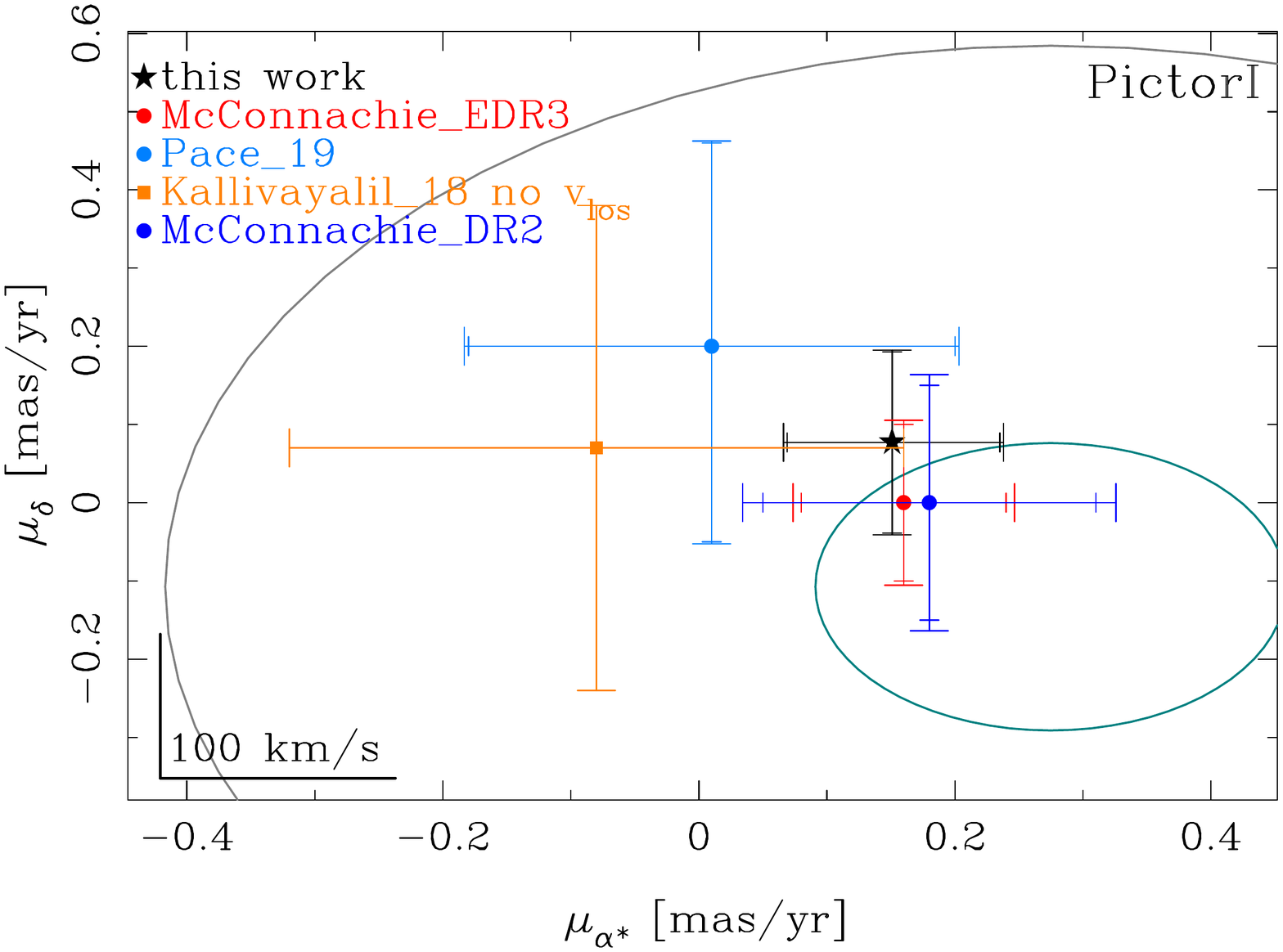}
      \includegraphics[width=0.30\textwidth,angle=0]{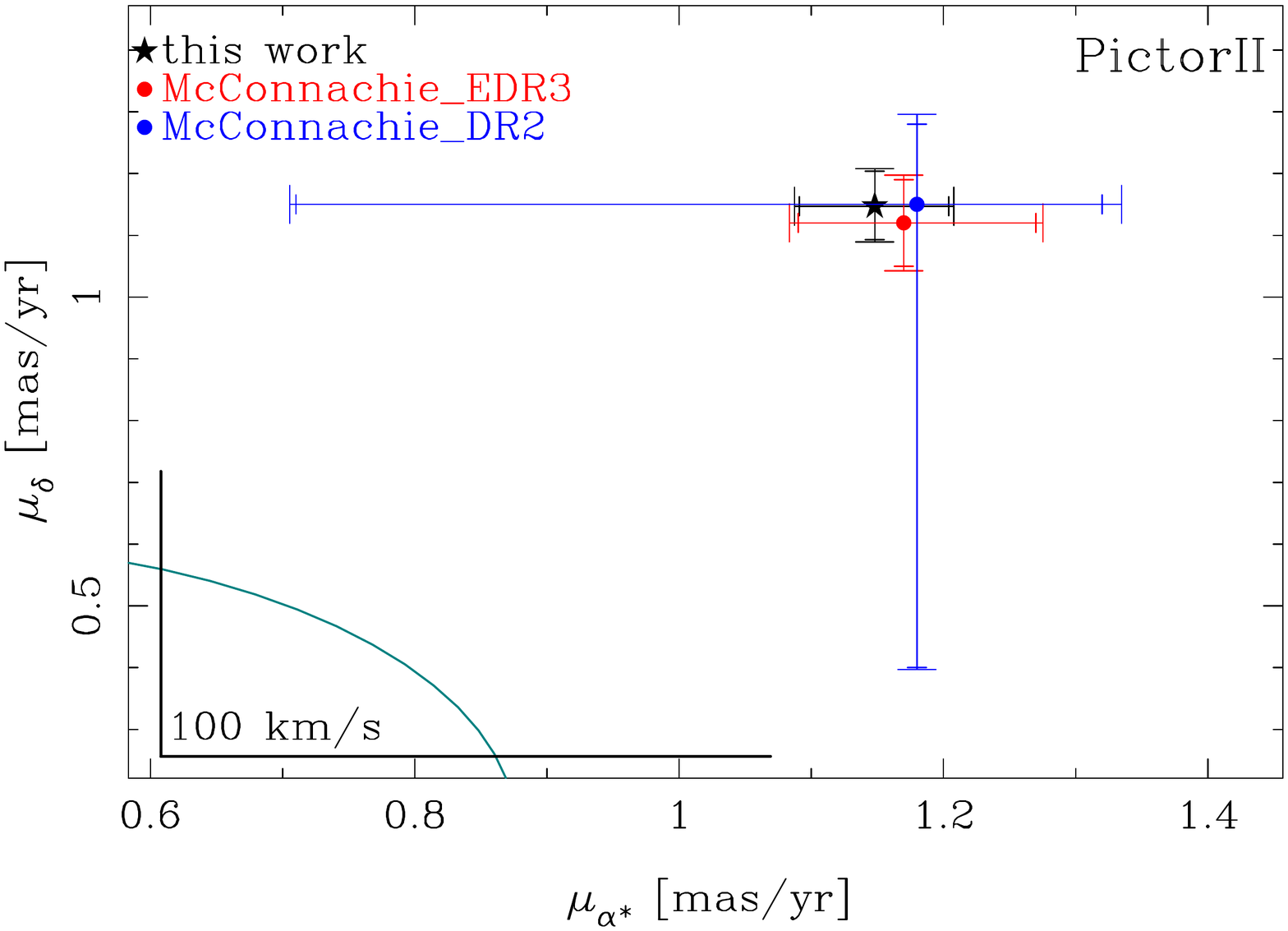}
            \includegraphics[width=0.30\textwidth,angle=0]{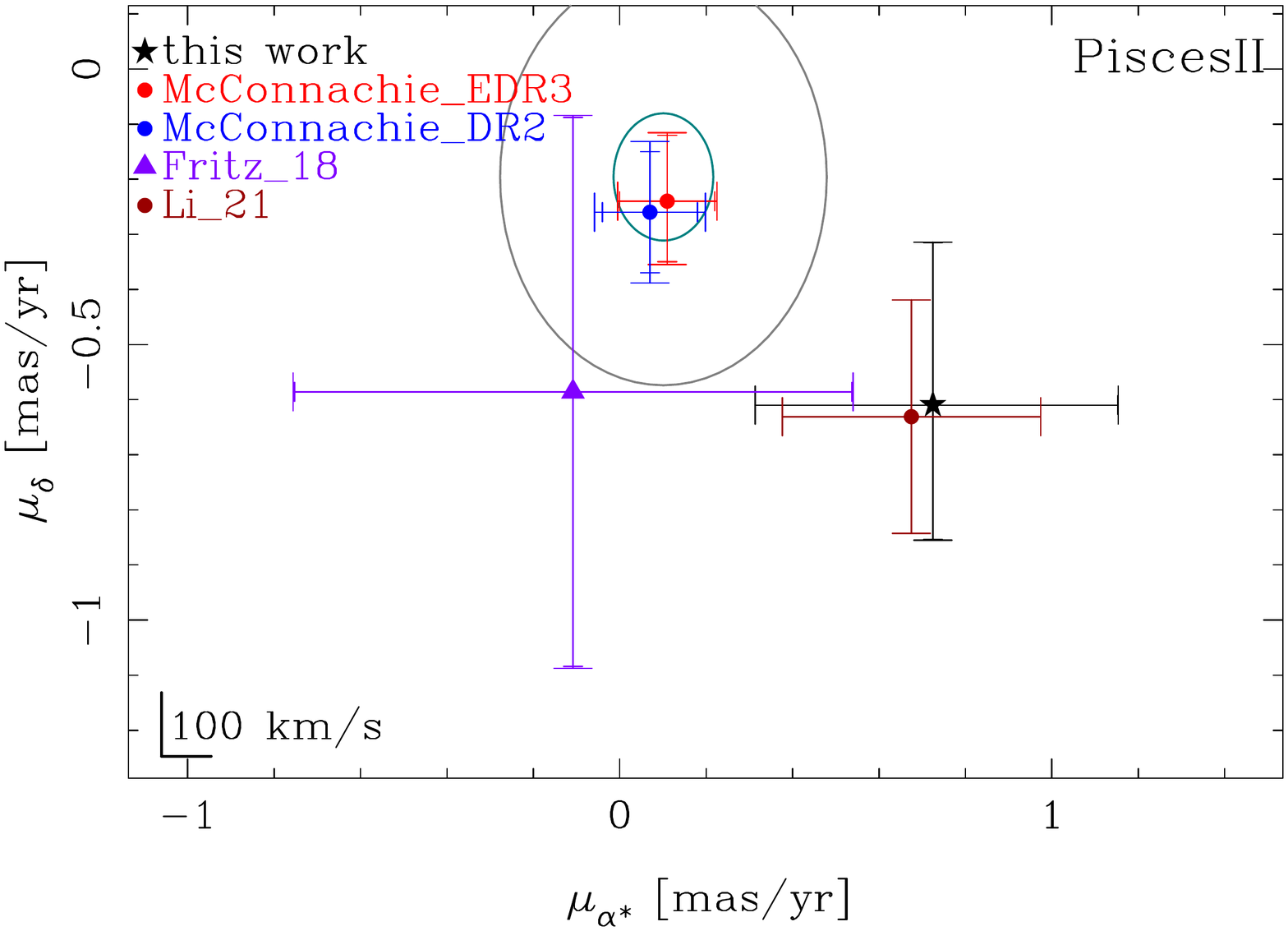}
            \includegraphics[width=0.30\textwidth,angle=0]{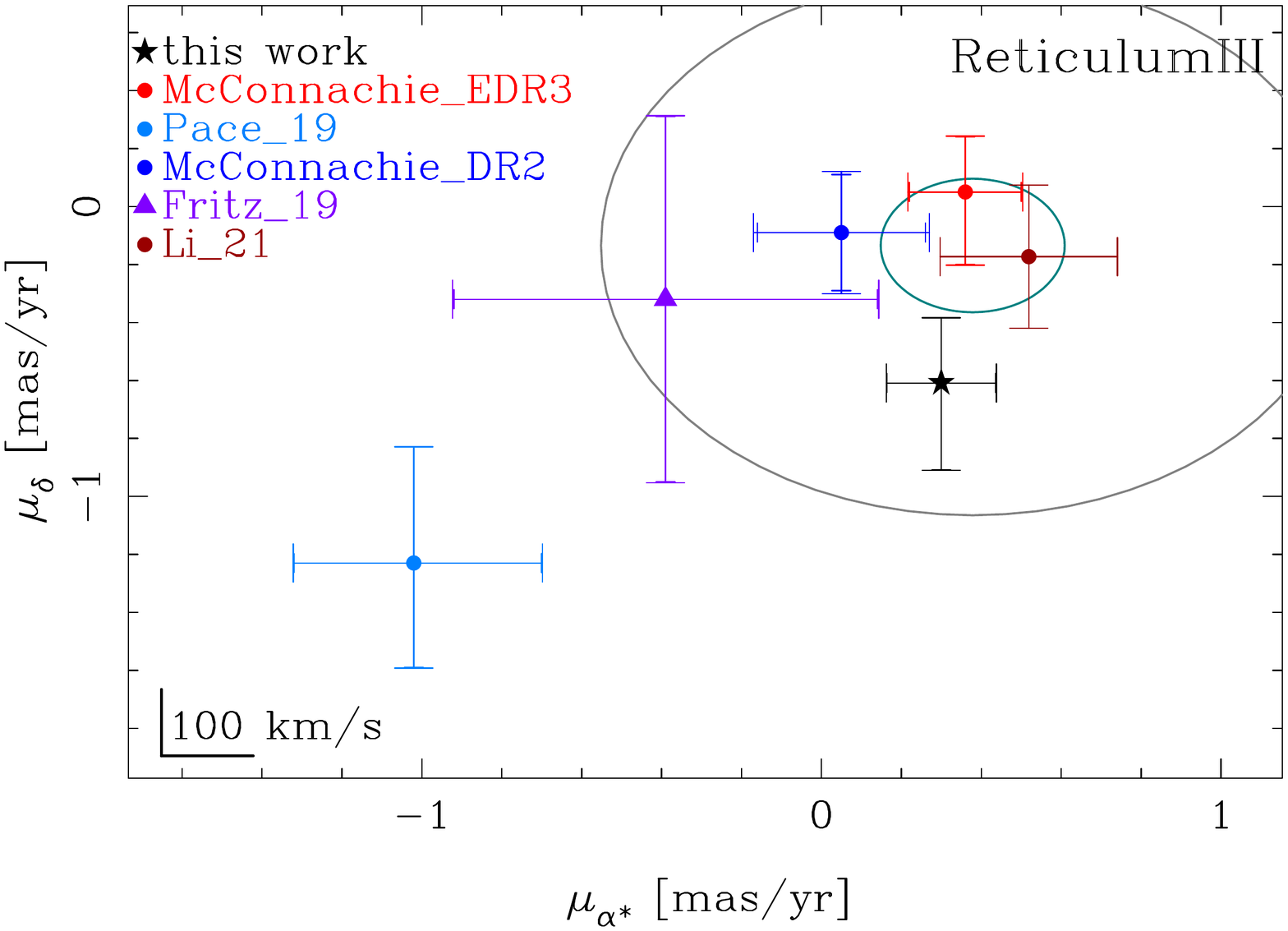}
\includegraphics[width=0.30\textwidth,angle=0]{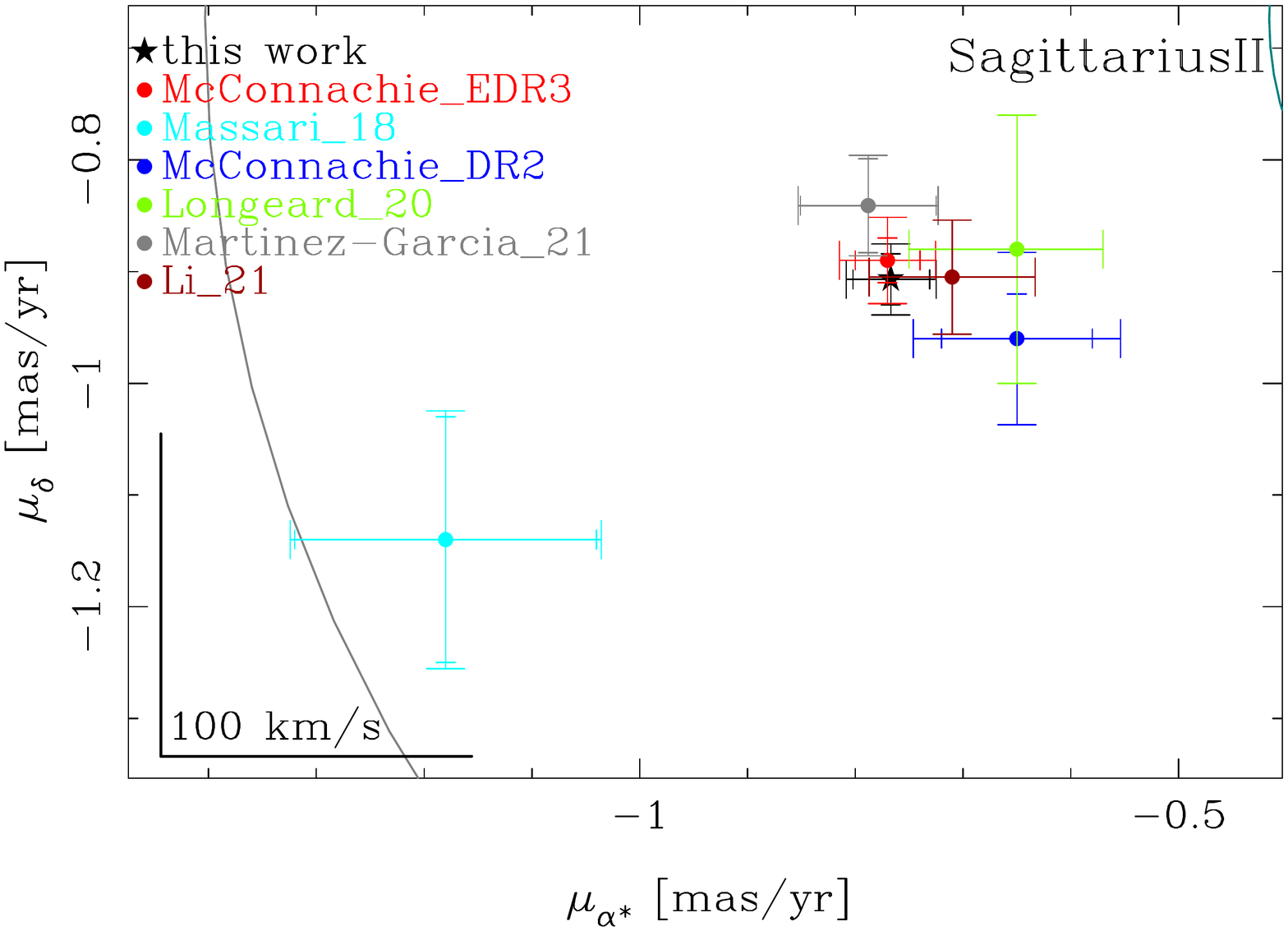}
\includegraphics[width=0.30\textwidth,angle=0]{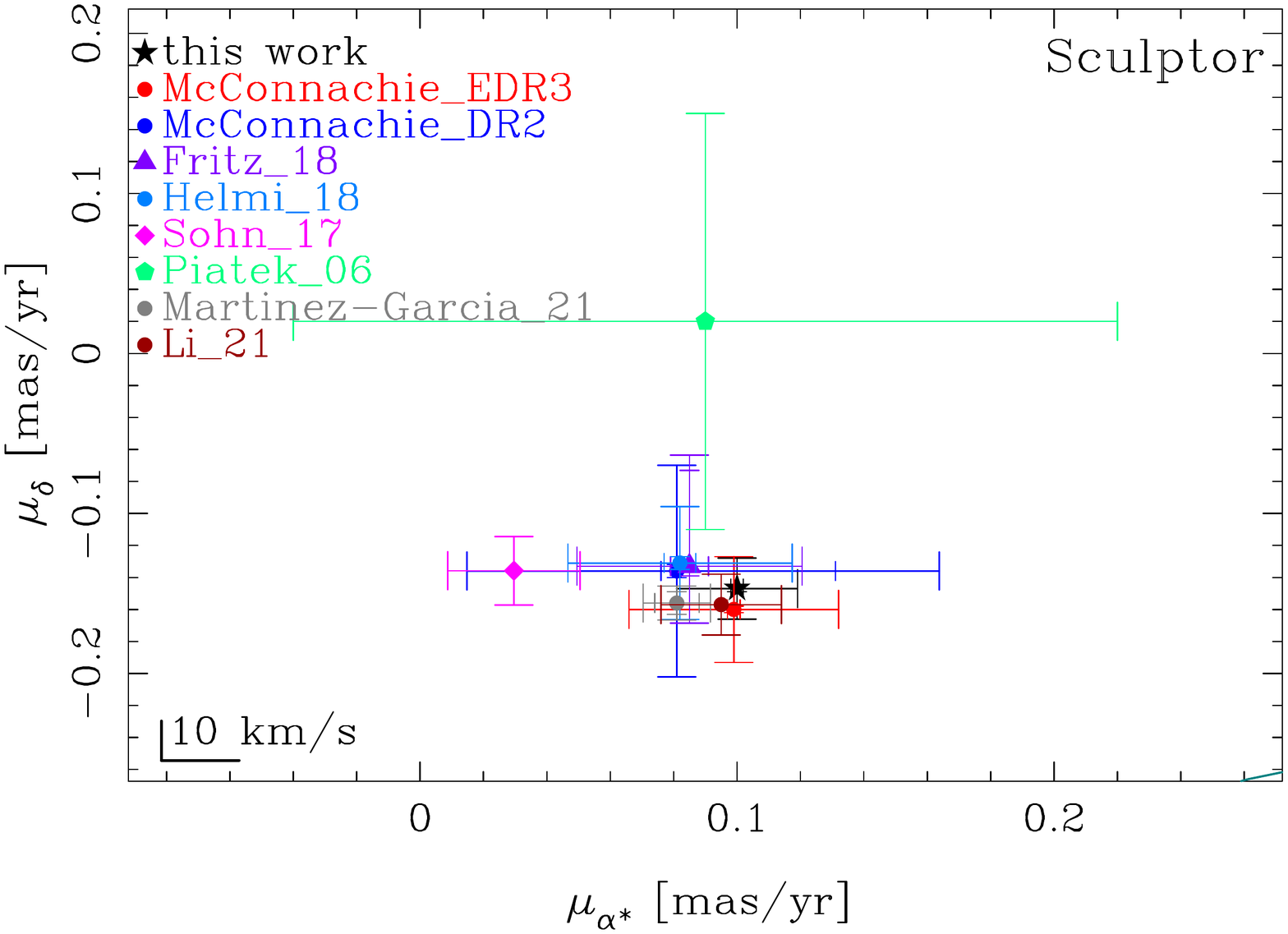}
\includegraphics[width=0.30\textwidth,angle=0]{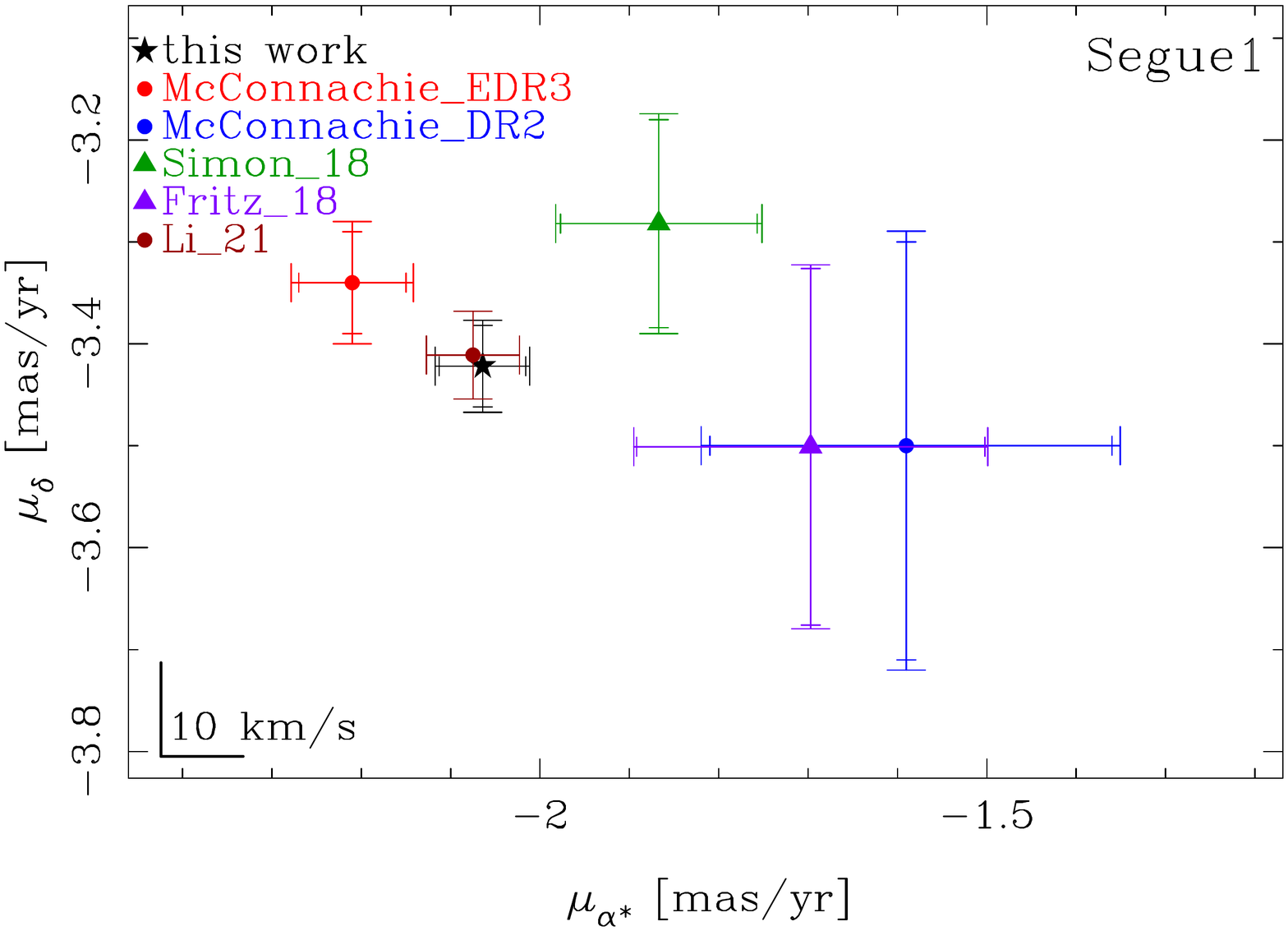}
\includegraphics[width=0.30\textwidth,angle=0]{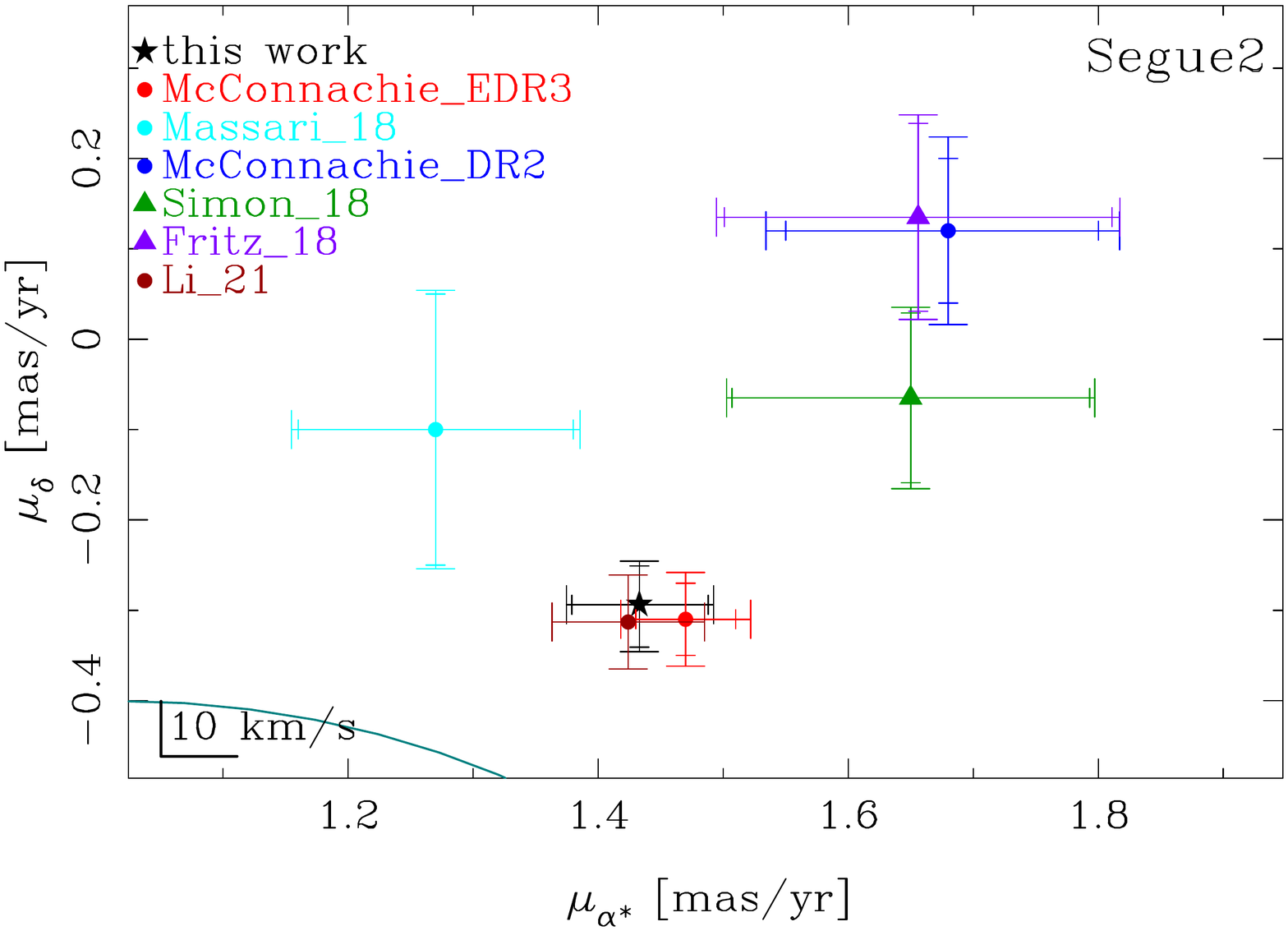}
\includegraphics[width=0.30\textwidth,angle=0]{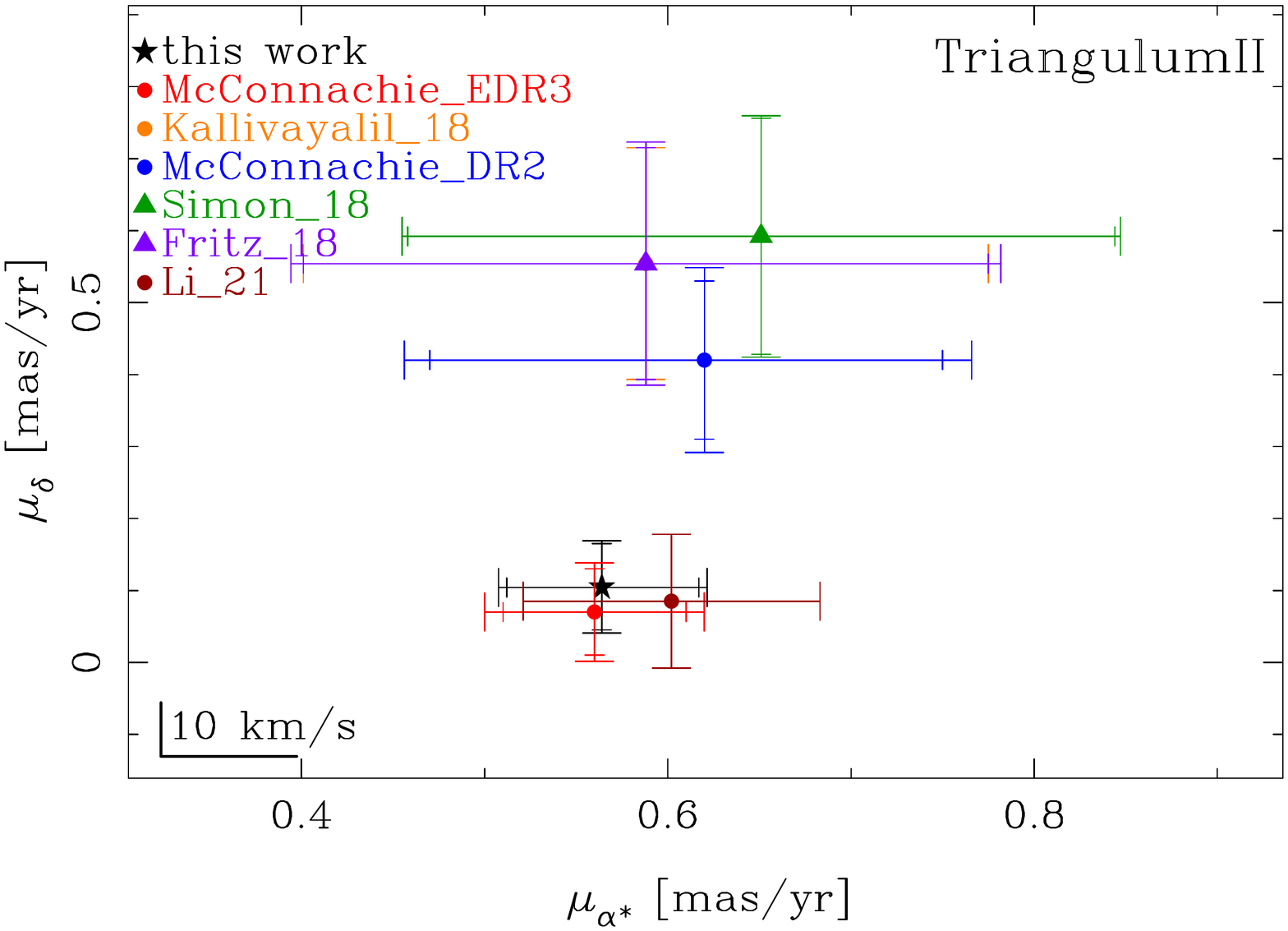}
\includegraphics[width=0.30\textwidth,angle=0]{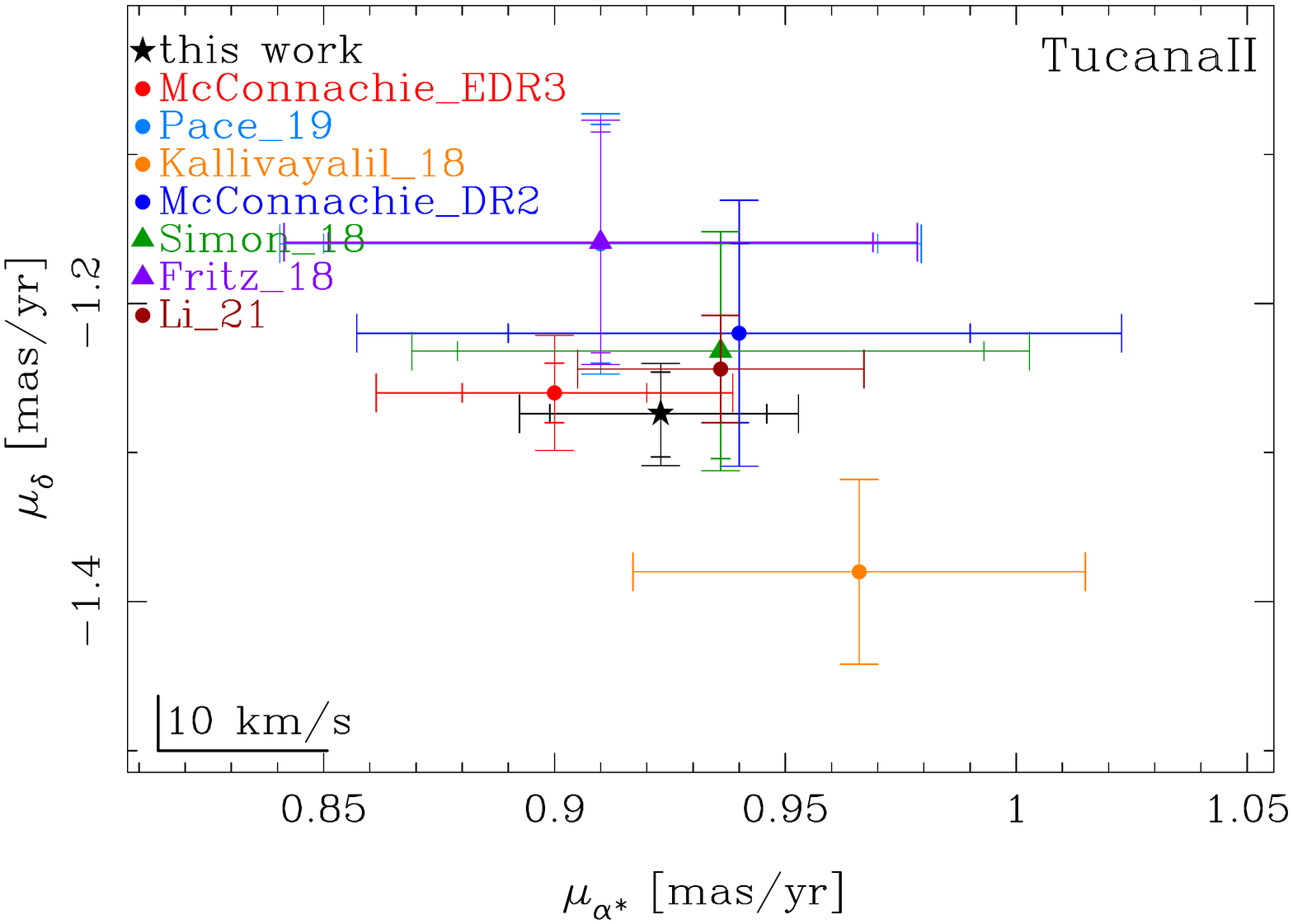}
      \caption{See Figure~\ref{fig:pms_lit}}
         \label{fig:pms_lit4}
   \end{figure*}

       \begin{figure*}
   \centering

\includegraphics[width=0.30\textwidth,angle=0]{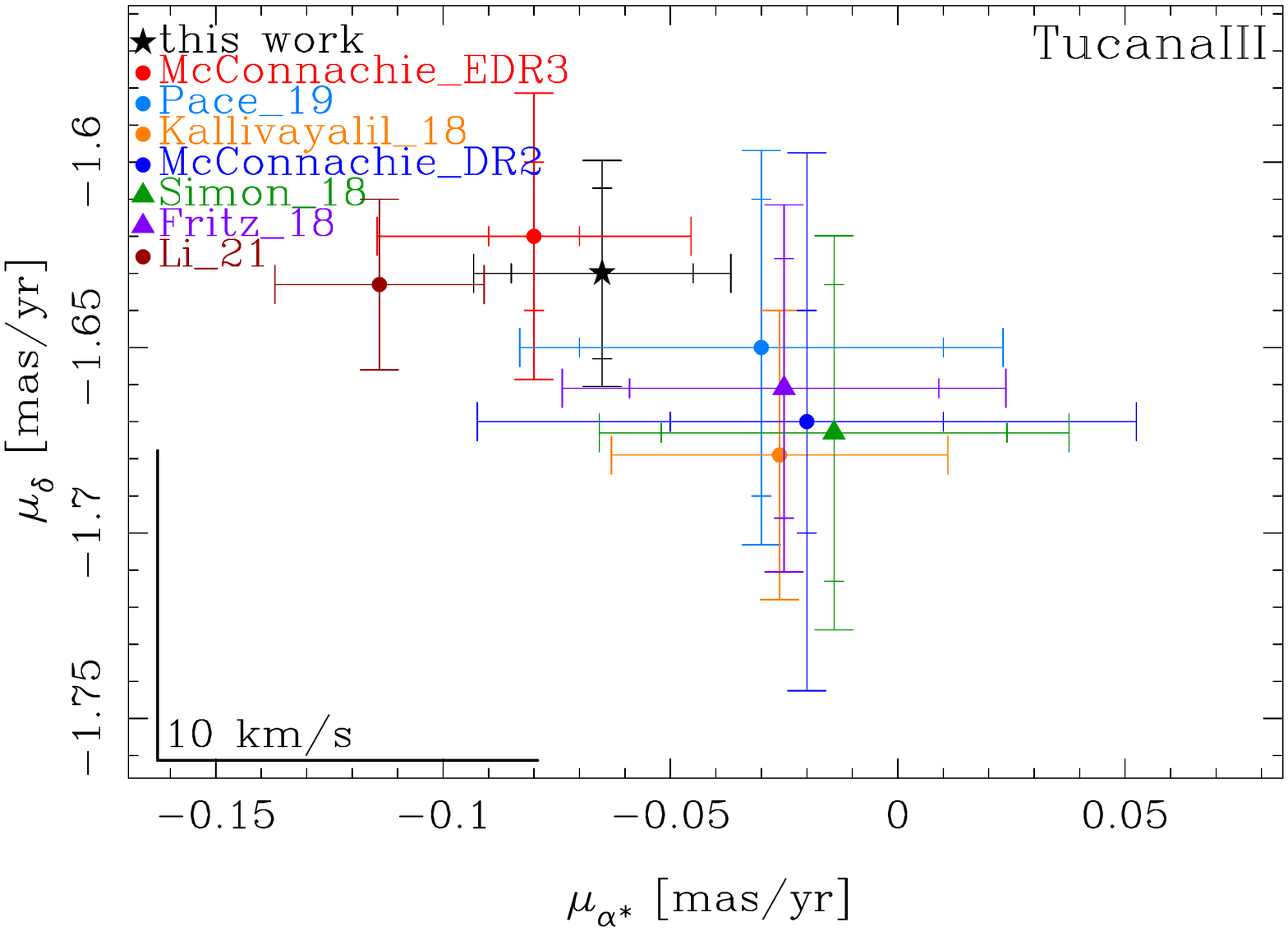}
\includegraphics[width=0.30\textwidth,angle=0]{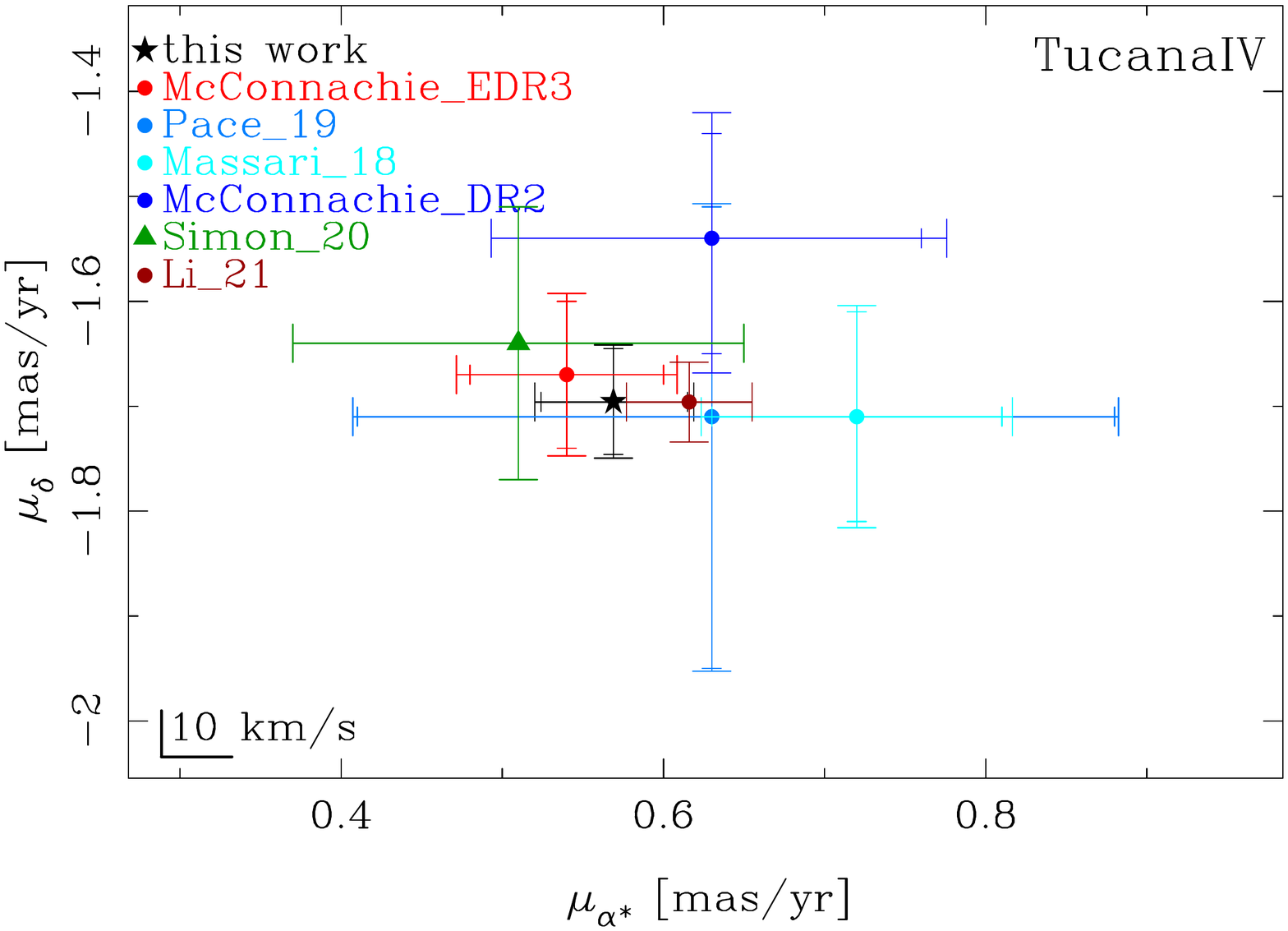}
   \includegraphics[width=0.30\textwidth,angle=0]{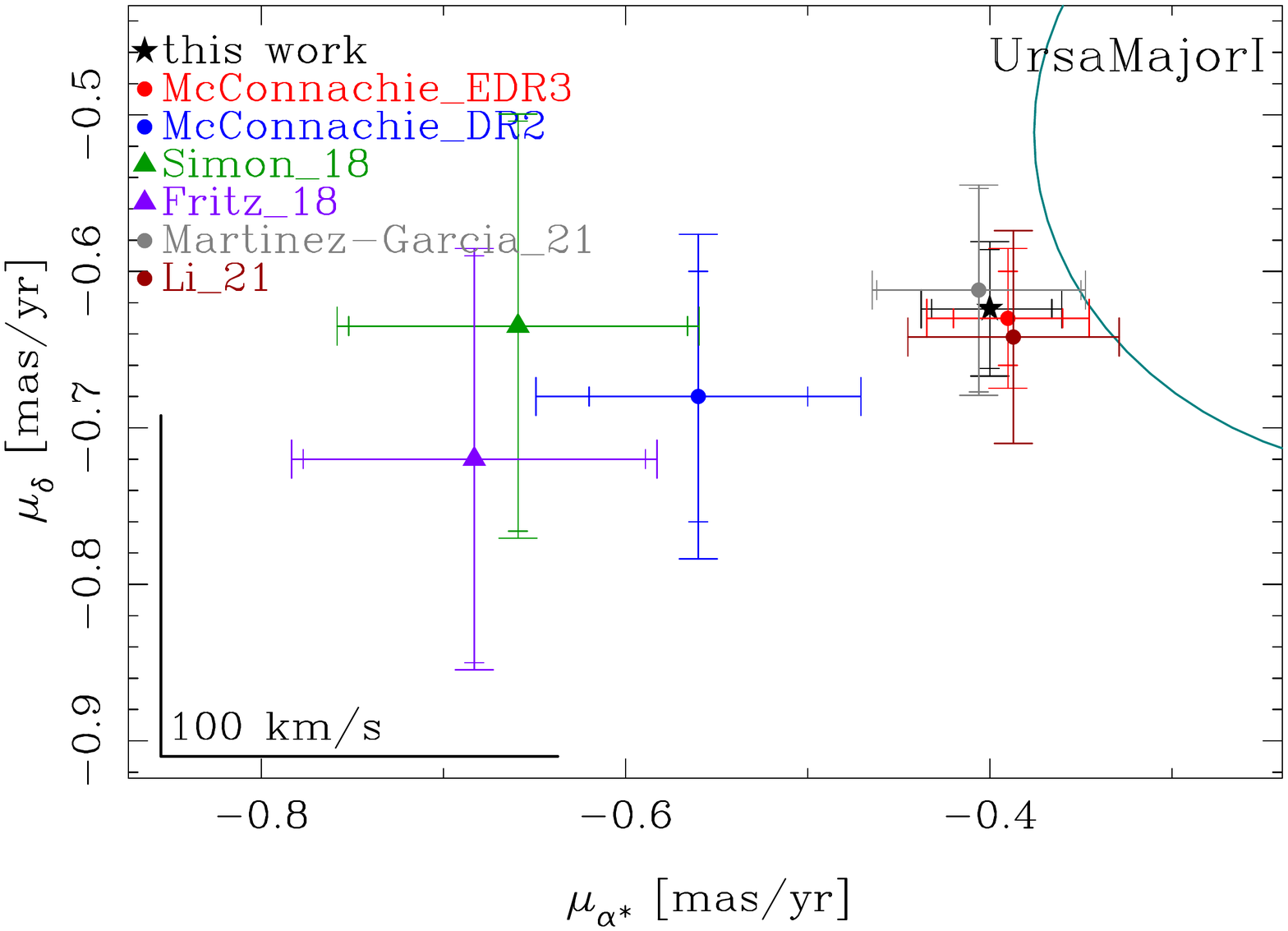}
      \includegraphics[width=0.30\textwidth,angle=0]{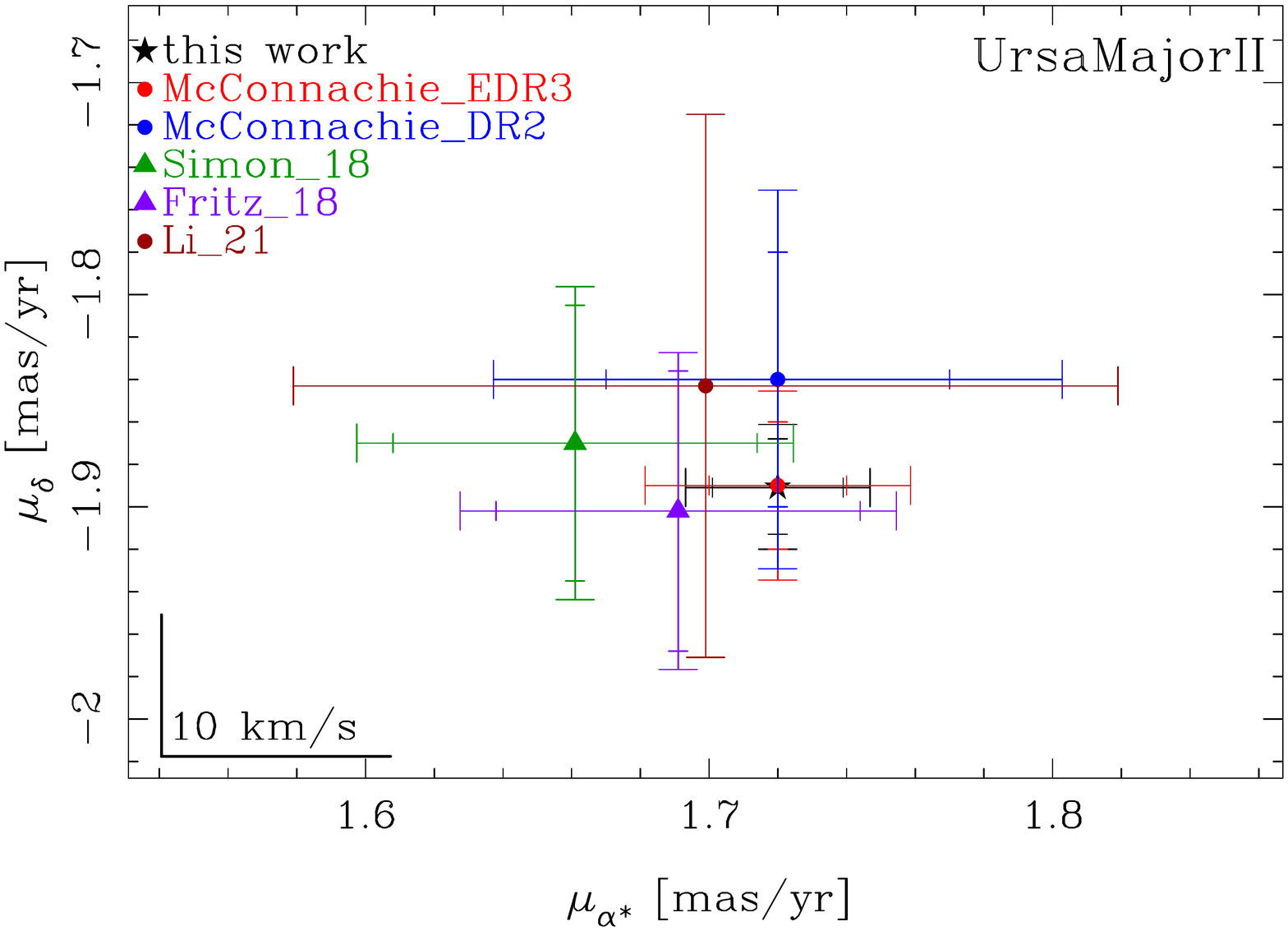}
            \includegraphics[width=0.30\textwidth,angle=0]{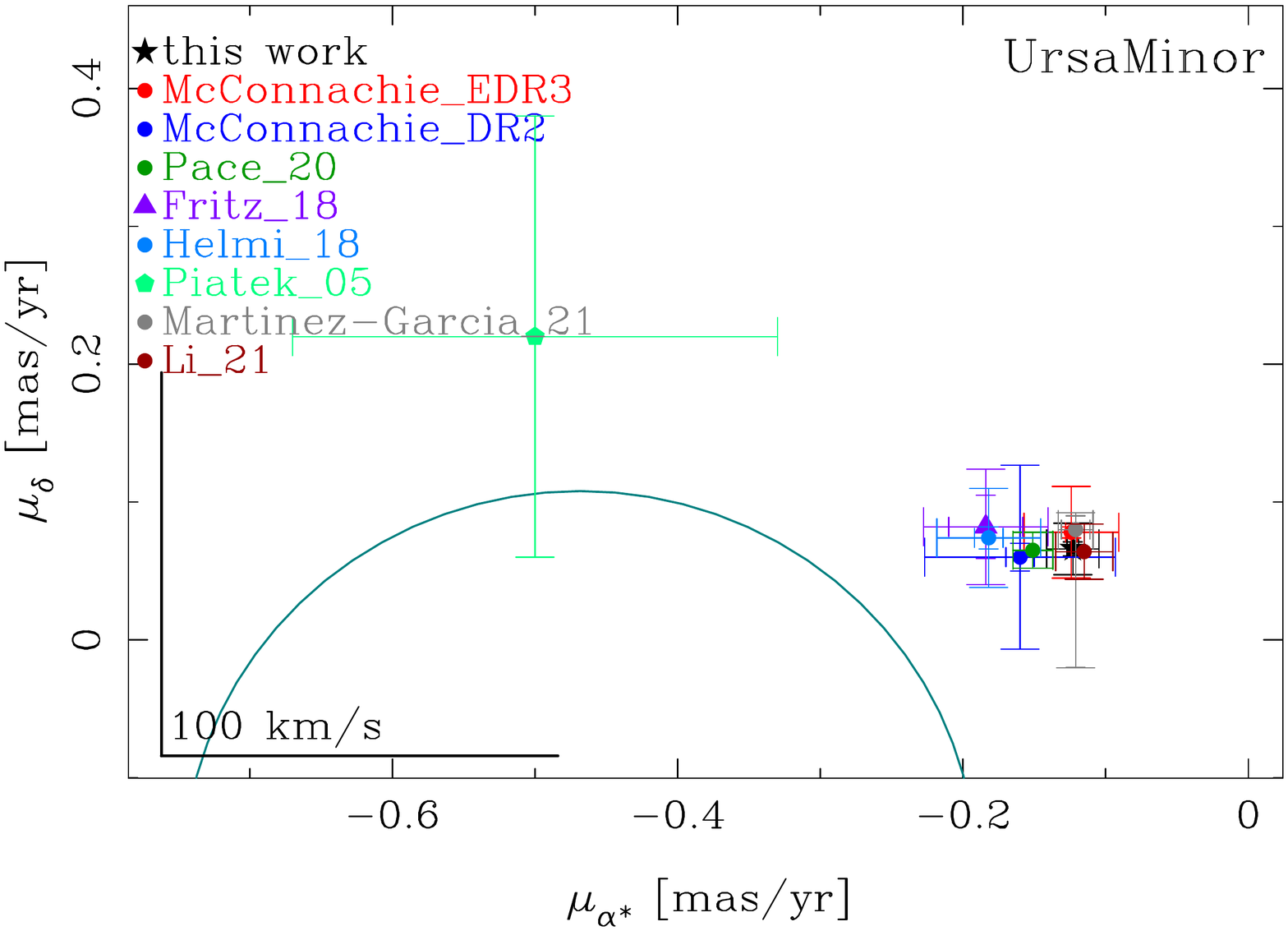}
\includegraphics[width=0.30\textwidth,angle=0]{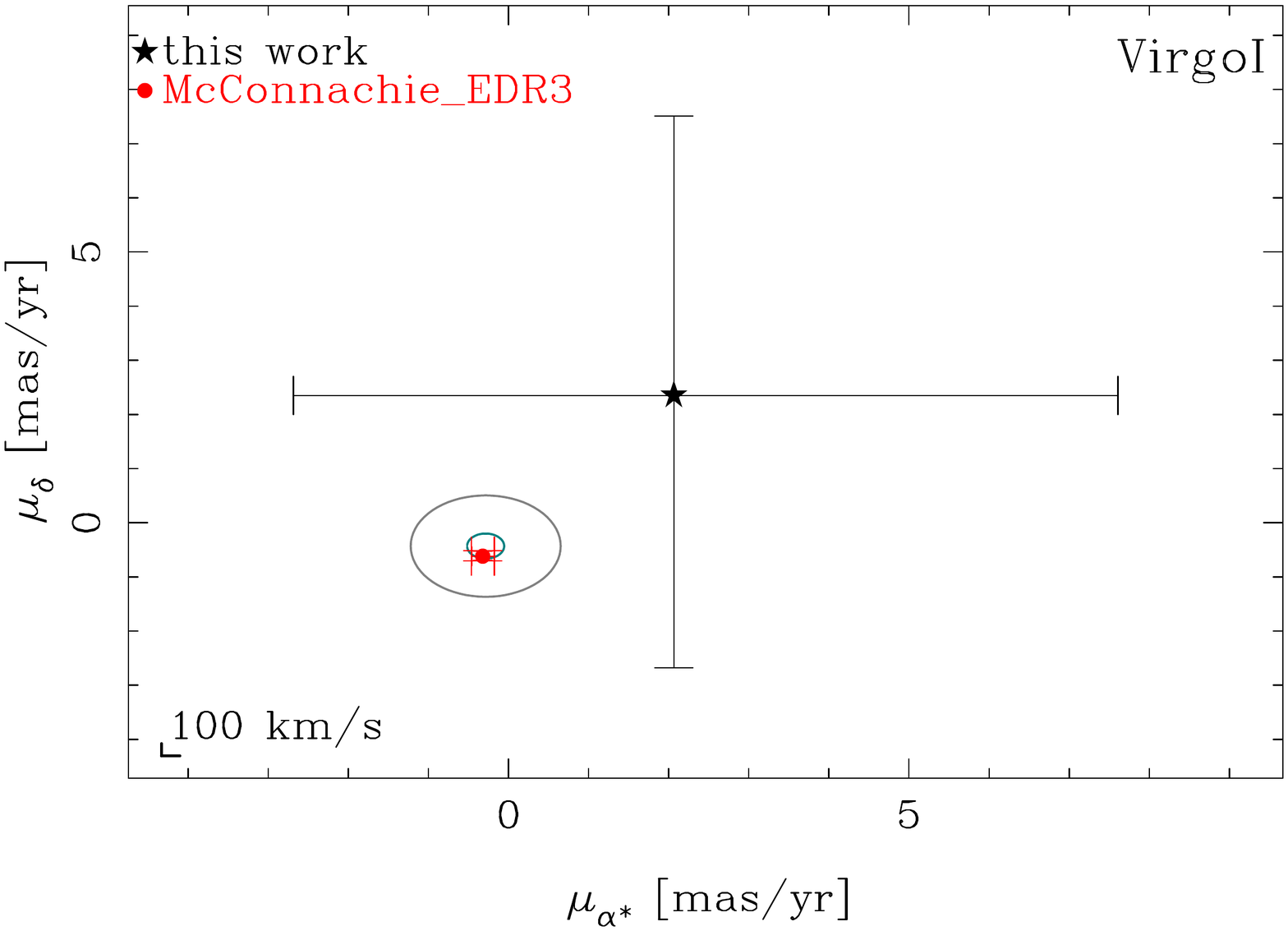}
\includegraphics[width=0.30\textwidth,angle=0]{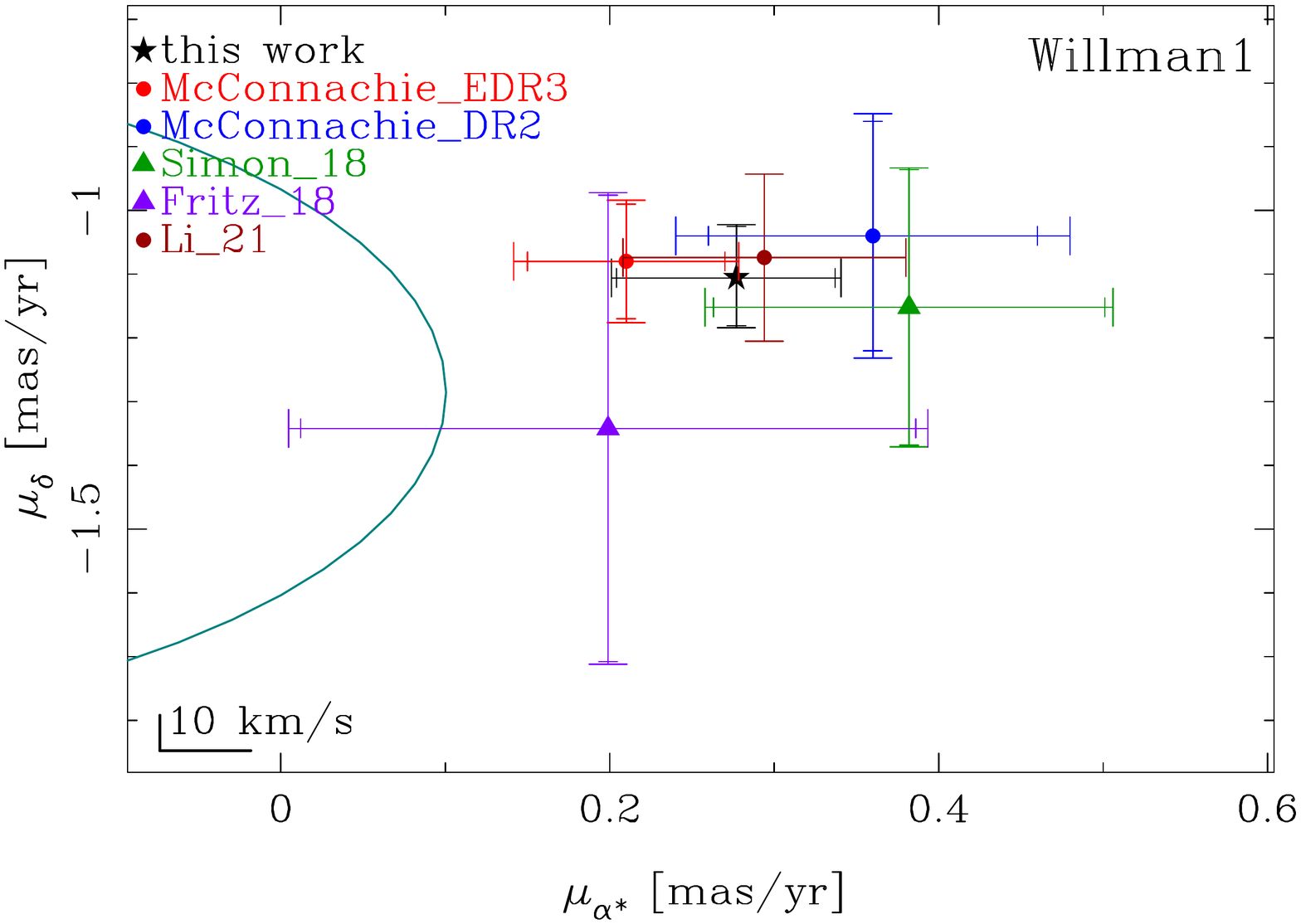}
      \caption{See Figure~\ref{fig:pms_lit}}
         \label{fig:pms_lit5}
   \end{figure*}

   \section{Plots on orbital histories}
Here we include plots showing the orbital evolution in the past 3 Gyr in the triaxial light MW potential with and without the inclusion of the LMC.

      \begin{figure*}
   \centering
\includegraphics[width=0.24\textwidth]{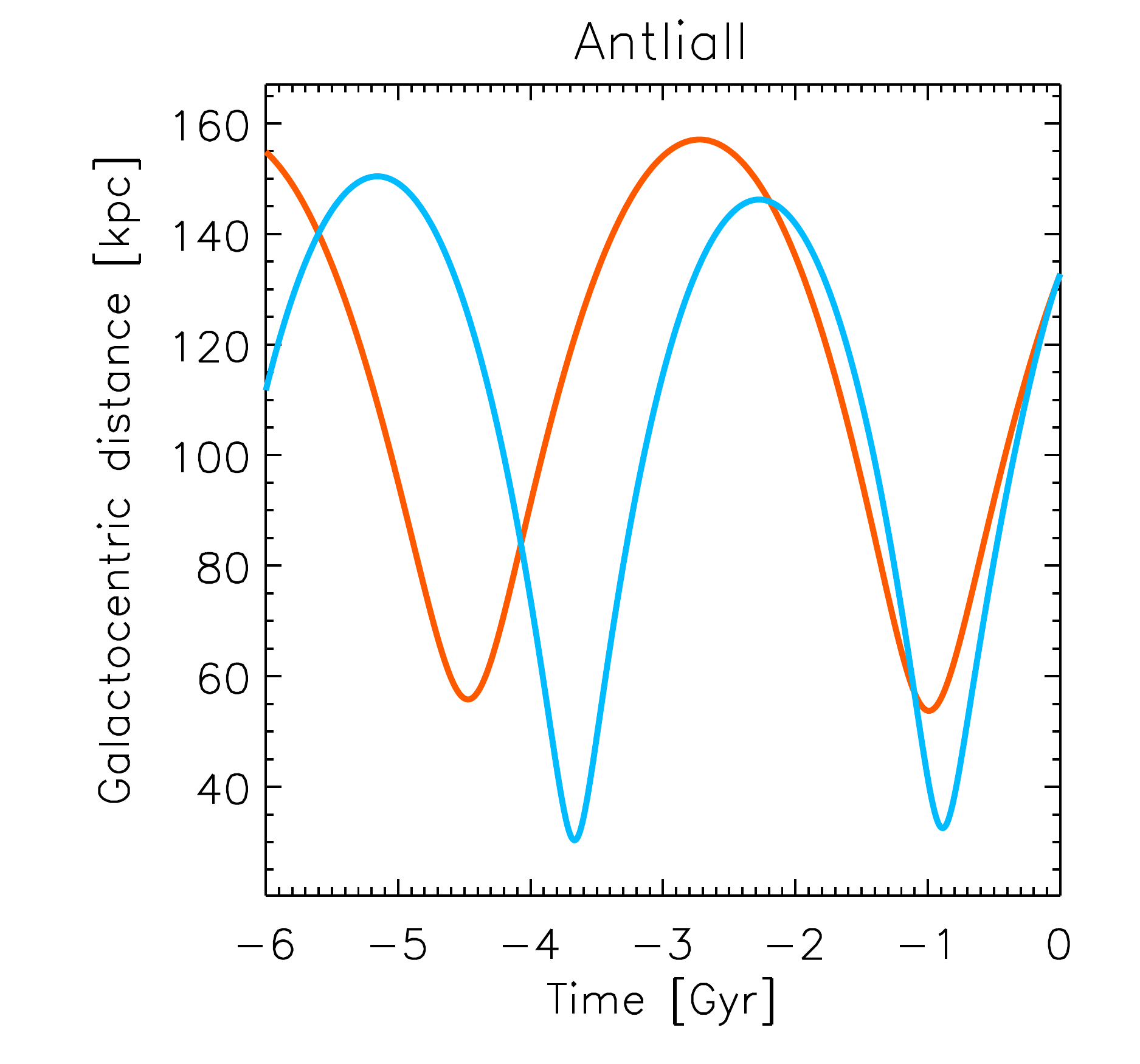}
\includegraphics[width=0.24\textwidth]{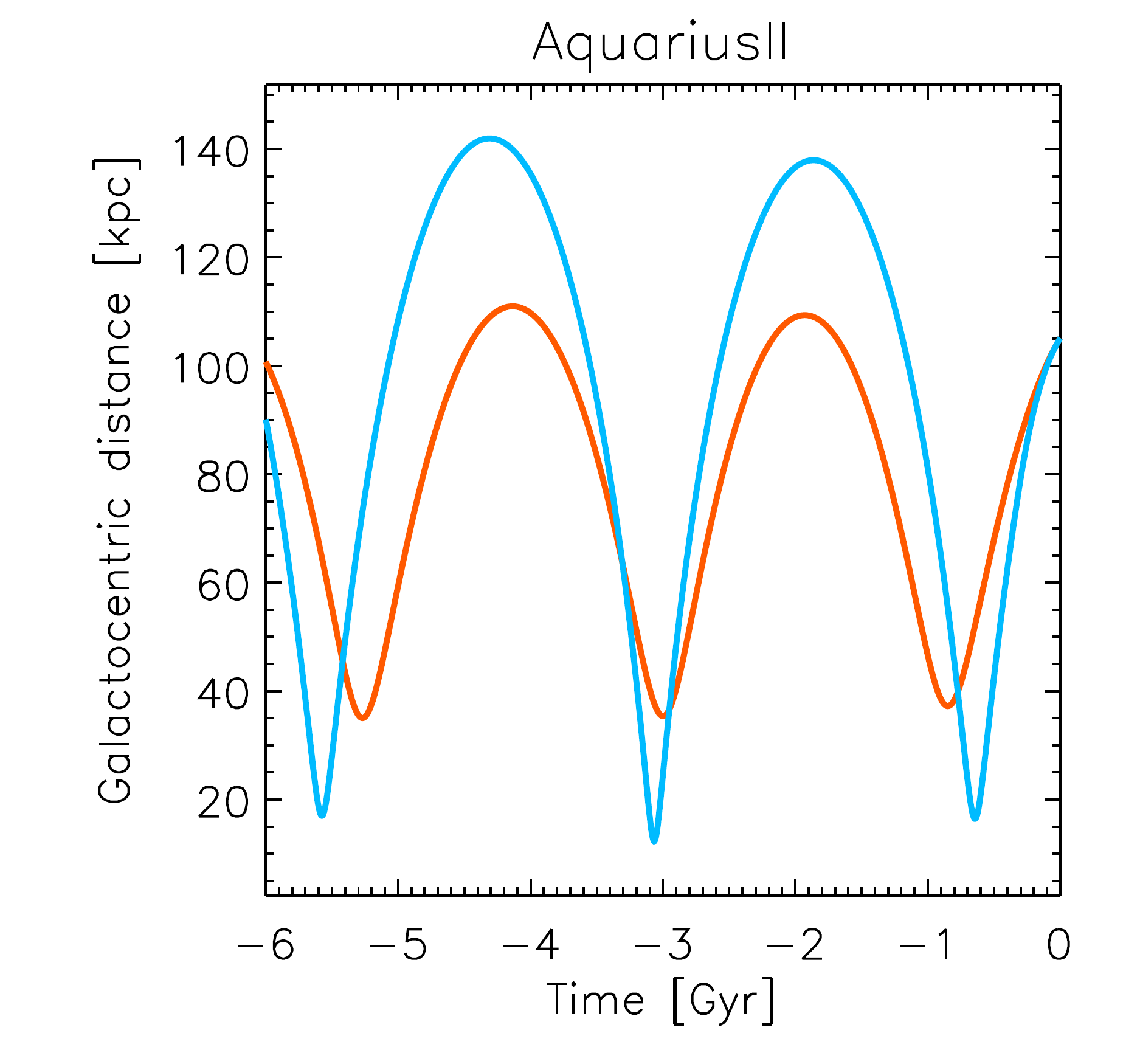}
\includegraphics[width=0.24\textwidth]{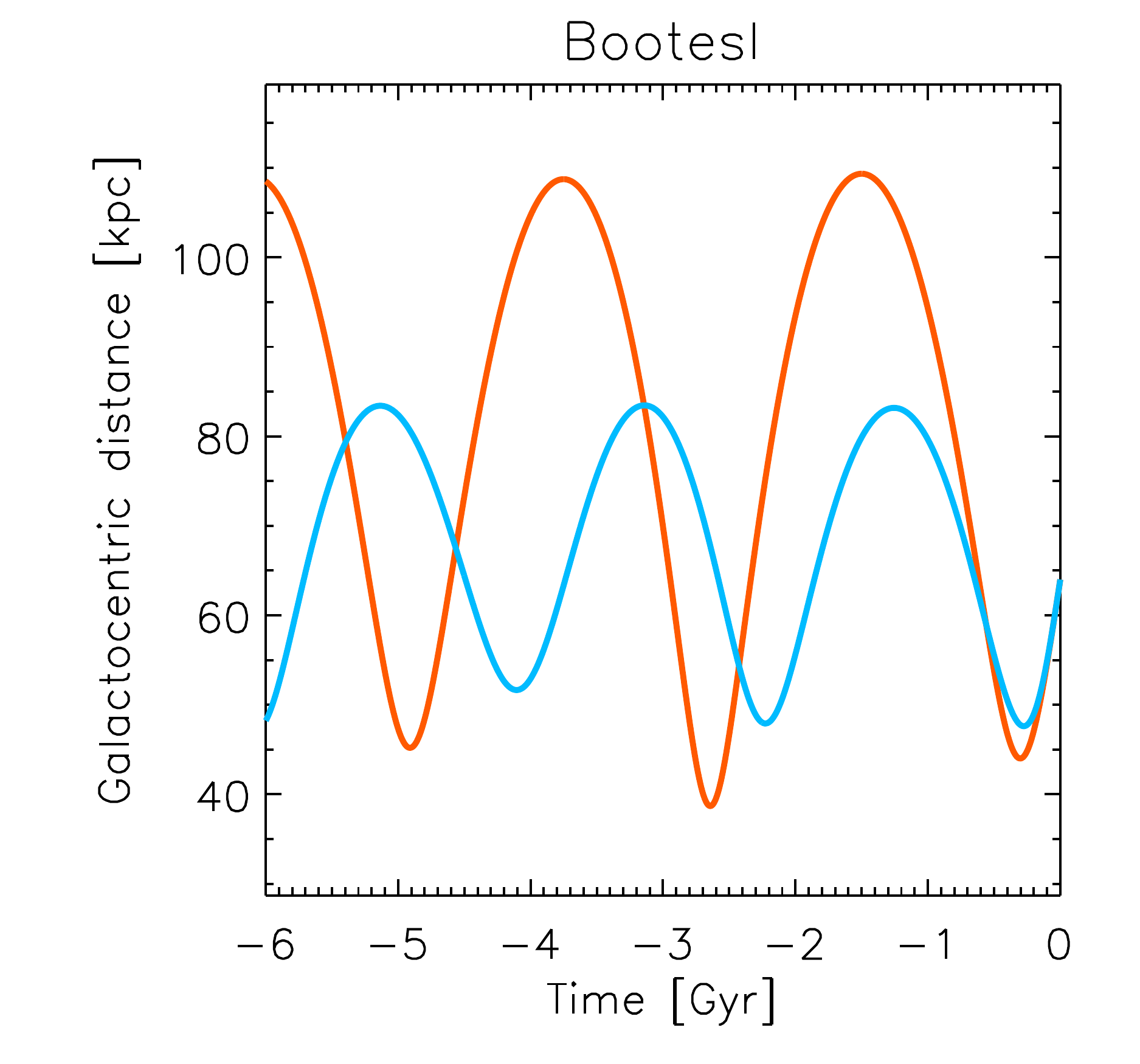}
\includegraphics[width=0.24\textwidth]{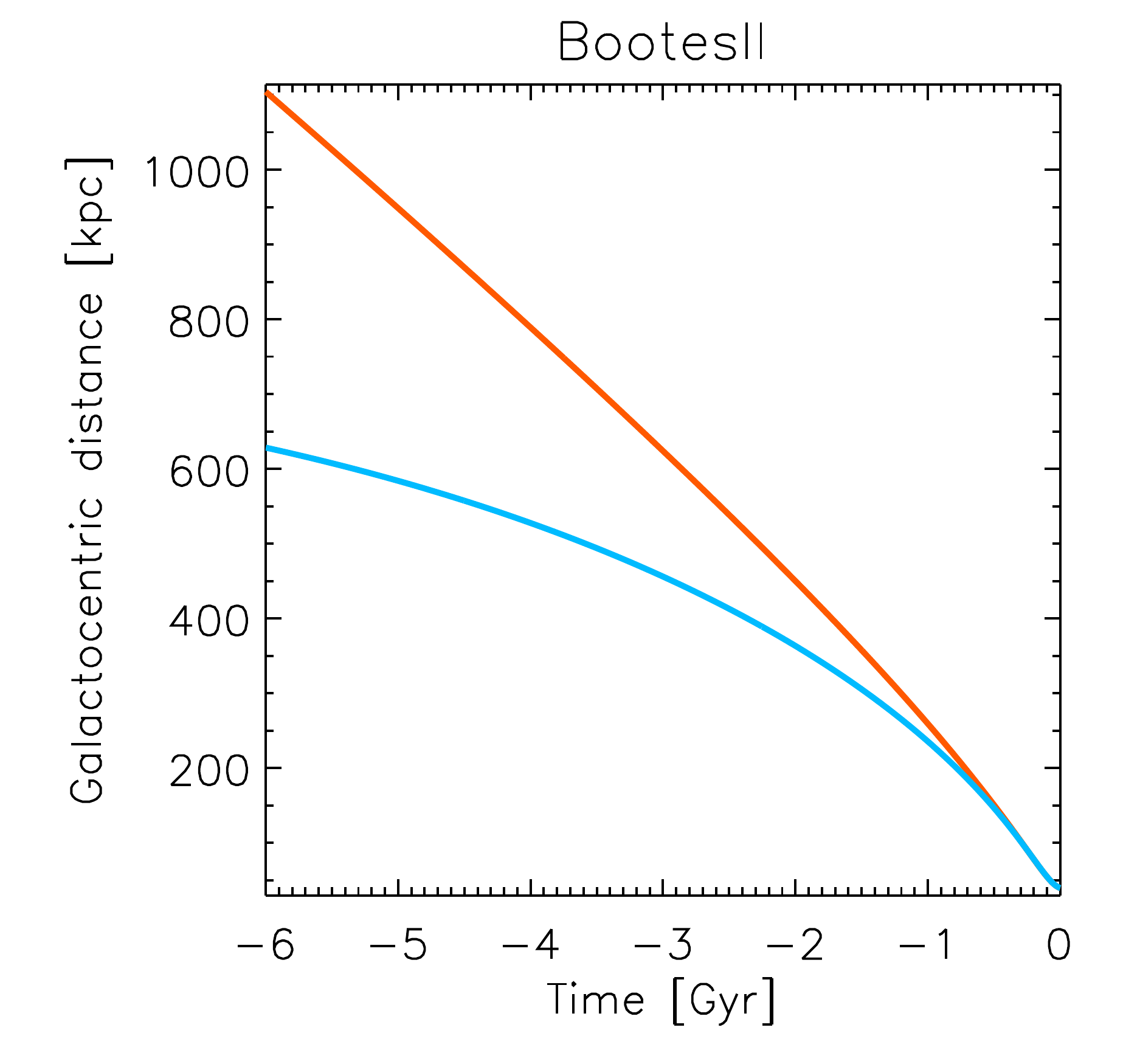}
\includegraphics[width=0.24\textwidth]{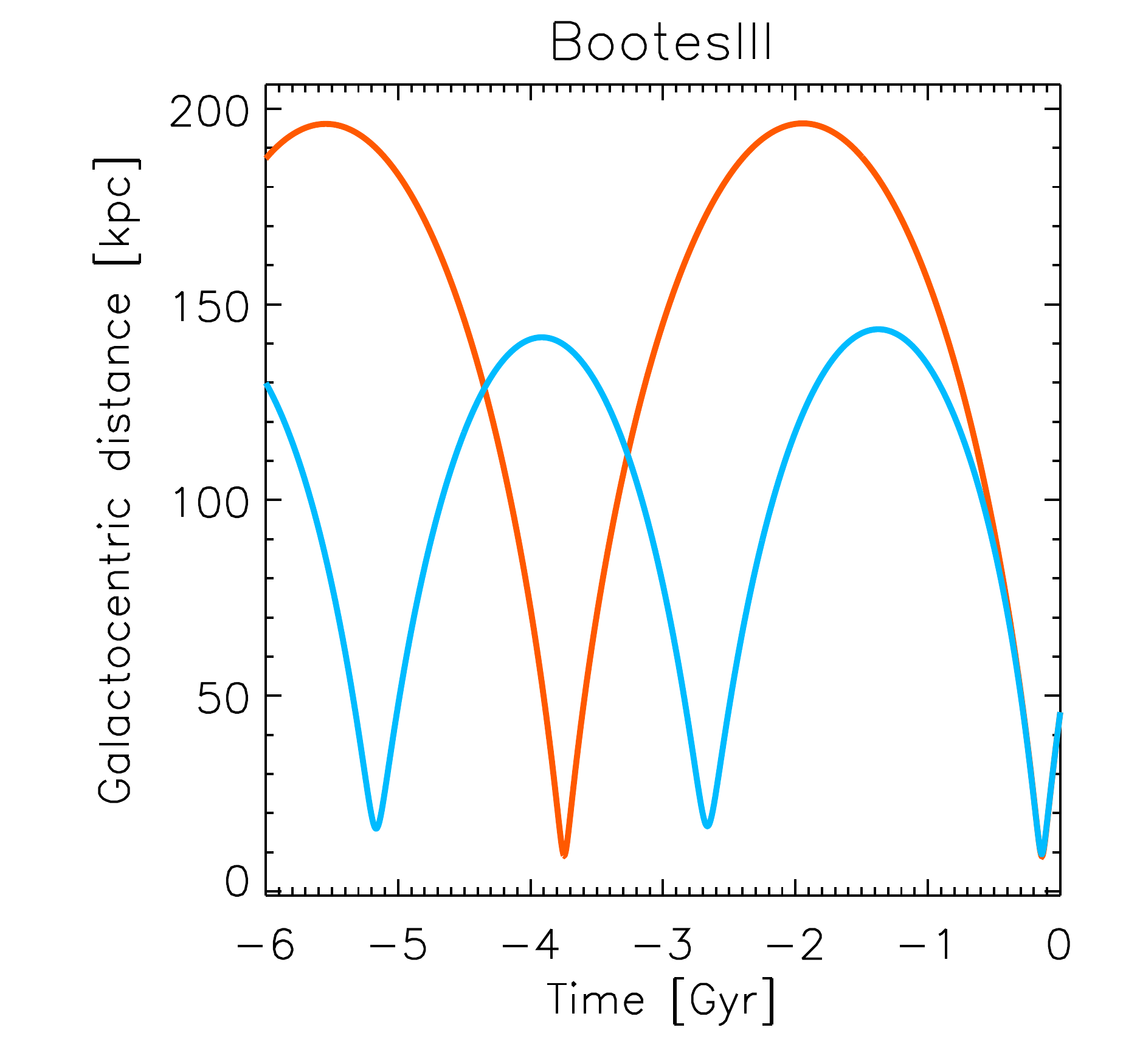}
\includegraphics[width=0.24\textwidth]{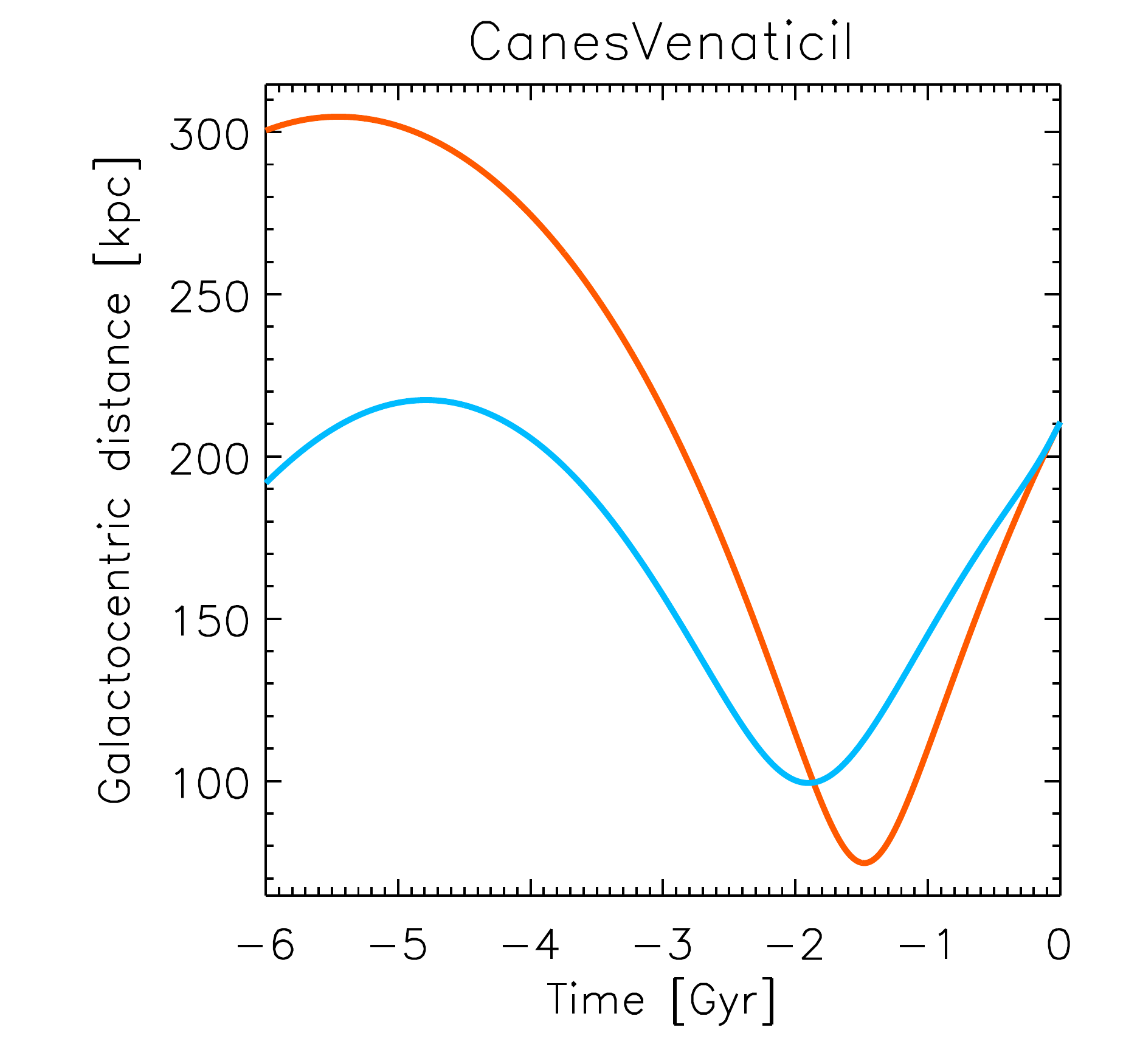}
\includegraphics[width=0.24\textwidth]{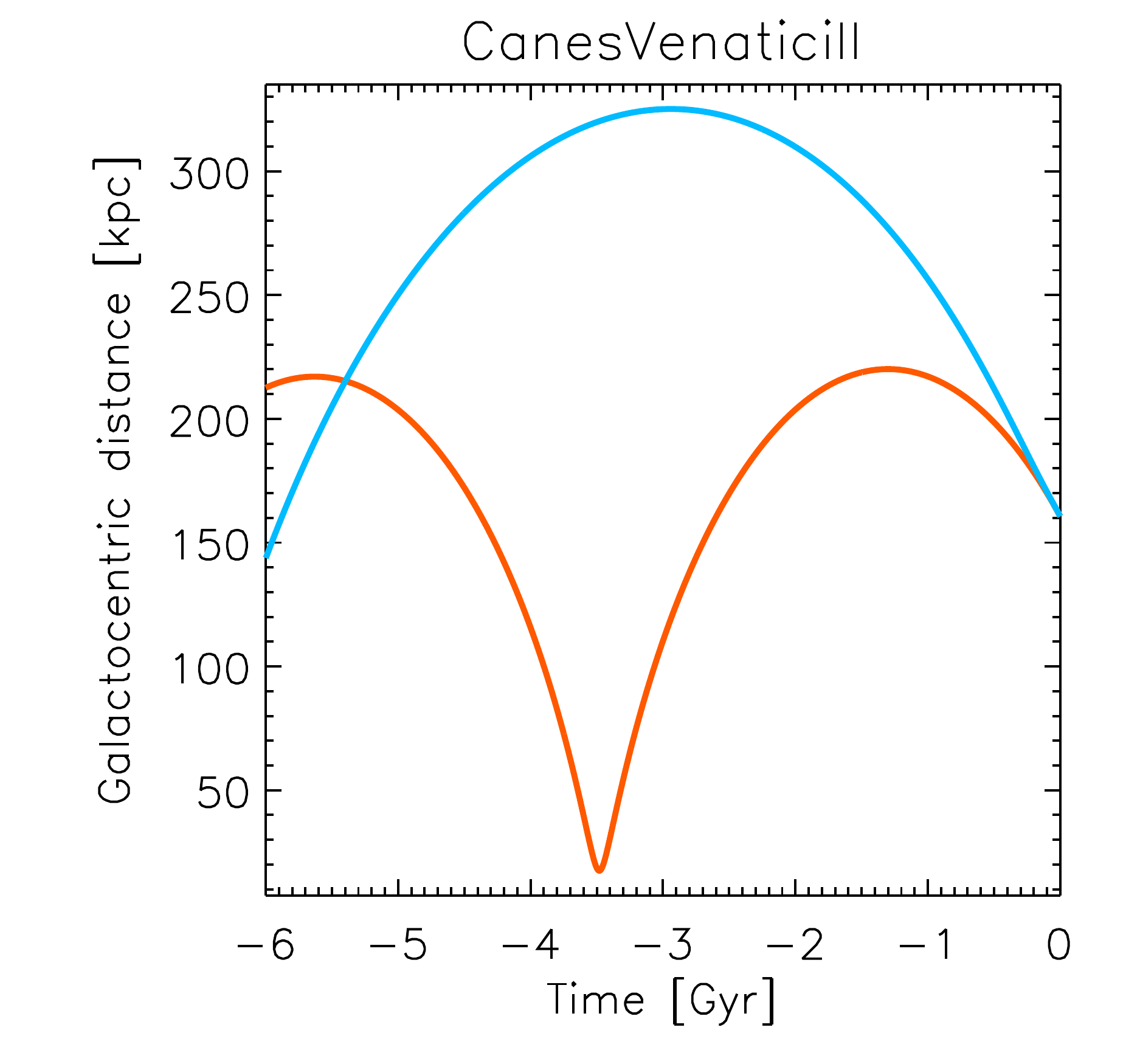}
\includegraphics[width=0.24\textwidth]{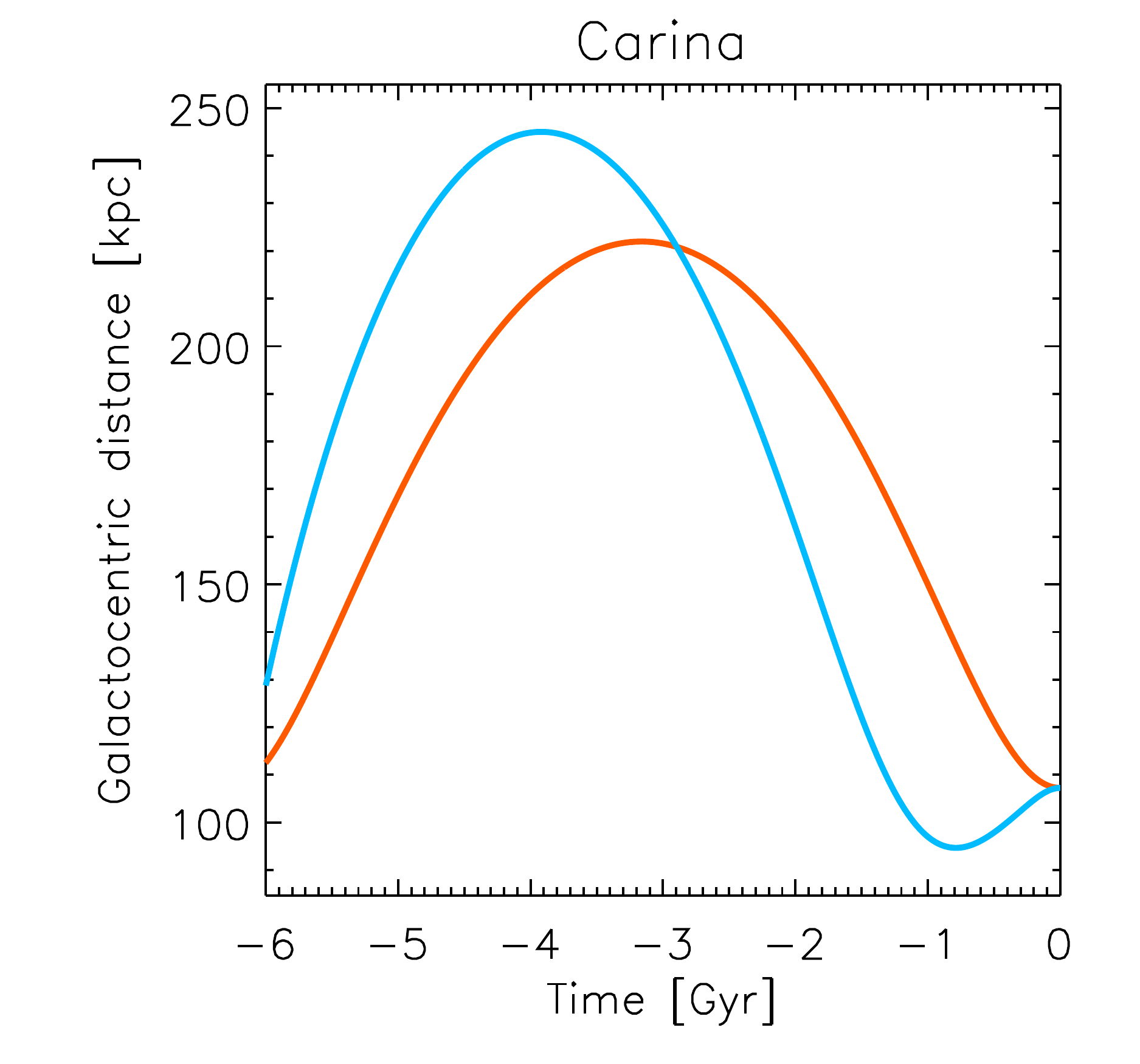}
\includegraphics[width=0.24\textwidth]{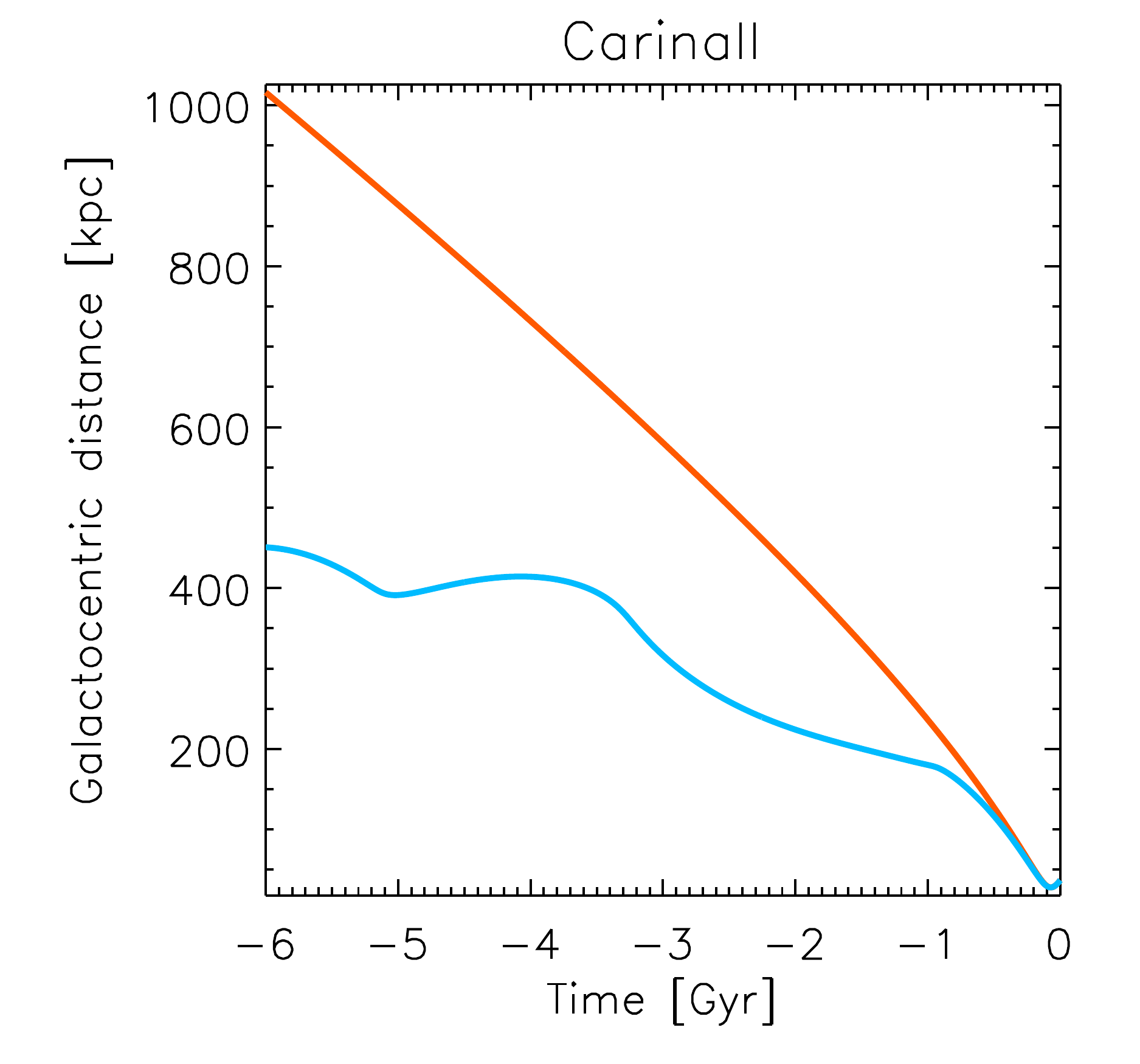}
\includegraphics[width=0.24\textwidth]{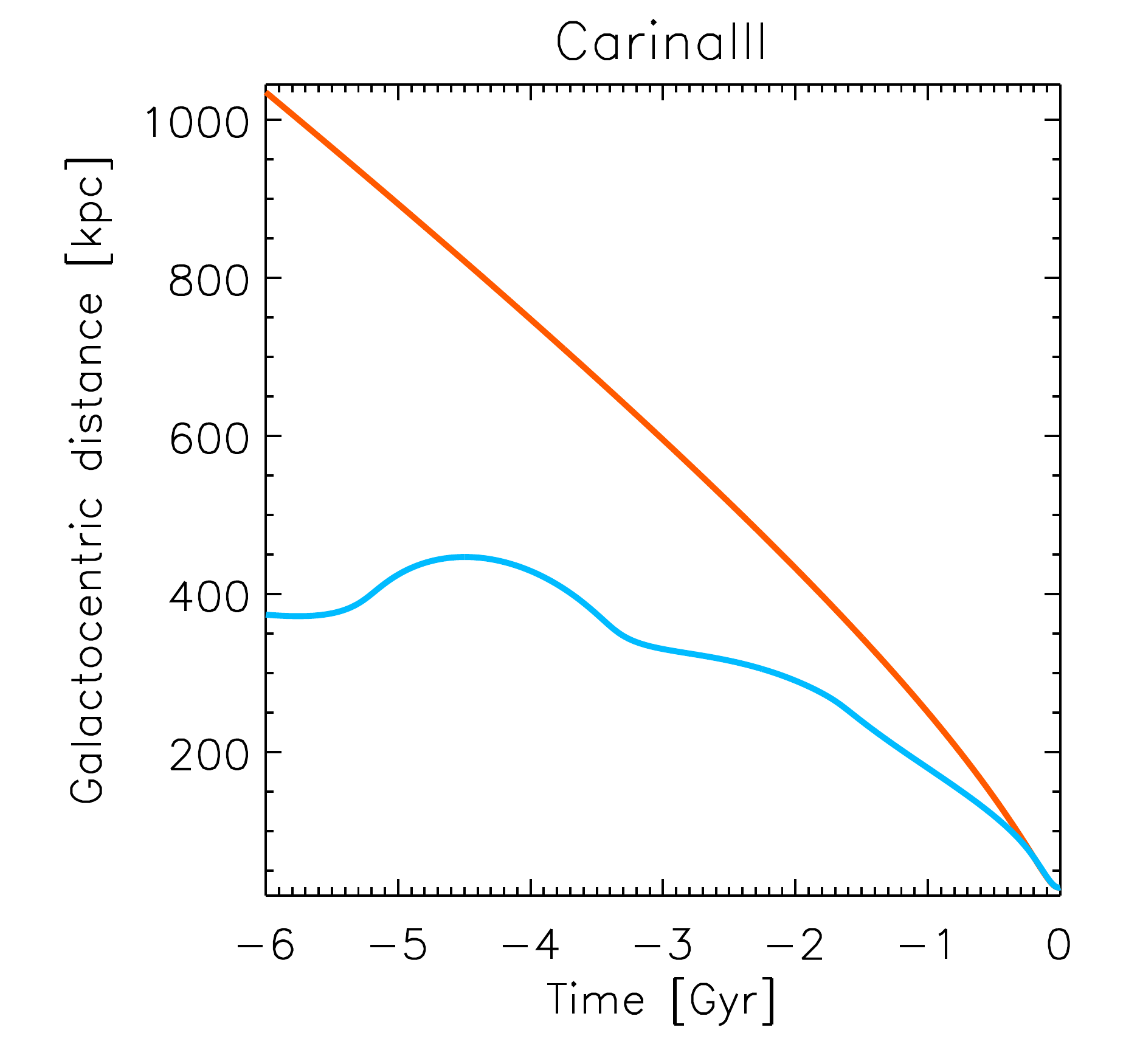}
\includegraphics[width=0.24\textwidth]{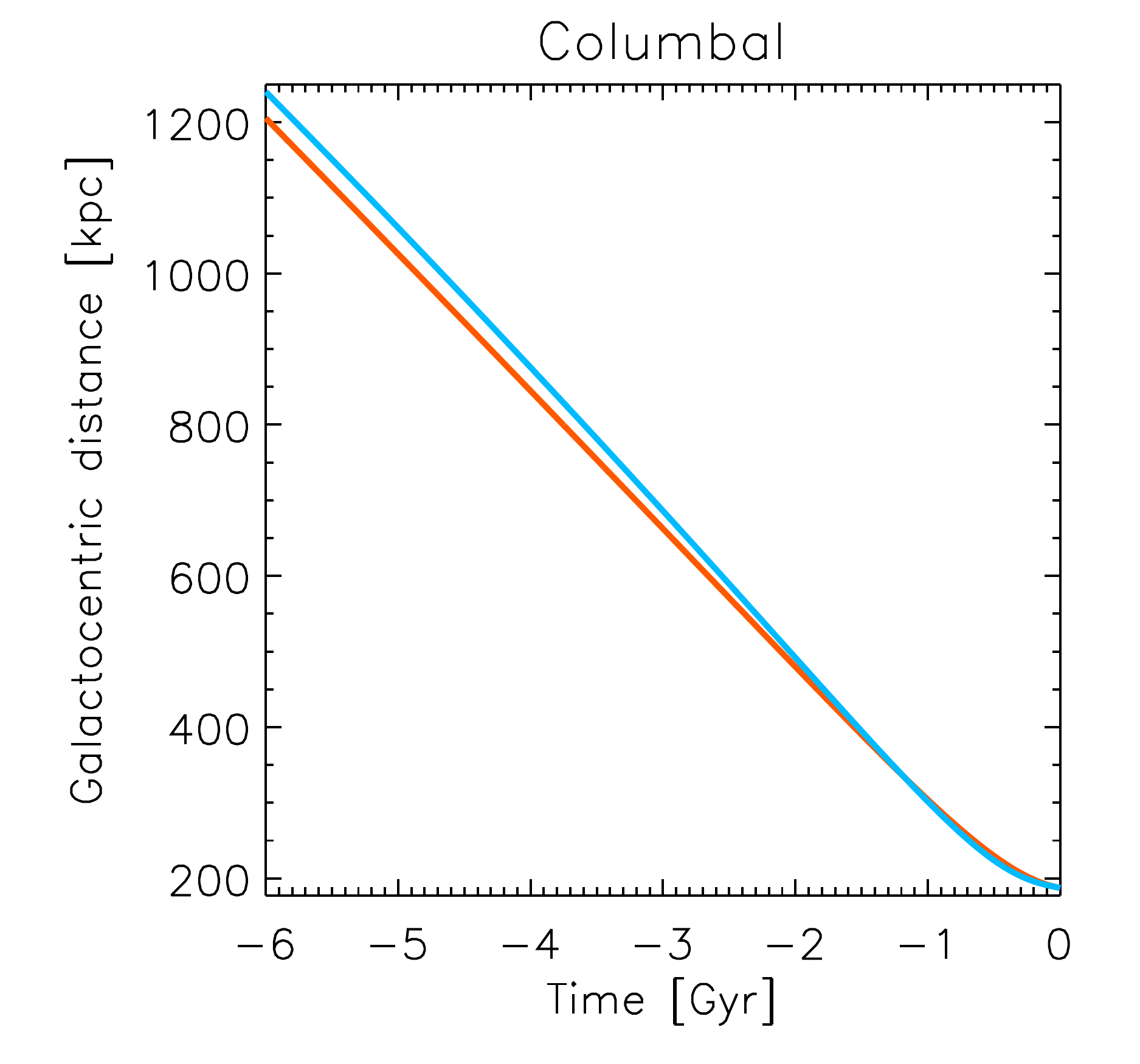}
\includegraphics[width=0.24\textwidth]{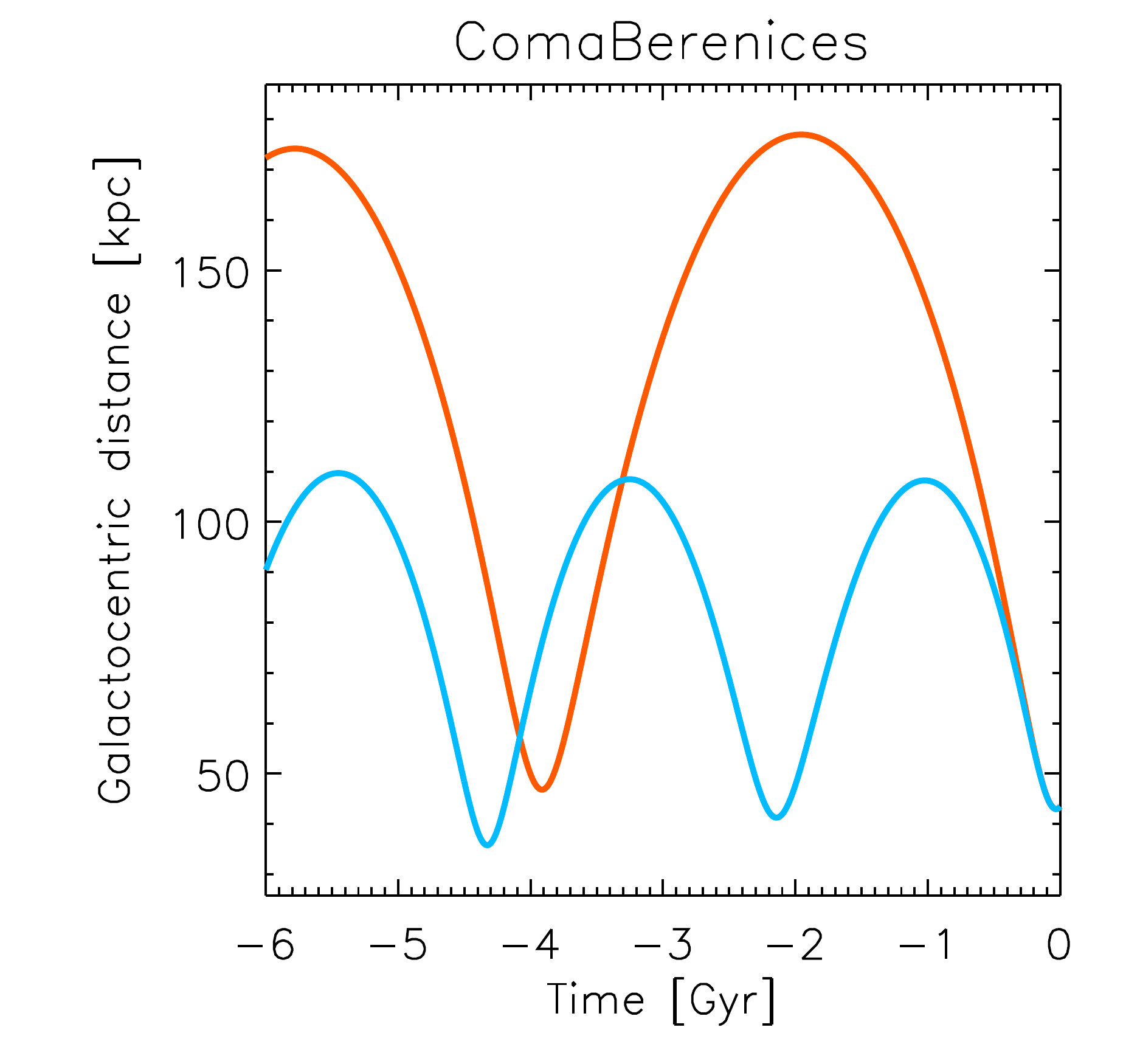}
\includegraphics[width=0.24\textwidth]{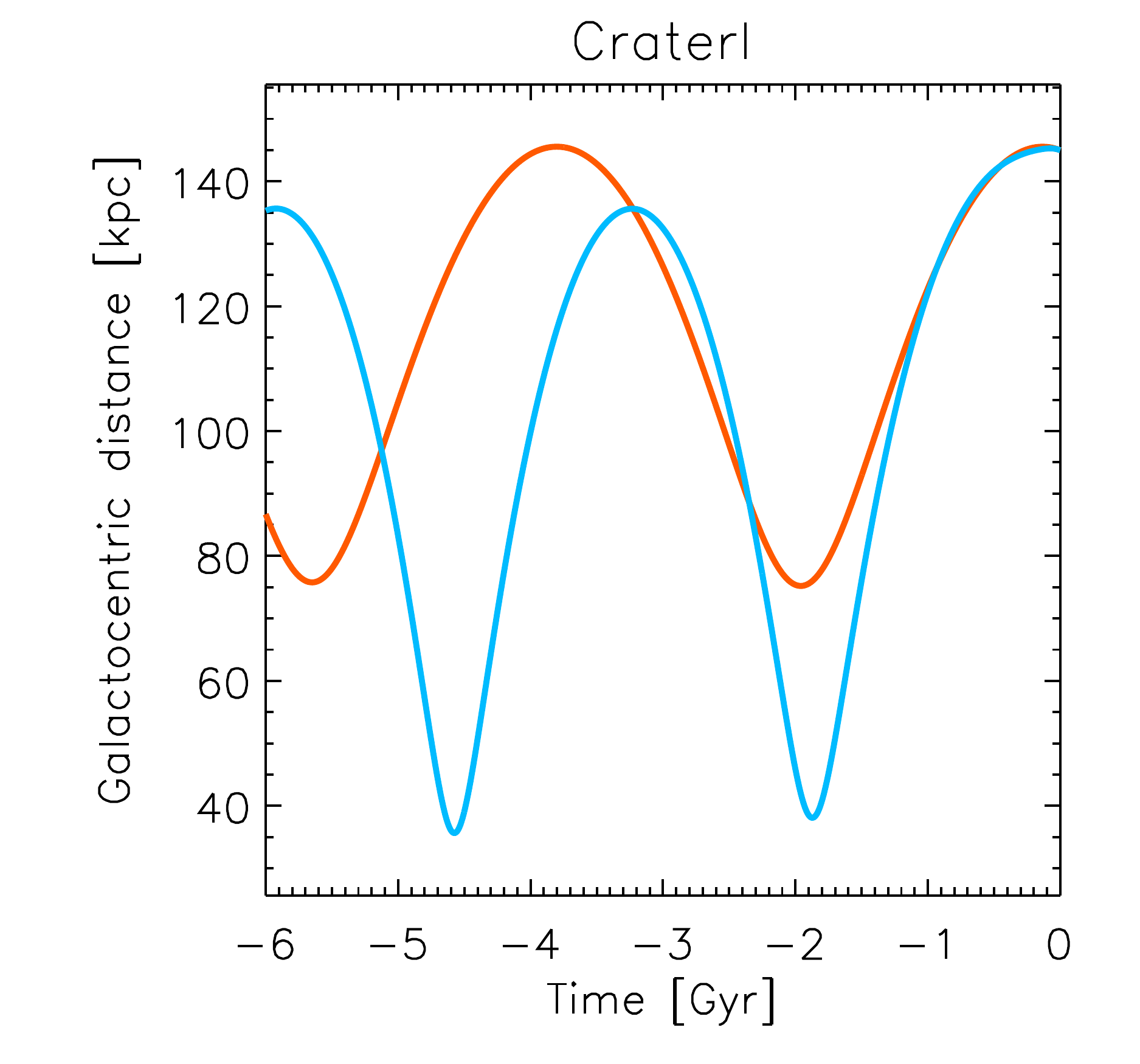}
\includegraphics[width=0.24\textwidth]{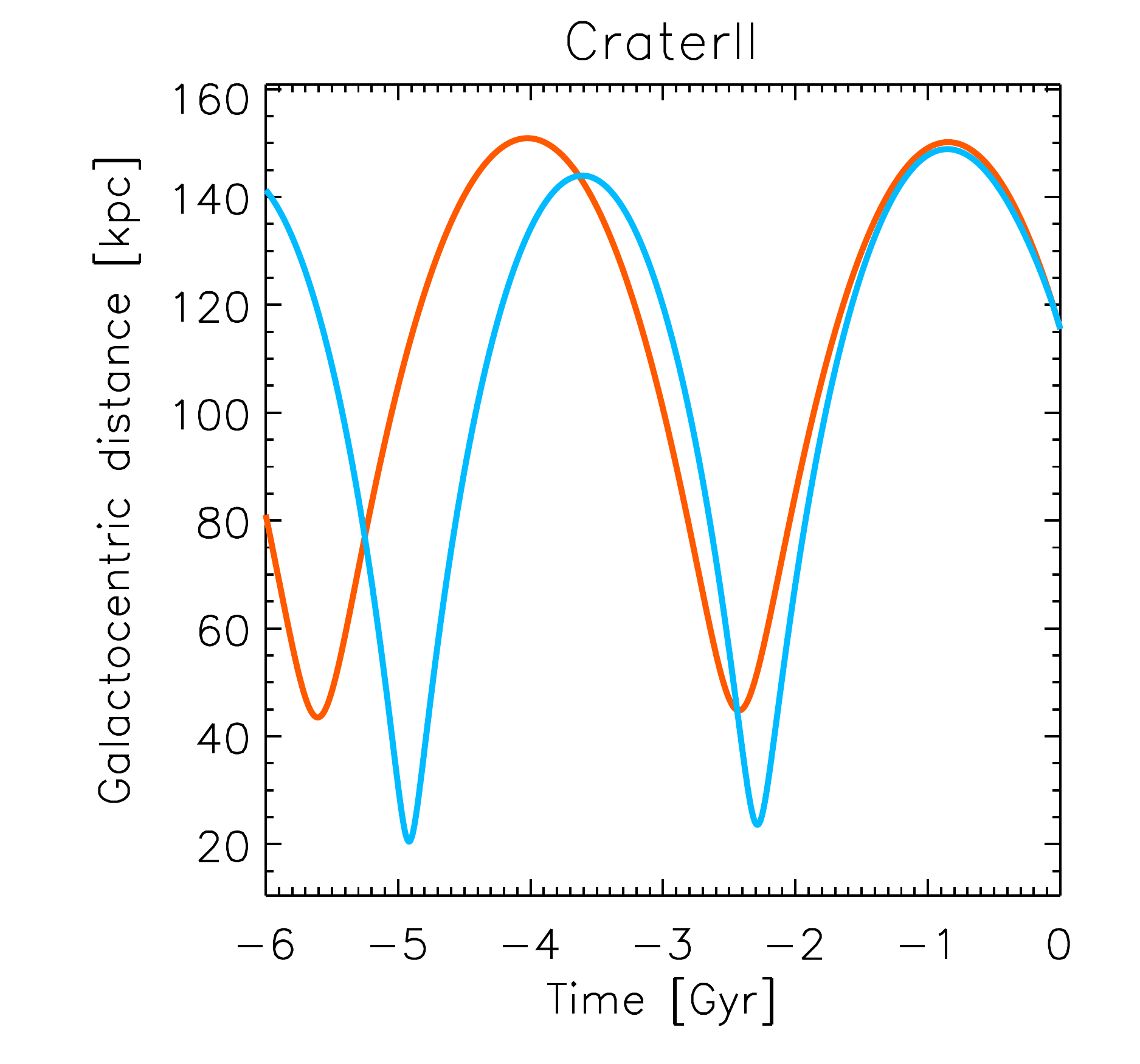}
\includegraphics[width=0.24\textwidth]{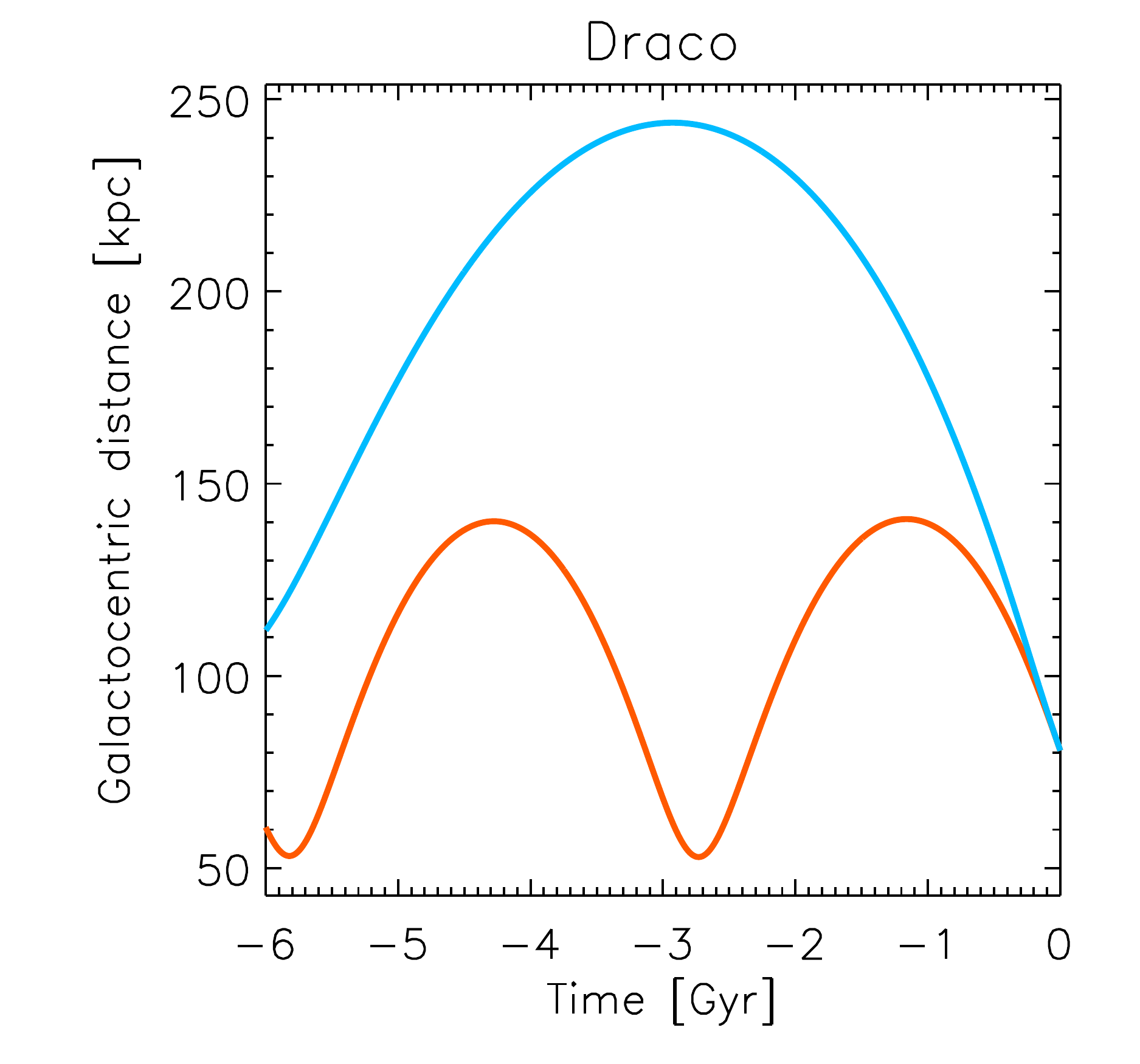}
\includegraphics[width=0.24\textwidth]{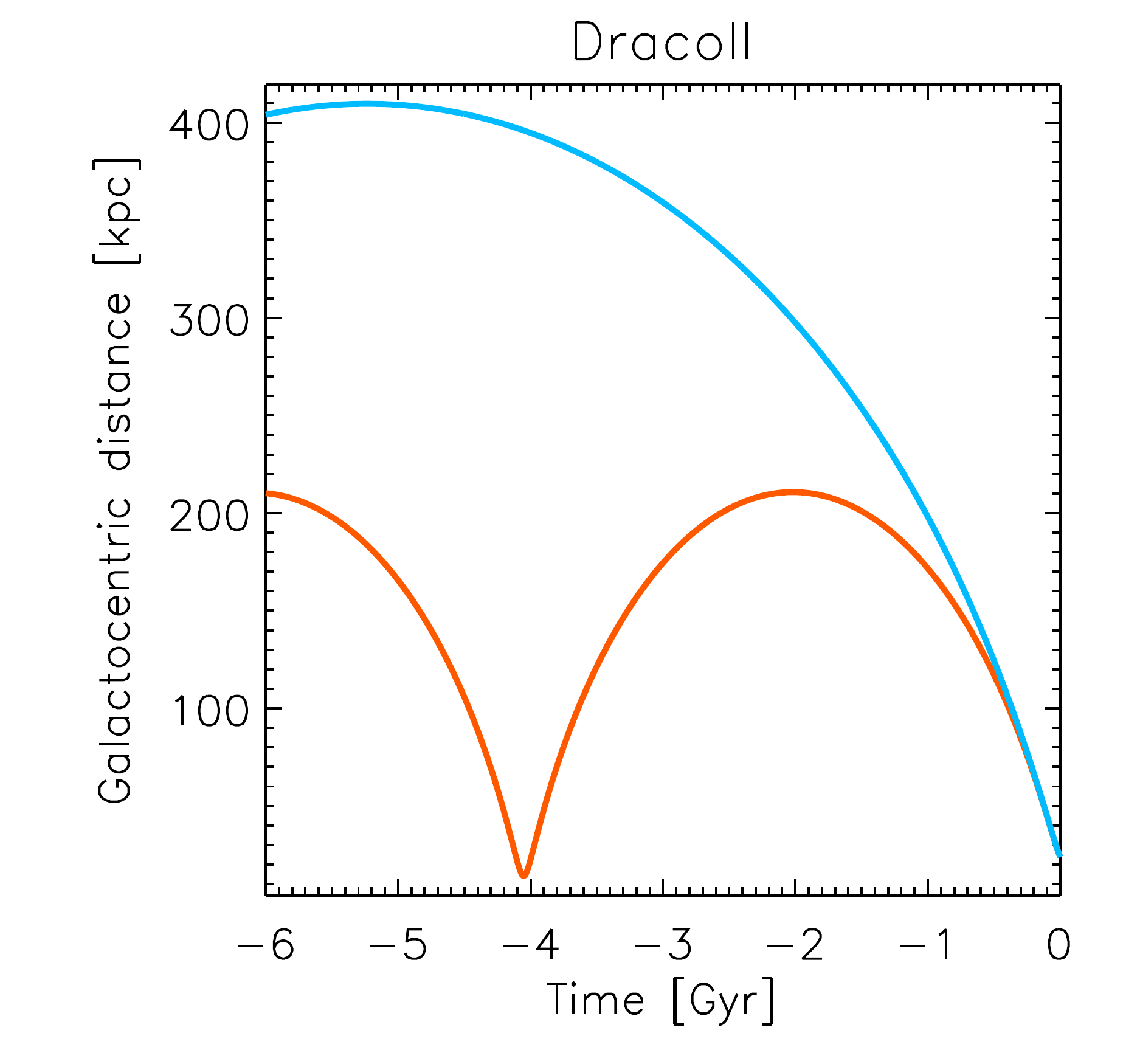}
\includegraphics[width=0.24\textwidth]{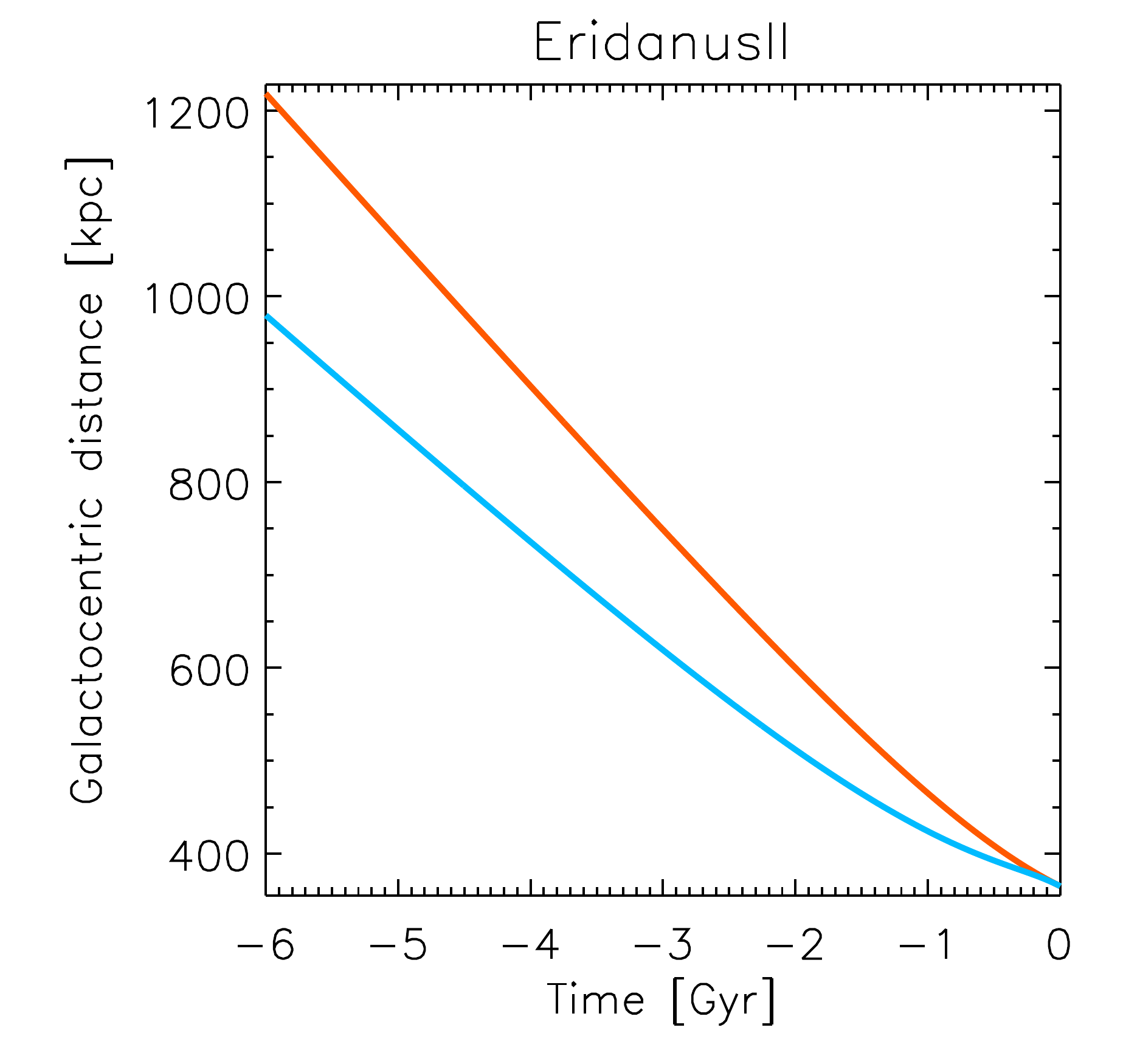}
\includegraphics[width=0.24\textwidth]{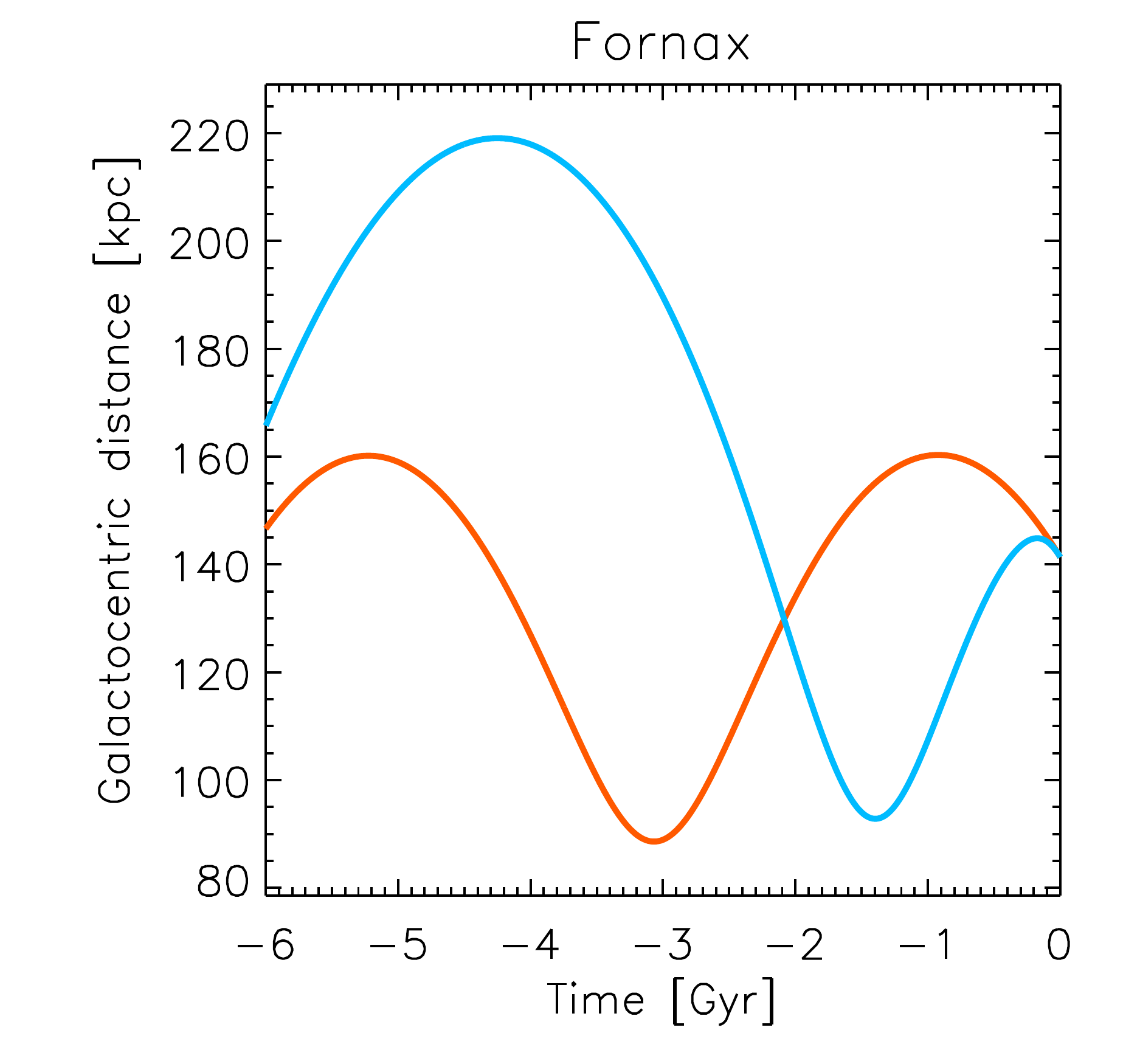}
\includegraphics[width=0.24\textwidth]{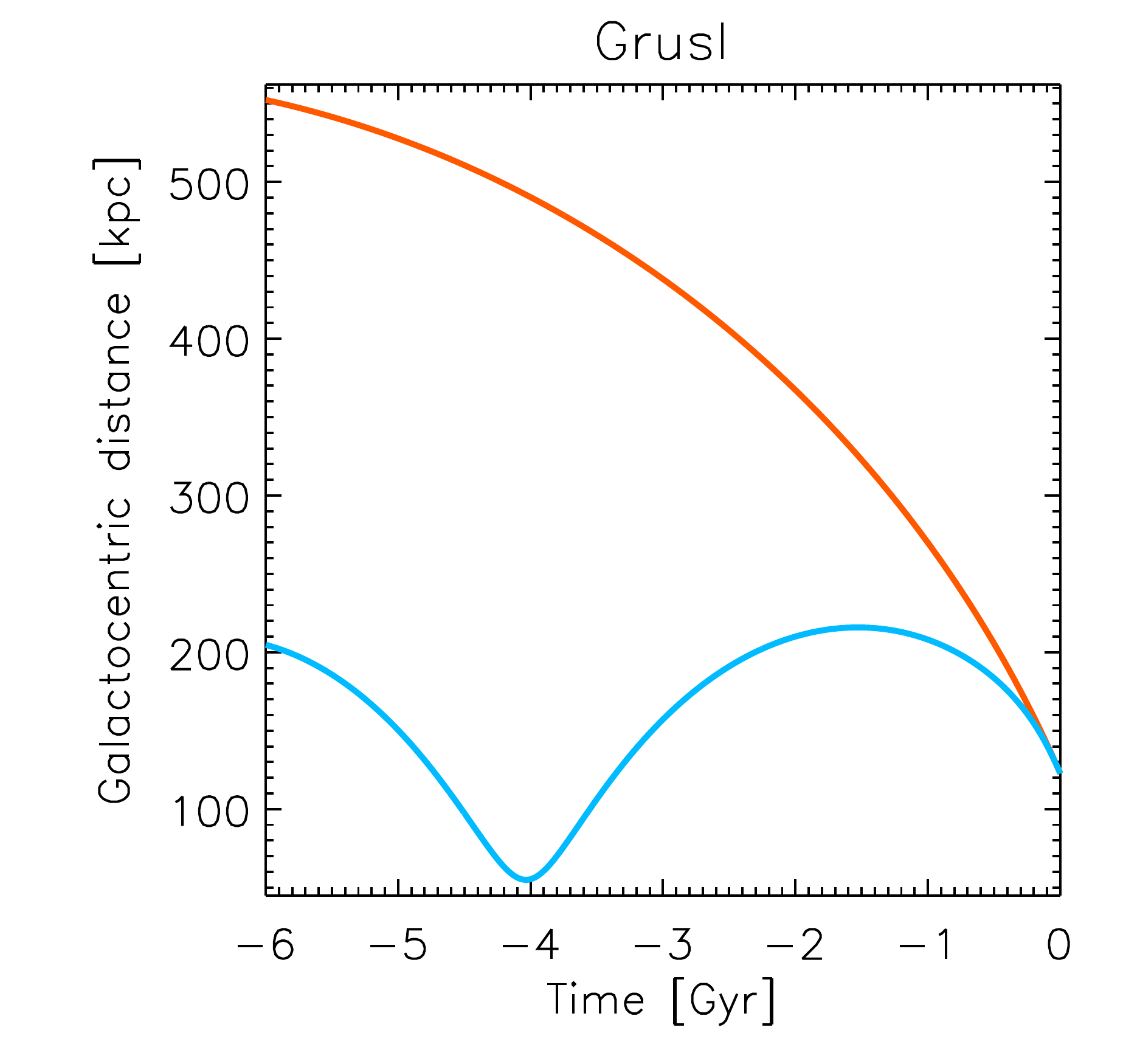}
\includegraphics[width=0.24\textwidth]{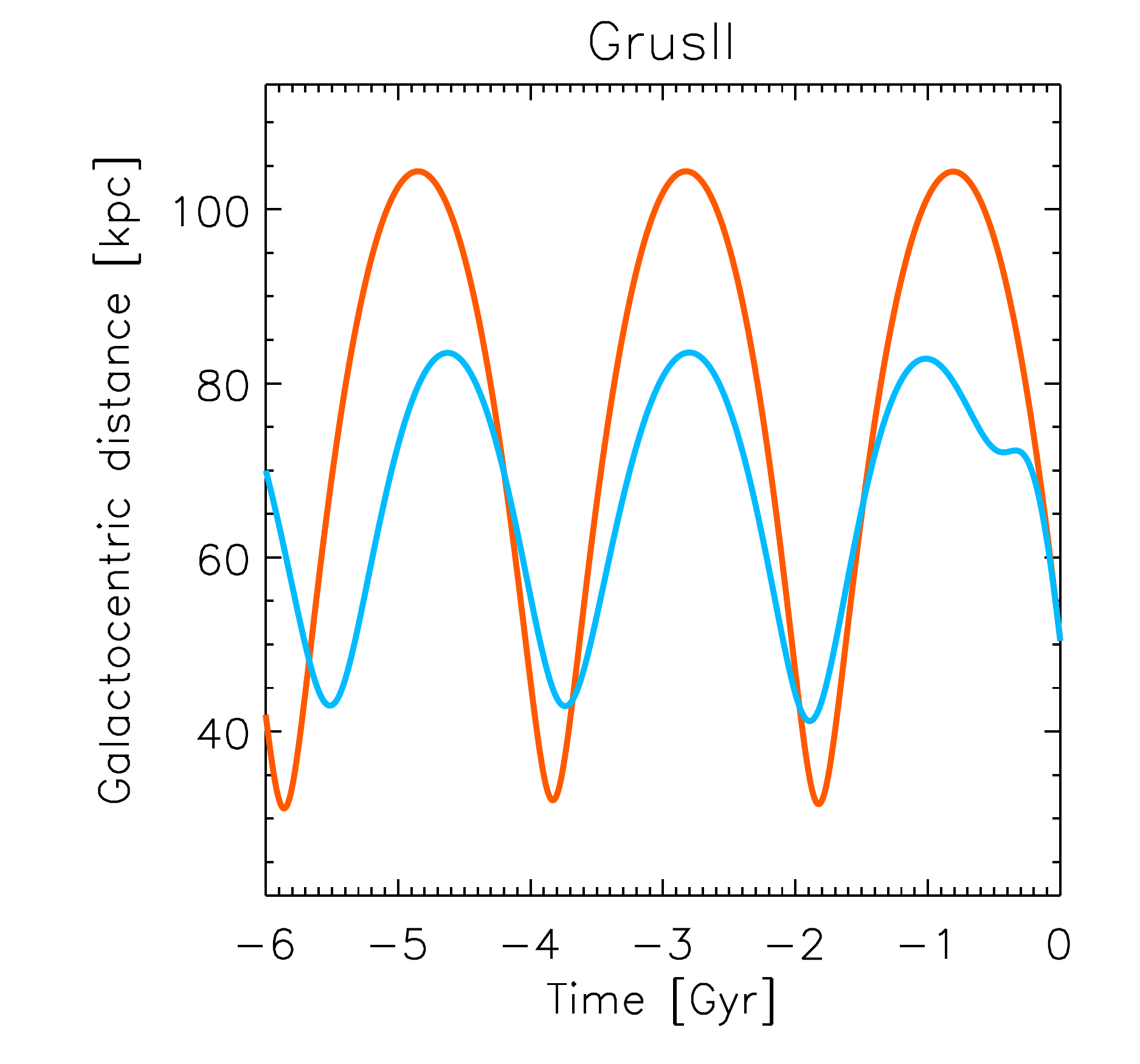}
       \caption{Distance from the MW center as a function of time for the past 6 Gyr (at present time t$=0$) for the  "perturbed" and "isolated" "Light MW" potentials (cyan and orange lines, respectively), i.e. in the case with and without the the infall of a massive LMC; the orbits are determined from the observed, error-free, bulk motions.}
         \label{fig:orbits}
   \end{figure*}

      \begin{figure*}
   \centering
\includegraphics[width=0.24\textwidth]{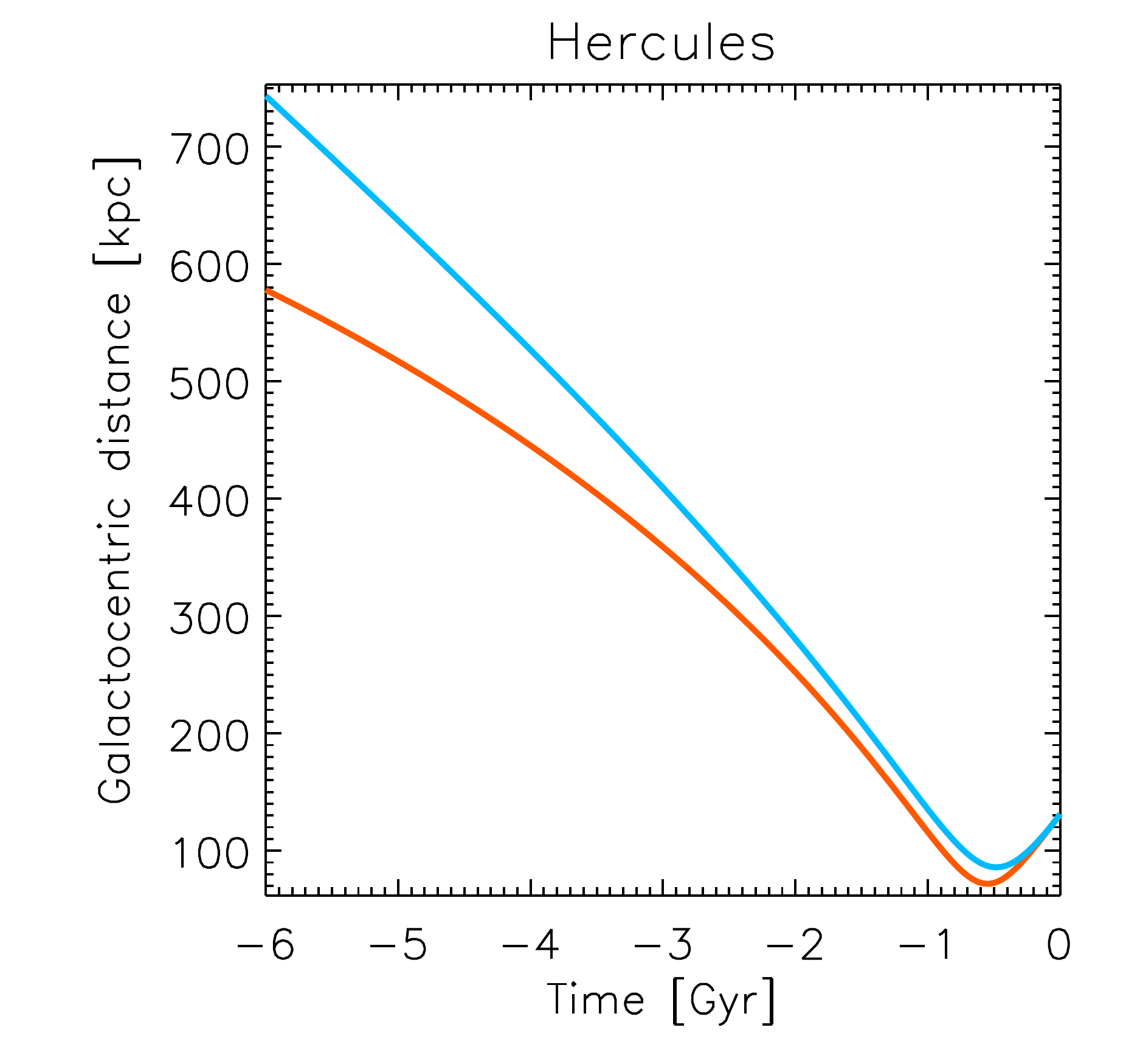}
\includegraphics[width=0.24\textwidth]{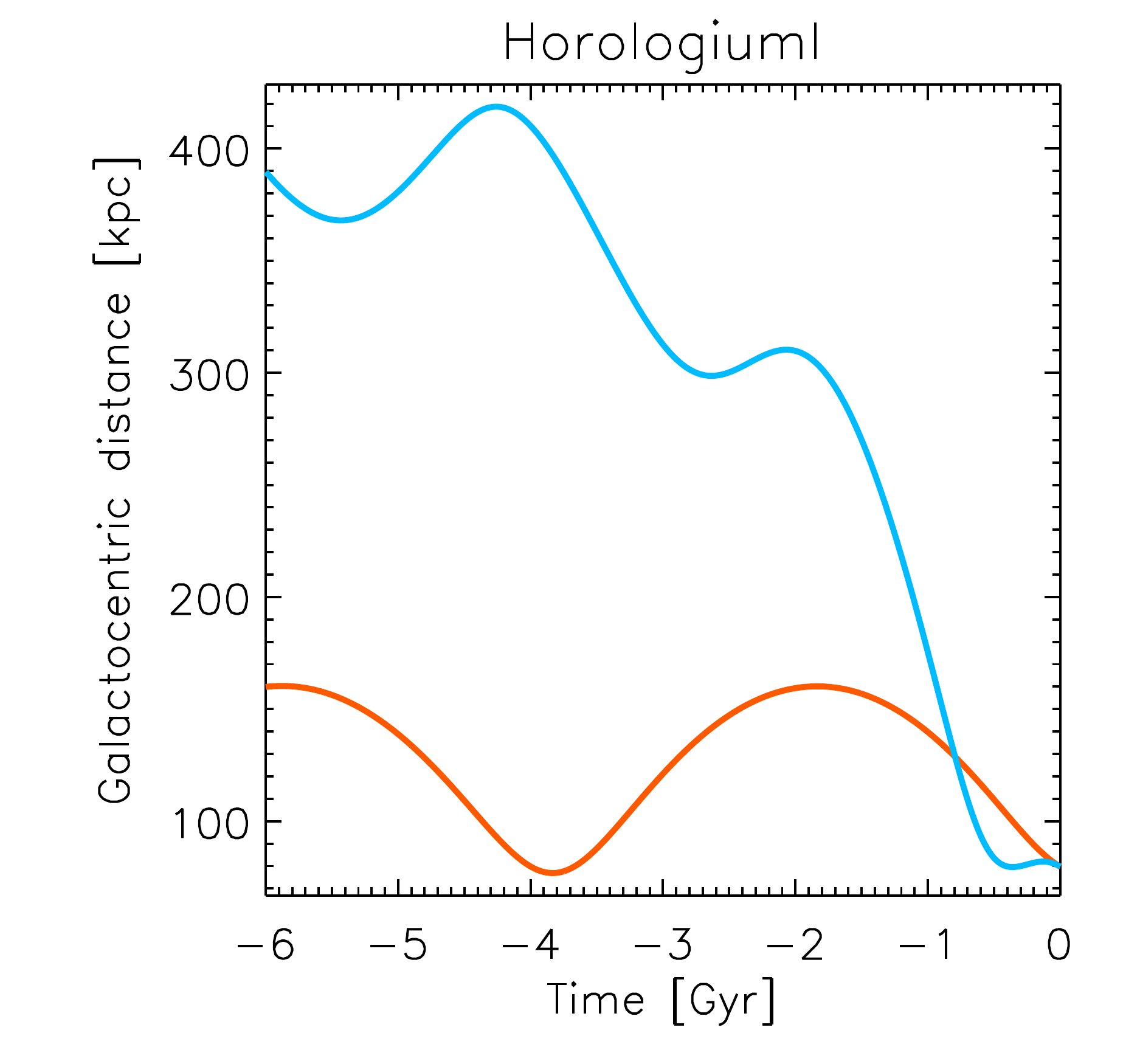}
\includegraphics[width=0.24\textwidth]{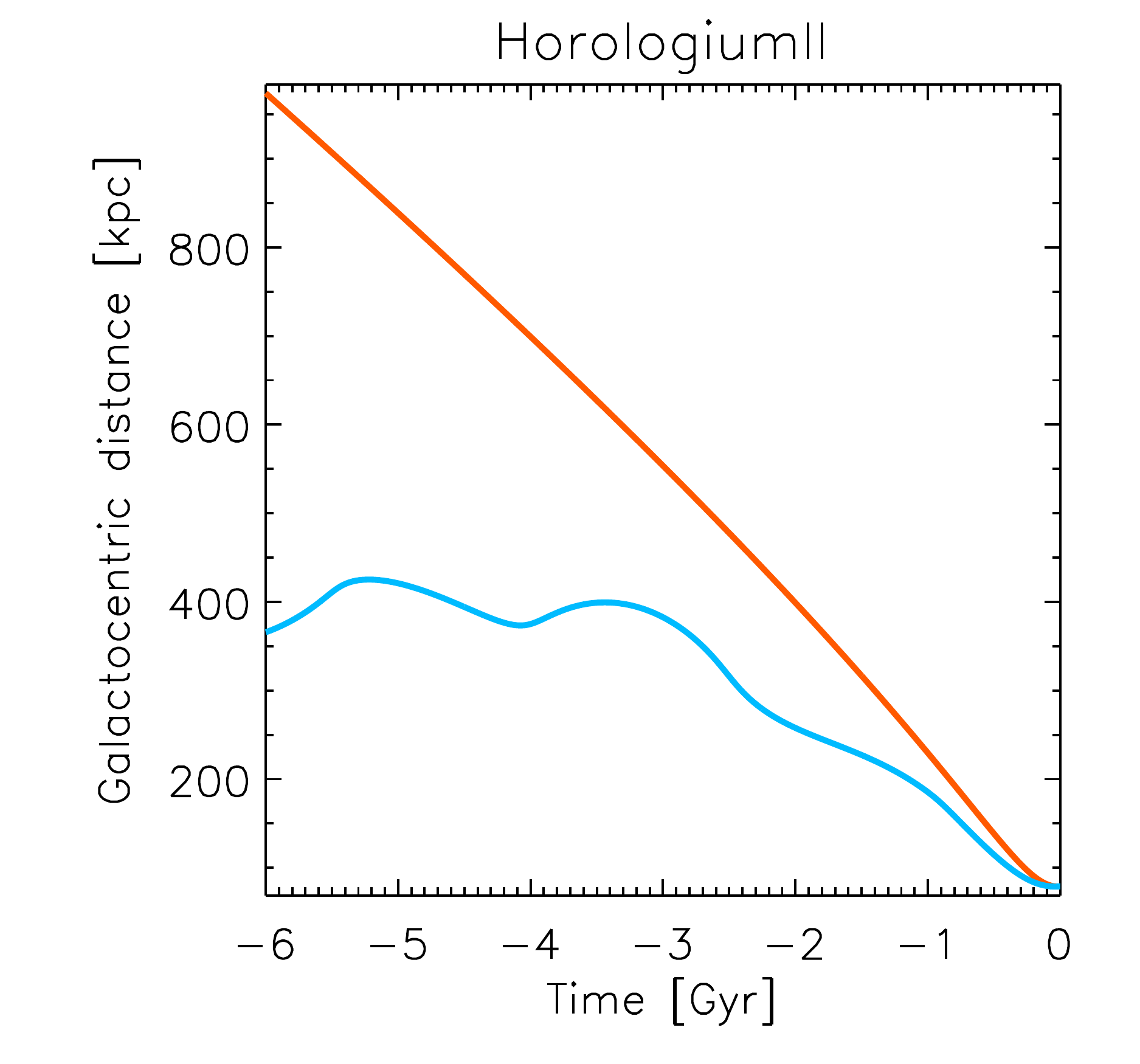}
\includegraphics[width=0.24\textwidth]{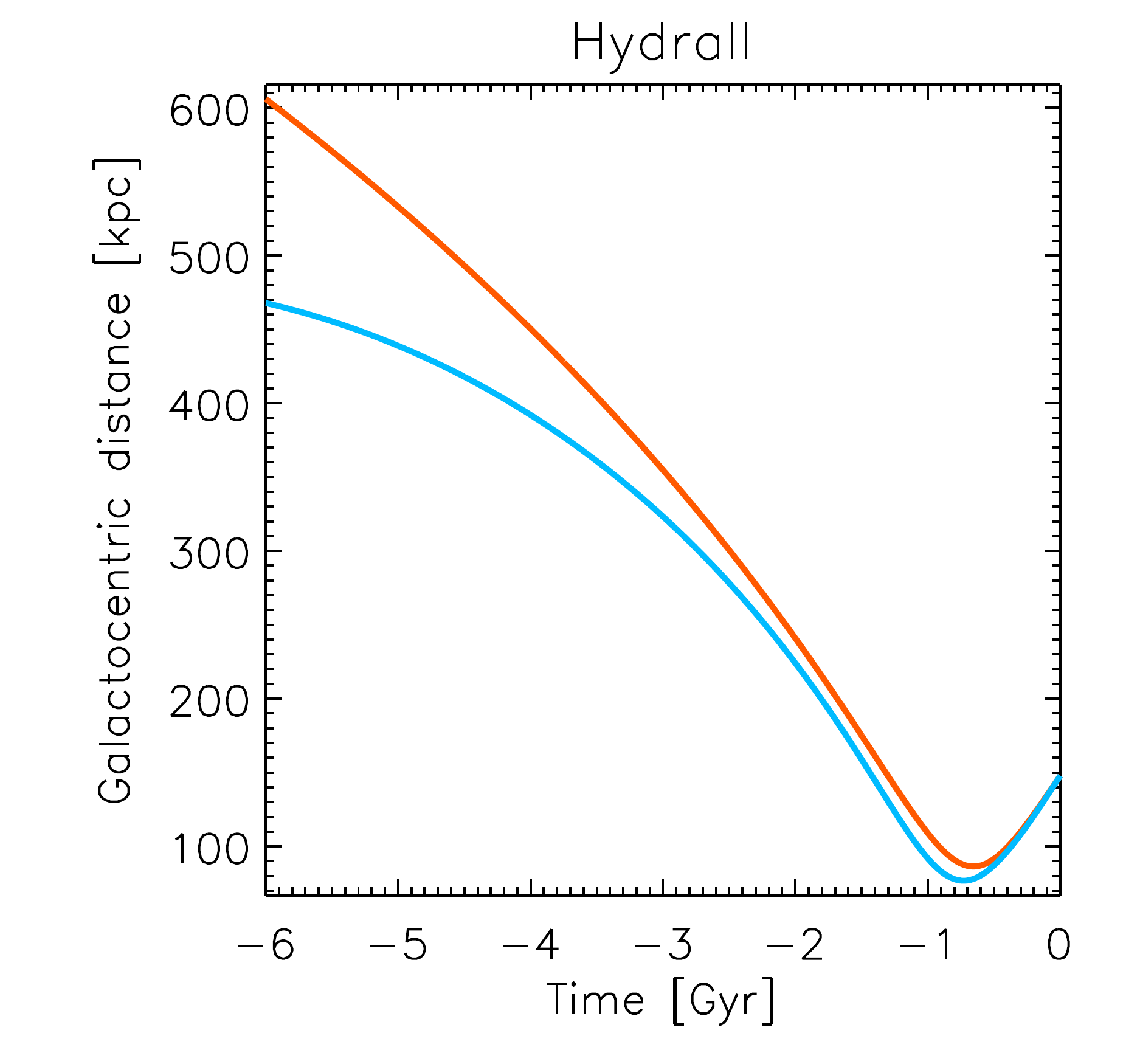}
\includegraphics[width=0.24\textwidth]{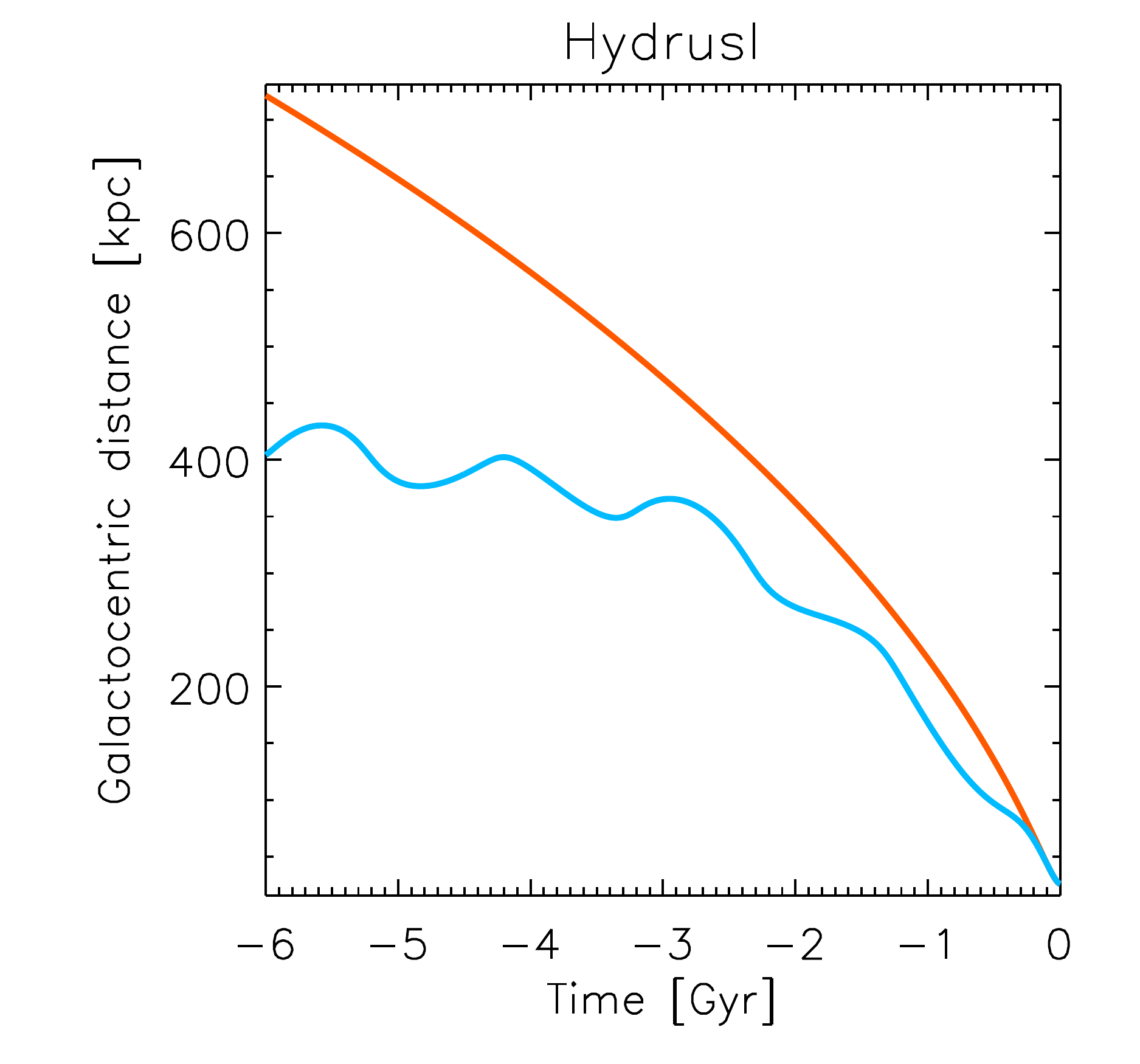}
\includegraphics[width=0.24\textwidth]{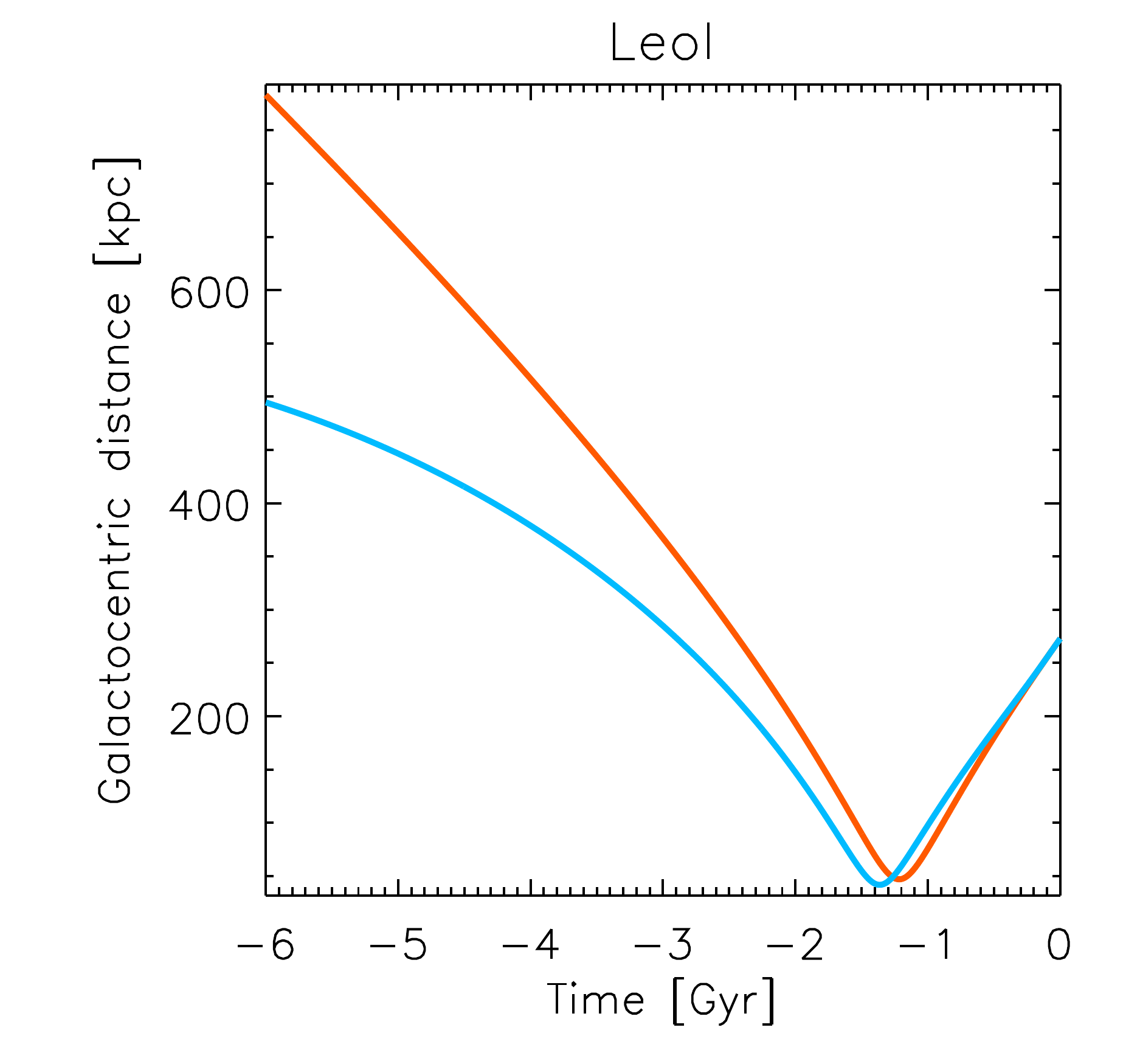}
\includegraphics[width=0.24\textwidth]{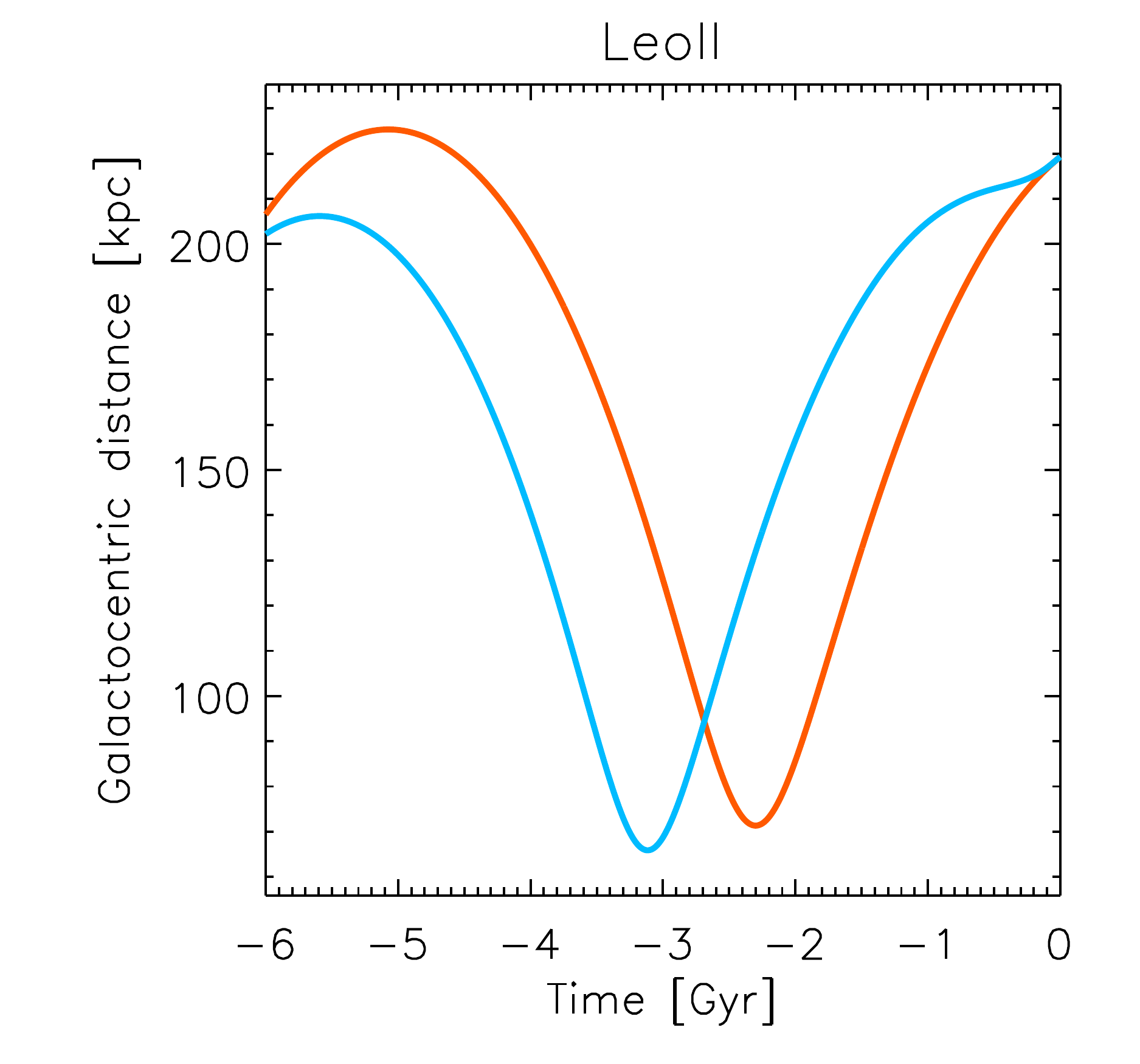}
\includegraphics[width=0.24\textwidth]{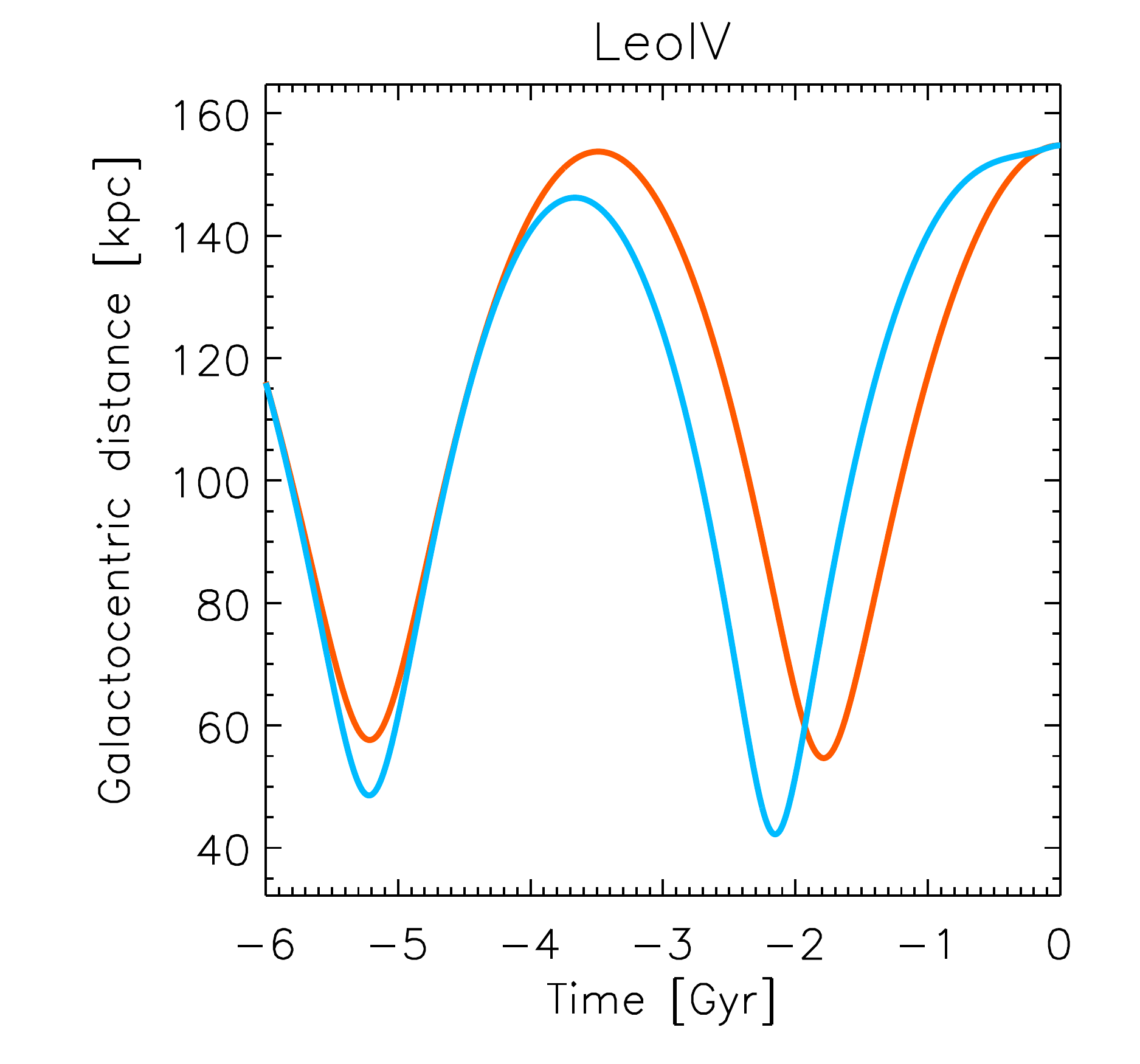}
\includegraphics[width=0.24\textwidth]{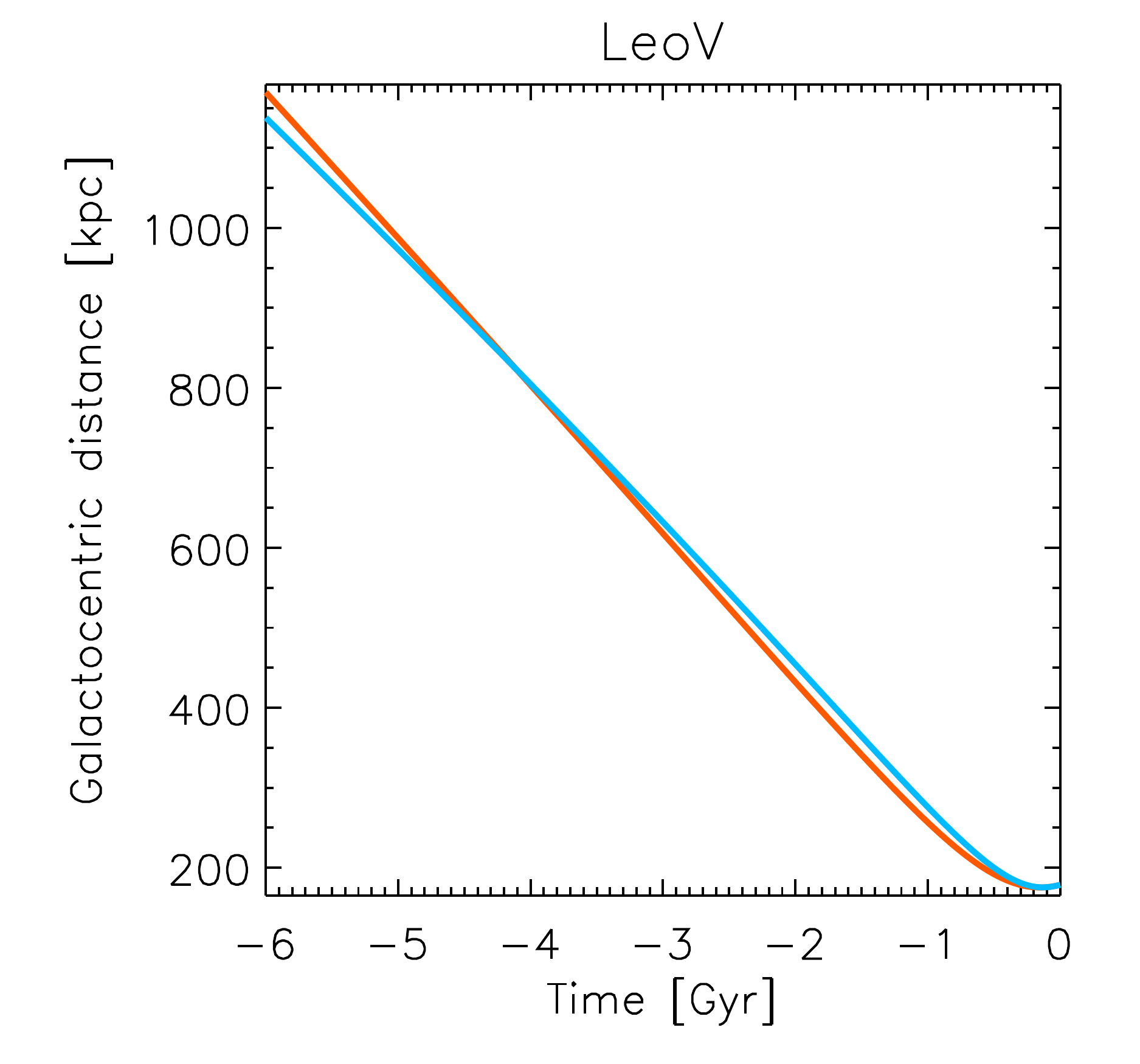}
\includegraphics[width=0.24\textwidth]{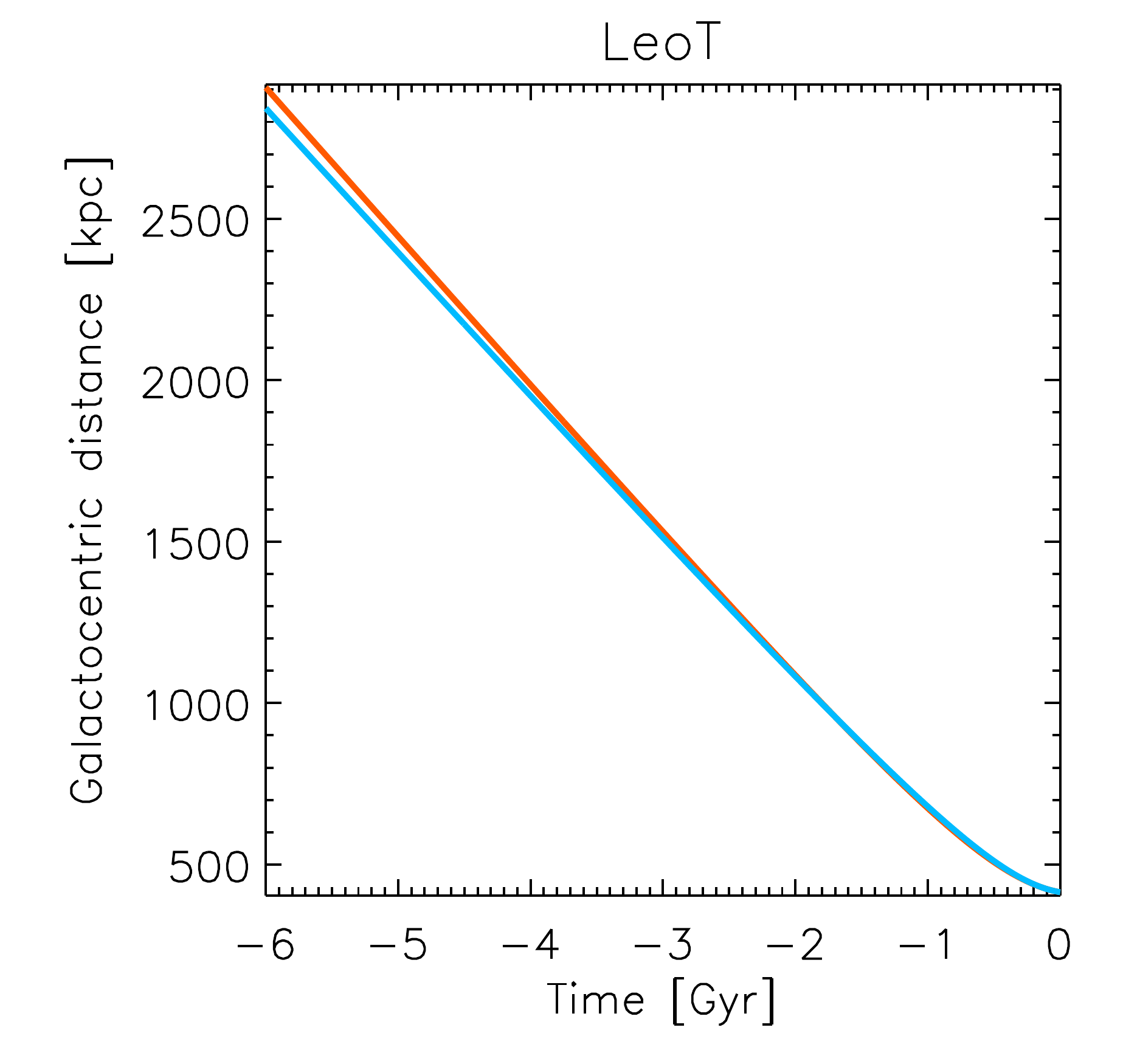}
\includegraphics[width=0.24\textwidth]{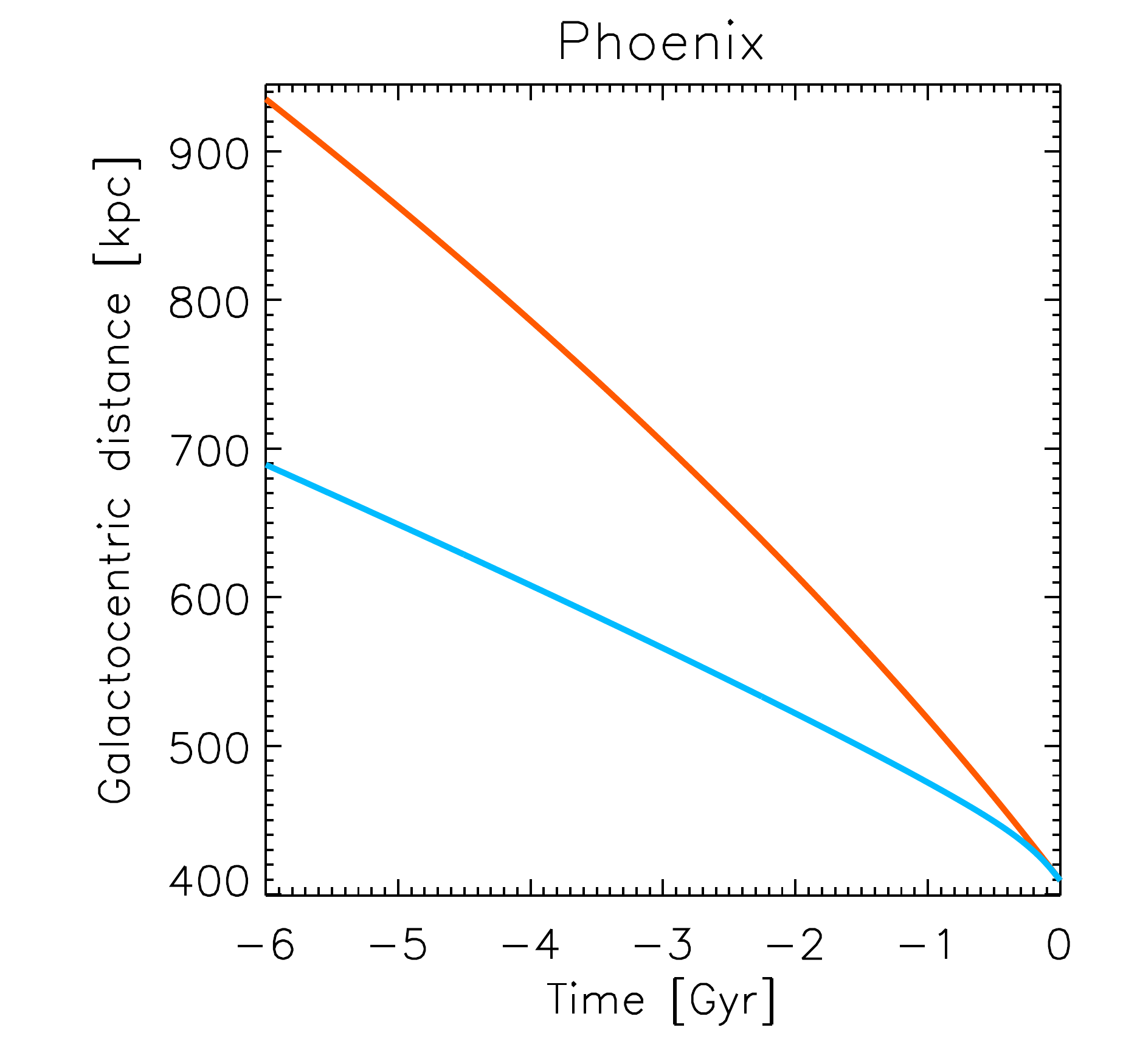}
\includegraphics[width=0.24\textwidth]{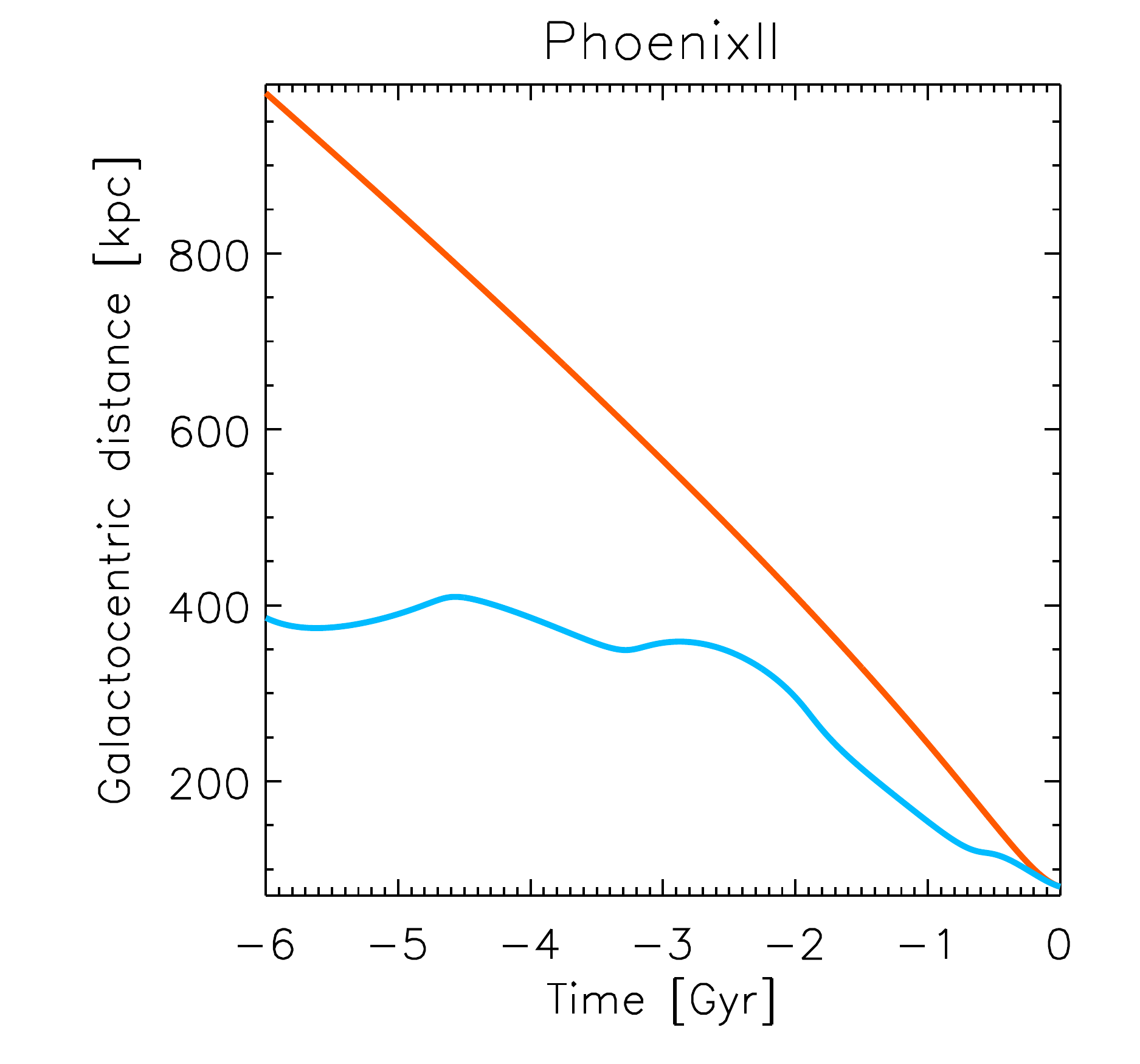}
\includegraphics[width=0.24\textwidth]{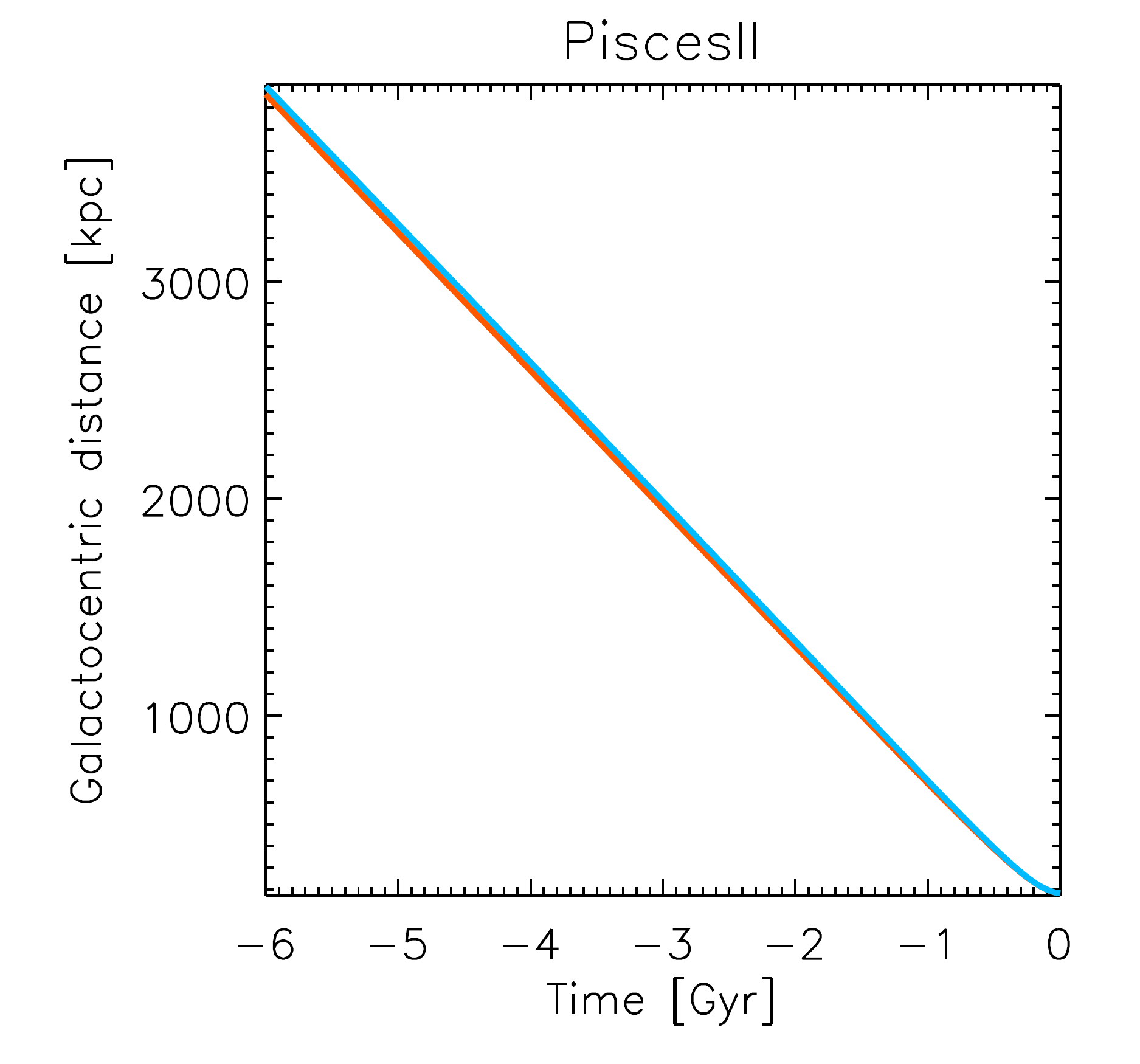}
\includegraphics[width=0.24\textwidth]{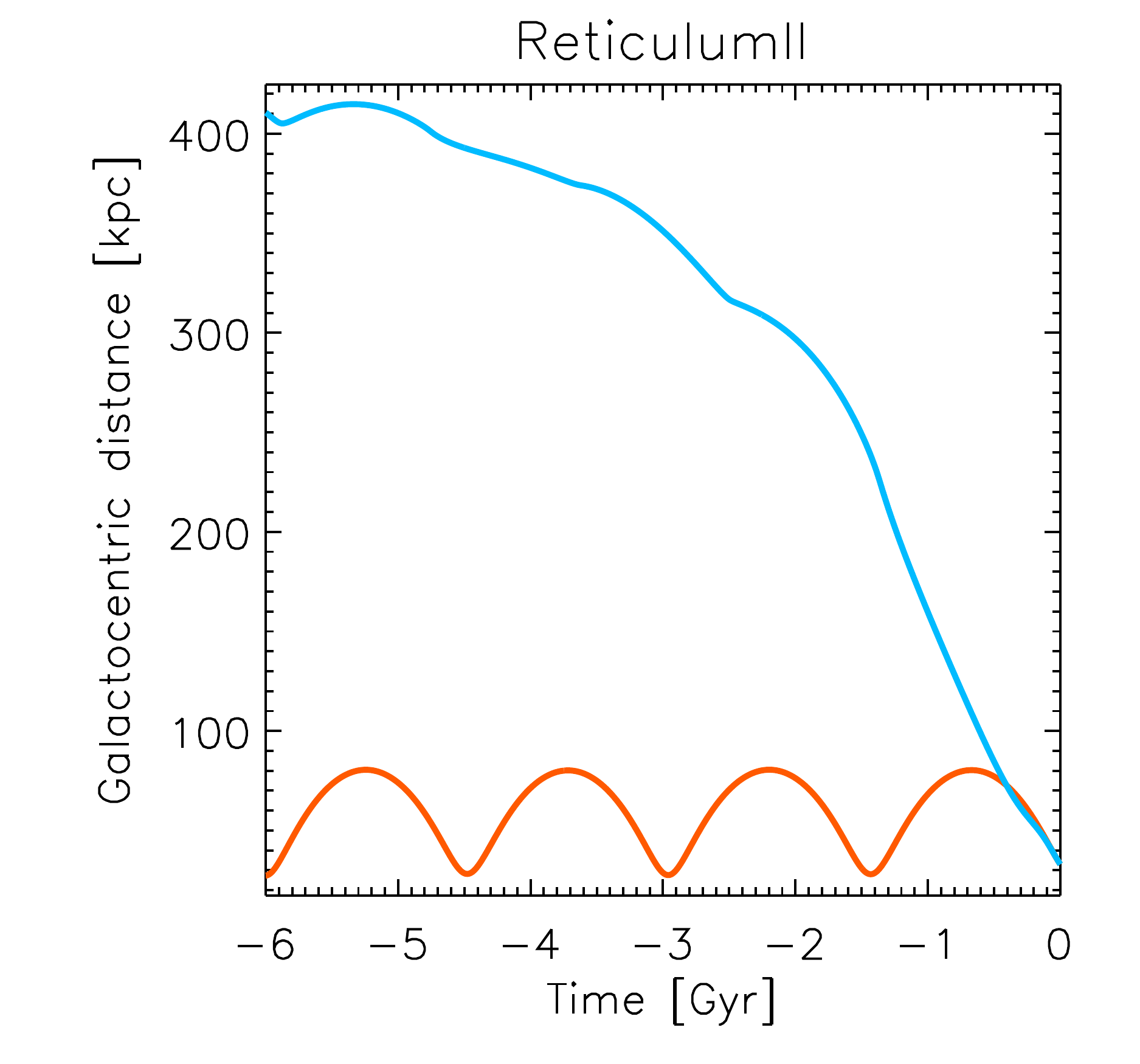}
\includegraphics[width=0.24\textwidth]{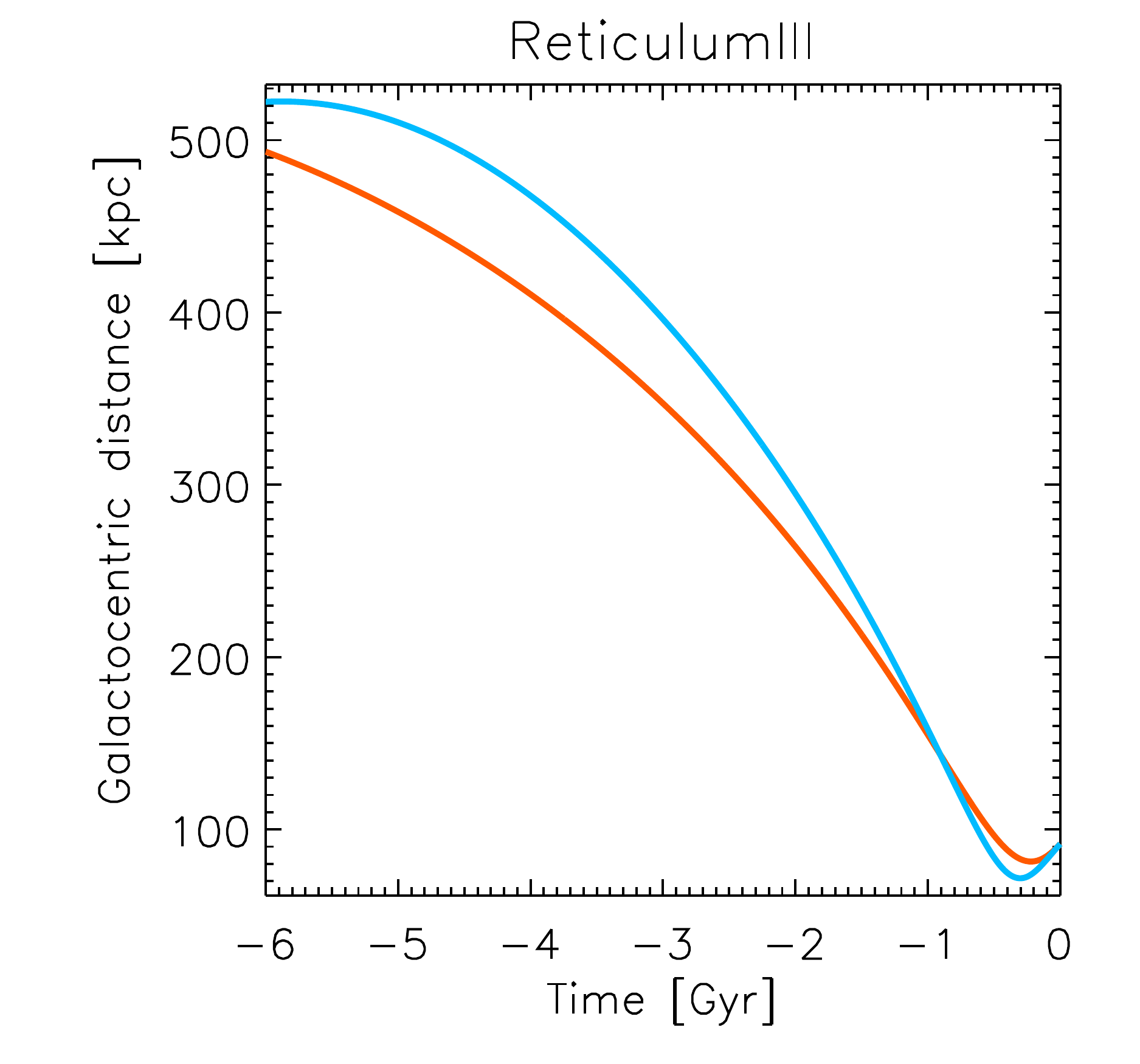}
\includegraphics[width=0.24\textwidth]{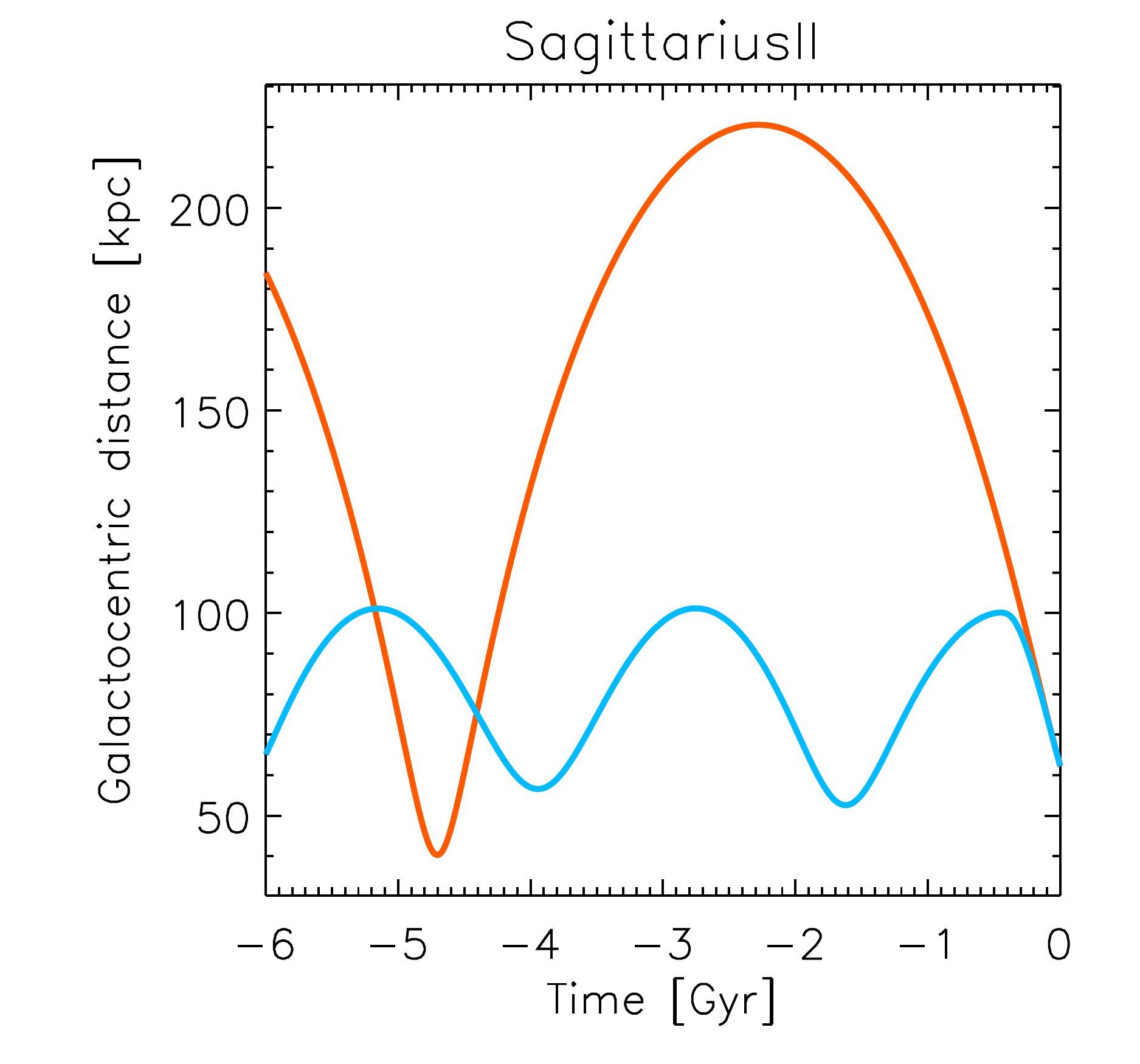}
\includegraphics[width=0.24\textwidth]{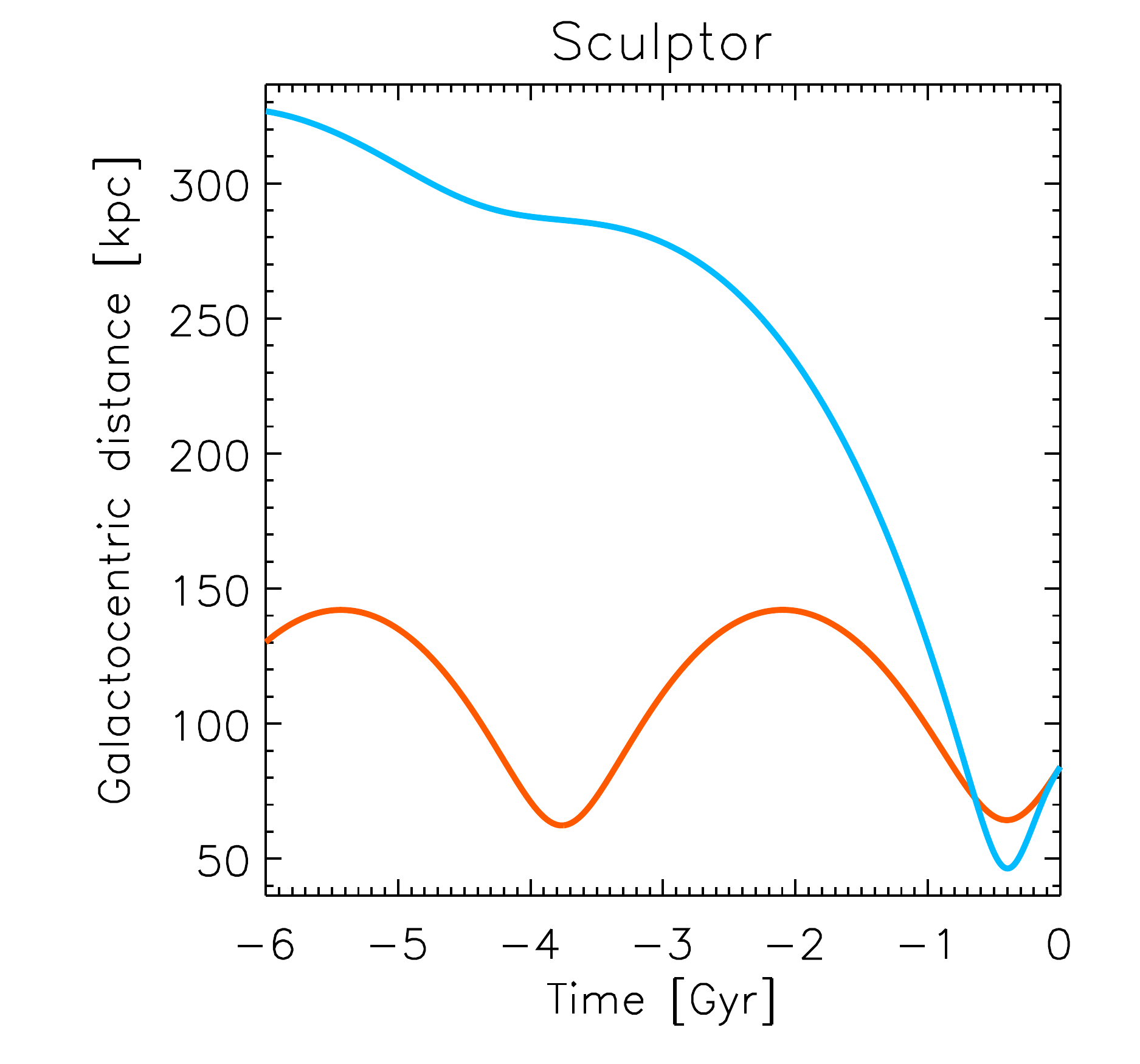}
\includegraphics[width=0.24\textwidth]{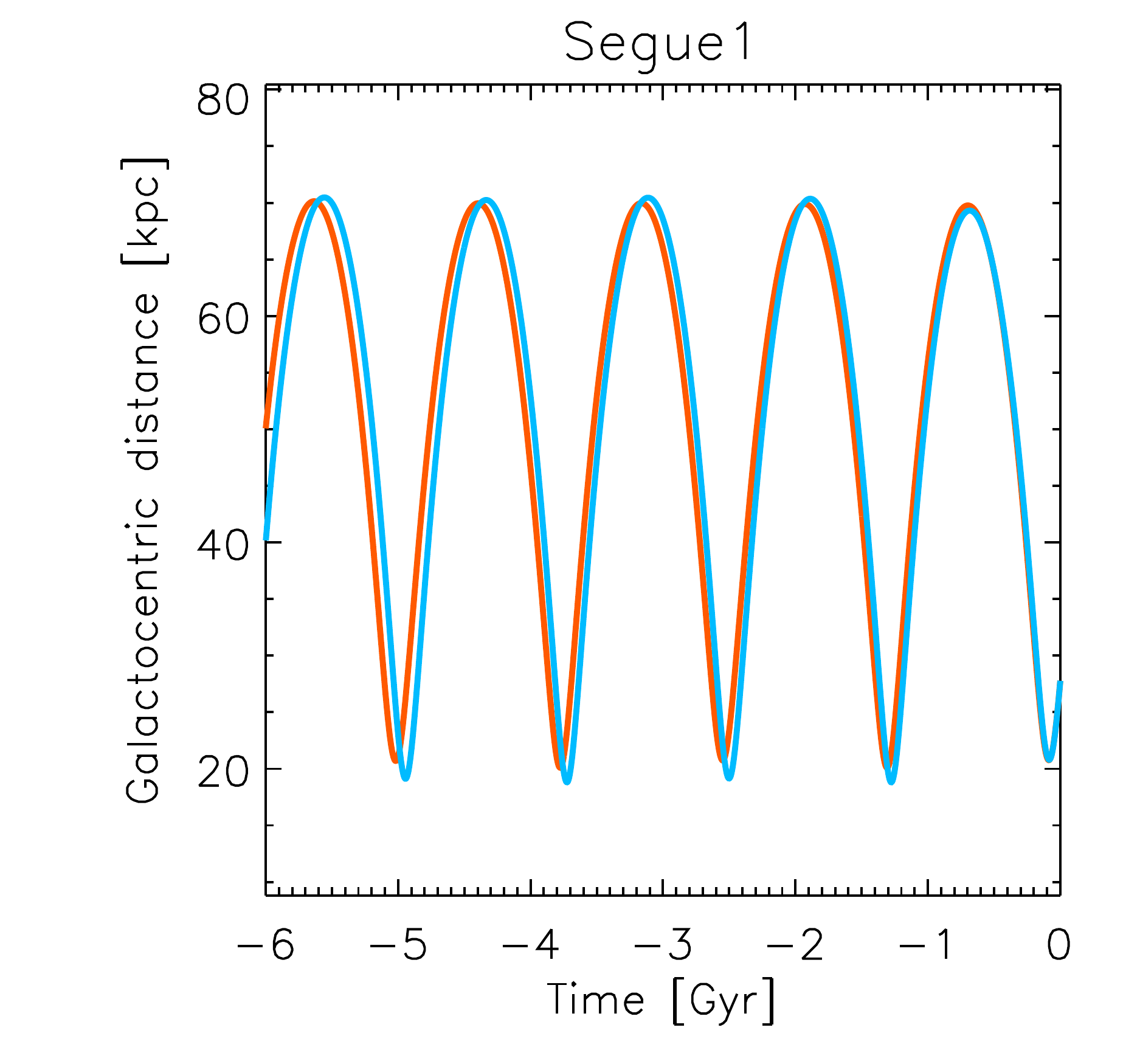}
\includegraphics[width=0.24\textwidth]{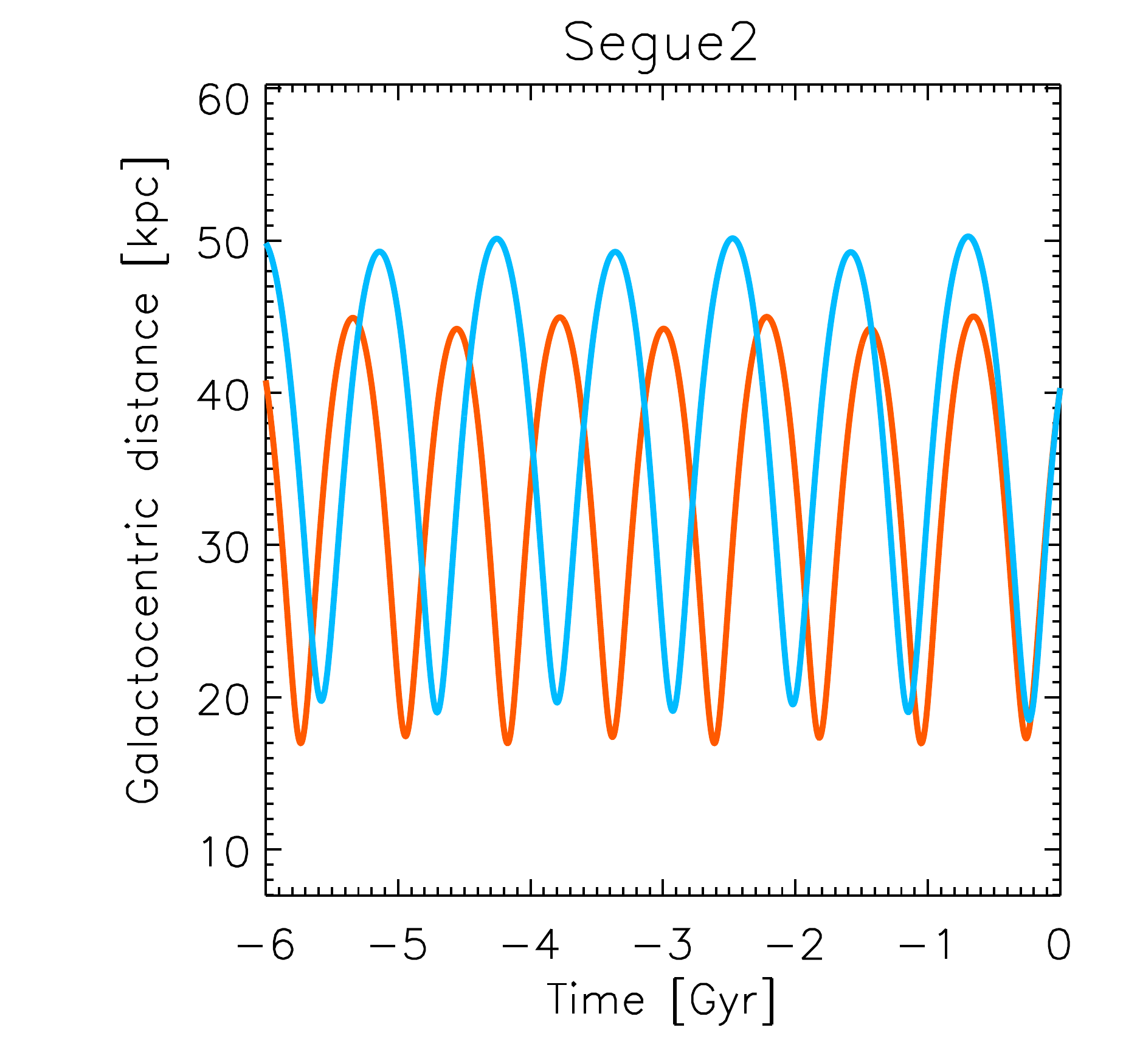}
\includegraphics[width=0.24\textwidth]{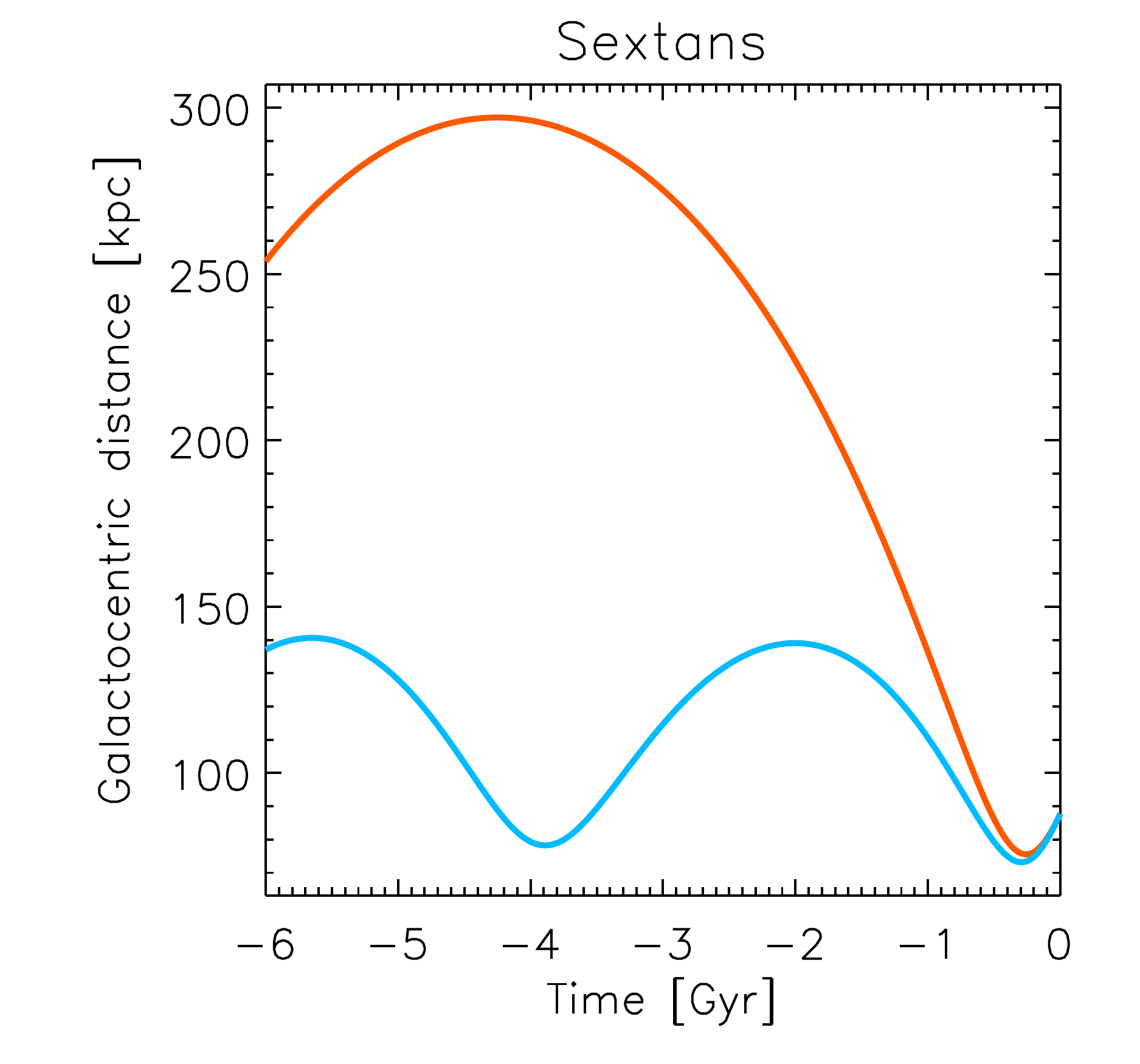}
       \caption{As \ref{fig:orbits}. For Pisces~II the orbits shown are obtained using the systemic motion from the inclusion of the spectroscopic information. }
         \label{fig:orbits2}
   \end{figure*} 
   
      \begin{figure*}
   \centering
\includegraphics[width=0.24\textwidth]{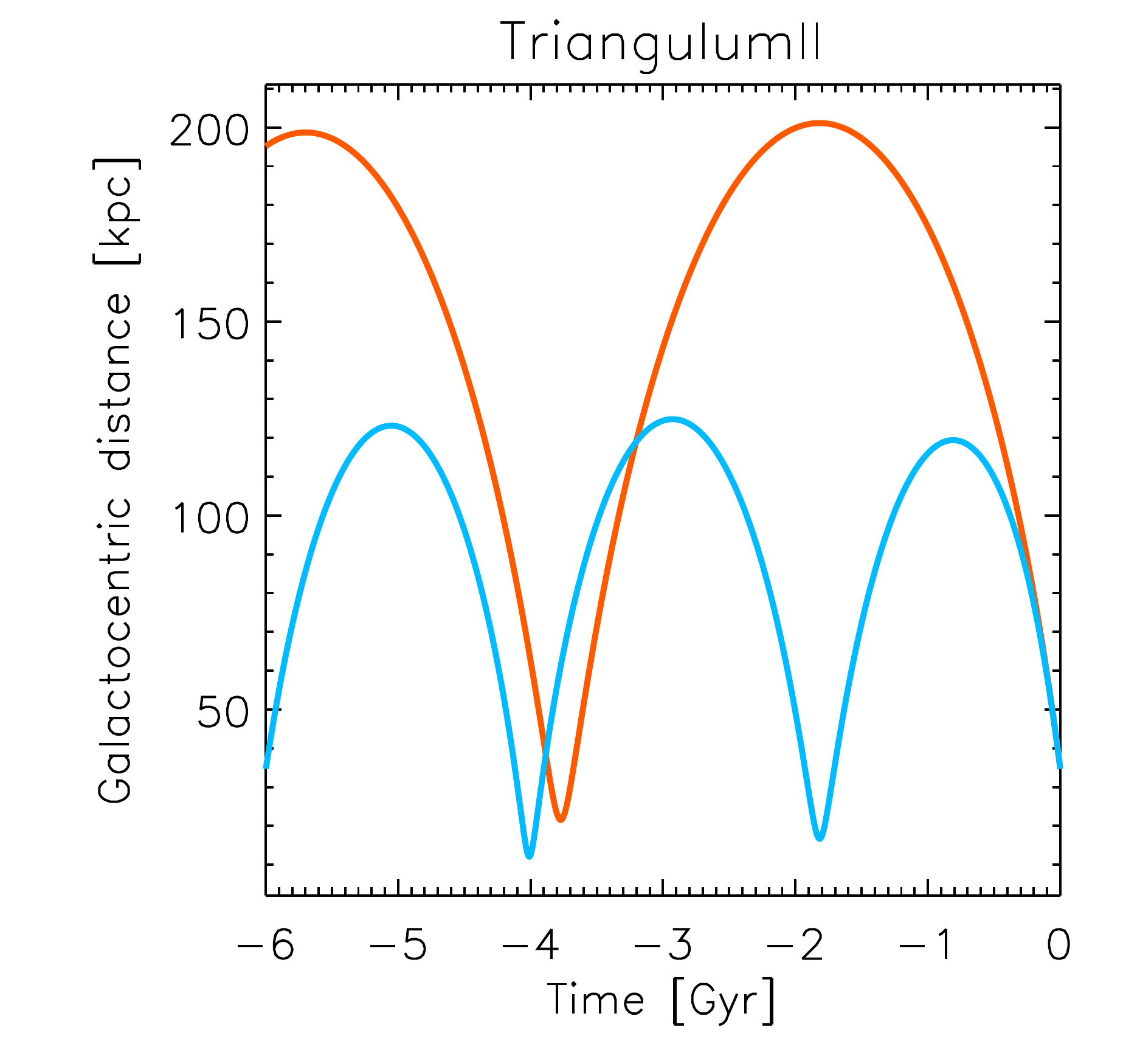}
\includegraphics[width=0.24\textwidth]{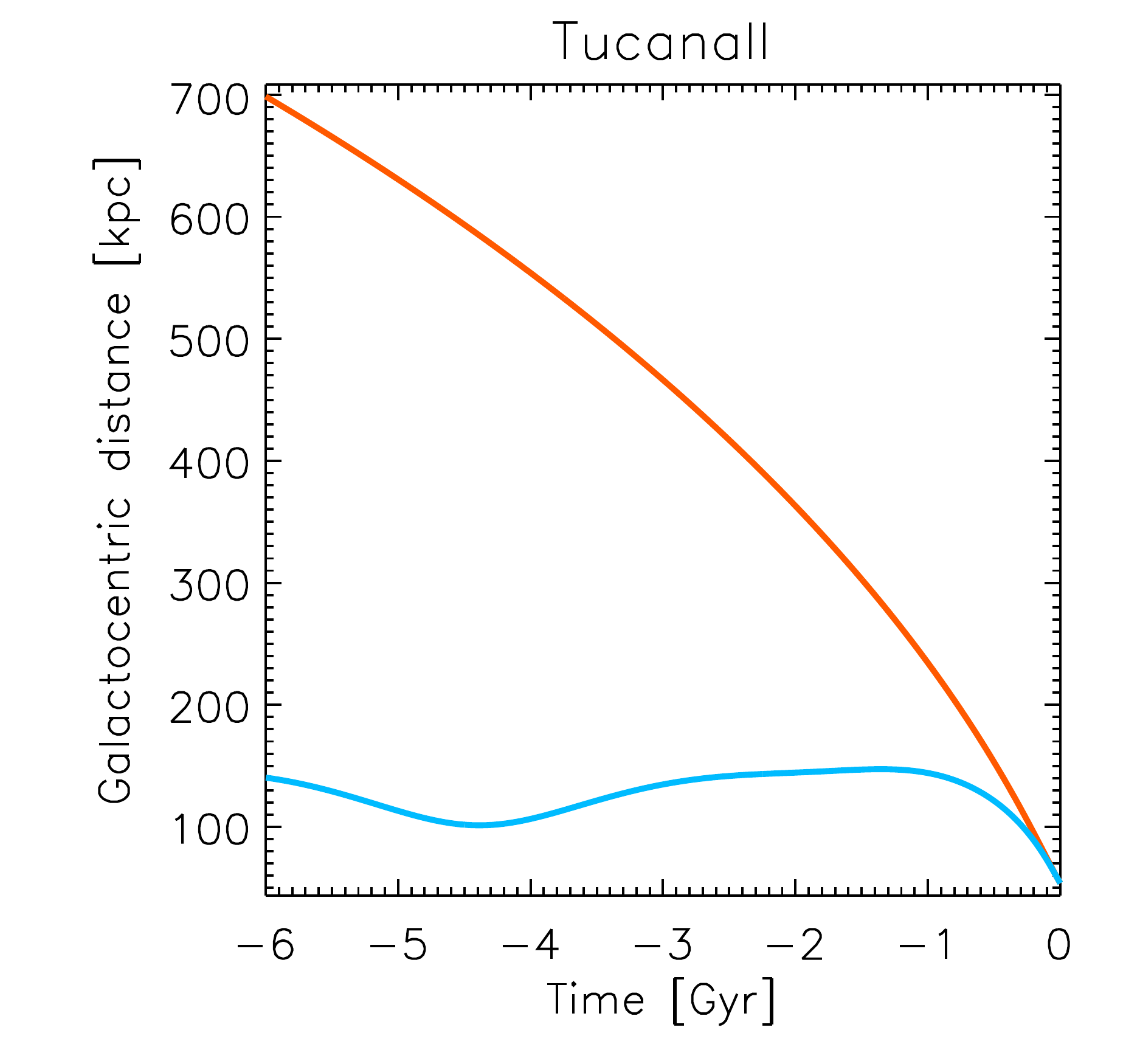}
\includegraphics[width=0.24\textwidth]{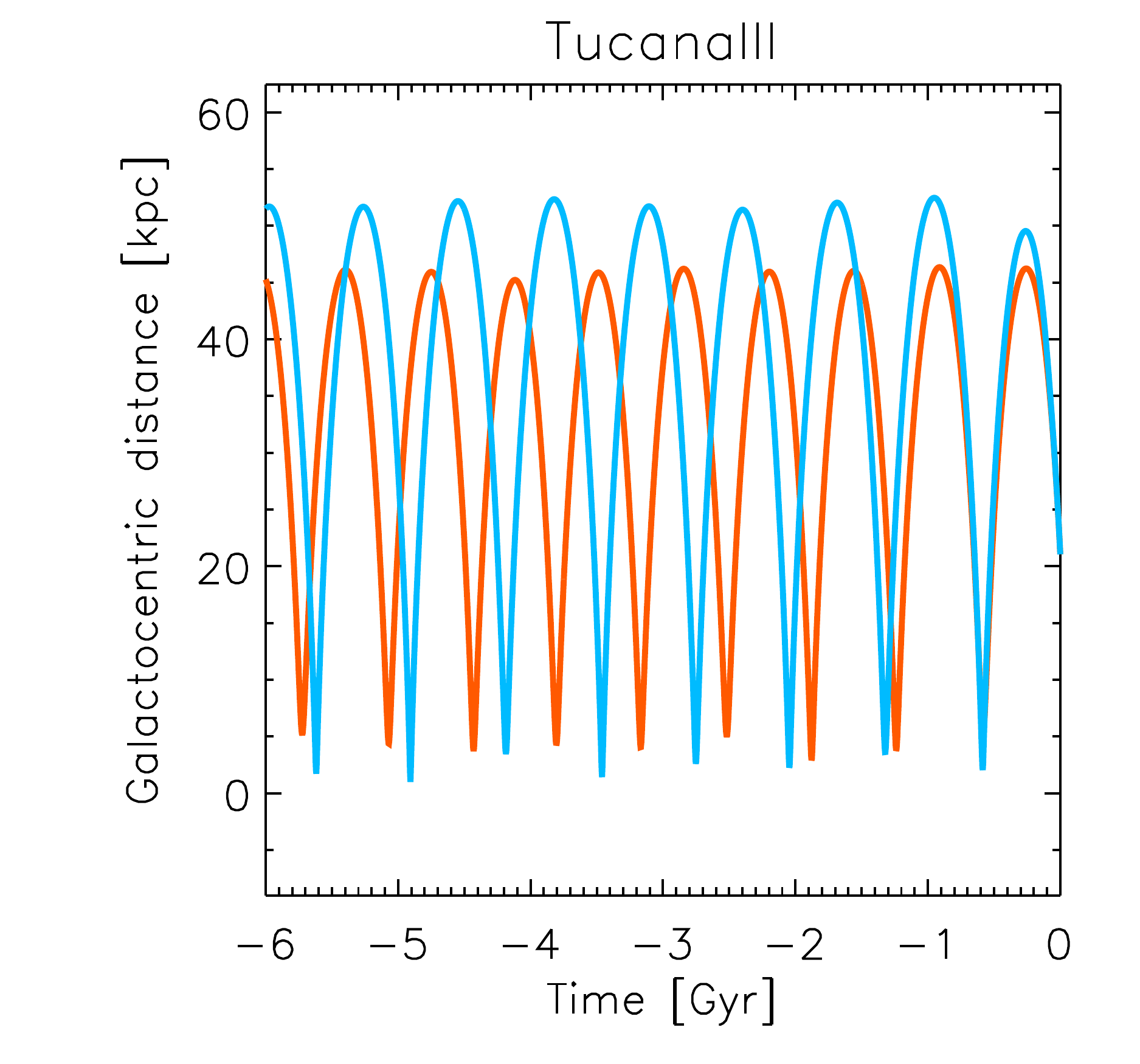}
\includegraphics[width=0.24\textwidth]{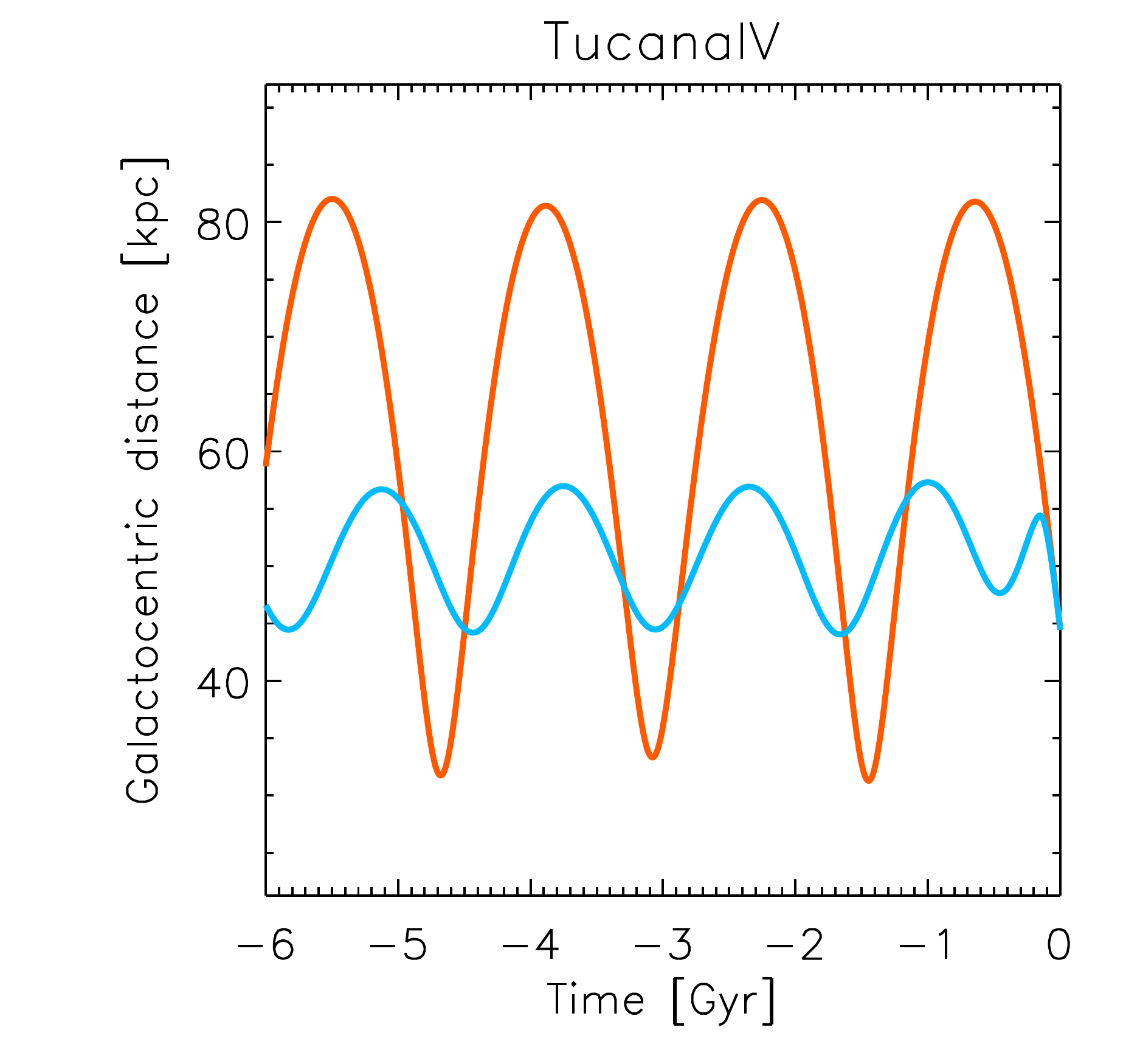}
\includegraphics[width=0.24\textwidth]{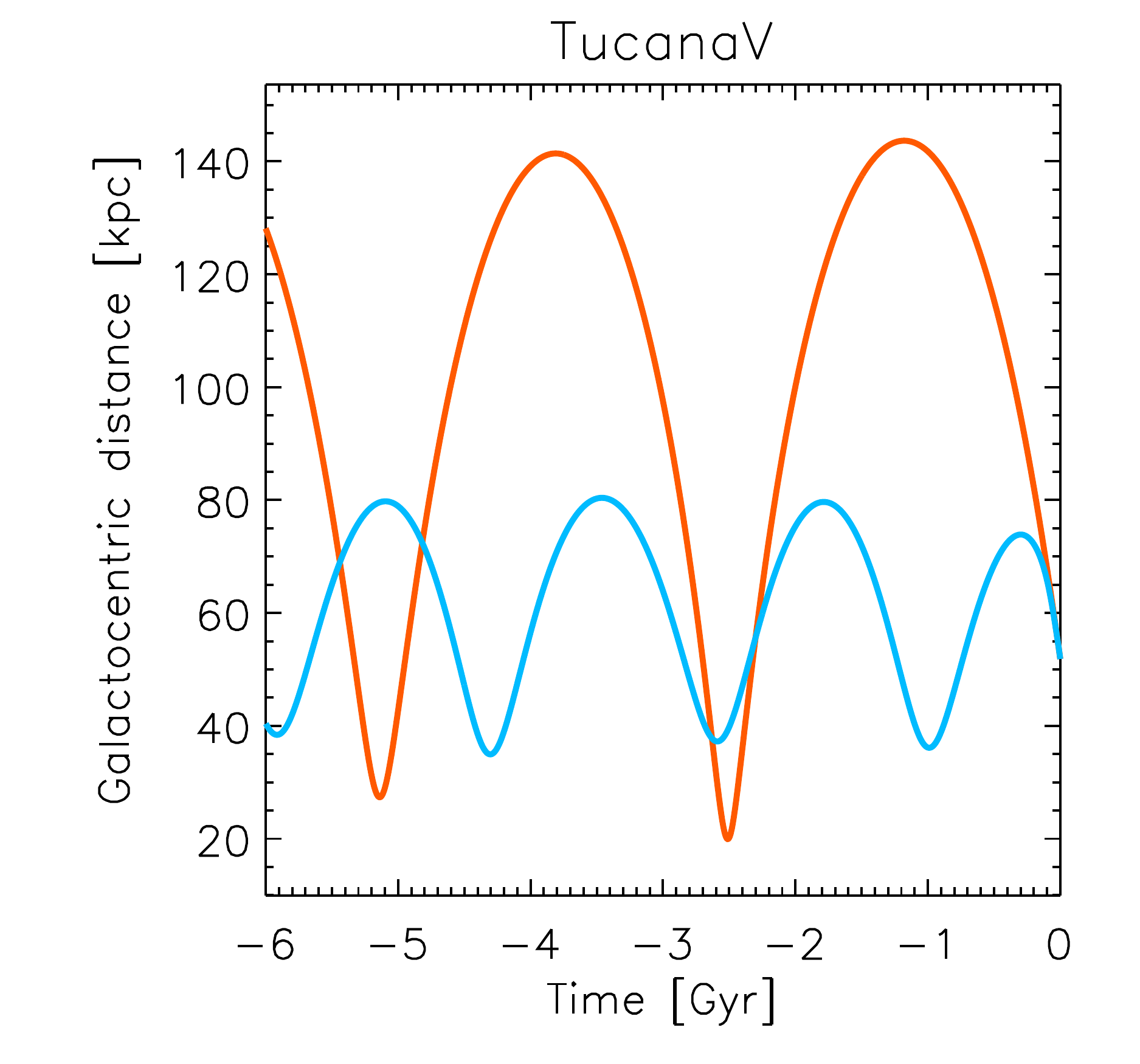}
\includegraphics[width=0.24\textwidth]{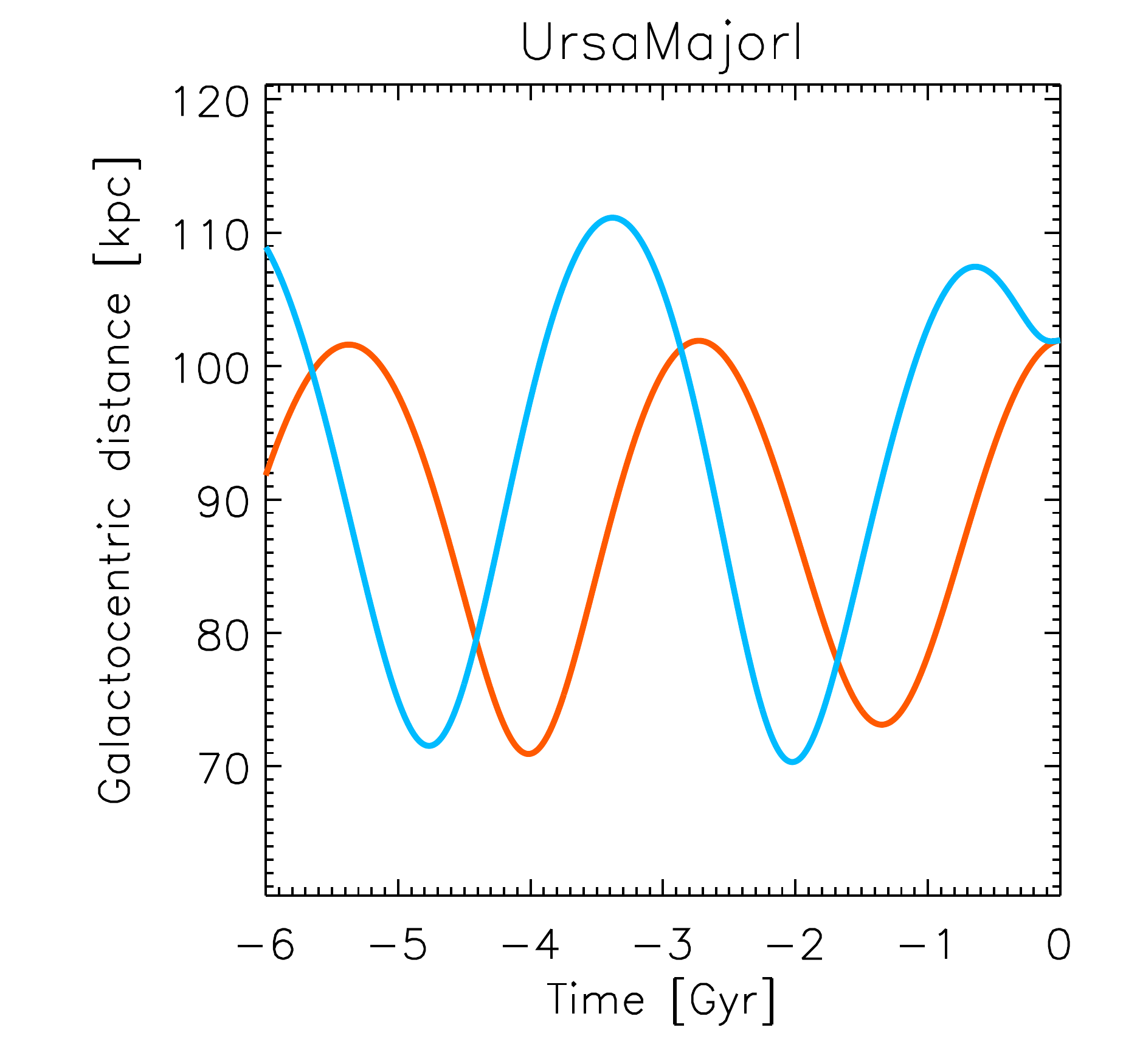}
\includegraphics[width=0.24\textwidth]{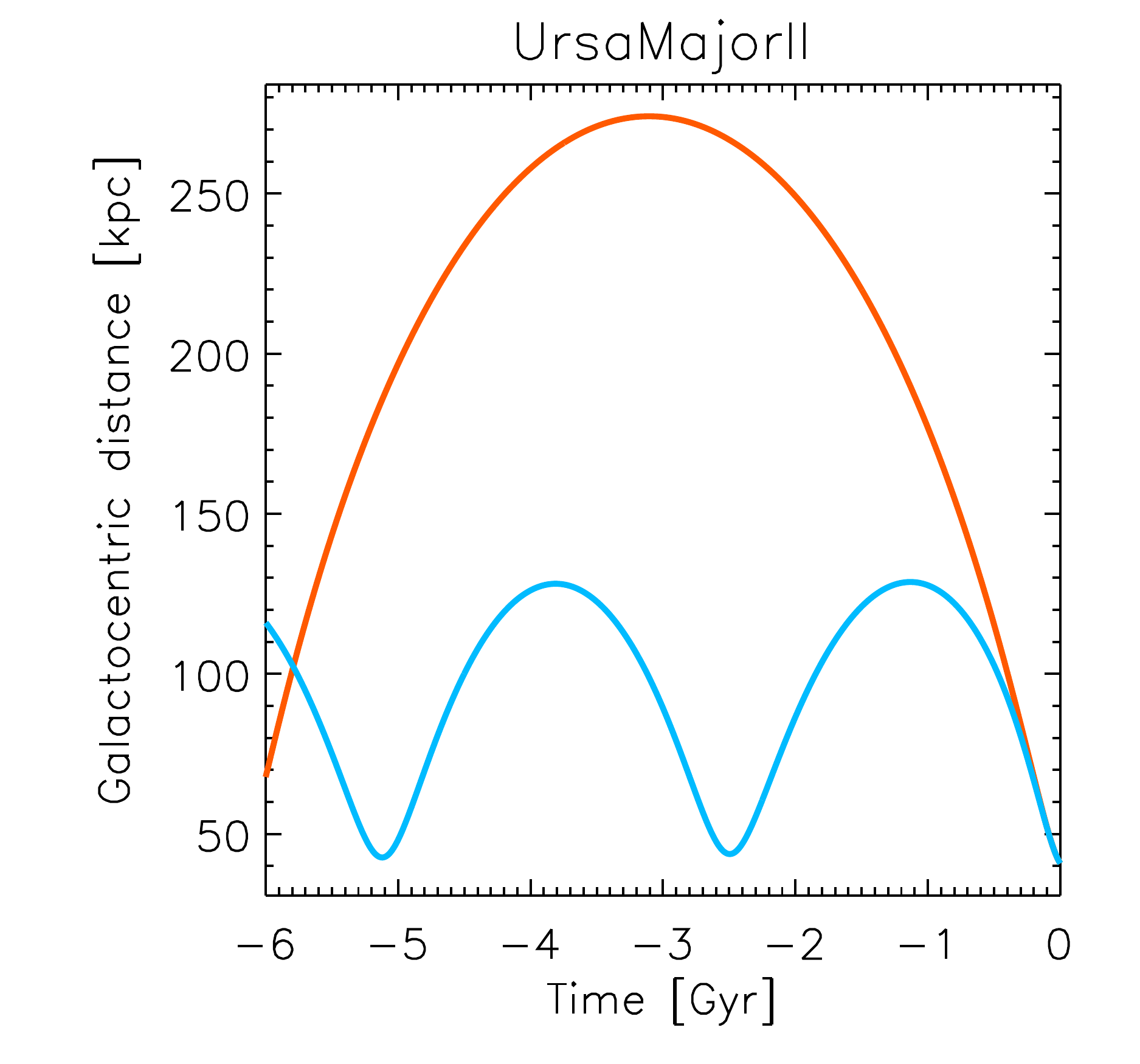}
\includegraphics[width=0.24\textwidth]{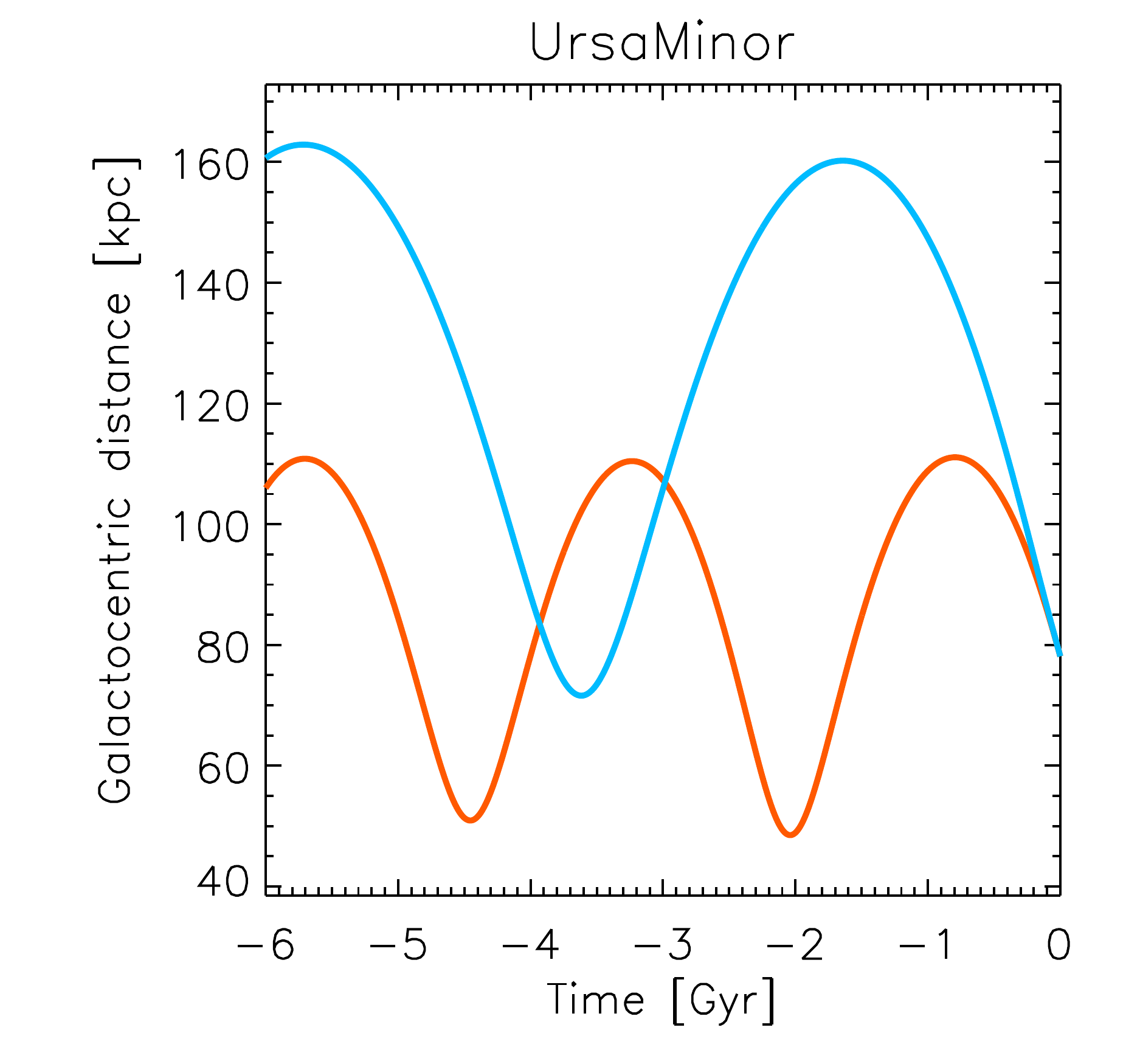}
\includegraphics[width=0.24\textwidth]{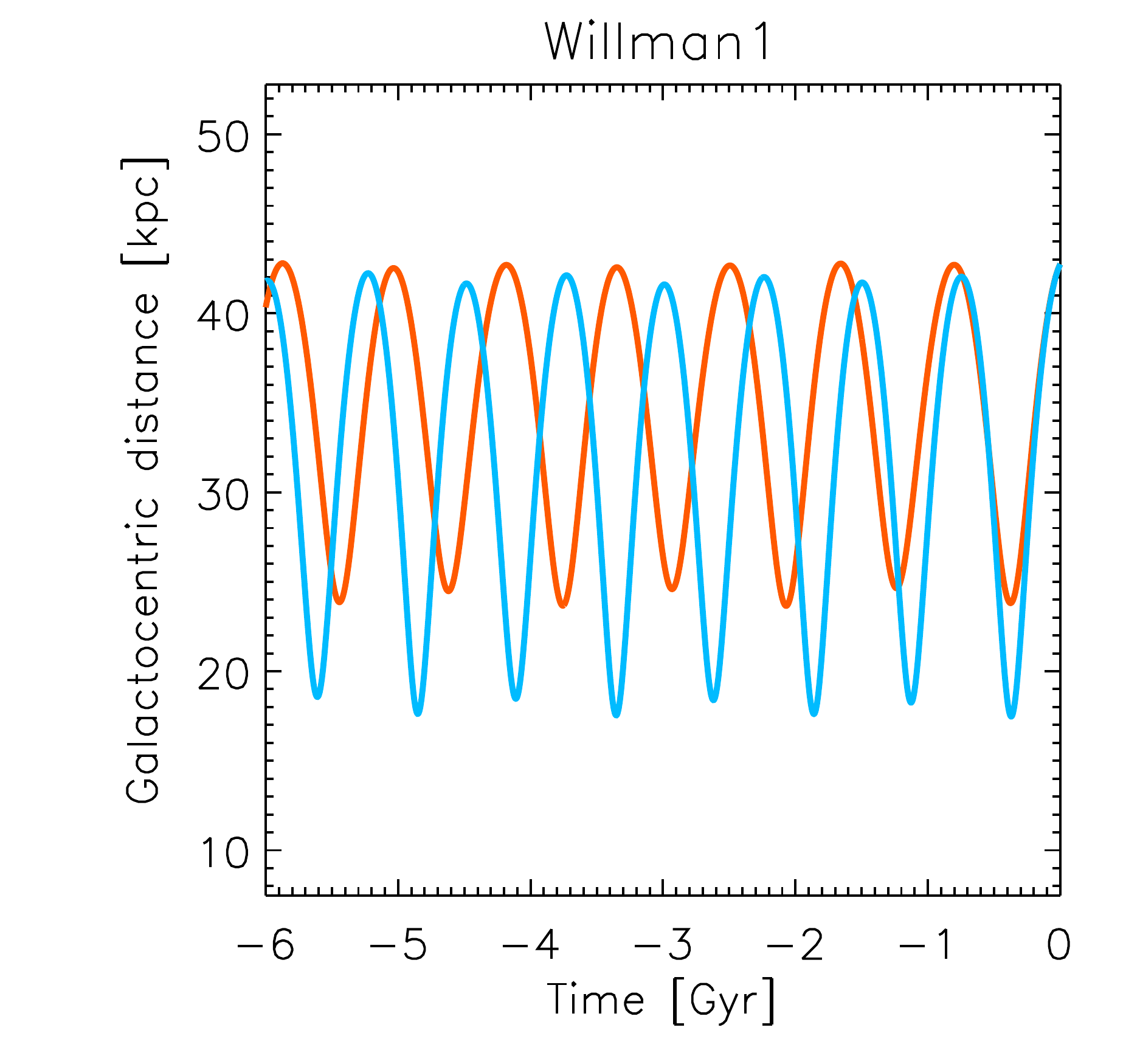}
\includegraphics[width=0.24\textwidth]{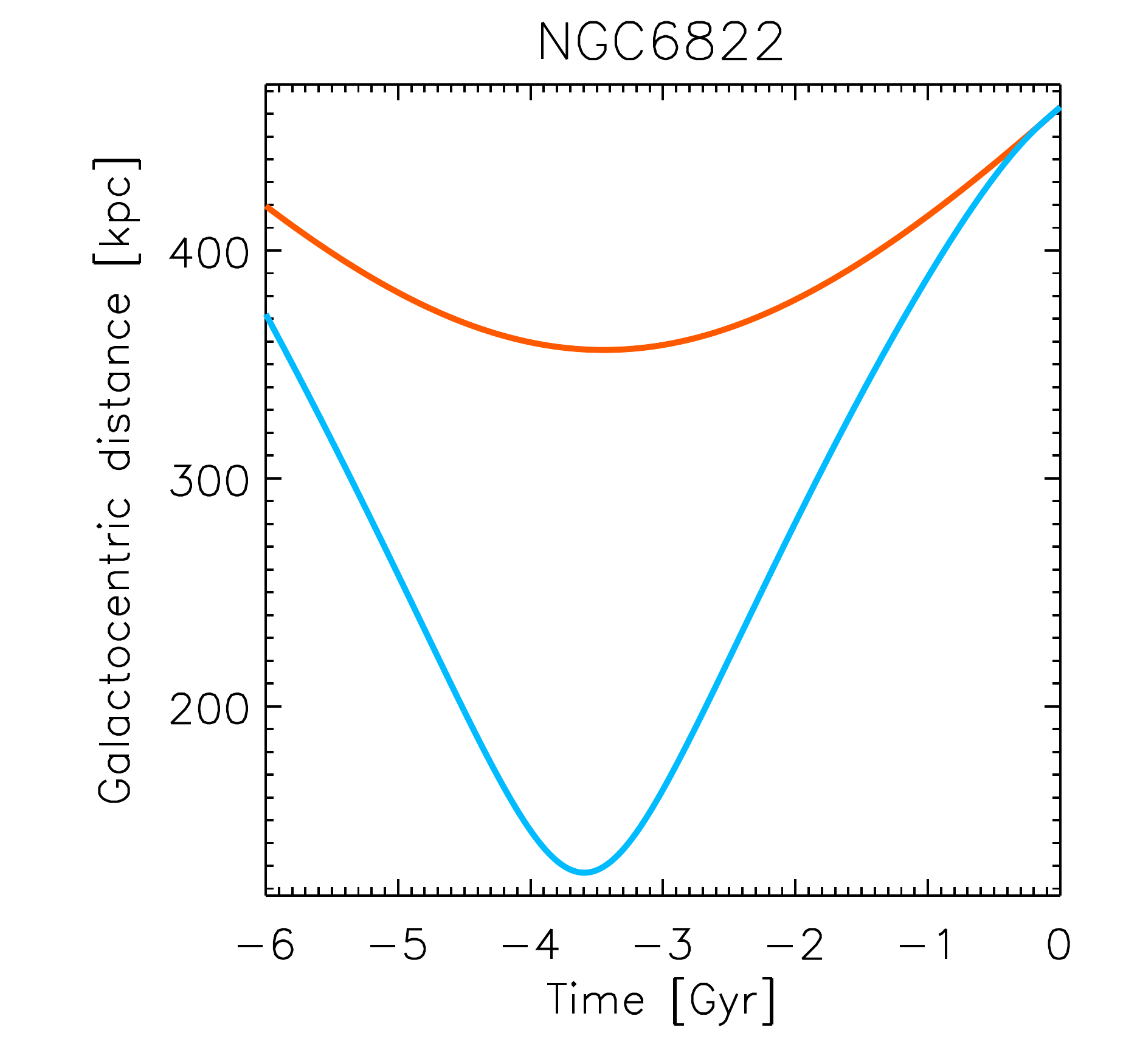}
        \caption{As Fig.~\ref{fig:orbits}. For Tucana~V the orbits shown are obtained using the systemic motion from the inclusion of the spectroscopic information.}
         \label{fig:orbits3}
   \end{figure*}

\end{appendix}

\end{document}